\documentclass[tightenlines]{osudissert96}
\usepackage{hyperref}
\usepackage[pagewise, mathlines]{lineno}
\usepackage[nottoc]{tocbibind}
\usepackage{booktabs}
\usepackage{setspace}
\usepackage{cite} 

\usepackage{fancyhdr}
\usepackage{tocloft}
\usepackage{type1cm}
\usepackage{courier}
\usepackage{url}
\makeatletter
\def\url@leostyle{%
  \@ifundefined{selectfont}{\def\UrlFont{\sf}}{\def\UrlFont{\small\ttfamily}}}
\makeatother
\usepackage{graphicx}
\usepackage{relsize}
\usepackage[latin1]{inputenc}
\usepackage{amsfonts}
\usepackage{amsmath}
\usepackage{amssymb}
\usepackage{amsthm}
\usepackage{mathrsfs}
\usepackage{bm}
\usepackage{caption}
\usepackage{xcolor}
\usepackage{xspace}

\newcommand{\La}{\langle}
\newcommand{\Ra}{\rangle}
\newcommand{\eq}{{\,=\,}}

\newcommand{\Tdec}{T_\mathrm{dec}}
\newcommand{\edec}{e_\mathrm{dec}}
\newcommand{\ecc}{\varepsilon}
\newcommand{\Tc}{T_\mathrm{c}}
\newcommand{\dNdy}{dN_\mathrm{ch}/dy}
\newcommand{\nc}{\newcommand}
\nc{\pt}{p_\mathrm{T}} \nc{\kt}{k_\mathrm{T}} \nc{\mt}{m_\mathrm{T}}
\nc{\Kt}{K_\mathrm{T}} \nc{\Mt}{M_\mathrm{T}} \nc{\pL}{p_\mathrm{L}}

\newcommand{\btu}{\bigtriangleup}
\newcommand{\eps}{\varepsilon}

\begin{document}
%
%

\author{Huichao Song}
\title{\LARGE{Causal Viscous Hydrodynamics \\[0.10in]  for Relativistic Heavy Ion Collisions }}
\unit{Graduate Program in Physics}
\advisorname{Ulrich W. Heinz}
\member{Richard J. Furnstahl}
\member{Michael A. Lisa}
\member{Junko
Shigemitsu}

%
%

\maketitle

%


%
%

\begin{abstract}

\startonehalfspace {\large The viscosity of the QGP is a presently
hotly debated subject. Since its computation from first principles
is difficult, it is desirable to try to extract it from experimental
data. Viscous hydrodynamics provides a tool that can attack this
problem and which may work in regions where ideal hydrodynamics
begins to fail.}

\startonehalfspace{\large This thesis focuses on viscous
hydrodynamics for relativistic heavy ion collisions. We first review
the 2nd order viscous equations obtained from different approaches,
and then report on the work of the Ohio State University group on
setting up the equations for causal viscous hydrodynamics in 2+1
dimensions and solving them numerically for central and noncentral
Cu+Cu and Au+Au collisions at RHIC energies and above. We discuss
shear and bulk viscous effects on the hydrodynamic evolution of
entropy density, temperature, collective flow, and flow
anisotropies, and on the hadron multiplicity, single particle
spectra and elliptic flow. Viscous entropy production and its
influence on the centrality dependence of hadron multiplicities and
the multiplicity scaling of eccentricity-scaled elliptic flow are
studied in viscous hydrodynamics and compared with experimental
data. The dynamical effects of using different versions of the
Israel-Stewart second order formalism for causal viscous fluid
dynamics are discussed, resolving  some of the apparent
discrepancies between early results reported by different groups.
Finally, we assess the present status of constraining the shear
viscosity to entropy ratio of the hot and dense matter created at
RHIC.}

\end{abstract}

%
%


%
%


%
%

\begin{acknowledgements}

\startonehalfspace{\large First I would like to express my deepest
gratitude to my advisor Prof. Ulrich Heinz, for his thoughtful
guidance during my graduate study in the past five years. I enjoy to
work with him very much. He led me to a fantastic research area,
which made me busy and fruitful; he provided me with all kinds of
opportunities to broaden my horizon and allow me to grow up as a
scientist. Especially, I thank him for his patience, encouragement
and continuous support even during the hardest times of my research
without which my early research would not have found its proper
direction. I also enjoy the research atmosphere he created for us,
through which I learned a lot from other group members. Here I
especially thank Evan for many technical discussions on
hydrodynamics, and Guang-You and Abhijit for discussions beyond
hydrodynamics. }

\startonehalfspace{\large For work covered in this thesis, I
gratefully acknowledgement the fruitful and enlightening discussions
with S. Bass, K. Dusling,  R. Fries, P. Huovinen, P. Kolb,  M. Lisa,
D. Molnar,  P. Petreczky, S. Pratt, P. Romatschke  and D. Teaney. I
especially thank K. Dusling and P. Romatschke for their close
collaborations during the code verification. I am also grateful for
many thought-stimulating discussions and debates with N. Demir, S.
Jeon, J. -Y. Jia, V. Koch, Y. Kovchegov, R. Lacey, R.  Neufeld, {J.
Noronha-Hostler}, J. Randrup, R. Snellings, P. Steinberg, A. Tang,
I. Vitev, {{X. -N. Wang}}, X. Zhe, and many others (This includes
numerous people in the RHIC community, whom I met at conferences,
workshops and seminars -- I apologize for not listing all of them
here individually). }

\startonehalfspace{\large I also thank my B. S. and M. S. thesis
advisors Prof. Guo-Mo Zeng and Prof. Yu-Xin Liu, who gave me early
scientific training and encouraged me to switch to new research
areas to further explore my potentials.}

\startonehalfspace{\large Last but not least, I thank my husband
Yulu Che for the life I could enjoy with him  together and for a
warm home where I could rest with peace of mind. I thank my parents
for their unconditional love and support.}

\end{acknowledgements}


%
%

\begin{center}
\startcontentspace
\tableofcontents
\startdoublespace
\end{center}
\pagebreak


%
%

\chapter{Introduction}
\section{The Quark Gluon Plasma and Relativistic Heavy Ion
Collisions}

Shortly after the discovery of the asymptotic freedom of QCD, people
realized that common nuclear matter, confining quarks and gluons
within individual protons and neutrons, could  transform into a new
de-confinement phase (at high energy densities) -- the quark gluon
plasma (QGP)~\cite{QGP-Lee:1974ma,QGP-Collins:1974ky}, a form of
matter that once existed in the very early universe or some variant
of which possibly still exists in the inner core of a neutron star.
It was theoretically conjectured that such extreme conditions can
also be realized on earth through colliding two heavy nuclei with
ultra-relativistic energies~\cite{QGP-Baumgardt:1975qv}, which
transform a fraction of the kinetic energies of the two nuclei into
heating the QCD vacuum within an extremely small volume. However, it
actually took more than 25 years of efforts to reach the threshold
for the QGP phase transition, from the first relativistic heavy ion
program at the Bevalac at LBL (with beam energies of up to $1
\mathrm{GeV/nucleon}$) in the early
1980s~\cite{Rev-Nagamiya:1982kn}, to the AGS at BNL (with center of
mass energies per nucleon pair $\sqrt{s} \sim 5 \mathrm{GeV}$) and
the SPS at CERN (with $\sqrt{s} \sim 17 \mathrm{GeV}$) in the late
1980s~\cite{Rev-Harris:1996zx}, to the RHIC at BNL (with $\sqrt{s} =
200 \mathrm{GeV}$), starting in
2000~\cite{Rev-Arsene:2004fa,Rev-Back:2004je,Rev-Adams:2005dq,Rev-Adcox:2004mh},
and to the inpending heavy ion program at the LHC at CERN (with
$\sqrt{s} = 5.5 \mathrm{TeV}$), which will probably start towards
the end of next year~\cite{Rev-Abreu:2007kv}.

At AGS and SPS energies and below, there was no unambiguous evidence
for  QGP formation, although a number of signals found at the SPS
strongly suggested the formation of a ``new state of
matter"~\cite{Rev-Heinz:2000bk}\footnote{\emph{The SPS measurements
 indicate that the plasma was already created there, but was
very short lived.}}. Only after the beginning of the RHIC program
during the summer of 2000, more and more evidence showed that the
QGP had been
discovered~\cite{Rev-Arsene:2004fa,Rev-Back:2004je,Rev-Adams:2005dq,Rev-Adcox:2004mh,
Gyulassy:2004vg,Gyulassy:2004zy,Muller:2006ee}. Before discussing
these QGP signatures, we will first explain the concept of the QGP
and how its features have become richer during the past few years.\\

\subsection{From weakly coupled QGP to strongly coupled QGP}

The quark gluon plasma is defined as a thermalized state of quarks
and gluons without color confinement. It was originally conceived as
a weakly coupled gas, motivated by the asymptotic freedom of QCD at
high energies. Using the statistics of relativistic massless
fermions and bosons, one obtains the equation of state (EOS) for a
free massless QGP gas~\cite{RHIC-Book}:
\begin{eqnarray}
p= \left[g_{g}+\frac{7}{8}(g_{q}+g_{\bar{q}})\right]
\frac{\pi^2}{90} T^4, \qquad \qquad e=
\left[g_{g}+\frac{7}{8}(g_{q}+g_{\bar{q}})\right] \frac{\pi^2}{30}
T^4.
\end{eqnarray}
Here, $g_{g}+\frac{7}{8}(g_{q}+g_{\bar{q}})$  is the total
degeneracy of the QGP. For gluons with 8 colors and 2 polarizations,
$g_{g}=8 \times 2$. For quarks with 3 colors, 2 spins and $N_f$
flavors, $g_{q}=3 \times 2 \times N_f$. The $7/8$ factor comes from
Fermi-Direc statistics. The above equations give the ideal EOS for a
non-interaction massless QGP, $p=e/3$. If one assumes that the phase
transition happens at $e \approx 1 \mathrm{GeV/fm^3}$, roughly seven
times that of normal nuclear matter, then one finds a critical
temperature of $T_c \approx 160 \ \mathrm{MeV}$ for a massless 2
flavor $(u, d)$ QGP.

In principle, the free energy and the EOS of the weakly coupled QGP
can be perturbatively calculated order by order in thermal QCD.
However, the QCD free energy shows bad convergence of the
perturbative expansion in powers of the strong coupling constant
$g$~\cite{Arnold:1994ps,Arnold:1994eb}. To solve this problem, one
needs to reorganize the perturbation theory. This led to the recent
developments of  hard thermal loop (HTL) perturbation theory, the so
called $\phi$-derivable approach,  dimensionally reduced screened
perturbation theory, and others (see the review
article~\cite{Andersen:2004fp} for details), all of which yield much
improved convergence for the QCD free energy and an improved EOS at
high temperature. Still, the calculation needs non-perturbative
input, due to the existence of a non-perturbative magnetic mass
scale $m_{mag} \sim g^2 T$, which enters the EOS at order
$g^6$~\cite{Linde:1978px,Linde:1980ts}. On the other hand, even the
most sophisticated weakly coupled QCD methods fail near the phase
transition where non-perturbative effects become dominant.

\begin{figure}[t]
\centering
\includegraphics[width=8.5cm,height=58mm, angle=0]{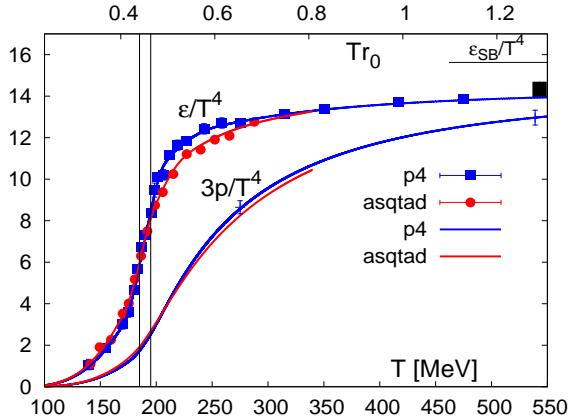}
\vspace{-3mm} \caption[Lattice EOS ]{Energy density and three times
the pressure as a function of temperature, calculated from lattice
QCD with asqtad and p4 action~\cite{Bazavov-EOS:2009zn}.
}\label{Lattice-EOS}
\end{figure}

Lattice QCD offers a non-perturbative approach to study the QCD
properties at finite temperature. Recent developments in this field
have made available  quite precise lattice calculations of the EOS
with almost physical quark
masses~\cite{Bazavov-EOS:2009zn,Cheng:2007jq}.
Fig~\ref{Lattice-EOS}. shows the Lattice EOS (at zero chemical
potential) from Bazavov et al., using two different improved
staggered fermion actions (the asqtad and p4 action,
respectively)~\cite{Bazavov-EOS:2009zn}. They found that both
deconfinement and chiral symmetry restoration happen in a narrow
temperature region: $ 185 \ \mathrm{MeV} <T_c< 195 \
\mathrm{MeV}$\footnote{\emph{Other groups found a lower transition
temperature $155 \ \mathrm{MeV} < T_c < 174 \
\mathrm{MeV}$~\cite{Aoki:2006br}}.}, which is indicated by the
narrow band in Fig~\ref{Lattice-EOS}.  At higher temperature $T >
300\ \mathrm{MeV}$, one finds that the lattice EOS approaches
85-90\% of the Stefan-Boltzmann limit for an ideal non-interacting
QGP gas. But this does not necessarily  mean that the QGP is already
weakly coupled there. From the EOS alone, lattice QCD simulations
can not distinguish a weakly coupled QGP from a strongly coupled
QGP. Indeed, the concept of a strongly coupled QGP, which behaves
like an almost perfect liquid with very low
viscosity~\cite{Heinz:2001xi,Gyulassy:2004vg,Gyulassy:2004zy,Shuryak:2004cy},
came from the strong collective flow observed in RHIC experiments
and its very successful descriptions by ideal
hydrodynamics~\cite{Kolb:2003dz,Huovinen:2003fa}. However, why the
QGP is strongly coupled at RHIC energies is still a theoretical
challenge\footnote{\emph{Some models trying to explain the strongly
coupled nature of the QGP at RHIC energies can be found in
Refs.~\cite{Shuryak:2003ty,Liao:2005pa}. Other authors explore the
possibility that the strongly coupled dynamics can be obtained
through weak coupling expansions using HTL
resummation~\cite{Ipp:2007zz,PRV-Peshier}.}}, and it is also unknown
whether the QGP created in the future LHC experiments will continue
to be a strongly coupled liquid or behave more like a weakly
coupled gas. \\

\subsection{Stages of a heavy ion collision and theoretical tools}

\textbf{\underline{Stages of a heavy ion collision}}\\[-0.10in]

In this subsection, we introduce different stages of a heavy ion
collision and the main theoretical tools to describe or simulate
these stages. After a discussion of these theoretical tools, we will
list several experimental probes and signatures to detect the
transient QGP phase.

The different stages for an ultra-relativistic heavy ion collision
($\sqrt{s} \gtrsim 100 \ \mathrm{GeV}$) are schematically
illustrated in Fig~\ref{HIC} and~\ref{Space-Time}. The initial stage
before the collision is followed after impact by a pre-equilibrium
stage, an expanding QGP and hadron resonance gas (HRG) stage  and a
final freeze-out/decoupling stage. In more detail, these stages are
characterized as follows~\cite{Heinz:2004qz}:

\begin{figure}[t]
\centering
\includegraphics[width=160mm,height=40mm,angle=0]{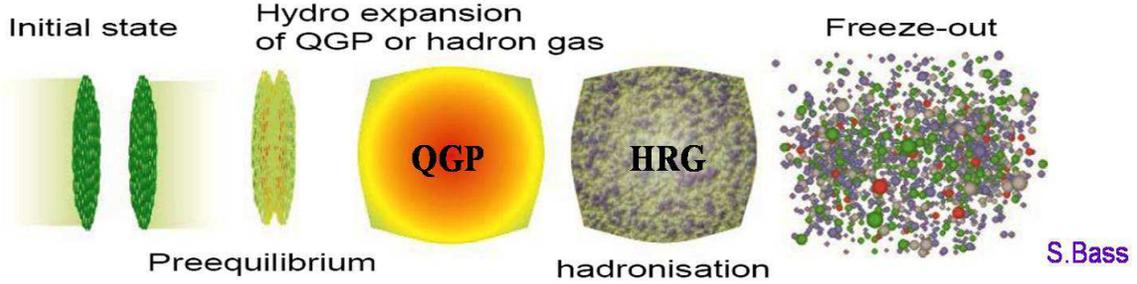}
\vspace{-3mm} \caption[Different stages for relativistic heavy ion
collisions ]{Different stages for relativistic heavy ion
collisions~\cite{StagesHIC}.}\label{HIC}
\end{figure}

\begin{enumerate}

\item Initial stage at $\tau<0$:  Two Lorentz contracted heavy nuclei
approach each other with more than 99.9\% of the speed of light. At
sufficiently high collision energy (possibly reached at RHIC), this
initial stage can be described by dense gluon walls known as the
Color Glass Condensate (CGC) (this is further explained in the
theoretical tools below).

\item  Pre-equilibrium stage and thermalization: The energetic collision of the two heavy nuclei
excites the QCD vacuum and produces a dense pre-equilibrium matter
consisting of quarks, anti-quarks and gluons. It takes around $1 \
\mathrm{fm/c}$ for the pre-equilibrium bulk matter to achieve local
thermalization and form the quark-gluon plasma. In this very early
pre-equilibrium stage, the primary collisions between fast partons
inside the colliding nuclei also generate ``hard probes" with either
large mass or large transverse momentum, such as heavy quark pairs
($c \bar{c}$ and $b \bar{b}$), pre-equilibrium real or virtual
photons, and very energetic quarks and gluons with large transverse
momentum (from which  jets are formed after hadronization).

\begin{figure}[b]
\centering \vspace{-5mm}
\includegraphics[width=90mm,height=52mm,angle=0]{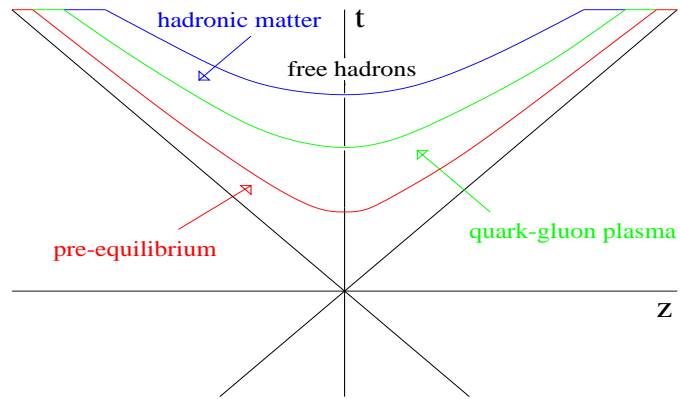}
\vspace{-8mm} \caption[The space-time picture of a heavy ion
collision]{The space-time picture of a heavy ion
collision~\cite{Satz:2002qj}.} \vspace{-0mm}\label{Space-Time}
\end{figure}

\item QGP expansion and hadronization: After
thermalization, the quark-gluon plasma (QGP),  driven by  thermal
pressure gradients, expands and cools down very quickly. After
reaching the critical temperature $T_c \simeq  170 \ \mathrm{MeV}$,
it hadronizes and turns into hadronic matter, consisting of a
mixture of stable and unstable hadrons and hadron resonances.
Hadronization happens continuously at the edge of the QGP fireball
during the whole QGP expansion period. In central Au+Au collisions
at RHIC, it takes around $10 \ \mathrm{fm/c}$ for the QGP fireball
to expand and fully convert to hadronic matter.

\item Hadronic expansion and decoupling: The hadronic matter continues to
expand until the system becomes very dilute. Then individual hadrons
decouple from the system (kinetic freeze-out) and free-stream to the
detector. Like the QGP hadronization process, the hadronic
decoupling happens continuously  at the edge of the fireball, where
the density is low. After complete hadronization, it takes another
$5-10 \ \mathrm{fm/c}$ for the hadronic matter to completely
freeze-out.
\end{enumerate}\vspace{3mm}

\textbf{\underline{Theoretical tools for different stages of a heavy
collisions}}\\[-0.10 in]

\begin{figure}[h]
\centering \vspace{1mm}
\includegraphics[width=150mm,height=60mm,angle=0]{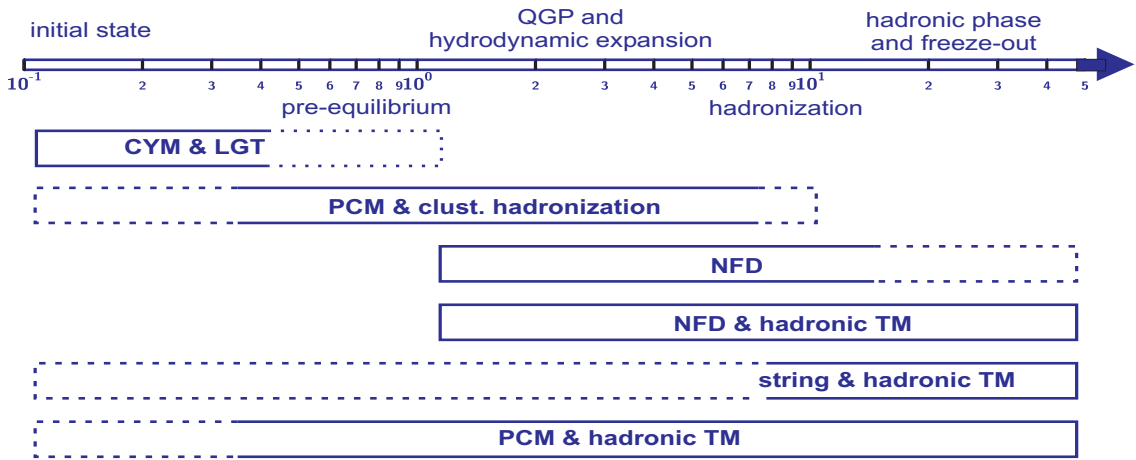}
\vspace{-1mm} \caption[Theory tools for RHIC and their range of
applicability]{Theory tools for RHIC and their range of
applicability~\cite{Bass:2001gb}. Solid bands denotes the safe range
to apply respective theoretical tools, while the dashed and dotted
bands refer to the region where the approach is still applied, but
may be questionable or unsafe.  CYM: Classical Yang-Mills
Theory~\cite{McLerran:1993ni,McLerran:1993ka}, LGT: Lattice Gauge
Transport~\cite{Poschl:1998px,Bass:1998nz}, PCM: Parton Cascade
Model~\cite{Geiger:1994he}, NFD: Nuclear Fluid Dynamics or
Hydrodynamics~\cite{Hydro-Clare:1986qj,Kolb:2003dz,Huovinen:2003fa}.
String and Hadronic TM: String and Hadronic Transport
Model~\cite{Sorge:1989dy,Bass:1998ca,Bleicher:1999xi}.
}\vspace{15mm} \label{Tools}
\end{figure}

There does not exists a unique theoretical tool to describe the
whole heavy ion collision process from the very beginning till the
end. The different energy and time scales during different collision
stages imply dramatic changes of the effective physical degrees of
freedom and their interactions, which thus require different tools
for their description. A complete description requires matching
these tools to each other,  generating so-called hybrid approaches.
For example, the results from the Color Glass Condensate (CGC)
theory can  be used as initial conditions for dynamical evolution
models such as hydrodynamics or the parton cascade. Hydrodynamics or
parton cascade models are then connected with a hadron cascade model
for a description of the late hadronic stage. The hadron cascade
model is equipped with an afterburner to generate quantum
statistical or final-state interaction induced 2-particle
correlations. Fig~\ref{Tools} shows the different theoretical
approaches and their ranges of applicability. Although plotted by S.
Bass as early as 2001, it still very nicely illustrates our present
understanding,  in spite of several new theoretical developments in
the past years\footnote{\emph{These new developments include new
ideas to understand
thermalization~\cite{Arnold:2003rq,Arnold:2004ti} as well as first
numerical implementations of hybrid approaches, such as
hydrodynamics+hadron
cascade~\cite{Hirano:2005wx,Nonaka:2006yn}.}}.\\

\textbf{Color Glass Condensate}: At very high collision energies,
particle production at mid rapidity probes the nuclear structure
functions in the small-$x$ regime ($x$ is the fraction between
parton momentum and beam momentum). It is well known that the gluon
distribution function $xG(x,Q^2)$ increases dramatically with
decreasing of $x$, for large enough probe resolution $Q^2$. When the
gluon density becomes high enough, two gluons start to recombine to
one. This leads to gluon saturation below some momentum scale
$Q_s^2$. At this momentum scale, each gluon mode has macroscopic
occupation number $\sim \frac{1}{\alpha_s}$, which is why this state
has been called a condensate. These small-$x$ gluons are generated
by radiation corrections from gluons at large $x$, whose natural
evolution time scale is Lorentz dilated. This time dilation is
transferred to the small-$x$ degrees of freedoms, making them evolve
very slowly compared to other natural time scales: their behavior is
``glassy". Considering these factors, together with the color nature
of gluons, this initial stage has been named Color Glass Condensate
(CGC)~\cite{McLerran:1993ni,McLerran:1993ka,Iancu:2003xm}. The
production of the pre-equilibrium secondary soft gluons is related
to breaking the phase coherence in this initial CGC wave
function~\cite{Lappi:2006fp}.\\[-0.05in]

\textbf{Parton cascade model}: After the initial parton production,
one needs to describe the pre-equilibrium stage and its
thermalization.  For dilute systems with incoherent parton
configurations,  the classical motion of on-shell partons can be
described by the Parton Cascade  Model
(PCM)~\cite{Geiger:1994he,Zhang:1997ej,Molnar:PRC00-PCM}, which
solves the Boltzmann equation with a leading order PQCD collision
term. Initially, the PCM was assumed to work for the pre-equilibrium
stage, the subsequent thermalization period  and the succeeding QGP
expansion stage. However, the typical thermalization time obtained
from PCM with 2-body ($gg \leftrightarrow gg$) PQCD scattering cross
sections is of the order of 5 fm/c~\cite{Molnar:PRC00-PCM}, which is
too long for the fast thermalization ($< 1 \ \mathrm{fm/c}$)
required by RHIC data~\cite{Heinz:2001xi,Heinz:2002un}. The recent
development of a PCM with $2 \leftrightarrow 3$ ($ gg
\leftrightarrow ggg $) processes leads to faster thermalization and
indeed reproduces the large observed elliptic
flow~\cite{Xu:PRC05-Thermal,Xu:PRL08-V2shear}. However, it is
applied to a dense system, which creates tension with the
assumptions under which the Boltzmann equation is valid. In a very
dense system (such as the very early pre-equilibrium stage at RHIC
and LHC energies) the partons scatter so frequently that they are no
longer on shell. To treat partons with off-shell energies requires
the use of quantum transport theory. The theoretical framework for
quark-gluon quantum transport was developed by Heinz more than 20
years
ago~\cite{Heinz:1983nx,Heinz:1984yq,Heinz:1985qe,Heinz-Elze:1989un},
but its numerical demands are exorbitant, and it has not yet been
implemented numerically.\\[-0.05in]

\textbf{Thermalization}: The formation of the quark gluon plasma
requires two aspects: local equilibrium (thermalization) and local
momentum isotropy. Generally, it is believed that the system
achieves momentum isotropy before completing thermalization, which
is assumed to happen at a time scale of $\sim 0.6 \ \mathrm{fm/c}$
at RHIC energies as indicated by the validity of ideal hydrodynamic
simulations~\cite{Heinz:2001xi,Heinz:2002un}. Elucidating the
mechanism for fast isotropization and thermalization has been a
theoretical challenge. Although much progress has been made during
the past years, the thermalization mechanism  is still not fully
understood. Recently, people realized that the Weibel instability,
which is known from traditional electromagnetic
plasmas~\cite{Weibel:1959}, might help to understand the fast
isotropization and thermalization of the
QGP~\cite{Mrowczynski:1988dz,Mrowczynski:1993qm,Arnold:2003rq,Arnold:2004ti}.
The presence of a Chromo-Weibel instability in the early parton
plasma, which is characterized by strong momentum anisotropies, has
been confirmed by numerical simulations in
1+1-dimensional~\cite{Arnold:2003rq,Arnold:2004ti,Rebhan:2004ur} and
3+1-dimensional~\cite{Arnold:2005vb,Rebhan:2005re} hard thermal loop
effective theory and in a 3+1-dimensional expanding
Glasma\footnote{\emph{Glasma is the name introduced for initially
highly coherent gluon matter that makes the transition from the
Color Glass Condensate to the Quark Gluon
Plasma~\cite{Lappi:2006fp}.}}
~\cite{Romatschke:2006nk,Romatschke:2005pm}. However, non-Abelian
saturation effects temper the exponential growth of the
Chromo-Weibel instability, leading to significantly larger time
scales even for isotropization  than those required for the fast
thermalization approximately observed at RHIC~\cite{Arnold:2005vb,Romatschke:2006nk}. \\[-0.05in]

\textbf{Hydrodynamics}: If thermalization is achieved and can be
locally maintained during the subsequent expansion, the further
evolution of the QGP and hadronic matter can be described by
hydrodynamics~\cite{Hydro-Clare:1986qj}. Hydrodynamics is a
macroscopic approach which describes the system by macroscopic
variables, such as local energy density, pressure, temperature and
flow velocity. It requires knowledge of the equation of state, which
gives a relation between pressure, energy and baryon density, but no
detailed knowledge of the microscopic dynamics. The simplest version
is ideal hydrodynamics~\cite{Kolb:2003dz,Huovinen:2003fa}, which
totally neglects viscous effects and assumes that local equilibrium
is always perfectly maintained during the fireball expansion.
Microscopically, this requires that the microscopic scattering time
is very much shorter than the macroscopic expansion (evolution) time
and that the mean free path is much smaller than the system size. If
this is not satisfied, viscous effects come in, and one can apply
viscous hydrodynamics as long as the deviation from local
equilibrium remains small~\cite{Heinz:2009xj,Teaney:2009qa}. If the
system is far away from equilibrium, one has to switch to a kinetic
theory
approach, such as parton~\cite{Geiger:1994he} or hadron~\cite{Bass:1998ca} cascade models. \\[-0.05in]

\textbf{Hadron cascade model and hybrid approaches}: The hadron
cascade model~\cite{Sorge:1989dy,Bass:1998ca,Bleicher:1999xi}, which
solves the Boltzmann equation for a variety of hadron species with
flavor-dependent cross-sections, is a successful tool to describe
the hadronic matter created at AGS and SPS energies. At these
collision energies, the hadron cascade model is initialized by a
superposition of hadrons and hadronic strings, produced in the
primary nucleon-nucleon collisions. At RHIC energies and above,
hybrid approaches that combine a parton cascade model or
hydrodynamics with a hadron cascade, provide a ``unified"
description of the evolution of the QGP and the succeeding hadronic
matter. However, some caution must be taken for the transition
between the models. Parton + hadron cascade hybrids must deal with
the problem of converting partons to hadrons without violating the
second law of thermodynamics (i.e. without losing entropy), and they
have difficulties to incorporate the change in the structure of the
QCD vacuum during the phase transition\footnote{\emph{Some
development try to solve this problem can be found
in~\cite{Lin:2004en,Zhang:2007rab}}}. Hydrodynamics + hadron cascade
hybrids~\cite{Hirano:2005wx,Nonaka:2006yn} can more easily
accommodate these, by employing a realistic EOS from lattice QCD.
One generally stops hydrodynamics at a switching temperature
slightly below $T_c$, converting the fluid to hadrons using a
Cooper-Frye prescription (see Chap.~3.4) for phase-space
distributions to generate (via Monte-Carlo) initial momentum and
spectra profiles for the hadron cascade simulations. However, this
procedure can not deal with a potential feed-back from cascade
hadrons to the hydrodynamics fluid along space-like parts of the
matching freeze-out hypersurface. \\

\subsection{QGP signatures }

Possible signals and probes for the quark-gluon plasma have been
investigated for around 30 years since the birth of the field. Such
signatures include: collective
flow~\cite{Ollitrault:1992bk,Reisdorf:1997fx,Kolb:2003dz,Huovinen:2003fa},
strangeness enhancement~\cite{Rafelski:1982pu,Sollfrank:1995bn},
charmonium suppression~\cite{Matsui:1986dk,Kluberg:2009wc}, thermal
photon and dilepton emission~\cite{McLerran:1984ay,Gale:2003iz}, jet
quenching~\cite{Wang:1991xy,Wang:1996yh,Gyulassy:2003mc,Kovner:2003zj},
critical fluctuations~\cite{Jeon:2003gk}, and others. Some of the
predicted signature were already found in earlier heavy ion
experiments at AGS and SPS energies~\cite{Rev-Heinz:2000bk}.
However, none of these signatures allow individually to prove QGP
formation, as they are contaminated by the dynamical evolution of
the fireball through various stages, usually from the very early
pre-equilibrium stage through (perhaps) a QGP phase to the late
hadronic stage. The combination of three observations at RHIC, that
finally convinced the community that the QGP has been successfully
created, were the measurements of strong anisotropic collective
flow, valence quark number scaling of the elliptic flow  $v_2$, and
jet quenching ~\cite{2008jna,Jacobs:2007dw}. This led to the
announcement in 2005 that the QGP had been discovered at
RHIC~\cite{Gyulassy:2004vg,Gyulassy:2004zy,2008jna,PefectLiquid}.
 \\[-0.05 in]

\textbf{\underline{Collective flow}:} \\[-0.10 in]

The hadron momentum spectra, their angular distribution (flow
patterns) and the particle yield ratios are primary observables for
the bulk medium created in heavy ion collisions. One of the main
discoveries of RHIC is that the medium displays  strong collective
dynamics~\cite{Adler:2002pu,Adler:2003kt,Adams:2003am,Adams:2003zg},
which, for the first time in the history of particle and nuclear
physics, could be quantitatively well described by ideal
hydrodynamics~\footnote{\emph{Although large collective flow was
also observed in heavy ion collisions at AGS and SPS energies, the
corresponding data can not be quantitatively described by
hydrodynamics models ~\cite{Agakichiev:2003gg}.}}. In the language
of hydrodynamics, the collective flow is driven by pressure
gradients,  thus providing access to the equation of state (EOS) of
the medium. Whereas the azimuthally averaged radial flow receives
contributions from all expansion stages, the anisotropic ``elliptic"
flow seen in non-central collisions is generated mostly during the
hot early stage, and thus provides the information about the QGP
phase, namely its thermalization and its EOS
~\cite{Kolb:2003dz,Huovinen:2003fa,Kolb:2000sd}.

\begin{figure}[t]
\centering
\includegraphics[width=75mm,height=60mm]{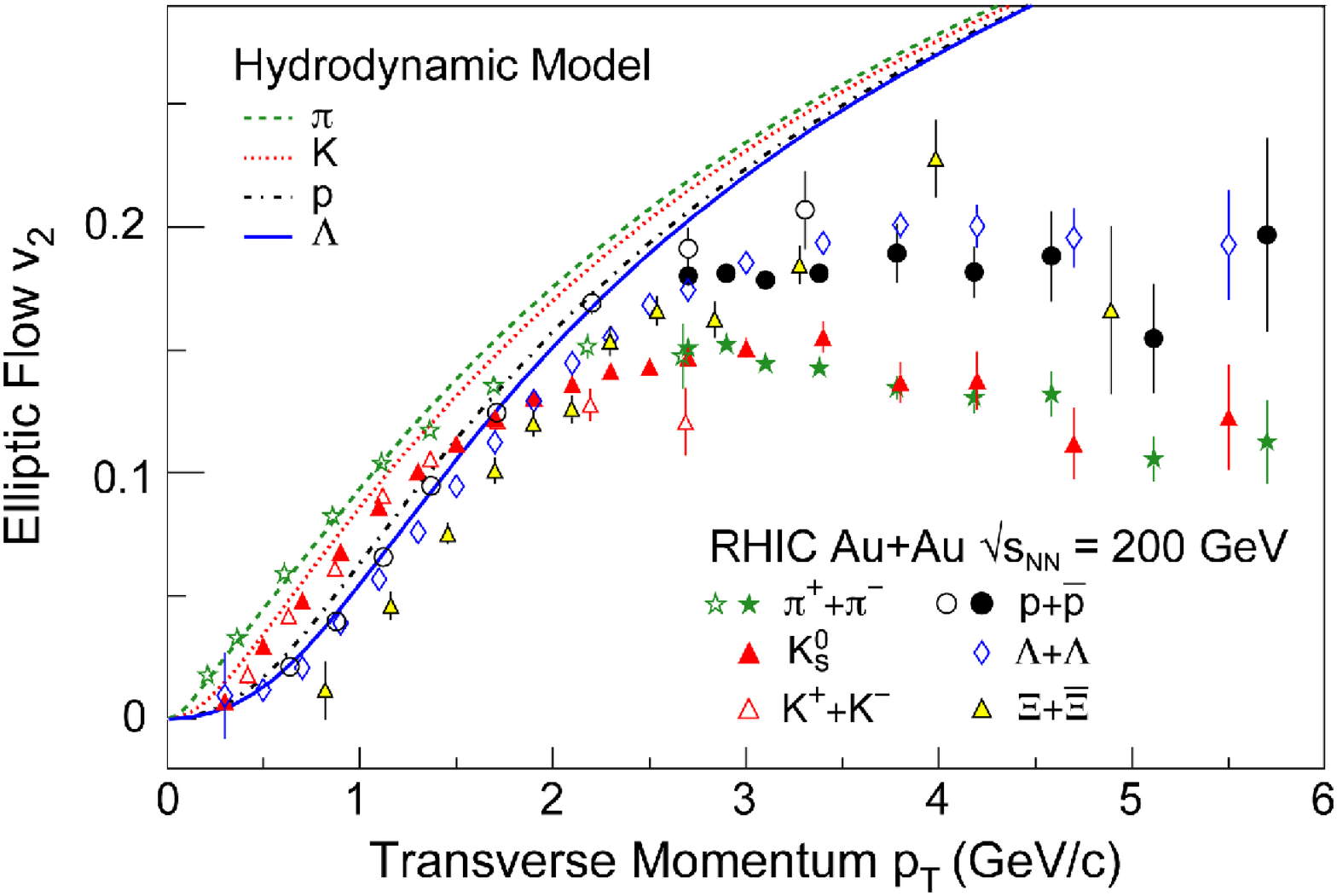}
\vspace{-0.3cm}\caption[$v_2(p_T)$: experiential  data vs. ideal
hydrodynamics predictions]{Elliptic flow $v_2$ for different hadron
species~\cite{Sorensen:2003kp,Adams:2003am,Adams:2005zg}, plotted as
a function of transverse momentum, compared with ideal hydrodynamics
predictions~\cite{Huovinen:2001cy,Heinz:2004pj}.} \label{V2-Ideal}
\end{figure}

The elliptic flow $v_2$ is defined as the 2nd Fourier coefficient of
the azimuthal distribution of hadron spectra:
\begin{eqnarray}
\label{cos2phi}
  v_2(p_T)=\langle\cos(2\phi_p)\rangle\equiv
  \frac{\int d\phi_p\,\cos(2\phi_p)\,\frac{dN}{dy\,p_T dp_T\,d\phi_p}}
                {\int d\phi_p\,\frac{dN}{dy\,p_T dp_T\,d\phi_p}}\,,\quad
\end{eqnarray}
where $\frac{dN}{dy\,p_T dp_T\,d\phi_p}$ is the angular distribution
of the transverse momentum ($p_T$) dependent spectra, and
$y=\frac{1}{2}\ln[(E+p_L)/(E-p_L)]$ is the rapidity of the
particles.

Fig~\ref{V2-Ideal} compares the experimental elliptic flow data
$v_2(p_T)$ with ideal hydrodynamic
predictions~\cite{Sorensen:2003kp,Adams:2003am,Adams:2005zg,Huovinen:2001cy}.
For $p_T < 1.5 \ \mathrm{GeV}$, where most (more than 98\%) of the
particles are produced, ideal hydrodynamics shows excellent
agreement with the experimental data and correctly predicts the
observed splitting for different hadron species. This strongly
indicates that the bulk of the matter is strongly coupled and
behaves like an almost perfect fluid~\cite{2008jna}.

For a successful description of the RHIC data, especially for the
elliptic flow, ideal hydrodynamic requires a fast thermalization of
the system, which must happen on a time scale of about $0.6 \
\mathrm{fm/c}$~\cite{Heinz:2001xi,Heinz:2002un}. Around that time,
the early matter has a peak temperature of $\sim 350 \ \mathrm{MeV}$
-- about twice  the QGP phase transition temperature. Although this
gives indirect evidence for QGP formation, the success of ideal
hydrodynamics is primarily evidence for the formation of a
\emph{thermalized}
new form of matter at $T \sim 2 T_c$~\cite{Heinz:2001xi,Gyulassy:2004vg,Gyulassy:2004zy}. \\[-0.05 in]

\textbf{\underline{Quark degrees of freedom and partonic collectivity}:} \\[-0.10 in]

The success of the statistical model in analyzing of the measured
hadron abundances~\cite{BraunMunzinger:2001ip,BraunMunzinger:2003zd}
gives additional evidence for the QGP phase transition. Particle
ratios including a large number of hadron species are well described
by a thermal model with just two parameters, $T$ and $\mu$, yielding
a chemical freeze-out temperature $T_{ch}=160-170 \ \mathrm{MeV}$,
which is approximately equal to the QCD phase transition temperature
as determined by lattice QCD simulations. This strongly indicates
that hadrons are born during the QGP to hadron phase transition, and
that flavor changing interactions (from inelastic collisions)
quickly cease right after the phase transition~\cite{Heinz:2004qz}.

\begin{figure}[t]
\centering \vspace{-0.1cm}
\includegraphics[width=150mm,height=62mm]{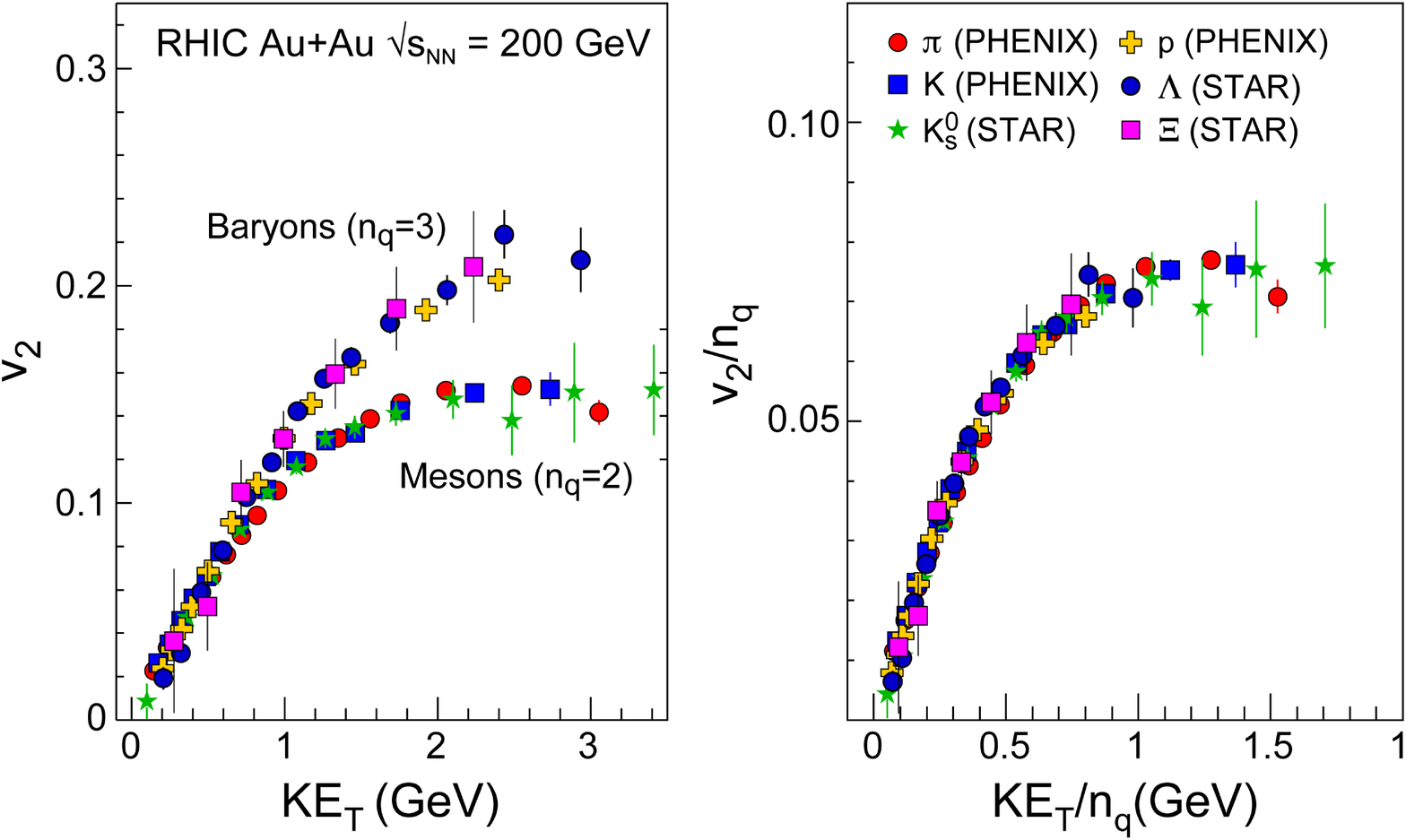}
\vspace{-0.3cm} \caption[Valence quark number scaling of
$v_2$]{left: The transverse kinetic energy $KE_T$ dependence of
$v_2$ for various hadron species. Mesons and baryons fail to
different universal curves respectively. right: $v_2$ and $KE_T$ are
scaled by the number of valence quarks. All of the hadron species
fail to the same universal curve: the differential elliptical flow
per quark.~\cite{Adare:2006ti} } \label{V2-Scale}
\end{figure}

Direct evidence for quark degrees of freedom in the newly formed
matter can be extracted from detailed measurements of the
$p_T$-differential elliptic flow $v_2 (p_T)$ for a large variety of
mesons and baryons. This observable, shown in Fig.~\ref{V2-Ideal},
shows a characteristic splitting between mesons and baryons at
intermediate transverse momentum $p_T=2-5 \ \mathrm{GeV}$. Even
though this is above the $p_T$ range where hydrodynamics is valid,
the underlying collective flow of the fireball still affects hadron
production at this $p_T$~\cite{Heinz:2004qz}. After replacing the
transverse momentum $p_T$ by the transverse kinetic energy
$KE_T=\sqrt{p_T^2+m^2}-m$, the mass splitting at $p_T< 1.5 \
\mathrm{GeV}$ disappears. Fig.~\ref{V2-Scale} left shows, as a
result of this procedure, two universal $v_2$ scaling curves for
baryons and mesons, respectively~\cite{Adare:2006ti}. Baryons, which
contain three valance quarks, show stronger $v_2$ at intermediate
transverse momentum than mesons, containing only two valance quarks.
This phenomenon can be well explained by the quark recombination
model~\cite{Fries:2003vb,Fries:2003kq,Greco:2003xt,Greco:2003mm}, in
which collectively flowing baryons and mesons are generated by the
coalescence of quarks that collectively flow with the medium.
According to the recombination model, the baryon and meson elliptic
flow coefficients are expressed in terms of the quark elliptic flow
in the following way~\cite{Molnar:2003ff}:
\begin{eqnarray}
  v_2^{(M)}(p_T)= 2  v_2^{(q)}(p_T/2); \qquad \qquad v_2^{(B)}(p_T)= 3
  v_2^{(q)}(p_T/3).
\end{eqnarray}
As the right panel in Fig.~\ref{V2-Scale} shows, a corresponding
rescaling of both $v_2$ and $KE_T$ by the number of valence quark (2
for mesons and 3 for baryons) leads to a universal $v_2$ scaling
curve for \emph{all} hadron species.  In short, the universal
valence quark number scaling $v_2$ suggests that the collectively
flowing matter directly involves quarks, and that the quark
collective properties are
transferred to those of hadrons by quark recombination. \\[-0.05in]

\textbf{\underline{Jet quenching}:}\\[-0.10in]

At the very beginning of a heavy ion collision, hard scatterings of
incoming quarks and gluons every now and then create a pair of
energetic fast partons with large transverse momentum. Each of the
two fast partons will finally fragment into a spray of hadrons,
forming what is called a jet. The rates for such hard process are
small but grow rapidly with increasing  collision energy. At RHIC
energies, the fast partons for the first time became sufficiently
abundant as useful probes for the hot medium. Their production rates
are well understood, both experimentally from pp collision at the
same energy or theoretically via PQCD. What makes them useful for
heavy ion collisions is that, before hadronization to jets, they
have to travel through the hot fireball medium formed in the
collision, which modifies their initial energy and momentum.

\begin{figure}[t]
\centering \vspace{-0.0cm}
\includegraphics[width=0.5 \linewidth, height=60mm]{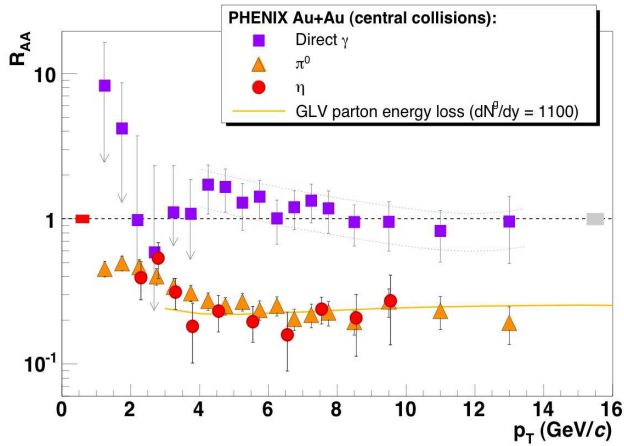}
\vspace{-0.3cm} \caption[Jet quenching in central Au+Au
collisions]{The nuclear modification factor factor $R_{AA}$ as a
function of transverse momentum $p_T$ for direct photons $\gamma$,
as well as $\pi^0$ and $\eta$ mesons in central Au + Au
Collisions~\cite{Adler:2006bv}. $R_{AA}$ is defined as the ratio of
the cross section per nucleon-nucleon collision measured in a heavy
ion collision divided by the cross section measured in $\mathrm{p  +
p}$ collisions. $R_{AA}=1$ if the heavy collisions can be viewed as
a simple superposition of $\mathrm{p  +  p}$ collisions.}
\label{Jet-Quench}
\end{figure}

\begin{figure}[t]
\centering
\includegraphics[width=73mm,height=60mm]{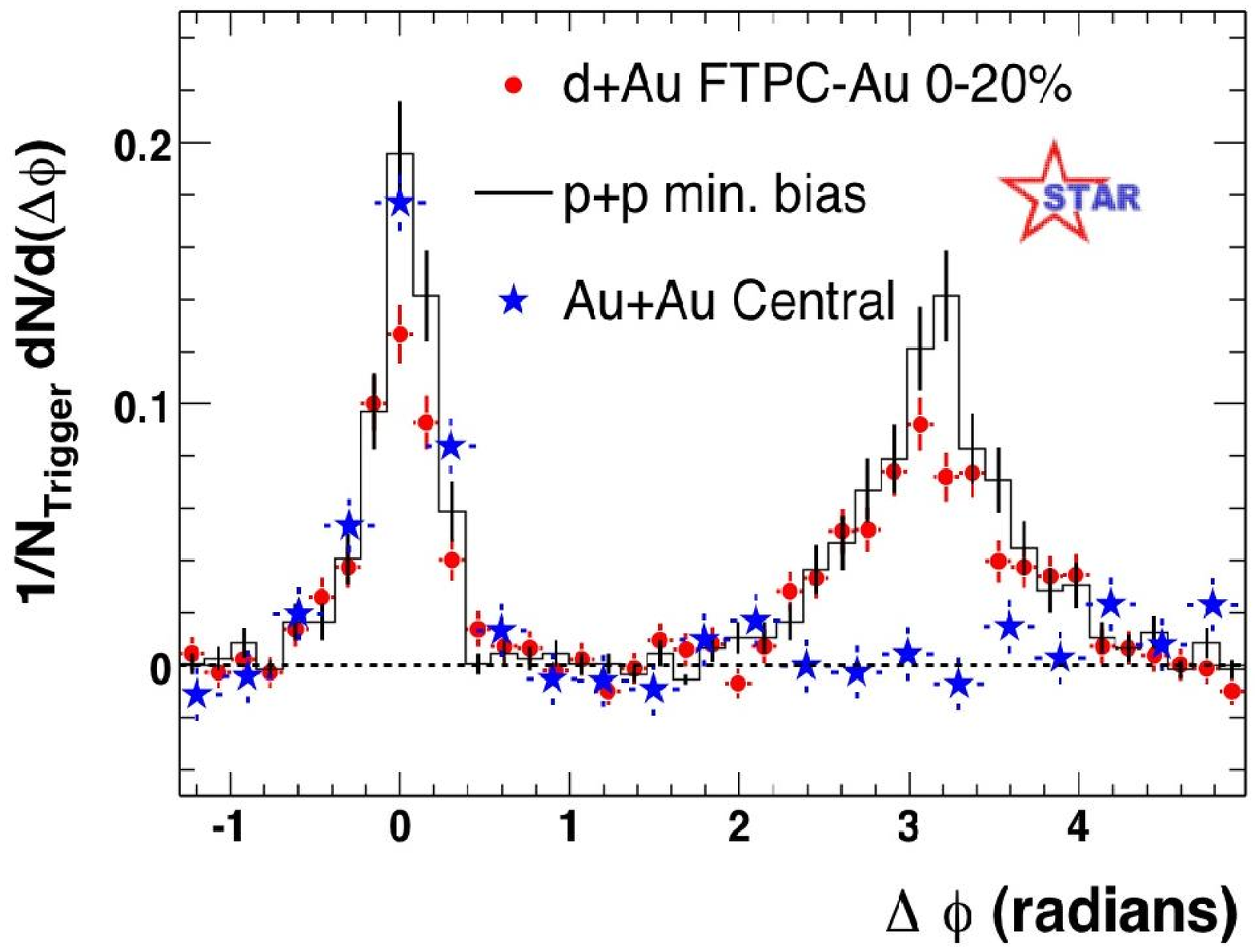}
\includegraphics[width=76mm,height=60mm]{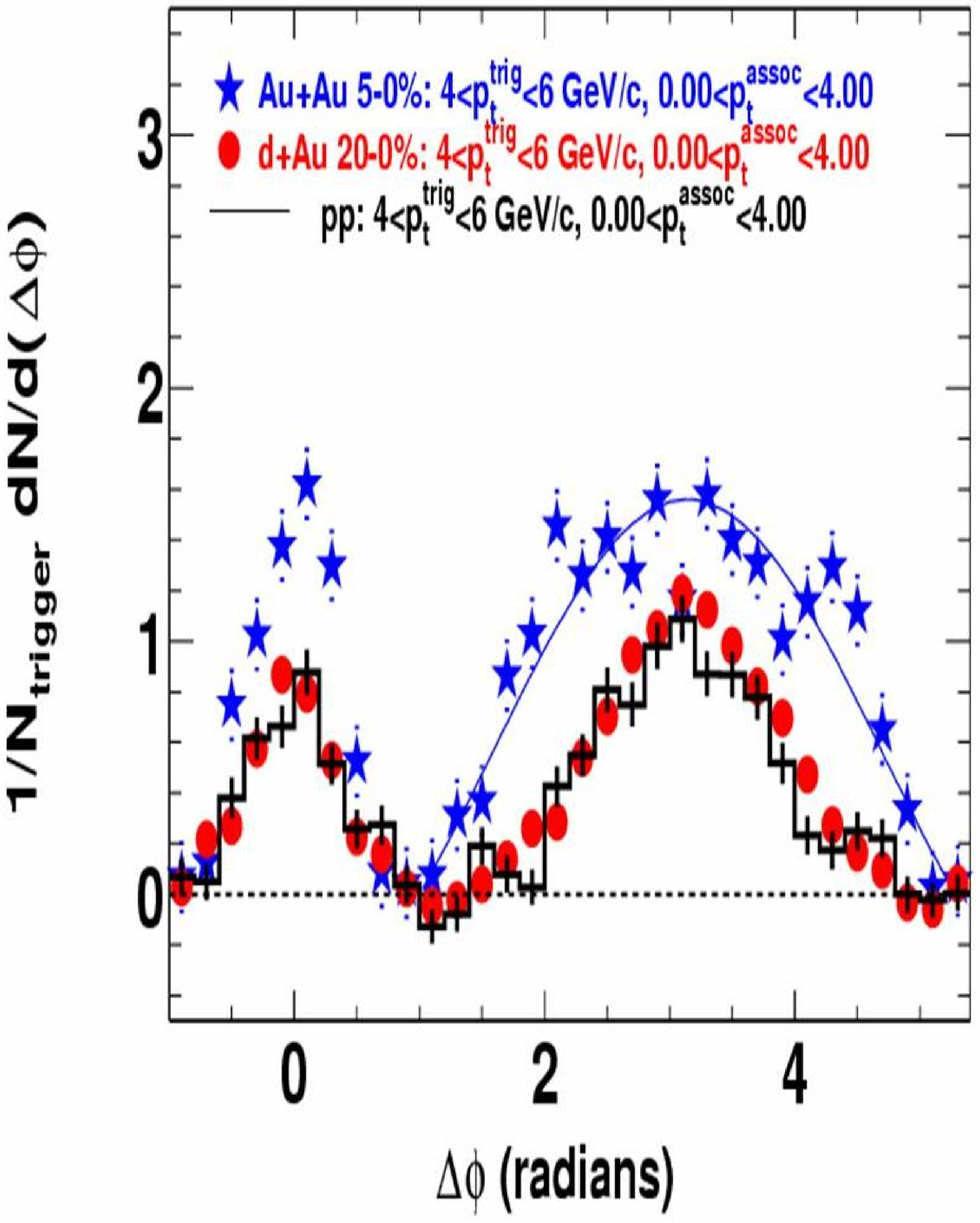}
\vspace{-0.3cm} \caption[Azimuthal angular correlations between high
momentum hadrons]{Azimuthal angular correlations between high
momentum hadrons in $\mathrm{p \ + \ p}$, d + Au, and central Au +
Au collisions. In both cases, the trigge particle of the pair has
high-momentum ($p_T > 4 \mathrm{GeV}$). Left: Associated particles
recoiling with high momenta ($p_T > 2 \mathrm{GeV}$) exhibit strong
suppression in Au + Au~\cite{Adams:2003im}. Right: Associated
particles with low recoil momenta ($p_T > 0.15 \mathrm{GeV}$) are
strongly enhanced in Au + Au collisions~\cite{Adams:2005ph}. }
\label{JetCor}
\end{figure}

One of the most exciting results from RHIC, right after the
discovery of strong elliptic flow, was the observation of a strong
suppression of high-$p_T$ hadrons in central Au+Au collision,
compared with the scaled results from the pp
collisions~\cite{Adler:2002xw,Adler:2003qi,Adams:2003kv}. The
experimentally observed suppression by a factor 4-5 agrees
qualitatively  with theoretical predictions for jet
quenching~\cite{Wang:1991xy,Gyulassy:1993hr,
Gyulassy:2003mc,Kovner:2003zj}, which argued that the QGP formation
could lead to large energy loss of fast partons by collision induced
gluon radiation and thus suppresses the production of high $p_T$
hadrons fragmented from such partons. The discovery of jet quenching
is illustrated in Fig~\ref{Jet-Quench}. The strong suppression of
pions and $\eta$ mesons is distinctly in contrast to that of direct
photons, which show no suppression since they interact only
electrodynamically and thus directly escape from the medium without
further interactions.

Angular correlations between a high-$p_T$ leading (trigger) hadron
with other energetic hadrons provide additional strong support for
the picture of significant parton energy loss in the QGP
medium~\cite{Wang:1996yh}. Energy momentum conservation requires
that fast partons are always generated in pairs, moving along
opposite directions in the pair center of mass frame. This leads to
back-to-back correlations between the resulting jets, as shown by
two peaks in the azimuthal angle correlation separated by $180^0$.
Such a back-to-back correlation is clearly seen in p+p and d+Au
collisions (see left panel of
Fig.~\ref{JetCor}~\cite{Adams:2003im,Adler:2002tq}). In central
Au+Au collisions, however, one sees only one such peak in the
direction of fast trigger hadron (blue stars in the left panel of
Fig.~\ref{JetCor}). This can be understood if one assumes that the
energetic parton pair is created near the surface of fireball: the
near-side outgoing parton quickly escapes from the medium and
fragments into the leading hadron and other softer hadrons
correlated in angle with the leading hadron, while its
inward-traveling partner at $180^0$ loses most of its energy through
interactions with the QGP medium and no longer contributes to
energetic hadrons in the away-side (recoiling)
direction~\cite{Wang:1996yh}. The energy carried by the away-side
fast parton is deposited in the medium, which leads to an
enhancement of soft hadron production in the away-side hemisphere,
as shown in the right panel of Fig~\ref{JetCor}~\cite{Adams:2005ph}.
Soon, people realized that the much broadened away-side correlations
may be related to a possible collective  ``hydrodynamic" response of
the medium to the energy and momentum deposition from the fast
parton, called Mach
Cone~\cite{CasalderreySolana:2004qm,CasalderreySolana:2006sq,Qin:2009uh}.\\

\subsection{``RHIC scientists serve up the perfect liquid "}
The 2007 NSAC Long Range Plane for Nuclear Physics States that ``The
experiments performed at RHIC since it began operation in 2000 have
provided spectacular evidence that the QGP does
exist"~\cite{2008jna}, but its properties are quite different from
the earlier expectation based on PQCD, which had suggested that the
QGP should behave like a dilute gas. In fact, the strong collective
flow together with the very good description from ideal
hydrodynamics indicates that the quark gluon plasma created at RHIC
is strongly coupled and behaves like a nearly perfect liquid with
very low viscosity~\cite{2008jna}.

The discovery of the  strongly coupled QGP is intellectually
exciting since it surprisingly connects super-string theory with
relativistic heavy ion experiments. Mapping strongly coupled quantum
field theories to weakly coupled gravity by the AdS/CFT
correspondence~\cite{Maldacena:1997re,Aharony:1999ti}, string
theorists have developed tools for gaining (at least) qualitative
insights into the strongly coupled QCD-like
systems~\cite{Son:2007vk,Gubser:2009sn} where traditional PQCD
methods fail to apply. Meanwhile,  insights from other strongly
coupled systems may help us to understand the nature of the strongly
coupled QGP. An example comes from  strongly interacting fermion
systems, created in optical traps at extremely low temperatures,
which show similar hydrodynamic
behavior~\cite{JET-Duke-fluid,JET-Duke-eta}. The advantage of such
cold atom experiments lies in the tunable coupling strength, via
Feshbach resonances, which allows to switch between strongly and
weakly coupled fermion systems and thus may help us to understand
the transition between strongly and weakly coupled QGP.

With these exciting developments in the past years, the next phase
of the RHIC physics program and of the incoming heavy ion program at
the LHC will focus on detailed investigations of the QGP, ``both to
quantify its properties and to understand precisely how they emerge
from the fundamental properties of QCD"~\cite{2008jna}. Fundamental
questions that need to be addressed include (see
Ref~\cite{Future-QGP} for details):

\begin{itemize}
  \item  \emph{What is the mechanism of the unexpectedly fast thermal
  equilibrium?} \\[-0.25in]

  \item  \emph{What is the initial temperature and thermal evolution of the
  produced matter?} \\[-0.25in]

  \item  \emph{What is the energy density and equation of state of the
  medium?} \\[-0.25in]

  \item  \emph{What is the viscosity of the produced matter?} \\[-0.25in]

  \item  \emph{Is there direct evidence for deconfinement,color
  screening, and a partonic nature of the hot dense medium?
  What is the screening length?} \\[-0.25in]

  \item  \emph{Is the chiral symmetry restored by QCD?} \\[-0.25in]

  \item  \emph{How does the new form of matter hadronize at the phase transition?}
\end{itemize}

In this thesis, I will concentrate on one of the above questions,
the viscosity of the QGP. To answer it requires continuous
progresses on both theoretical and experimental sides, especially
the development of viscous hydrodynamics and  future high statistics
flow measurements for a variety of identified hadrons species. This
thesis will focus on viscous hydrodynamics for relativistic heavy
collisions\footnote{\emph{This thesis mainly covers work done and
results obtained by the
author~\cite{Heinz:2005bw,Song:2007fn,Song:2007ux,Song:2008si,Heinz:2008qm,Song:2008hj,Song:2009Bulk}.
Results from other groups published over the same time period can be
found in
Ref~\cite{Baier:2006um,Baier:2006gy,Romatschke:2007jx,Romatschke:2007mq,Luzum:2008cw,
Luzum:2009sb,Dusling:2007gi,Dusling:2008xj,Dusling:2009bc,
Molnar:2008xj,Huovinen:2008te,Chaudhuri:2007qp,Chaudhuri:2008sj,Chaudhuri:2008je}.
We will refer to these papers for comparison when needed or
useful.}}. The transport coefficients (such as shear viscosity and
bulk viscosity) are free parameters in viscous hydrodynamic
calculations. The hope is that by tuning these parameters for best
fits of a sufficiently large set of sensitive experimental
observables, the QGP viscosity can be extracted phenomenologically
from experimental data. This requires not only the development of a
practical and accurate viscous hydrodynamics code, but a careful
investigation of other ingredients (EOS, initial and final
conditions, etc.), which may affect the viscosity-sensitive
observables. Some of these issues will be addressed in this thesis.
(Of course, the QGP transport coefficients are not just simple
numbers, but depend on the thermodynamics properties of the
fireball, which evolve with time). We should therefore use, as much
theoretical knowledge as available, to constrain the the temperature
dependence of these variables, at least at the qualitative level. In
the rest of this chapter we will briefly review the present status
of theoretical understanding and knowledge of the transport
coefficients for QCD matter and other strongly coupled systems.\\

\section{Transport coefficients for non-relativistic fluids}
There are several transport coefficients that characterize the
internal "friction" in a fluid. For example, the shear viscosity
(dynamic viscosity) $\eta$ measures the fluid's resistance to flow,
the bulk viscosity (volume viscosity) $\zeta$ measures the fluid's
resistance to expansion, and the thermal conductivity (heat
conductivity) $\lambda$ measures the fluid's ability to conduct
heat.  In this section, we will discuss shear and bulk viscosity in
some detail, since they will be used in later parts
of this thesis.\\[-0.05in]

\textbf{\underline{Shear viscosity}}\\[-0.10in]

Classically, the shear viscosity is defined as the ratio between the
friction force $F$ per area and the transverse flow gradient
$\nabla_y v_x$,
\begin{eqnarray}
\frac{F}{A}=\eta \nabla_y v_x.
\end{eqnarray}

\begin{figure}[t]
\vspace{-7mm}
\includegraphics[bb=75 40 610 710,width=60mm,height=85mm,angle=90]{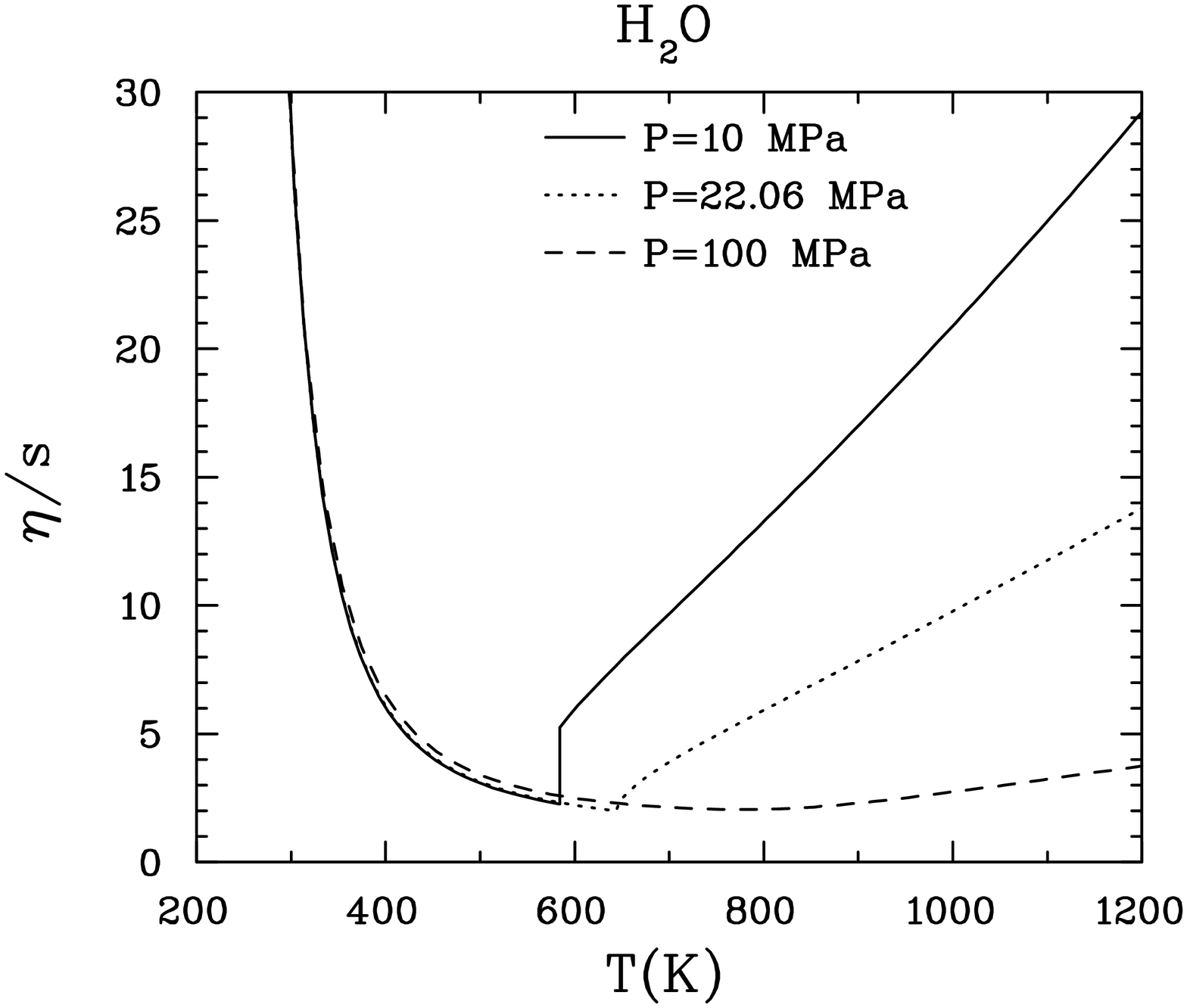}
\includegraphics[width=70mm,height=51mm]{./Figs/Chap1/etaS-HeN.eps}
\vspace{-3mm}
 \caption[$\eta/s$ for common fluids ]{left: $\eta/s$
as a function of temperature for water at different pressures
~\cite{Kapusta-Csernai:2006zz}. right: $\eta/s$ for helium, nitrogen
and water ~\cite{Kovtun:2004de}.} \label{Fig-etas-Fluid}
\end{figure}

Microscopically, shear viscosity is associated with momentum
transfer between particles in different fluid cells. For a dilute
gas of non-relativistic particles, the shear viscosity $\eta$ and
shear viscosity to entropy density ratio $\eta/s$ are estimated
as~\cite{Molecular-theory}:
\begin{eqnarray}
\eta \sim  n m \bar{v}  l_{\mathrm{mfp}} \sim \frac{m
\bar{v}}{\sigma}, \qquad   \qquad   \frac{\eta}{s}  \sim \frac{m
\bar{v}}{\sigma n},
\end{eqnarray}
where $n$ is the particle density, $m$ is the particle mass,
$\bar{v}$ is the mean velocity, $l_{\mathrm{mfp}}=\frac{1}{n\sigma}$
(where $\sigma$ is the transport cross section) is the mean free
path, and the entropy density $s$ is proportional to the particle
density $s \sim n$. The shear viscosity of a non-relativistic dilute
gas increases with temperature $\eta \sim \sqrt{m T}$ (taking
$\bar{v} \sim \sqrt{T/m}$) and is approximately independent of
density\footnote{\emph{This estimate agrees qualitatively with
results derived from the Chapman-Enskog
expansion~\cite{Molecular-theory}.}}. The $1/n$ density dependence
of $\eta/s$ translates into a pressure dependence of $\eta/s$, as
shown in the left panel of Fig~\ref{Fig-etas-Fluid}. Using the ideal
gas EOS $p=n k_B T$, one finds that, for fixed pressure, $\eta/s$
increases with temperature approximately as
\begin{eqnarray}
\eta/s\sim T^{3/2}\label{eta-1}
\end{eqnarray}
in the high temperature gas phase.

For a liquid, the momentum transfer between different fluid cells
involves the motion of voids. The density of voids decreases with
temperature, which leads to an increase of shear viscosity due to
the growth of mean free path. Detailed
calculation~\cite{Kinetic-theory-liquid} shows that
\begin{eqnarray}
\eta \simeq h n e^{E/(k_B T)}   \qquad \frac{\eta}{s} \sim h
e^{E/(k_B T)} \label {eta-2},
\end{eqnarray}
where $E$ is an activation energy and $h$ is the Planck constant,
and again $s \sim n$. This result shows that for a liquid $\eta/s$
increases with decreasing temperature.

Taking the liquid phase and gas phase together, Eq.(\ref{eta-1}) and
Eq.(\ref{eta-2}) imply that $\eta/s$ is likely to reach a minimum
during the liquid-gas phase transition. This is confirmed by
experimental results shown in Fig~\ref{Fig-etas-Fluid} which plots
in the right panel $\eta/s$ as a function of temperature for three
different fluids (helium, nitrogen and water)~\cite{Kovtun:2004de},
and for water at different pressures in the left
panel~\cite{Kapusta-Csernai:2006zz}. In all of these cases,
$\eta/s$ reaches a minimum near the phase transition.\\[-0.05in]

\textbf{\underline{Bulk viscosity}}\\[-0.10in]

In non-relativistic fluid dynamics, the bulk viscosity $\zeta$ is
defined as a combination of shear viscosity $\eta$ and volume
viscosity\footnote{\emph{The volume viscosity (also known as
``second viscosity") $\nu$ measures the resistance to expansion; it
determines the dynamics of a compressible fluid.}} $\nu$:
$\zeta=\nu+\frac{2}{3}\eta$. Bulk viscosity vanishes for a scale
invariant system. For a dilute monatomic gas, experimental data and
kinetic theory show that bulk viscosity is vanishingly small or zero
~\cite{Gas-Dynamics}. The bulk viscosity of a dilute diatomic gas,
however, has an appreciable value due to the exchange of energy
between translational and rotational degrees of freedom during
collisions ~\cite{Non-Uniform-Gas}. In a dense gas, bulk viscosity
is associated with the internal friction force arising from the
change of volume at constant shape~\cite{BulkVis-Past-Present}. In a
liquid, bulk viscosity is related to the rearrangement of molecules
during acoustic compression and
rarefaction~\cite{BulkVis-Past-Present}.

Experimentally, bulk viscosity is generally determined from
measuring the sound absorption coefficient. In contrast to shear
viscosity, there are no comprehensive bulk viscosity data sets for
many varieties of fluids over a wide range of temperature and
density, and the existing bulk viscosity data have large error bars.

The phase transition behavior of the bulk viscosity of spherical
molecules was studied by Meier, Laesecke and Kabelac using
molecular-dynamics simulations with Lennard-Jones potentials. They
found that the bulk viscosity shows a peak in the vicinity of the
gas-liquid phase transition~\cite{Meier:2005} (see also Fig.6 and
related comments in Ref.~\cite{Kapusta:2008vb}).\\

In short, the shear (bulk) viscosity to entropy ratio reaches a
minimum (maximum) near phase transitions for common non-relativistic
fluids. As we will see in Sec.~\ref{Sec-VIS-QCD}, the same holds for
relativistic QCD matter. This is not a rigorous statement from first
principles calculations, but appears to be supported by specific
examples of both
experimental data and theoretical results.\\

\section[Transport coefficients for relativistic fluids: weak vs. strong coupling]
{Transport coefficients for relativistic fluids: weak coupling vs.
strong coupling}

Relativistic hydrodynamics can be viewed as an effective theory for
quantum field systems at large distance and time scales. The
dynamics of long wavelength, low frequency fluctuations are
characterized by transport coefficients (shear viscosity $\eta$,
bulk viscosity $\zeta$ and heat conductivity $\lambda$, electric
conductivity $\sigma$ etc.). For a relativistic fluid, the transport
coefficients define the leading order corrections of the energy
momentum tensor $T_{\mu\nu}$ and charge current $N_{\mu}$ from their
local equilibrium forms. For example, in the local rest frame the
stress tensor reads at leading order in gradients:
$T_{ij}=p_{eq}(e)\delta_{ij}-\eta(\partial_i u_j +
\partial_i u_j+\frac{2}{3}\delta_{ij} \partial_k u_k u_j) -\zeta
\delta_{ij} \mathrm{\nabla} \cdot u$. In this section, we will
briefly review methods for calculating these transport coefficients
(especially shear and bulk viscosity) in both weak and strong
coupling regimes.

For a weakly coupled system, the transport coefficients can be
calculated from kinetic theory or from linear response theory. The
kinetic theory approach
~\cite{Jeon:1992kk,Jeon:1994if,Arnold:2000dr,Arnold:2003zc,Defu:2005hb,York:2008rr}
starts from local equilibrium distribution functions and expands the
full distribution function order by order in terms of gradients of
the four-velocity, temperature and chemical potential. The viscosity
coefficients are associated with the first order terms in velocity
gradients, and can be determined from the Boltzmann equations if the
collision term is known explicitly (they depend on the transport
cross section in the collision term).

In linear response theory, the transport coefficients can be
rigorously expressed by Kubo formulae~\cite{Kubo:1957}, that relate
the transport coefficients to the slope of associated spectral
functions at zero frequency. For example,
\begin{eqnarray}
\eta=\lim_{\omega\rightarrow 0} \lim_{\mathbf{k}\rightarrow
0}\frac{\rho^{12,12}(\omega,\mathbf{k})}{2\omega}, \qquad
\zeta=\frac{1}{9}\lim_{\omega\rightarrow 0}
\lim_{\mathbf{k}\rightarrow
0}\frac{\rho^{ii,jj}(\omega,\mathbf{k})}{2\omega},
\label{etaZeta-dif}
\end{eqnarray}
where the spectral functions are given by the imaginary part of the
retarded green functions, $\rho(\omega,\mathbf{k})=2\mathrm{Im}G_R(
\omega,\mathbf{k})$, for certain components of the energy-momentum
tensor:
\begin{eqnarray*}
G_R^{\mu\nu,\kappa\sigma}( \omega,\mathbf{k})=-i\int d^4x
e^{i(\omega t-\mathbf{k} \mathbf{x})} \langle \left[
T^{\mu\nu}(\mathbf{x},t) T^{\kappa\sigma} (0,0) \right] \Ra.
\end{eqnarray*}

The Kubo formulae can be applied to both weakly and strongly coupled
systems (see Chap.~1.5) and can be generalized for lattice
simulations. In lattice QCD, the spectral function $\rho(\omega)$ is
obtained from the Matsubara (imaginary time) Green function $G_E$,
instead of the retarded Green function $G_R$. (Both of them have
spectral representation in terms of $\rho$~\footnote{\emph{$G_R$ is
obtained from $G_E$ by replacing $i \omega_n$ with $\omega-p_0 + i
\varepsilon$:
\begin{eqnarray*}
G_E(i \omega_n)=\int\frac{d
\omega}{2\pi}\frac{\rho(\omega)}{\omega-i \omega_n}, \qquad
G_R(p_0)=-i \int\frac{d \omega}{2\pi}\frac{\rho(\omega)}{\omega-p_0
-i \varepsilon}.
\end{eqnarray*}
Here $\omega_n =2 \pi n T (n \in \mathbb{Z} )$ are bosonic Matsubara
frequencies, and T is the temperature of the medium.}}. ) The
calculation of shear and bulk viscosity on the
lattice~\cite{Karsch:1986cq,Nakamura:2004sy,Meyer:2007ic,Meyer:2007dy}
starts with the calculation of the corresponding imaginary time
correlators of the energy momentum tensor in Euclidean time $0 \leq
\tau < 1/T$, which is related to the spectral function by
\begin{eqnarray}
G_E(\tau)=T \sum_n e^{-i \omega_n \tau}\int\frac{d
\omega}{2\pi}\frac{\rho(\omega)}{\omega-i \omega_n} =\int \frac{d
\omega}{2\pi}\frac{\cosh[\omega(\tau-1/(2T))] }{\sinh[\omega/(2T)]}
\rho(\omega). \label{lattice-dif}
\end{eqnarray}
In principle, one can obtain the spectral function $\rho(\omega)$ by
inverting eq.(\ref{lattice-dif}), and then use
eq.(\ref{etaZeta-dif})  to calculate the transport coefficients. In
practice, one can only compute a finite number of points for
$G_E(\tau)$ on the lattice, which makes the unique extraction of an
analytical spectral function $\rho(\omega)$ impossible. Generally,
one makes an ansatz for $\rho(\omega)$ with a small number of
parameters, motivated by (usually somewhat model-dependent)
considerations on the expected behavior of $\rho(\omega)$ at small
and large $\omega$, and then fits these parameters to the Monte
Carlo
data~\cite{Karsch:1986cq,Nakamura:2004sy,Meyer:2007ic,Meyer:2007dy}.\\

\section{Shear and bulk viscosity for QCD matter}
\label{Sec-VIS-QCD}

\textbf{\underline{Shear viscosity}}\\[-0.10in]

\begin{figure}[t]
\includegraphics[width=55mm,height=78mm,angle=90,clip=]{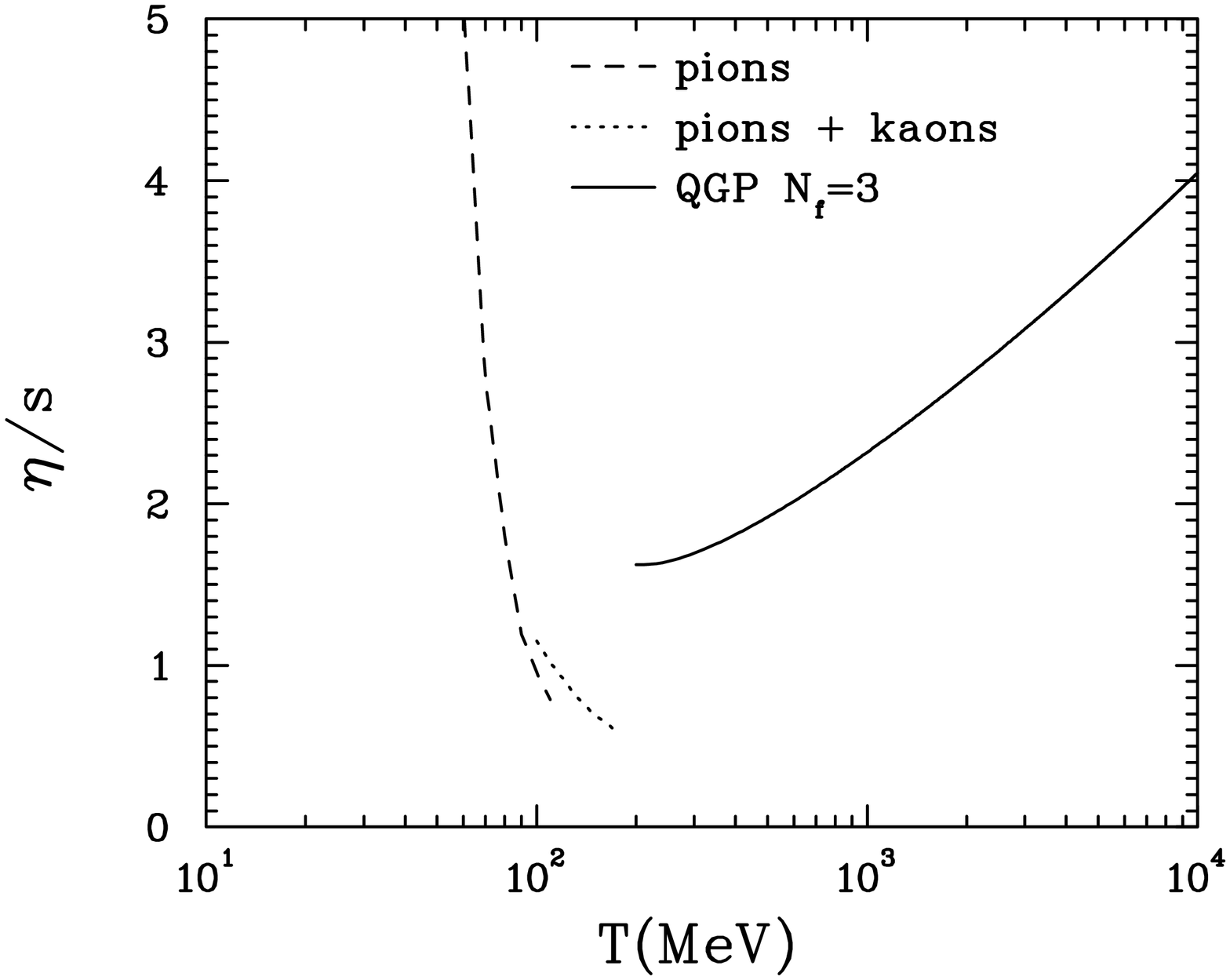}
\hspace{3mm}
\includegraphics[width=0.45\linewidth,height=55mm,clip=]{./Figs/Chap1/etaS-Nasser.eps}
\vspace{-3mm} \caption[$\eta/s$ for QGP and hadron gases ]{left:
$\eta/s$ for low temperature hadronic phase and high temperature QGP
phase~\cite{Kapusta-Csernai:2006zz}. right: $\eta/s$ for hadron
resonance gas in chemical and kinetic
equilibrium~\cite{Demir:2008tr}.} \label{Fig-etas-QCD}
\end{figure}

Attempts to calculate the QGP shear viscosity using weakly coupled
QCD started more than 20 years
ago~\cite{Hosoya:1983xm,Hosoya:1983id}. Using kinetic theory in the
relaxation time approximation and using a simple perturbative
estimate for the latter, one found that the shear viscosity of the
QCD matter behaves as $\eta \sim \frac{T^3}{\alpha_s^2
\mathrm{ln}(1/\alpha_s)}$~\cite{Hosoya:1983xm,Hosoya:1983id}. A
first calculation of the leading logarithmic contribution from the
Boltzmann equation was performed in~\cite{Baym:1990uj}, with a
complete result finally published in~\cite{Arnold:2000dr}. A full
leading order calculation that also computed the coefficient under
logarithm was performed by Arnold, Moore and Yaffe using effective
kinetic theory in the hard thermal loop
approximation~\cite{Arnold:2003zc}. For a weakly coupled QGP with
three massless quark flavors, the shear viscosity to entropy ratio
is~\cite{Arnold:2003zc}:
\begin{eqnarray}
\frac{\eta}{s}=\frac{5.12}{g^4 \ln(2.42/g)}. \label{eta-QCD}
\end{eqnarray}
This result is shown as the solid line in the left panel of
Fig.~\ref{Fig-etas-QCD}, using the two-loop renormalization group
expression for the running coupling
$g(T)$~\cite{Kapusta-Csernai:2006zz}. (Note that using
Eq.\ref{eta-QCD} with a running coupling is phenomenological and
beyond the order of accuracy at which Eq.\ref{eta-QCD} was derived.)

In principle, the QGP shear viscosity can also be non-perturbatively
calculated by lattice QCD using Kubo formulae. However, such
calculations are highly non-trivial in the standard Monte-Carlo
simulations due to the large noise to signal ratio for the relevant
operators and the ill-posed inversion problem for the spectral
function $\rho (\omega)$, given the finite number of data points for
the Euclidean correlators~\cite{Meyer:2007ic}. A first attempt to
calculate the shear viscosity of SU(3) gluonic matter on an $8^3
\times 4$ lattice comes from the pioneering work of Karsch and
Wyld~\cite{Karsch:1986cq}, using a three-parameter ansatz for the
spectral function $\rho (\omega)$. This method was later implemented
by Nakamura and Sakai to calculate both shear and bulk viscosity on
larger lattices ($16^3 \times 8$ and $24^3 \times 8$) with improved
Iwasaki action~\cite{Nakamura:2004sy}. They found that the shear
viscosity to entropy density ratio is less than one, $\eta/s < 1$,
and that the bulk viscosity is zero within errors for the
temperature range $1.4\leq T/T_c \leq1.8$. A new evaluation of the
shear viscosity for SU(3) gluon dynamics was performed by
Meyer~\cite{Meyer:2007ic} using a two-level
algorithm~\cite{Meyer:2003hy}, which dramatically improves the
statistical accuracy of the relevant Euclidean correlators. He
derived a robust upper bound $\eta/s<1.0$ for the shear viscosity
entropy ratio and estimated that:
\begin{eqnarray}
\eta/s = \left\{ \begin{array}{l@{~~~}l}
   0.134(33) & (T=1.65T_c), \\
 0.102(56) & (T=1.24T_c).
          \end{array} \right.
\end{eqnarray}

Early calculations of the shear viscosity for the hadronic matter
started from a theory of massless pions in the low-energy chiral
limit, giving~\cite{Prakash:1993bt}
\begin{eqnarray}
\frac{\eta}{s}=\frac{15}{16\pi}\frac{f_\pi^4}{T^4},
\end{eqnarray}
where $f_\pi= 93\ \mathrm{MeV}$ is the pion decay constant. The
ratio $\eta/s$ diverges at zero temperature; together with the
logarithmic perturbative results discussed above, $\eta/s$ is seen
to reach a minimum near the QGP phase transition. A more detailed
calculation of
 $\eta/s$  for hadronic matter with both pions and kaons can be found in
Ref.~\cite{Prakash:1993bt}; it goes beyond the chiral approximation
and includes intermediate resonances such as the $\rho$ meson (see
also~\cite{Davesne:1995ms}). The left panel of
Fig.~\ref{Fig-etas-QCD} shows these two results, which qualitatively
agree with each other at low temperature but gradually deviate from
each other above $T > 100\ \mathrm{MeV}$ due to kaon excitations.

Recently, $\eta/s$ for a hadron resonance gas including a large set
of hadron species up to 2 GeV was extracted from
UrQMD\footnote{\emph{UrQMD: Ultra-relativistic Quantum Molecular
Dynamics, a hadron cascade
model~\cite{Sorge:1989dy,Bass:1998ca,Bleicher:1999xi}.}}
simulations, using the Kubo formula ~\cite{Demir:2008tr}.
Fig.~\ref{Fig-etas-QCD} shows the extracted $\eta/s$ as a function
of temperature $T$ for a chemically equilibrated hadron resonance
gas with zero chemical potential. Below $T<100 \mathrm{MeV}$,
$\eta/s$ quickly rises with decreasing temperature, in qualitative
agreement with the chiral pion result~\cite{Prakash:1993bt}. Between
100 MeV and 160 MeV, $\eta/s$ saturates at a value $\sim 1$. For a
hadron resonance gas out of chemical equilibrium, a similar tendency
was demonstrated in Fig.~4 of Ref.~\cite{Demir:2008tr}, where it is
shown that at fixed temperature $T$, a non-zero baryon chemical
potential reduces $\eta/s$. For $\mu_B/T=3.0-3.4$ (an
unrealistically large chemical potential for RHIC energies),
$\eta/s$ can decrease to 0.3-0.4 near the phase transition, and the
plateau structure between $100 \ \mathrm{MeV}$ and 160 MeV is
replaced by a tendency that $\eta/s$ continues to decrease with
increasing $T$.

It is worthwhile to point out that Ref.~\cite{NoronhaHostler:2008ju}
argued that including Hagedorn states\footnote{\emph{Hagedorn states
are highly unstable, massive hadronic resonances that exist close to
$T_c$. These states are not included in the above UrQMD
simulation.}} can significantly reduce $\eta/s$ in the hadronic
phase near $T_c$ since the highly degenerate Hagedorn states could
dramatically increase the entropy density. The authors of
Ref.~\cite{NoronhaHostler:2008ju} investigated Hagedorn state
effects on $\eta/s$ for a gas of pions and
nucleons~\cite{Itakura:2007mx} and for a hadron-resonance gas with
excluded volume corrections~\cite{Gorenstein:2007mw} and found that
near $T_c$, $\eta/s$ can be significantly reduced to values close to
the KSS bound $\eta/s = 1/4\pi$
(see Chap.~1.5).\\[-0.05in]

\textbf{\underline{Bulk viscosity}}\\[-0.10in]

At sufficiently high temperature, the equation of state of a
massless QGP satisfies $p=e/3$ which, on the classical level, leads
to a vanishing bulk viscosity for the weakly coupled QGP. However,
quantum corrections break the conformal symmetry and result in a
non-zero bulk viscosity. Detailed calculations from Arnold, Dogan
and Moore~\cite{Arnold:2006fz} showed:
\begin{eqnarray}
\zeta=\frac{A \alpha_s^2 T^3}{\ln[\mu^* / m_D]}\, . \label{bulk-QCD}
\end{eqnarray}
Here, $A$, $\mu^*$ and $m_D^2$ depend on the number of quark flavors
in QCD. For QCD with three massless quark flavors, $A=0.657$,
$\mu^*=7.77 T$ and $m_D^2=1.5 g^2 T^2$ . Relating Eq.
(\ref{bulk-QCD}) with the shear viscosity calculated within weakly
coupled QCD~\cite{Arnold:2000dr,Arnold:2003zc} one finds an
approximate, but simple relationship between shear and bulk
viscosity:
\begin{eqnarray}
\zeta \simeq 15\eta (c_s^2-1/3)^2 . \label{bulk-Weak}
\end{eqnarray}
While this relation does not hold exactly~\cite{Arnold:2006fz}, it
gives the right order of magnitude of $\zeta/\eta$.

\begin{figure}[t]
\includegraphics[width=55mm,height=78mm,angle=90]{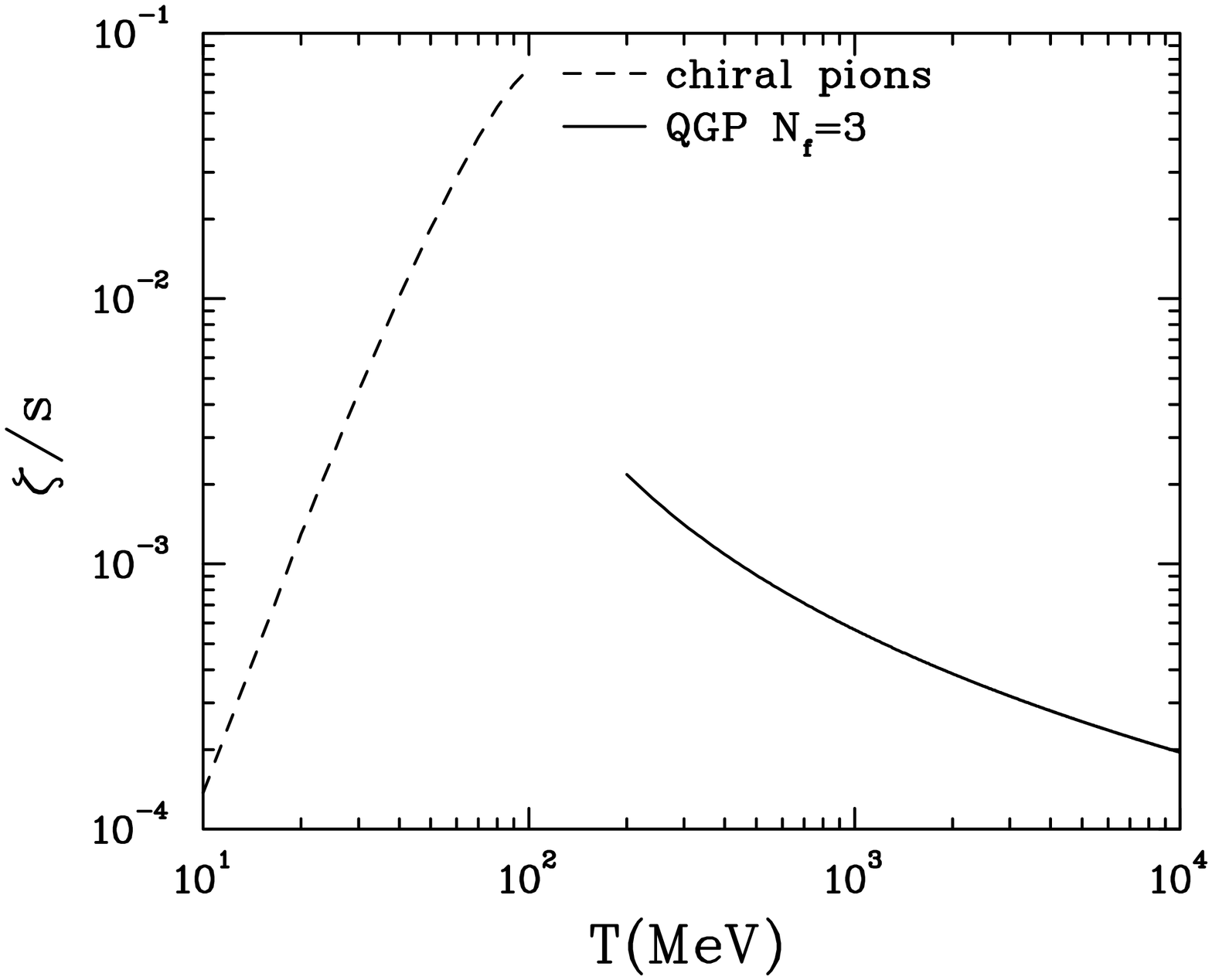}
\hspace{3mm}
\includegraphics[width=55mm,height=78mm,angle=90]{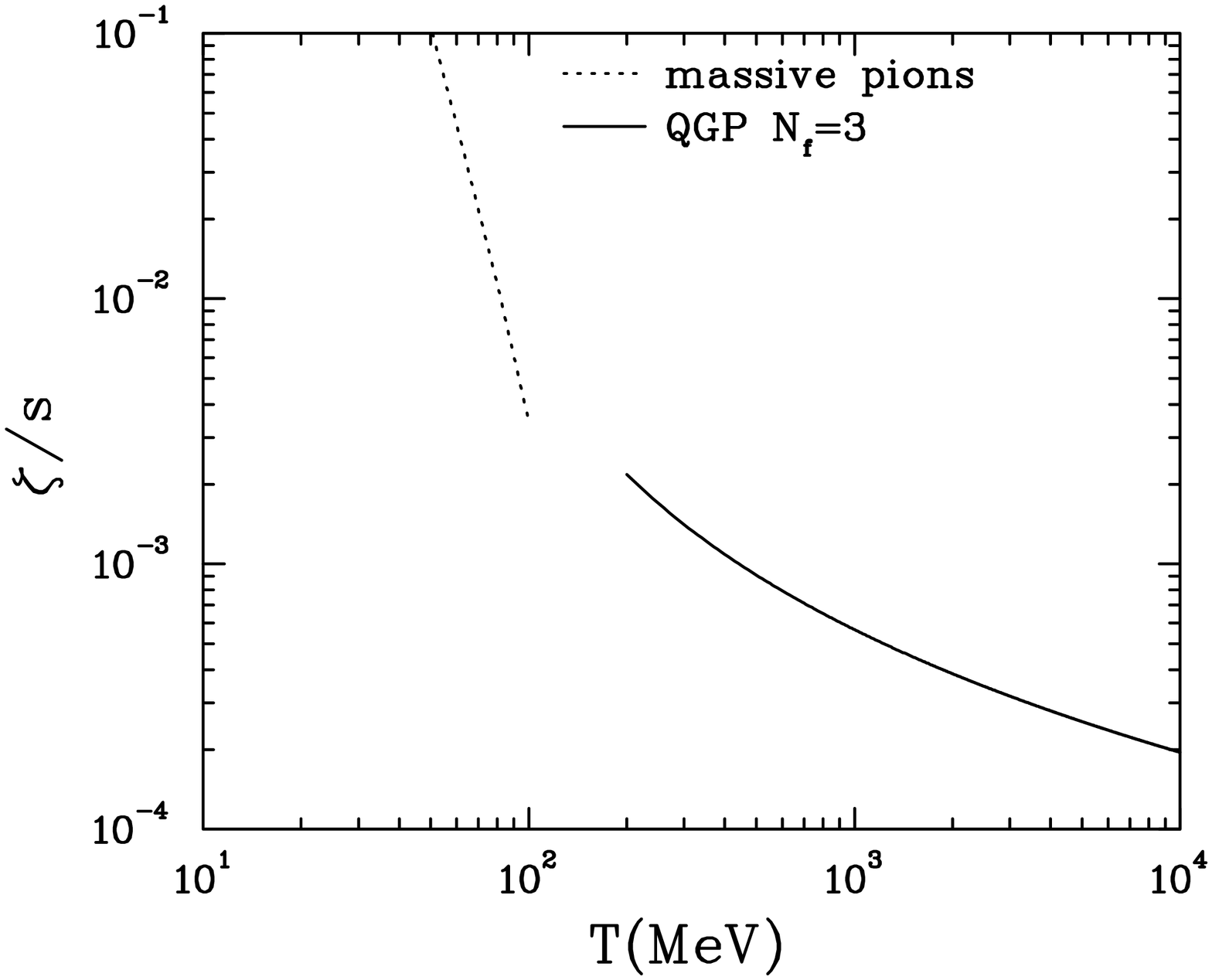}
\vspace{-3mm} \caption[$\zeta/s$ for QGP and hadron gases ]{left:
$\zeta/s$ for high temperature QGP and for massless (left) and
massive (right) pions~\cite{Kapusta:2008vb}.} \label{Fig-Zetas-QCD}
\end{figure}

The bulk viscosity of interacting massless pions was calculated by
Chen and Wang in the framework of kinetic theory, combined with
chiral perturbation theory. They found that $\zeta/s$ monotonically
increases with temperature below $T=120 \ \mathrm{MeV}$:
$\frac{\zeta}{s} \sim \frac{T^4}{f_\pi^4}$. The bulk viscosity for
massive pions was calculated by Prakash et al. in the 1990's, using
kinetic theory with experimental elastic scattering cross sections
as inputs~\cite{Prakash:1993bt,Davesne:1995ms}. In contrast to the
case of massless pions, one finds that $\zeta/s$ is a decreasing
function of temperature. The bulk viscosity to entropy density ratio
$\frac{\zeta}{s}$ for massless and massive poins, together with the
result for a weakly coupled QGP, are illustrated in
Fig.~\ref{Fig-Zetas-QCD}.

The behavior of bulk viscosity near the QCD phase transition was
first investigated by Paech and Pratt within the linear sigma
model~\cite{Paech:2006st}. Their work showed qualitatively that
$\zeta/s$ reaches a maximum near the phase transition. This behavior
was confirmed in later research by Kharzeev et
al.~\cite{Kharzeev:2007wb,Kharzeev:2007wb} and
Meyer~\cite{Meyer:2007dy}, respectively. Using low energy theorems,
Kharzeev et al. connected the bulk viscosity with the lattice
interaction measure $e-3p$ and extracted a temperature dependent
$\zeta/s$ from lattice data for pure SU(3) gluon
dynamics~\cite{Kharzeev:2007wb} and for full
QCD~\cite{Karsch:2007jc}. Meyer directly extracted $\zeta/s$ from
lattice QCD simulations by calculating the corresponding correlation
function for a pure SU(3)gluon plasma. He showed that the peak value
of $\zeta/s$ near $T_c$ can reach as high as
0.73~\cite{Meyer:2007dy} (a value that is around 10 times larger
than the string theory estimates from holographical
models~\cite{Gubser:2008yx,Gubser:2008sz}). Both of these methods
involve an ansatz for the spectral function. However, the
parameterized spectral functions used by these authors  were
challenged by Moore and Saremi~\cite{Moore:2008ws}, who examined the
behavior of the spectral function with analytical methods  both near
the QCD phase transition
and in the weak coupling regime.\\

\section[Shear viscosity and bulk viscosity from $N=4$ SYM and
AdS/CFT]{Shear viscosity and bulk viscosity from $N=4$ SYM and the
AdS/CFT correspondence}

The transport coefficients calculated from weakly coupled QCD are
valid at sufficiently high temperature where the running coupling
constant becomes small. However, the temperature reached in
relativistic heavy ion collisions is not very high (top RHIC
energies: $T \sim 350 \ \mathrm{MeV}$, and LHC energies: $T \sim 600
\ \mathrm{MeV}$). It is thus questionable to directly apply the
weakly coupled QCD results at LHC/RHIC temperatures. Some insights
for this problem can be gained from studies of $N=4$ supersymmetric
Yang-Mills theory (SYM), where the shear viscosity can be calculated
in both the strong coupling~\cite{Policastro:2001yc,Buchel:2004di}
and weak coupling regimes~\cite{Huot:2006ys}. Without too much
effort, one can parameterize the behavior at intermediate coupling
by interpolating between the two limits~\cite{Huot:2006ys}.

For $N=4$ SYM theory with infinite $N_c$ and large, but finite
t'Hooft coupling $g^2 N_c$ (strongly coupling regime), $\eta/s$ was
calculated by Buchel, Liu and Starinets (using the gauge/gravity
correspondence, see below). They found~\cite{Buchel:2004di}
\begin{eqnarray}
\frac{\eta}{s}=\frac{1}{4\pi} \left[1+ \frac{135 \zeta(3)}{8 (2
g^2N_c)^{3/2}} + \ldots \right].
\end{eqnarray}
The $\eta/s$ of $N=4$ SYM theory in the weakly coupled regime was
investigated by the McGill group using the kinetic theory approach.
They found a much smaller $\eta/s$ than what was previously found in
weakly coupled QCD at the same value of the coupling constant.
However,  good agreement can be achieved after re-scaling each
result by a combination of Debye screening mass and the number of
degrees of freedom in the theory. Extending this mapping between
weakly coupled QCD and weakly coupled N=4 SYM to lower temperature
or into the strongly coupled regime, one finds that  $\alpha_s =0.5$
in QCD corresponds to $\lambda =4.7$ in N=4 SYM, where $\eta/s$ from
N=4 SYM is still relatively large, of the order of
1~\cite{Huot:2006ys}.

The finite temperature version of the gauge/gravity (AdS/CFT)
correspondence~\cite{Witten:1998qj,Witten:1998zw}, which maps finite
temperature gauge field theory at strong coupling onto weakly
coupled gravity in a curved space with a black whole, provides a new
method to study strongly coupled systems and helps us to gain
insights for the properties of a strongly coupled
QGP\footnote{\emph{The AdS/CFT correspondence can not be directly
applied to QCD since QCD is not a Conformal Field Theory, and its
gravity dual (if it exists) is not known.}}. One example is the
shear viscosity in the strong coupling limit. In the language of
black holes gravity, the shear viscosity, defined by the Kubo
formula, is connected with the absorption cross section of the black
hole: $\eta \propto \lim_{\omega \rightarrow 0} \sigma_{BH} =A$,
where the last equality comes from the general theorem on black
holes~\cite{Das:1996we} which states that the cross section
$\sigma_{BH} =A$ is equal to the horizon area $A$ for a broad class
of black holes. On the other hand, the horizon area $A$ also
represents the entropy of the black hole: $s_{BH}=\frac{A}{4 G}$.
This shows that the shear viscosity to entropy ratio $\eta/s$ is a
constant~\cite{Natsuume:2007qq}. The constant was found by Kovtun,
Son, and Starinets~\cite{Kovtun:2004de}, who showed that
\begin{eqnarray}
\frac{\eta}{s}=\frac{1}{4\pi}
\end{eqnarray}
is a universal value for a large class of black holes or branes.
Given that the corresponding dual gauge field theories are very
different, they conjectured that $1/4\pi$ is an absolute low bound
for $\eta/s$, now known as the KSS bound\footnote{\emph{Some
possibilities to violate the KSS bound are discussed
in~\cite{Buchel:2008vz,Brigante:2008gz}.}}.

The bulk viscosity vanishes in conformal field theories such as
$N=4$ SYM. To apply string theory techniques to the calculation of
the bulk viscosity in the strongly coupled regime thus requires
investigating  non-conformal deformations of the original AdS/CFT
correspondence. Some recent developments can be found in
Ref~\cite{Gubser:2008yx,Gursoy:2009kk,Karch:2006pv,Kajantie:2006hv}.
Based on various holographical model computations, Buchel proposed a
lower bound for the bulk viscosity to entropy
ratio~\cite{Buchel:2007mf}:
\begin{eqnarray}
\frac{\zeta}{s}\geq 2 (\frac{1}{3}-c_s^2) \frac{\eta}{s}.
\end{eqnarray}
(Note the different dependence on $\frac{1}{3}-c_s^2$ from the
weakly coupled result~(\ref{bulk-Weak})). Using a scalar potential
tuned to reproduce the lattice equation of state, Gubser et al.
calculated the bulk viscosity via gravity duals of non-conformal
gauge theories~\cite{Gubser:2008yx,Gubser:2008sz}. They found that
$\zeta/s$ rises sharply near $T_c$, with
$\zeta/s|_{T_c} \simeq 0.05 $.\\

\section{Notation and outline of this thesis}

Throughout thesis we adopt units in which $\hbar = c = k_{\rm B} =
1$. The metric tensor is always taken to be $g^{\mu \nu} =
\mbox{diag} (+1,-1,-1,-1)$. $\Delta^{\mu\nu}=g^{\mu\nu}-u^\mu u^\nu$
is the projector onto the space transverse to the fluid velocity
$u_\mu$, $\Delta^{\mu\nu} u_\mu=0$. The partial derivative
$\partial^{\mu}$ can then be decomposed as:
\begin{eqnarray*}
\partial^{\mu} = \nabla^\mu + u^{\mu} D, \qquad \qquad \qquad \qquad
\qquad \qquad
\end{eqnarray*}
where, $\nabla^\mu=\Delta^{\mu \nu} \partial_{\nu}$ is the
transverse component of the partial derivative $\partial_{\mu}$, and
$D=u^{\mu}\partial_{\mu}$ is the corresponding longitudinal
component. In the fluid rest frame, $D$ reduces to the time
derivation and $\nabla^\mu$ reduces to the spacial gradient. One
therefore also uses the denotation $\dot{f}=Df$.

We also frequently use the symmetric, anti-symmetric and the $\La \
\Ra$ brackets in Chap.~2, following
Ref.~\cite{Muronga:2001zk,Muronga:2003ta,
Muronga:2004sf,Israel1976}:
\begin{eqnarray*}
A_{(\mu} B_{\nu)} &=&\frac{1}{2}\left(A_\mu B_\nu+A_\nu
B_\mu\right),
\\
A_{[\mu} B_{\nu]} &=&\frac{1}{2}\left(A_\mu B_\nu-A_\nu
B_\mu\right),
\\
A_{\langle \mu} B_{\nu\rangle} &=&
\frac{1}{2}\left(\Delta^\alpha_\mu \Delta^\beta_\nu +
\Delta^\alpha_\nu \Delta^\beta_\mu-\frac{2}{3} \Delta^{\alpha \beta}
\Delta_{\mu \nu}\right) A_\alpha B_\beta.
\end{eqnarray*}

Using the above notations, the commonly used local fluid rest frame
variables in dissipative viscous hydrodynamics are expressed in
terms of the energy momentum tensor $T^{\mu\nu}$, charge current
$N^{\mu}$ and entropy current $S^{\mu}$ as following (see Chap.~2.1
and Ref.~\cite{Muronga:2001zk,Muronga:2003ta, Muronga:2004sf} for
details):
\begin{eqnarray*}
n & \equiv & \makebox[1.5in][l]{$u_\mu N^\mu$} \mbox{net density of charge};\hspace{3.0cm}\\
V^\mu &\equiv& \makebox[1.5in][l]{$\btu^\mu_\nu N^\nu$} \mbox{net flow of charge};\\
\eps &\equiv& \makebox[1.5in][l]{$u_\mu T^{\mu\nu} u_\nu$} \mbox{energy density};\\
p + \Pi &\equiv& \makebox[1.5in][l]{$\displaystyle -1/3
\btu_{\mu\nu} T^{\mu\nu}$ } \mbox{p: thermal pressure, $\Pi$: bulk
pressure;
}\\
\pi^{\mu\nu} &\equiv& \makebox[1.5in][l]{$T^{\langle\mu\nu\rangle}$
}\mbox{shear stress
tensor};\\
W^\mu &\equiv&  \makebox[1.5in][l]{$u_\nu T^{\nu\lambda}
\btu^\mu_\lambda $}
\mbox{energy flow};\\
q^\mu &\equiv& \makebox[1.5in][l]{$W^\mu - h V^\mu$}\mbox{heat flow};\\
s &\equiv& \makebox[1.5in][l]{$u_\mu\, S^\mu $}\mbox{entropy density};\\
\Phi^\mu &\equiv& \makebox[1.5in][l]{$\btu^\mu_\nu S^\nu $}
\mbox{entropy flux}.
\end{eqnarray*}

In the 1st and 2nd order formalisms for viscous hydrodynamics
(Chap.~2) one also frequently encounters the following scalar and
tensors constructed from the gradients of the four velocity $u^\mu$
:
\begin{eqnarray*}
\theta &\equiv& \makebox[3.in][l]{$\partial \cdot u$}
\mbox{expansion rate},\\
\nabla ^{\La \mu}u ^{\nu \Ra} &\equiv& \makebox[3.in][l]{$\sigma
^{\mu\nu}=\frac{1}{2}(\nabla^\mu u^\nu+\nabla^\nu
u^\mu)-\frac{1}{3}\nabla^{\mu\nu}\partial_\alpha u^\alpha$}
\mbox{velocity stress tensor}, \\
\Omega^{\mu\nu}  &\equiv& \makebox[3.in][l]{$- \nabla^{[\mu}
u^{\nu]}$} \mbox{vorticity tensor}.
\end{eqnarray*}

Note that the above notations are for Cartesian coordinates $x^\mu
\eq(t,x,y,z)$. In Chap.~3, we change to curvilinear coordinates
$x^m\eq(\tau,x,y,\eta)$ for the convenience of describing a
longitudinally boost invariant system (i.e. viscous hydrodynamics in
2+1 dimensions). To express the above variables in $(\tau,x,y,\eta)$
coordinates, one can notationally  replace $\mu, \nu$ by $m, n$, but
must also replace the partial derivative $\partial_\mu$ by the
covariant derivative $d_m$, see Chap.~3.1 and Appendix A.~1 for detail.\\[0.15in]

\textbf{\underline{Outlines:}} \\[-0.10in]

This thesis focuses on viscous hydrodynamics for relativistic heavy
ion collisions. Chapter~2 outlines the 2nd order formalism for
viscous hydrodynamics, obtained from different approaches. Chapter~3
sets us up for numerical calculations in (2+1)-dimension, assuming
exact longitudinal boost invariance. The discussion there includes
explicit transport equations in (2+1)-dimension, the initial
conditions, the EOS, final conditions and the transport coefficients
used in viscous hydrodynamic simulations. Generic shear and bulk
viscosity effects are studied in Chapter~4 and Chapter~6,
respectively, by comparing runs with ideal and viscous hydrodynamics
using identical initial and final conditions.  Chapter~5 focuses on
the system size dependence of the shear viscous effects and
investigates the multiplicity scaling of the elliptic flow
coefficient $v_2$ for both ideal and viscous hydrodynamics. In
Chapter~7, we report on some of the most recent developments in
viscous hydrodynamics, including the search for the optimized I-S
equations for numerical implementations and recent efforts on code
verifications among different groups. Chapter~8 assesses the current
uncertainties  for extracting the QGP viscosity from experimental
data and briefly comments on some future directions for viscous
hydrodynamics. Chapter~9 briefly outlines other methods for
estimating the QGP shear viscosity. Our conclusions are summarized
in Chapter~10.
\chapter[Relativistic Viscous Hydrodynamics: the Formalism]{Relativistic Viscous Hydrodynamics\\
\hspace{2cm} -- the Formalism} In this chapter we introduce the
formalism for viscous (dissipative) hydrodynamics. Initial attempts
to formulate relativistic dissipative fluid dynamics started as a
relativistic generalization of the Navier-Stokes (N-S)
equations~\cite{Eckart,Landau}. Unfortunately, it turned out that
the relativistic N-S equations are unsuited for numerical
implementation since they developed instabilities due to
exponentially growing modes, in particular at high frequencies, and
violate causality (the N-S formalism will be briefly described in
Chap.~2.2.). These difficulties are largely avoided in the 2nd order
formalism developed 30 years ago by Israel and
Stewart~\cite{Israel:1976tn, Israel1976, Israel:1979wp},
 which goes beyond the so-called 1st order Navier-Stokes approach
 by expanding the entropy current to 2nd order in
dissipative flows, replacing the instantaneous identification of the
dissipative flows with their driving forces multiplied by some
transport coefficient (as is done in Navier-Stokes theory) by a
kinetic equation that evolves the dissipative flows rapidly but
smoothly towards their Navier-Stokes limit. The deduction of the I-S
equation from the entropy current expansion and the 2nd law of
thermodynamics will be introduced in Chap.~2.3. The I-S formalism
can also be derived from the Boltzmann
equation~\cite{Baier:2006um,Betz:2008me,York:2008rr}, by expanding
the distribution function around local equilibrium. The most general
form of 2nd order viscous hydrodynamics was derived
in~\cite{Baier:2007ix,York:2008rr} for systems with conformal
symmetry and in~\cite{Betz:2008me} for systems without such symmetry
(i.e. which also feature bulk viscosity and heat conductivity). Some
of this will be discussed in Chap.~2.5.  All  versions of the I-S
formalism directly solve evolution equations for the dissipative
flows. In contrast, another type of 2nd-order formalism, developed
by the Ottinger and Grmela (O-G formalism)~\cite{OG1,OG2,OG3}, uses
evolution equations for auxiliary fields, rather than the
dissipative flows (see Chap.~2.6). The standard dissipative flows
can be approximately reconstructed from the auxiliary fields~\cite{PRV-Dusling}. \\

\section{From ideal to viscous hydrodynamics}
%

Hydrodynamics is a macroscopic tool to describe the expansion of the
QGP and hadronic matter. It starts from the conservation laws for
the conserved charge currents $N^{\mu}_i (x)$ and the energy
momentum tensor $T^{\mu \nu}(x)$ ($x$ denotes the 4-dimensional
space-time coordinates $(t, \mathbf{x})$)~\cite{Rischke:1998fq}:
\begin{subequations}
\label{eq-conserv}
\begin{eqnarray}
&&\partial_\mu N^{\mu}_i (x)=0\, ,\quad i=1,...,k; \hspace{3cm}
\label{eq-conserv-a}\\
&&\partial_\mu T^{\mu \nu}(x)=0\, . \label{eq-conserv-b}
\end{eqnarray}
\end{subequations}

For simplicity, one sets $k=1$ and only considers the conserved net
baryon number current. This leaves $14$ independent variables: $10$
from the symmetric energy momentum tensor  $T^{\mu \nu}$ and $4$
from the charge current $N^{\mu}$. However, this system  of
equations is unclosed since (\ref{eq-conserv}) only offers $5$
independent equations. To solve this problem, one needs either to
reduce the number of independent variables through physically
assumptions, or to provide more equations.

Ideal hydrodynamics~\cite{Kolb:2003dz,Huovinen:2003fa} solves this
problem via the first route. By assuming perfect local thermal
equilibrium, one can decompose the charge current and energy
momentum tensor as follows: 
\begin{subequations}
\label{ideal-decomp}
\begin{eqnarray}
 &&N^{\mu}=nu^{\mu}\, ,
 \label{ideal-decomp-a}\\
 &&T^{\mu \nu}=e u^{\mu}u^{\nu}-p\Delta^{\mu \nu}, \quad (\Delta^{\mu \nu}=g^{\mu
 \nu}-u^{\mu}u^{\nu})\, ,
 \label{ideal-decomp-b}
\end{eqnarray}
\end{subequations} 
where $n(x)$, $e(x)$, $p(x)$ are the local net baryon density,
energy density and pressure, respectively, and $u^{\mu}(x)$ is the
4-flow velocity which is time like and normalized:
$u^{\mu}u_{\mu}=1$. The above ideal fluid decomposition reduces the
14 independent variables into  6 (1 each for $e$, $n$ and $p$ and 3
independent components in $u^\mu$). With one additional input -- the
equation of state (EOS) $p=p(n,e)$ -- the system of equations is
closed and can be solved to simulate the evolution of the system.
Using the fundamental thermodynamic identity $e+p = Ts+\mu n$, it is
easy to show that, in the absence of shocks (i.e. discontinuity in
$e$, $n$ or $p$), Eqs.~(\ref{eq-conserv}) conserve entropy,
$\partial_\mu S^\mu =0$, with $S^\mu = s u^\mu$.

In classical kinetic theory, $N^{\mu}$ and  $T^{\mu \nu}$ are
associated with the microscopic phase-space distribution function
$f(x,p)$ as follows~\cite{Rischke:1998fq,Kinetic-Theory80}:
\begin{subequations}
\label{Kinetic-def}
\begin{eqnarray}
&&N^{\mu} = \sum_i n_i\int \frac{d^3p}{E}p^{\mu}f_i(x,p)\, , \hspace{4cm}\\
&&T^{\mu \nu}=\sum_i \int \frac{d^3p}{E}p^{\mu}p^{\nu}f_i(x,p)\, .
\end{eqnarray}
\end{subequations}
Here $n_i$ are the baryon charges per particle of species $i$
(particles and anti-particles count as sperate species). Plugging in
the equilibrium distribution function $f_{i,\ eq}(x,p)=
\frac{1}{e^{[p\cdot u(x)+\mu_i (x)]/T(x)}\pm 1}$ (where $\mu_i (x)$
and $T(x)$ are the local chemical potential of particle species $i$
and the temperature, respectively) into eq.(\ref{Kinetic-def}), one
directly obtains the ideal fluid decomposition (\ref{ideal-decomp}).
Local thermal equilibrium is thus the basic assumption behind ideal
hydrodynamics. In kinetic theory, it requires that the microscopic
mean free path is much smaller than the system size and the
microscopic collision time scale is much shorter than the
macroscopic evolution time scale. The concept of local thermal
equilibrium is, however, more general and can be formulated without
recourse to kinetic theory. Ideal fluid dynamics always applies if
local thermal equilibrium is ensured.

If microscopic processes are not fast enough to satisfy the above
conditions, the system is no longer in local equilibrium. For a
near-equilibrium system, the distribution function can be decomposed
into an equilibrium part plus a small deviation:
\begin{eqnarray}
f(x,p)=f_{eq}(x,p)+\delta f(x,p)\, ,\qquad \qquad (\delta f \ll f)
\hspace*{-0.5cm} \label{nonequil-distrib}
\end{eqnarray}
Putting eq.~(\ref{nonequil-distrib}) into eq.~(\ref{Kinetic-def})
one finds
\begin{subequations}
\label{NTmunu}
\begin{eqnarray}
&&N^{\mu}=N^{\mu}_{eq} +\delta N^{\mu}, \hspace*{5.3cm} \\
&&T^{\mu \nu}= T^{\mu \nu}_{eq} + \delta T^{\mu \nu},
\end{eqnarray}
\end{subequations}
where the dissipative flows $\delta N^{\mu}$ and $\delta T^{\mu
\nu}$ are generated by the non-equilibrium contribution $\delta f$.
By imposing the ``Landau matching conditions"~\cite{Landau} for an
arbitrary frame $u^\mu$: $u_\mu \delta N^{\mu}=0 $ and $u_\mu \delta
T^{\mu \nu} u_\nu =0 $, one associates  $N^{\mu}_{eq}$ and $ T^{\mu
\nu}_{eq}$ with the equilibrium definitions of net baryon density
$n$ and  energy density $e$ and pressure $p$ ($n= N^{\mu} u_{\mu}$,
$e= u_{\nu} T^{\mu \nu} u_{\mu}$). In other words, these matching
conditions fix the temperature and chemical potential of the
equilibrium distribution $f_{eq}$ in (\ref{nonequil-distrib}) such
that $e$ and $n$ are defined in terms of $f_{eq}$ in the standard
way. Then $N^{\mu}$, $T^{\mu \nu}$ can be fully decomposed
as~\cite{Rischke:1998fq}:
\begin{subequations}
\label{tensor-decomp-1}
\begin{eqnarray}
 &&N^{\mu}= nu^{\mu}+V^{\mu}, \\
 &&T^{\mu \nu}= eu^{\mu}u^{\nu}-
p\Delta^{\mu \nu}-\Pi\Delta^{\mu \nu}+\pi^{\mu
\nu}+W^{\mu}u^{\nu}+W^{\nu}u^{\mu},
\end{eqnarray}
\end{subequations}
where $V^{\mu}$, $\Pi$, $\pi^{\mu \nu}$ and $W^{\mu}$ are called
dissipative or viscous flows. More specifically, $V^\mu=\Delta^{\mu
\nu} N_\nu$ describes a baryon flow  in the local rest frame, and
$W^\mu=\Delta^{\mu \nu}T_{\nu\alpha}u^{\alpha}\equiv\frac{e+p}{n}
V^\mu +q^\mu$ (where $q^\mu$ is the "heat flow vector") describes an
energy flow in the local rest frame. $W^{\mu}$,  $V^{\mu}$ and
$q^{\mu}$ are all transverse to the frame $u^{\mu}$:
$W^{\mu}u_{\mu}=0$, $V^{\mu}u_{\mu}=0$ and $q^{\mu}u_{\mu}=0$, so
each of them has 3 independent components. Since $q^{\mu}$ is
defined through $V^{\mu}$ and $W^{\mu}$, the 3 vectors leave 6
independent components altogether. $\Pi=-\frac{1}{3}\Delta_{\mu \nu}
T^{\mu \nu}-p$ is the viscous bulk pressure, which contributes to
the trace of energy momentum tensor.
 $\pi^{\mu \nu}=T^{\La \mu \nu \Ra}\equiv
[\frac{1}{2}(\Delta^{\mu \sigma} \Delta^{\nu \tau}+\Delta^{\nu
\sigma} \Delta^{\mu \tau})-\frac{1}{3}\Delta^{\mu \nu}
\Delta^{\sigma \tau}]T_{\tau\sigma}$ (where the expression $\La \mu
\nu \Ra$ is the shorthand for traceless and transverse to $u_\mu$
and $u_\nu$ as defined by the projector in square brackets) is
called the shear stress tensor. $\pi^{\mu\nu}$ is traceless and
transverse to $u^{\nu}$, $\pi^{\mu}_{\mu}=0$ and $\pi^{\mu\nu}
u_{\nu}=0$, and it is symmetric (like $T^{\mu\nu}$); it thus has 5
independent components.

The Landau matching conditions, however, leave the choice of frame
$u^\mu$ unconstrained. Two commonly used frames are the Eckart
frame~\cite{Eckart} and the Landau frame~\cite{Landau}. The Eckart
frame sets $u^\mu$ parallel to the charge flow $N^{\mu}$. As a
consequence, $V^{\mu}$ disappears and the energy flow vector reduces
to the heat flow vector $W^\mu=q^\mu$. However, for a system with
very small or vanishing net baryon number, which is approximately
realized in nuclear collisions at top RHIC energy, this frame is ill
defined~\cite{Gyulassy85}. Therefore, this thesis and other
theoretical viscous hydrodynamics frameworks aimed at RHIC
physics~\cite{Muronga:2001zk,Muronga:2003ta,Muronga:2004sf,Heinz:2005bw,
Muronga:2006zw,Muronga:2006zx} adopt the second choice -- the Landau
frame. In the Landau frame, $u^\mu$ is taken parallel to the
4-velocity of energy flow ($\propto T^{\mu \nu}u_\nu$), which leads
to a zero value for $W^\mu $. Then the tensor decomposition shown in
eq.~(\ref{tensor-decomp-1}) simplifies as follows:
\begin{subequations}
\label{tensor-decomp}
\begin{eqnarray}
 &&N^{\mu}= nu^{\mu}-\frac{n}{e+p} q^{\mu}, \\
 &&T^{\mu \nu}= eu^{\mu}u^{\nu}-
(p+\Pi)\Delta^{\mu \nu}+\pi^{\mu \nu}.
\end{eqnarray}
\end{subequations}
For the full evolution of all components of $T^{\mu\nu}$ and
$N^\mu$, we need in addition to the 5 conservation law equations
(\ref{eq-conserv}) additional equations for $q^\mu$,
$\Pi$ and $\pi^{\mu\nu}$.\\

%
\section{Navier-Stokes formalism}
%
The Navier-Stokes (N-S) formalism comes from the relativistic
generalization of the non-relativistic N-S equations, which impose
linear relationships between the dissipative flows and the
corresponding thermodynamic forces~\cite{Eckart,Landau}:
\begin{subequations}
\label{N-S}
\begin{eqnarray}
&&q^{\nu}=-\lambda \frac{n T^2}{e+p}\nabla ^{\nu} \left (\frac{\nu}{T}\right)\equiv \lambda X^{\nu}, \qquad\\
&&\Pi=-\zeta \theta \equiv  \zeta X, \\
&&\pi^{\mu \nu}=2\eta \nabla ^{\La \mu}u ^{\nu \Ra} \equiv 2\eta
X^{\mu\nu}.
\end{eqnarray}
\end{subequations}
The (positive) ``transport" coefficients $\zeta$, $\lambda$ and
$\eta$ are the  bulk viscosity, heat conductivity and shear
viscosity, respectively.  The scalar, vector and tensor
thermodynamical forces on the right hand side are explicitly given
by $X\equiv-\theta=-\partial \cdot u$, $X^{\nu}\equiv\frac{\nabla
^{\nu}}{T}-\dot{u}^{\nu}=-\frac{nT}{e+p}\nabla ^{\nu}\frac{\mu}{T}$
and $X^{\mu\nu}\equiv\nabla ^{\La \mu}u ^{\nu\Ra}=\sigma
^{\mu\nu}=\frac{1}{2}(\nabla^\mu u^\nu+\nabla^\nu
u^\mu)-\frac{1}{3}\Delta^{\mu\nu}\partial_\alpha u^\alpha$.

Unfortunately the N-S equations violate
causality~\cite{Hiscock:1983zz,Hiscock:1985zz,Hiscock:1987zz} and
lead to infinite speed of signal propagation: if the thermodynamic
forces turn off suddenly, the dissipative flows will also disappear
instantaneously. This conflicts with the fact that a macroscopic
process caused by microscopic scattering is delayed by a relaxation
time  comparable to the kinetic scattering time scale. Connected
with the acausality problem in the N-S formalism are numerical
instabilities~\cite{Hiscock:1983zz,Hiscock:1985zz,Hiscock:1987zz},
which render it useless for numerical simulations. Both of these two
problems are avoided in the 2nd order formalism which
is introduced in the following sections.\\

%
\section{Israel-Stewart formalism from the 2nd law of
thermodynamics}
%
The 2nd order formalism for relativistic dissipative fluids was
first obtained by Israel and Stewart in the late 1970's,
generalizing earlier non-relativistic work by I.
$\mathrm{M\ddot{u}ller}$~\cite{Muller1967}. They gave two
derivations, a macroscopic one based on the 2nd law of
thermodynamics, derived by expanding the entropy current up to the
2nd order in deviations from local equilibrium, and a microscopic
one based on a near-equilibrium expansion of the Boltzmannn
equation. Detailed presentations can be found in the original
articles of Israel and Stewart~\cite{Israel:1976tn,Israel:1979wp}
and in recent articles by
Muronga~\cite{Muronga:2001zk,Muronga:2003ta,Muronga:2004sf,Muronga:2006zw,Muronga:2006zx},
who first brought this work to the attention of the RHIC community.

In a relativistic equilibrium system, the entropy current is written
as:
\begin{eqnarray}
S_{eq}^{\mu}=p(\alpha, \beta)\beta ^{\mu}- \alpha N_{eq}^{\mu}+\beta
_{\nu}T_{eq}^{\mu \nu}, \label{Entropy-equil}
\end{eqnarray}
where $\alpha$ and $\beta$ are related to the local temperature $T$
and chemical potential $\mu$ by $\alpha =\frac{\mu}{T}$, $\beta
=\frac{1}{T}$. For a near-equilibrium system, one can generalize the
entropy current by the following off-equilibrium expansion:
\begin{eqnarray}
S^{\mu}=p(\alpha, \beta)\beta ^{\mu}-  \alpha (N_{eq}^{\mu}+\delta
N^{\mu})+\beta _{\nu}(T_{eq}^{\mu \nu}+ \delta T^{\mu
\nu})+Q^{\mu}(\delta N^{\mu},\delta T^{\mu \nu}),
\end{eqnarray}
where, in addition to the first order corrections $\delta N^{\mu}$
and $\delta T^{\mu\nu}$ appearing in the middle terms, the last term
$Q^{\mu}$ includes second and higher order corrections in terms of
$\delta N^{\mu}$ and $\delta T^{\mu \nu}$. After some algebra, the
entropy production rate can be written as:
\begin{eqnarray}
\partial_{\mu} S^{\mu}=\delta N^{\mu} \partial_{\mu} \alpha + \delta T^{\mu
\nu} \partial_{\mu} \beta_{\nu} +\partial_{\mu} Q^{\mu}.
\end{eqnarray}
After rewriting $\delta N^{\mu}$ and $\delta T^{\mu \nu}$ in terms
of  the corresponding dissipative flows $\Pi$, $q^{\mu}$ and
$\pi^{\mu\nu}$, expressing $\partial_{\mu} \alpha$ and
$\partial_{\mu} \beta_{\nu}$ through the thermodynamics forces
$X^\mu$ and $X^{\mu\nu}$ defined in eq.~(\ref{N-S}), this can be
further recast into:
\begin{eqnarray}
T\partial_{\mu} S^{\mu}=\Pi X-q^{\mu}X_{\mu} + \pi^{\mu\nu}
X_{\mu\nu}+ T \partial_{\mu} Q^{\mu}\geq 0.\label{2nd-law-2}
\end{eqnarray}
The inequality on the right implements the 2nd law of thermodynamics
$\partial \cdot S \geq 0$. Note that the first three terms on the
r.h.s. are first order, while the last term is higher order in the
dissipative flows.

The first order theory~\cite{Eckart,Landau} neglects the second
order and higher order contributions and sets $Q^{\mu}=0$. By
postulating linear relationships between the dissipative flows and
the thermodynamic forces with non-negative coefficients described by
the N-S equations (\ref{N-S}), the inequality $\partial \cdot S \geq
0$ is automatically satisfied:
\begin{eqnarray}
T\partial_\mu S^\mu=\frac{\Pi^2}{\zeta}-\frac{q_\mu q^\mu}{2\lambda
T}+\frac{\pi_{\mu\nu}\pi^{\mu\nu}}{2 \eta} \geq 0. \label{2nd-Law}
\end{eqnarray}
Note that, since $u^\mu q_\mu=0$, $q_\mu$ is space-like, $q^\mu
q_\mu<0$.   The N-S equations thus corresponds to a  1st order
expansion of the entropy current, which is why the Navier-Stokes
formalism is known as 1st order viscous hydrodynamics.

The 2nd order theory of Israel and
Stewart~\cite{Israel:1976tn,Israel:1979wp,Muronga:2001zk,Muronga:2003ta}
keeps $Q^\mu$, more precisely, it keeps all terms that are second
order in the dissipative flows:
\begin{eqnarray}
Q^{\mu}=-(\beta_0 \Pi^2-\beta_1 q^{\mu}q_{\mu}+\beta_2 \pi_{\mu\nu}
\pi^{\mu\nu})\frac{u^{\mu}}{2T}-\frac{\alpha_0 \Pi q^{\mu}}{T}+
\frac{\alpha_1 \pi^{\mu\nu}  q_{\nu}}{T},
\end{eqnarray}
where $\beta_0$, $\beta_1$, $\beta_2$ and $\alpha_0$, $\alpha_1$ are
phenomenological expansion coefficients. After a bit of algebra, the
entropy production rate $T\partial\cdot S$ is now written
as~\cite{Muronga:2003ta,Muronga:2004sf}
\begin{eqnarray}
T\partial_\mu S^\mu &=& -\Pi\left[\theta+\beta_0\dot{\Pi} +
{1\over2} T\partial_\mu \left({\beta_0\over T}u^\mu\right)\Pi
-\alpha_0\nabla_\mu q^\mu\right]\nonumber\\
{}&& -q^\mu\left[\nabla_\mu\ln T-\dot{u}_\mu-\beta_1\dot{q}_\mu -
{1\over2} T\partial_\nu \left({\beta_1\over T}u^\nu\right)q_\mu
-\alpha_0\nabla_\nu\pi^\nu_\mu -\alpha_1\nabla_\mu \Pi\right]\nonumber\\
{}&&+\pi^{\mu\nu}\left[\sigma_{\mu\nu}-\beta_2\dot{\pi}_{\mu\nu}+
{1\over2}T\partial_\lambda \left({\beta_2\over T}u^\lambda \right)
\pi_{\mu\nu} +\alpha_1 \nabla_{\La\nu}q_{\mu\Ra}\right] .
\end{eqnarray}
Again, one can ensure the second law of thermodynamics by writing
$T\partial_\mu S^\mu$ in the form of eq.~(\ref{2nd-Law}). This leads
to the following 2nd order viscous equations for the dissipative
flows:
\begin{subequations}
\label{dissi-trans}
\begin{eqnarray}
&&\dot{\Pi}=-\frac{1}{\tau_{\Pi}}\bigg[\Pi+\zeta \theta -l_{\Pi q}
\nabla_{\mu} q^{\mu}+\Pi \zeta T
\partial_{\mu} \big( \frac{\tau_\Pi u^{\mu}}{2\zeta T} \big) \bigg],
\label{dissi-trans-a}
\\
&&\Delta^{\mu}_{\nu}\dot{q}^{\nu}=-\frac{1}{\tau_{q}}\bigg[q_{\mu}+\lambda
\frac{n T^2}{e+p}\nabla ^{\mu}\frac{\nu}{T} + l_{q \pi}\nabla_\nu
\pi^{\mu\nu}+ l_{q \Pi}\nabla^\mu \Pi - \lambda T^2 q^{\mu}
\partial_{\mu}\big( \frac{\tau_q u^{\mu}}{2\lambda T^2}
\big)\bigg],\qquad\quad
\label{dissi-trans-b}
\\
&&\Delta ^{\mu \alpha} \Delta ^{\nu \beta}\dot{ \pi}_{\alpha \beta}
=-\frac{1}{\tau_{\pi}}\bigg[\pi^{\mu\nu}-2\eta \nabla ^{\La \mu}u
^{\nu\Ra}-l_{\pi q} \nabla^{\La\mu} q^{\nu\Ra}+ \pi_{\mu\nu} \eta T
\partial_{\alpha} \big( \frac{\tau_\pi u^{\alpha}}{2 \eta T} \big)
\bigg], \label{dissi-trans-c}
\end{eqnarray}
\end{subequations}
where $\beta_0$, $\beta_1$ and $\beta_2$ have been replaced by the
relaxation times $\tau_{\Pi}\equiv  \zeta \beta_0$, $\tau_{q}\equiv
\lambda T \beta_1$ and $\tau_{\pi}\equiv 2 \eta \beta_2$,
respectively, and the mixing coefficients $\alpha_0$, $\alpha_1$
have also been rewritten into $l_{\Pi q}=\zeta \alpha_0$, $l_{q \Pi
}=\lambda T \alpha_0$, $l_{q \pi }=\lambda T \alpha_1$ and $l_{\pi
q}=2\eta \alpha_1$. The relaxation times replace the instantaneous
identification between dissipative flows and thermodynamics forces
in the N-S equations. They ensure the causality of the theory and
the numerical stability~\cite{Hiscock:1983zz,Romatschke:2009im} as
long as they are not too small~\cite{Romatschke:2009im}. The viscous
hydrodynamics based on the I-S formalism is thus also known as
\emph{causal viscous hydrodynamics}\footnote{\emph{The projectors
$\Delta^{\mu\nu}$ on the r.h.s. of (\ref{dissi-trans} b, c) ensure
that the transversality to the flow $u^{\mu}$ of $q_{\mu}$ and
$\pi_{\mu\nu}$ is preserved during the time evolution.}}.

In this phenomenological approach, both the viscous coefficients and
the corresponding relaxation times are arbitrary free parameters,
which must be determined from  other theoretical consideration or
extracted from experimental data.\\

\section{Israel-Stewart formalism from kinetic theory}
The kinetic theory approach starts from a Taylor expansion of the
distribution function $f$ around its local equilibrium form:
\begin{eqnarray}
&& \label{f-dis}
f=f_0 \left[1+ (1\pm f_0)\delta f\right],  \qquad \delta f \ll 1,\\
\mathrm{with,}\hspace{1.3cm} &&\delta f({\bf x},t,{\bf
p})=\epsilon({\bf x},t) + \epsilon_{\lambda}({\bf x},t) p^\lambda+
\epsilon_{\lambda \nu}({\bf x},t) p^\lambda p^\nu +O(p^3).\qquad
\qquad \qquad  \nonumber
\end{eqnarray}
Note that the scalar, vector and tensor coefficients $\epsilon$,
$\epsilon_{\lambda}$, $\epsilon_{\lambda \nu}$ are functions of
space-time, which, through kinetic theory, can be related to the
dissipative flows. After putting eq.~(\ref{f-dis}) into the kinetic
definition (\ref{Kinetic-def}) of $T^{\mu\nu}$ and integrating out
the momentum degrees of freedoms, one finds that specific
combinations of $\epsilon$, $\epsilon_{\lambda}$, $\epsilon_{\lambda
\nu}$ correspond to bulk pressure $\Pi$, heat flow $q_\lambda$ and
shear stress tensor $\pi_{\lambda \nu}$. For simplicity, we consider
a fluid with only shear viscosity, neglecting bulk viscosity and
heat conductivity, which simplifies the expansion of $\delta f$ as
\begin{eqnarray}
\delta f({\bf x},t,{\bf p})= \epsilon_{\lambda \nu}({\bf x},t)
p^\lambda p^\nu,
\end{eqnarray}
with $ \epsilon_{\lambda \nu}$ being traceless. Then, the shear
stress tensor $\pi^{\mu\nu}$ is directly related to
$\epsilon^{\mu\nu}$ after integrating out the momentum degrees of
freedom in eq.~(\ref{Kinetic-def}). For a massless Boltzmann gas
$f_0=\exp{(- p_\mu u^\mu/T)}$, one finds~\cite{Baier:2006um}
\begin{eqnarray}
\epsilon^{\mu\nu}=\frac{1}{2T^2(e+p)} \pi^{\mu\nu},
\end{eqnarray}
with small quantum statistical corrections for fermions or
bosons~\cite{Dusling:2007gi}.

 In the kinetic theory approach, the macroscopic
conservation laws $\partial_\mu N^\mu=0$ and $\partial_\mu
T^{\mu\nu}=0$ and the 2nd order I-S equations can be obtained from
the Boltzmann equation $p^\mu d_{\mu} f({\bf x},t,{\bf p}) =
{\mathcal C}(x)$ after integrating out the momentum degrees of
freedom, $\int d\omega \equiv \int \frac{d^3 p}{(2\pi)^3\, p_0}$,
with different combinations of momentum $p^\mu$ as weighting:
\begin{eqnarray}
&&\int\ d\omega p^\mu \partial_\mu f = \int d\omega\ {\mathcal C}\,
,
    \label{Boltz1}\\
&&\int d\omega\ p^\mu p^\alpha \partial_\mu f = \int d\omega\
p^\alpha {\mathcal C}\, ,
    \label{Boltz2}\\
&&\int d\omega\ p^\mu p^{\alpha} p^{\beta} \partial_\mu f = \int
d\omega\ p^{\alpha} p^{\beta} {\mathcal C}\, .
    \label{Boltz3}
\end{eqnarray}
Here ${\mathcal C}(x)$ is the collision term whose precise form is
related to the interaction matrix elements between particles. For a
theory with conserved charges and with interactions that conserve
energy and momentum: $\int d\omega\ {\mathcal C}=0$ and $\int
d\omega\ p^\alpha {\mathcal C}=0$. Eqs.~(\ref{Boltz1}, \ref{Boltz2})
then lead to $\partial_\mu N^\mu=0$ and $\partial_\mu T^{\mu\nu}=0$,
respectively. After employing the relaxation time approximation for
the collision term:
\begin{eqnarray}
{\mathcal C}=-p_\mu u^\mu \frac{f-f_{0}}{\tau_{\pi}}\, .
\end{eqnarray}
and integrating out $\int d\omega$ in eq.~(\ref{Boltz3}) one finds
the 2nd order I-S equations~\cite{Baier:2006um,Baier:2007ix}:
\begin{eqnarray}
\pi^{\mu \nu}+\tau_\pi \left[\Delta^\mu_\alpha \Delta^\nu_\beta D
\pi^{\alpha \beta}+ \frac{4}{3}\pi^{\mu\nu}\nabla_\alpha u^\alpha-2
\pi^{\phi (\mu}\Omega^{\nu)}_{\ \phi} +\frac{\pi^{\phi \La\mu}
\pi^{\nu\Ra}_\phi}{2 \eta}\right] =\eta \nabla^{\La\mu} u^{\nu\Ra}\,
, \label{I-S-Kin}
\end{eqnarray}
where $\Omega_{\mu\nu}$ is the vorticity tensor,
$\Omega_{\mu\nu}=\Delta^{\alpha}_\mu \Delta^{\beta}_\nu
\partial_{[\beta} u_{\alpha]}$. In contrast to the macroscopic
 approach of Chap.~2.3, the relaxation time $\tau_\pi$ and
shear viscosity $\eta$ here are no longer independent from each
other. One finds $\tau_\pi=\frac{6\eta}{sT}$ for a massless
Boltzmann gas,
$\tau_\pi=\frac{6\eta}{4p}=\frac{(3+\frac{T}{s}\frac{dS}{dT})\eta}{sT}$
for a massive Boltzmann gas, and $\tau_\pi \simeq
1.024\frac{6\eta}{sT}$ for a massless Bose-Einstein
gas~\cite{Baier:2006um}. Actually, the shear viscosity $\eta$ can
also be self-consistently calculated in the kinetic theory approach,
after explicitly writing out the collision term. For purely gluonic
matter, where interactions are dominated by $2 \rightarrow 2$
scattering processes with HTL cross sections, one finds: $\eta =
\frac{960}{\pi^7} \zeta^2(5) T^3 \frac{(4\pi)^2}{g^4 \ln(4
\pi/g^2)}$~\cite{Baier:2006um}.

Based on consistent expansions to the 2nd order in Knudsen
number\footnote{\emph{Knudsen number is defined as microscopic mean
free path over macroscopic scale of hydrodynamics, $K\equiv
l_{mfp}/L_{hydro}$.}}, Betz, Henkel and Rischke~\cite{Betz:2008me}
deduced the complete set of I-S equations from the Boltzmann
equation for the bulk pressure $\Pi$, shear stress tensor
$\pi^{\mu\nu}$ and heat flow vector $q^{\mu}$  via Grad's
14-Momentum method~\cite{Grad14}:
\begin{subequations}
\label{dissi-trans2}
\begin{eqnarray}
\Pi & = & \Pi_{\rm NS} - \tau_\Pi\, \dot{\Pi} \nonumber \\
& + &  \tau_{\Pi q}\, q \cdot \dot{u}
 - \ell_{\Pi q}\, \partial \cdot q
 - \zeta \, \hat{\delta}_0\, \Pi\, \theta \nonumber \\
& + &  \lambda_{\Pi q} \, q \cdot \nabla \alpha
 +  \lambda_{\Pi \pi} \, \pi^{\mu \nu} \sigma_{\mu \nu}\;,
\label{eq:PiIS} \\[0.10in]
q^\mu  & = & q^\mu_{\rm NS}  - \tau_q \, \Delta^{\mu \nu}\dot{q}_\nu
\nonumber \\
& - &  \tau_{q \Pi}\, \Pi\, \dot{u}^\mu - \tau_{q \pi}\, \pi^{\mu
\nu}\, \dot{u}_\nu \nonumber
 +    \ell_{q \Pi}\, \nabla^{\mu} \Pi  -
\ell_{q \pi}\, \Delta^{\mu \nu}\, \partial^\lambda \pi_{\nu \lambda}
+  \tau_q \, \Omega^{\mu \nu} \, q_\nu
 - \frac{\kappa}{\beta}\, \hat{\delta}_1\, q^\mu \, \theta \nonumber \\
& - &  \lambda_{qq}\, \sigma^{\mu \nu}\, q_\nu +  \lambda_{q \Pi}\,
\Pi \, \nabla^\mu \alpha +  \lambda_{q \pi}\, \pi^{\mu \nu}\,
\nabla_\nu \alpha\; ,
\label{eq:qIS}\\[0.10in]
\pi^{\mu \nu} & = & \pi^{\mu \nu}_{\rm NS}  -
\tau_\pi\, \dot{\pi}^{\La \mu \nu\Ra}  \nonumber \\
& + &   2\, \tau_{\pi q}\, q^{\La\mu} \dot{u}^{\nu\Ra}
 +  2 \, \ell_{\pi q}\, \nabla^{\La\mu} q^{\nu\Ra}
 +  2\, \tau_\pi\, \pi_\lambda^{\hspace*{0.1cm}\La\mu}
\Omega^{\nu\Ra \lambda}
 -  2\, \eta\, \hat{\delta}_2\, \pi^{\mu \nu}\, \theta \nonumber \\
& - &    2\, \tau_\pi \, \pi_\lambda^{\hspace*{0.1cm}\La\mu}
\sigma^{\nu\Ra \lambda}-  2\, \lambda_{\pi q}\, q^{\La\mu}
\nabla^{\nu\Ra} \alpha + 2\, \lambda_{\pi \Pi}\, \Pi\, \sigma^{\mu
\nu}\;.
\end{eqnarray}
\end{subequations}
They found that the form of the I-S equations is independent of the
choice of frame (Eckart frame vs. Landau frame), whereas the
coefficients $\tau_\Pi,\, \tau_q,\, \tau_\pi$,  $\tau_{\Pi q},\,
\tau_{q \Pi} , \, \tau_{q \pi}$ etc. are frame dependent and are
complicated functions of temperature and chemical potential.  The
first terms on r.h.s of each equation, corresponding to the
Navier-Stokes first order approximation in Chap.2.2, are of first
order in the Knudsen number. All other term on r.h.s are of second
order in the Knudsen number. Compared with eq. (\ref{dissi-trans})
from the macroscopic approach, the Knudsen number expansion yields
more terms for the 2nd order viscous equations. Some of these terms
do not contribute to entropy production and thus can not manifest
themselves in the 2nd law of thermodynamics; some others are
suspected to correspond to higher order corrections in the entropy
expansion approach~\cite{Betz:2008me}.\\

\section{Israel-Stewart formalism from conformal symmetry constraints}
For conformally invariant theories, the action $S[\phi,g^{\mu\nu}]$,
as a functional of the external metric $g^{\mu\nu}$, is invariant
under Weyl transformations: $g_{\mu\nu} \rightarrow e^{-2 \omega }
g_{\mu\nu}$. As a consequence, the energy momentum tensor
$T^{\mu\nu}$ is traceless and transforms as $ T^{\mu\nu} \rightarrow
e^{6 \omega } T^{\mu\nu}$ under  Weyl rescaling in 4 dimensions.
Correspondingly, the shear stress tensor also transforms as
\begin{eqnarray}
\pi^{\mu\nu} \rightarrow e^{6 \omega }\pi^{\mu\nu}. \label{Weyl}
\end{eqnarray}

The above Weyl rescaling for the shear stress tensor $\pi^{\mu\nu}$
gives a constraint for the possible form of the viscous equations
for  $\pi^{\mu\nu}$, if one assumes that the conformal invariance is
preserved throughout  the evolution of the
system~\cite{Baier:2007ix}. It is easy to prove that the 1st order
N-S equation $\pi^{\mu\nu}=2\eta \nabla^{\La \mu}u^{\nu \Ra}$
naturally satisfies this constraint, since the shear viscosity and
velocity tensor transform as $\eta \rightarrow e^{3 \omega} \eta $
and $\nabla^{\La \mu}u^{\nu \Ra} \rightarrow e^{3 \omega}
\nabla^{\La \mu}u^{\nu \Ra}$, respectively.

The 2nd order viscous equations can be generalized from the 1st
order one by adding  2nd order corrections: $\pi^{\mu\nu}=2\eta
\nabla^{\La \mu}u^{\nu\Ra}+ \mathrm{(2nd \ order \ corrections)}$,
satisfying  the conformal constraint (\ref{Weyl}) and the
transversality and tracelessness properties: $\pi^{\mu\nu} u_\mu=0$
and $\pi^{\mu}_\mu=0$~\cite{Baier:2007ix}. The last two constraints
allow for 8 possible 2nd order corrections~\cite{Baier:2007ix}:
\begin{eqnarray}
&D^{\La \mu} \ln \epsilon\, D^{\nu\Ra} \ln \epsilon, \quad D^{\La
\mu} D^{\nu\Ra} \ln \epsilon, \quad \nabla^{\La \mu} u^{\nu \Ra}
\left(\nabla_\alpha u^\alpha\right), \quad P^{\mu \nu}_{\alpha
\beta}\ \nabla^{\La \alpha} u^{\gamma \Ra} g_{\gamma \delta}
\nabla^{\La \delta} u^{\beta \Ra}&\nonumber\\
&P^{\mu \nu}_{\alpha \beta}\ \nabla^{\La \alpha} u^{\gamma \Ra}
g_{\gamma \delta} \Omega^{\beta \delta},\quad P^{\mu \nu}_{\alpha
\beta}\ \Omega^{\alpha \gamma} g_{\gamma \delta} \Omega^{\beta
\delta},\quad u_\gamma R^{\gamma \La \mu \nu \Ra \delta}
u_\delta,\quad R^{\La \mu \nu \Ra}\, ,&
\end{eqnarray}
where  $R^{\alpha \mu\nu \beta}$ is  the  Riemann tensor and
$R^{\mu\nu}$ is the Ricci tensor. However, only 5 combinations of
the above terms transform homogeneously under Weyl transformation.
This determines the form of the 2nd order viscous equations, leaving
5 independent coefficients~\cite{Baier:2007ix}:
\begin{eqnarray}
\pi^{\mu\nu} &=& 2 \eta \nabla^{\langle \mu} u^{\nu\rangle} -
\tau_\pi \left[ \Delta^\mu_\alpha \Delta^\nu_\beta
D\pi^{\alpha\beta}
 + \frac 4{3} \pi^{\mu\nu}
    (\nabla_\alpha u^\alpha) \right] \nonumber\\
  &&\quad
  + \frac{\kappa}{2}\left[R^{\La \mu\nu \Ra}+2 u_\alpha R^{\alpha\La \mu\nu \Ra \beta}
      u_\beta\right]\nonumber\\
  && -\frac{\lambda_1}{2\eta^2} {\pi^{\La \mu}}_\lambda \pi^{\nu \Ra \lambda}
  -\frac{\lambda_2}{2\eta} {\pi^{\La \mu}}_\lambda \Omega^{\nu \Ra \lambda}
  - \frac{\lambda_3}{2} {\Omega^{\La \mu}}_\lambda \Omega^{\nu \Ra \lambda}\,
  .\
  \label{conformal}
\end{eqnarray}
Note that the $\kappa$ term is related to curved spaces and vanishes
in flat space. The conformal symmetry constraint requires that
$\Delta^\mu_\alpha \Delta^\nu_\beta D\pi^{\alpha\beta}$ must combine
with the term $\frac 4{3} \pi^{\mu\nu} (\nabla_\alpha u^\alpha)$
such that the additional derivatives of the function $\omega(x)$ in
the Weyl transformation  cancel between the two terms. The term
$\frac 4{3} \pi^{\mu\nu} (\nabla_\alpha u^\alpha)$ term can be
written in different ways as discussed in Chap. 2.7 below.

While the conformal symmetry constraint approach provides the
general form of the 2nd order viscous equations, it does not give
the 2nd order coefficients $\tau_\pi$, $\kappa$, $\lambda_1$,
$\lambda_2$ and $\lambda_3$. They must be determined through
explicit calculation. Such calculations were performed for weakly
coupled QCD by York and Moore~\cite{York:2008rr} and for strongly
coupled N=4 SYM theory in~\cite{Baier:2007ix}. Table \ref{Coef}
summarizes the results. It is reasonable to assume that the strong
 and weak coupling results bracket the value for these
coefficients in
QCD at realistic values for the coupling strength. \\

\begin{table}[h]
\begin{center}
\begin{tabular}{ccccccc}
\hline\hline  Approaches & $\tau_\pi$ & $\lambda_1$ & $\lambda_2$ &
$\lambda_3$ & $\kappa $\\[0.05 in]
\hline 1) N=4 SYM  \hspace{6mm} &  $\frac{2 - \ln 2}{2 \pi T}$ &
$\frac{\eta}{2 \pi T}$ & $-\ln{2} \frac{\eta}{\pi T}$ & 0&
$\frac{\eta}{\pi T}$\\[0.05 in]
\hline 2) Kinetic theory  & $\frac{5.0\ldots 5.9}{T}\frac{\eta}{s}$
& $\frac{4.1\ldots5.2}{T} \frac{\eta^2}{s}$ & $-2 \eta \tau_\pi$
& 0& ${\cal O}(g^{-4})T^2$\\[0.05 in]
\hline\hline
\end{tabular}
\end{center}
\vspace*{-7mm} \caption[The coefficients for the 2nd order I-S
equations] {2nd order Israel-Stewart Coefficients in the strong
coupling~\cite{Baier:2007ix} and weakly coupling~\cite{York:2008rr}
limits.} \label{Coef}
\end{table}

\section{\"Ottinger-Grmela formalism}
 The I-S formalism discussed in Chap 2.3-2.5 directly deals with
the dissipative flows ($\pi^{\mu\nu}$, $\Pi$ and $q^\mu$) and their
viscous equations.  In contrast, the \"Ottinger-Grmela  formalism
starts with an auxiliary field $c^{\mu\nu}$ and its evolution
equation (O-G equation). The details for the \"Ottinger-Grmela
formalism can be found in~\cite{OG1,OG2,OG3}. In numerical
implementation, one solves the O-G equation first and then relates
the auxiliary field $c_{\mu\nu}$ to the commonly used dissipative
flows in the subsequent spectra calculations~\cite{Dusling:2007gi}.

If taking the O-G equation and directly replacing the auxiliary
filed $c^{\mu\nu}$ by the dissipative flows, one obtains an
effective I-S equation in the \"Ottinger-Grmela approach. For
vanishing bulk viscosity and heat conductivity, the effective I-S
equations for shear stress tensor $\pi^{\mu\nu}$ was deduced by
Dusling and Teaney, and has the following
form~\cite{PRV-Dusling,TECHQM-Vis,Dusling:2009zz}:
\begin{eqnarray}
\pi^{\mu\nu} &=& 2 \eta \nabla^{\langle \mu} u^{\nu\rangle} -
\tau_\pi \left[ \Delta^\mu_\alpha \Delta^\nu_\beta
D\pi^{\alpha\beta}
 + [\frac 4{3}+\frac 2{3}-\frac{2 \beta}{\alpha} ]\pi^{\mu\nu}
    (\nabla_\alpha u^\alpha) \right] \nonumber\\
  && -\frac{\tau_\pi}{\eta} {\pi^{\La\mu}}_\lambda \pi^{\nu\Ra\lambda}
  - \tau_\pi {\pi^{\La \mu}}_\lambda \Omega^{\nu\Ra \lambda}
  ,\
  \label{O-G-2nd}
\end{eqnarray}
Where, the third order and higher order terms have been dropped.
$\alpha$ and $\beta$  are free parameters here. Generally, one uses
$\alpha$ to fix the relaxation time $\tau_\pi=\eta/p\alpha$ and sets
$\beta=\alpha/3$ to ensure
conformal invariance (see eq.~(\ref{conformal}) ).\\

\section{Final comments on the state of the formalism}

Comparing the 2nd order equations (\ref{dissi-trans-c}),
(\ref{I-S-Kin}), (\ref{dissi-trans2}c), (\ref{conformal}) and
(\ref{O-G-2nd}) superficially, one finds as the common terms that
are obtained in all of the different approaches only the following:
\begin{eqnarray}
\Delta ^{\mu \alpha} \Delta ^{\nu \beta} D \pi_{\alpha \beta}
=-\frac{1}{\tau_{\pi}}\bigg[\pi^{\mu\nu}-2\eta \nabla ^{\La \mu}u
^{\nu\Ra} \bigg]. \label{Simplifed-I-S}
\end{eqnarray}
This so-called \emph{simplified I-S equation}, was  used in our
early calculations~\cite{Song:2007fn,Song:2007ux} and in Chaudhuri's
work~\cite{Chaudhuri:2007qp,Chaudhuri:2008sj,Chaudhuri:2008je}.
However this equation does not preserve the conformal symmetry for a
conformal fluid (i.e. a fluid that consists of massless degrees of
freedom). To fix this problem one needs to add an additional term
$\frac 4{3} \tau_\pi \pi^{\mu\nu} (\nabla_\alpha u^\alpha)$ as
pointed out in Ref~\cite{Baier:2007ix}:
\begin{eqnarray}
\Delta ^{\mu \alpha} \Delta ^{\nu \beta} D \pi_{\alpha \beta}
=-\frac{1}{\tau_{\pi}}\bigg[\pi^{\mu\nu}-2\eta \nabla ^{\La \mu}u
^{\nu\Ra} + \frac 4{3} \tau_\pi \pi^{\mu\nu} (\nabla_\alpha
u^\alpha) \bigg].  \label{full-I-S}
\end{eqnarray}
In Ref.~\cite{Song:2008si} we called this equation the \emph{full
I-S equation}. The last term in eq.~(\ref{full-I-S}) is clearly
illustrated in Eqs.~(\ref{I-S-Kin}), (\ref{conformal}) and
(\ref{O-G-2nd})\footnote{\emph{Eq.(\ref{dissi-trans2})} also
contains a term of $\pi^{\mu\nu} \theta=\pi^{\mu\nu} (\nabla_\alpha
u^\alpha)$, but leaves an undetermined coefficient in front of it.
}. At first sight, eq.(\ref{dissi-trans-c}) seems to miss this term,
but this naive impression is incorrect.  For a conformally symmetric
fluid, the temperature $T$ is the only scale in the problem and
therefore $\eta\sim s\sim T^3$ and $\tau_\pi\sim 1/T$, hence $\eta
T/\tau_\pi \sim T^5$. In this limit, The last term in
eq.~(\ref{dissi-trans-c}) can then be rewritten as:
\begin{equation}
\label{full-IS-Var}-\frac{1}{2}\pi^{\mu\nu} \frac{\eta T}{\tau_\pi}
d_\lambda\left(\frac{\tau_\pi}{\eta T}u^\lambda\right) =
\frac{1}{2}\pi^{\mu\nu}\bigl(5 D(\ln T)-\theta\bigr) =  -\frac
4{3}\pi^{\mu\nu} (\nabla_\alpha u^\alpha).
\end{equation}
The last equality follows from the 2nd order hydrodynamic
equation~\cite{Romatschke:2007mq}, showing that, up to terms of
sub-leading order, the last terms in eq.~(\ref{dissi-trans-c}) and
eq.~(\ref{full-I-S}) agree.

The ``full I-S" equation, with slight variations, such as using the
replacement eq.~(\ref{full-IS-Var}), has been established as the
standard viscous equation for numerical
implementations~\cite{Song:2008si,Song:2009Bulk,Romatschke:2007mq,
Luzum:2008cw,Luzum:2009sb,Huovinen:2008te}. It has received
additional support by the observation made in~\cite{Song:2008si}
that it appears to minimize the dependence of the hydrodynamic
evolution on the value of the microscopic relaxation time
$\tau_\pi$. While the variations related to the use of
eq.~(\ref{full-IS-Var})  are equivalent for conformal systems, they
differ in principle for systems with an EOS that breaks conformal
invariance, e.g. through a phase transition. These differences turn
out to be negligible in practice~\cite{Song:2008si}. For a conformal
theory with vanishing chemical potentials, 4 other terms can be
added to the right hand side of the ``full I-S" equation, with
additional coefficients $\lambda_1$, $\lambda_2$, $\lambda_3$, and
$\kappa$ ($\kappa=0$ in Minkowski space), as shown in
eq.~(\ref{conformal}) \cite{Baier:2007ix}. These terms include
couplings to the vorticity tensor \cite{Baier:2007ix} which turns
out to be small in heavy-ion collisions if the initial longitudinal
velocity profile is boost invariant \cite{Romatschke:2007mq}. Even
more terms arise in a kinetic theory derivation that does not assume
conformal symmetry and includes the effects from bulk viscosity and
heat conductivity, as shown by
Eqs.~(\ref{dissi-trans-c},~\ref{dissi-trans2}). Their coefficients
can be obtained from kinetic theory \cite{Betz:2008me,York:2008rr}
at weak coupling or from the AdS/CFT correspondence at strong
coupling \cite{Baier:2007ix}(see Table.\ref{Coef}). The so far
accumulated numerical evidence \cite{Song:2008si,Luzum:2008cw} (see
Chap.~7.3 and~7.4) suggests that, except for the last terms in
Eqs.~(\ref{full-I-S}), all other second order terms are unimportant
in practice, but a systematic study that confirms this beyond doubt
remains outstanding.
\chapter[Causal Viscous Hydrodynamics in 2+1 Dimensions]{Causal Viscous Hydrodynamics \\
\hspace{3cm} in 2+1 Dimensions}

Although the Israel-Stewart 2nd order formalism for relativistic
dissipative fluid dynamics was established thirty years ago, its
numerical implementations for describing relativistic heavy ion
collisions started only quite
recently~\cite{Muronga:2001zk,Muronga:2003ta,Muronga:2004sf,Heinz:2005bw,
Muronga:2006zw,Muronga:2006zx}. Currently all of the available
viscous hydrodynamics codes are restricted to either 1+1
dimensions~\cite{Baier:2006um,Baier:2006gy,Romatschke:2007jx} or 2+1
dimensions~\cite{Song:2007fn,Song:2007ux,Song:2008si,Song:2009Bulk,
Romatschke:2007mq,Luzum:2008cw, Luzum:2009sb,Dusling:2007gi,
Molnar:2008xj,Chaudhuri:2007qp,Chaudhuri:2008sj,Chaudhuri:2008je}.
Here the first number indicates the number of spatial dimensions in
which the code solves numerically for the evolution of the
hydrodynamic fields, and the ``+1" stand for time. All existing
codes assume longitudinal boost invariance, which allows to treat
the longitudinal expansion along the beam direction analytically.
This is a good approximation at RHIC and LHC energies for particle
production near midrapidity. (1+1)-d codes assume additionally
azimuthal symmetry around the beam axis, allowing spatial expansion
in the radial direction. With them, we can only simulate heavy ion
collisions with zero impact parameter. To describe non-central
collisions, and in particular anisotropic (elliptic) flow, requires
a (2+1)-d code, which allows for anisotropic expansion in the two
dimensions transverse to the beam.

During the past years, the OSU group developed a (2+1)-dimensional
viscous hydrodynamic code to describe the space-time evolution of
the QGP and subsequent hadronic matter in the two dimensional
transverse plane -- VISH2+1 (Viscous Israel-Stewart Hydrodynamics in
2+1
dimensions)~\cite{Song:2007fn,Song:2007ux,Song:2008si,Song:2009Bulk}.
In this chapter, we will summarize the crucial ingredients for the
numerical calculations. These include the explicit form of the
viscous hydrodynamic equations in 2+1 dimensions (Chap.~3.1), the
initial conditions (Chap.~3.2), the equation of state (Chap.~3.3)
and the freeze-out condition, including the freeze-out procedure and
the calculation of spectra (Chap.~3.4).  We also discuss the inputs
for shear viscosity, bulk
viscosity and the relaxation times (Chap.~3.5).\\

%
\section{Viscous hydrodynamic equations in 2+1 dimensions}
%
For ease of numerical implementation, the current viscous
hydrodynamic calculations focus on boost-invariant systems, realized
by assuming a specific "scaling" velocity profile $v_z=z/t$ along
the beam direction. As shown by Bjorken~\cite{Bjorken:1982qr}, this
profile is a solution of the hydrodynamic equations (ideal or
viscous) if the initial conditions are independent of the
longitudinal reference frame (boost invariance), i.e. do not depend
on space-time rapidity $\eta$. After implementing the Bjorken
approximation, the (3+1)-d viscous hydrodynamics reduces to (2+1)-d
viscous hydrodynamics with boost invariance. Currently, all of the
existing (2+1)-d viscous hydrodynamic
codes~\cite{Song:2007fn,Song:2007ux,Song:2008si,Luzum:2008cw,
Luzum:2009sb,Dusling:2007gi} also assume zero net baryon density and
zero heat conductivity. These conditions are approximately realized
in experiments at top RHIC and LHC energies and simplify the
implementation further by eliminating the need to solve for the
flows of baryon number and heat.

Longitudinally boost-invariant systems  are conveniently described
in curvilinear  coordinates $x^m\eq(\tau,x,y,\eta)$, where
$\tau\eq\sqrt{t^2{-}z^2}$ is the longitudinal proper time,
$\eta\eq\frac{1}{2}\ln\bigl(\frac{t{+}z}{t{-}z}\bigr)$ is the
space-time rapidity, and $(x,y)$ are the usual Cartesian coordinates
in the plane transverse to the beam direction $z$. In this
coordinate system, the transport equations for the full energy
momentum tensor $T^{\mu\nu}$ are written
as~\cite{Heinz:2005bw,Song:2007ux}:
\begin{subequations}
\label{transport-T}
\begin{eqnarray}
&&\partial_\tau \widetilde{T}^{\tau\tau}
 +\partial_x (v_x\widetilde{T}^{\tau\tau})
 +\partial_y (v_y\widetilde{T}^{\tau\tau}) = {\cal S}^{\tau\tau},
\\
&&\partial_\tau \widetilde{T}^{\tau x}
 +\partial_x (v_x\widetilde{T}^{\tau x})
 +\partial_y (v_y\widetilde{T}^{\tau x}) = {\cal S}^{\tau x},
\\
&&\partial_\tau \widetilde{T}^{\tau y}
 +\partial_x (v_x\widetilde{T}^{\tau y})
 +\partial_y (v_y\widetilde{T}^{\tau y}) = {\cal S}^{\tau y}.
\end{eqnarray}
\end{subequations}
Here $\widetilde{T}^{m n}\equiv\tau(T_0^{mn}{+}\pi^{mn}-\Pi
\Delta^{mn})$, $T_0^{mn}\eq{e}u^mu^n{-}p\Delta^{mn}$ being the ideal
fluid contribution, $u^m\eq(u^\tau, u^x, u^y,
0)=\gamma_\perp(1,v_x,v_y,0)$ is the flow profile (with
$\gamma_\perp\eq\frac{1}{\sqrt{1{-}v_x^2{-}v_y^2}}$), and
$g^{mn}\eq\mathrm{diag}(1,-1,-1,-1/\tau^2)$ is the metric tensor for
our coordinate system. The source terms ${\cal S}^{mn}$ on the right
hand side of Eqs.~(\ref{transport-T}) are given explicitly as
\begin{subequations}
\label{Source2+1}
\begin{eqnarray}
&&{\cal S}^{\tau\tau}=-(p+\Pi) -\tau^2\pi^{\eta\eta}
                   -\tau\partial_x(p v_x{+}\Pi v_x{+}\pi^{x\tau}{-}v_x\pi^{\tau\tau})
\nonumber\\
&&\hspace*{4.5cm}  -\,\tau\partial_y(p v_y{+}\Pi v_y{+}\pi^{y\tau}{-}v_y\pi^{\tau\tau}),
\label{S00}
\\
&&{\cal S}^{\tau x}=-\tau\partial_x(p{+}\Pi{+}\pi^{xx}{-}v_x\pi^{\tau x})
                   -\tau\partial_y(\pi^{xy}{-}v_y\pi^{\tau x}),
\label{S01}
\\
&&{\cal S}^{\tau y}=-\tau\partial_x(\pi^{xy}{-}v_x\pi^{\tau y})
                   -\tau\partial_y(p{+}\Pi{+}\pi^{yy}{-}v_y\pi^{\tau y}).
\label{S02}
\end{eqnarray}
\end{subequations}
The transport equations for the shear pressure tensor and bulk
pressure in 2+1 dimensions are written
as~\cite{Song:2007ux,Song:2008si}:
\begin{subequations}
\label{PimnPI2+1}
\begin{eqnarray}
&&D\tilde{\pi}^{mn}=-\frac{1}{\tau_{\pi}}(\tilde{\pi}^{mn}{-}2\eta\tilde{\sigma}^{mn})
   -(u^m\tilde{\pi}^{nk}{+}u^n\tilde{\pi}^{mk}) Du_k
   -\frac{1}{2}\tilde{\pi}^{mn} \frac{\eta T}{\tau_\pi}
       d_k\left(\frac{\tau_\pi}{\eta T}u^k\right),\hspace*{1.0cm}\\
&&D\Pi=-\frac{1}{\tau_{\Pi}}(\Pi{-}\zeta\theta)
   -\frac{1}{2}\Pi \frac{\eta T}{\tau_\Pi}
       d_k\left(\frac{\tau_\Pi}{\eta T}u^k\right),
\end{eqnarray}
\end{subequations}
Here, we have written out the explicit (2+1)-d form for the full
``I-S equations" described in Chap. 2.7. The expressions for
$\tilde{\sigma}^{mn}$ and $\tilde{\pi}^{mn}$ are found in
Eqs.~(\ref{tilde-pi}, \ref{tilde-sigma}) in Appendix~\ref{appa1};
they differ from $\pi^{mn}$ in Eqs.~(\ref{S00}-\ref{S02}) and
$\sigma^{mn}$ given in Ref.~\cite{Heinz:2005bw} by a Jacobian
$\tau^2$ factor in the $(\eta\eta)$-component:
$\tilde{\pi}^{\eta\eta}\eq\tau^2\pi^{\eta\eta}$,
$\tilde{\sigma}^{\eta\eta}\eq\tau^2\sigma^{\eta\eta}$. This factor
arises from the curved metric where the local time derivative
$D\eq{u^m}d_m$ must be evaluated using covariant derivatives $d_m$.
Since $u^\eta\eq0$, no such extra Jacobian terms arise in the
derivative $Du_k$ in the first line of Eq.~(\ref{PimnPI2+1}). In
more detail,
$D{\eq}{u^\tau}\partial_\tau{\,+\,}u^x\partial_x{\,+\,}u^y\partial_y$,
$\theta=\partial\cdot u\eq\partial_\tau u^\tau + \partial_x u^x +
\partial_y u^y +\frac{u^\tau}{\tau}$
and $\sigma^{mn}\eq\nabla^{\left\langle m\right.} u^{\left.n
\right\rangle}$ $\eq\frac{1}{2}(\nabla^m u^n{+}\nabla^n
u^m)-\frac{1}{3} \Delta^{mn}\theta$ (with
$\nabla^m\eq\Delta^{ml}d_{l}$).

Even though several components of the symmetric shear pressure
tensor $\pi^{mn}$ are redundant \cite{Heinz:2005bw} on account of
its tracelessness and transversality to the flow velocity $u^m$,
VISH2+1 propagates all 7 non-zero components and uses the
tracelessness and transversality conditions as checks of the
numerical accuracy \cite{Song:2007fn}. We find them to be satisfied
with an accuracy of better than $1-2\%$ everywhere except for the
fireball edge where the $\pi^{mn}$ are very small and the error on
the transversality and tracelessness constraints can become as large
as 5\%.\\

\section{Initial conditions}
\label{sec2c}
The initialization of a hydrodynamic simulation requires a starting
time $\tau_0$ and initial profiles for the energy momentum tensor
$T^{\tau \tau}$, $T^{\tau x}$ and $T^{\tau y}$, which are given by
the initializations for the energy density, velocity and stress
stress tensor $\pi^{mn}$. Through all of this thesis, we set
$\tau_0=0.6 \ \mathrm{fm/c}$, following the ``standard"
thermalization time used in most ideal hydrodynamic
simulations~\cite{Kolb:2003dz}, and we use zero initial transverse
flow velocity. We have not explored the need for rescaling $\tau_0$
when including viscosity; this awaits a careful comparison with the
experimental data. The initializations for energy density and
$\pi^{mn}$ profiles are
described below.\\[-0.05in]

\subsection{Initializations for the energy density}

\textbf{\underline{Glauber model initialization}}\\[-0.10in]

A simple Glauber model initialization assumes that the initial
energy density in the transverse plane is proportional to the
wounded nucleon density \cite{Kolb:1999it,Song:2007ux}:

\hspace{-2cm}
\begin{eqnarray}
\label{Glauber}
 && e_0(x,y;b)=K n_\mathrm{WN}(x,y;b)
\\
&& = K \biggl\{T_A\bigl(x{+}{\textstyle\frac{b}{2}},y)\biggl[1-
       \biggl(1-\frac{1-\sigma T_B\left(x{-}\frac{b}{2},y\right)}{B}\biggr)^B
       \biggr]
\nonumber\\
&& \quad\ \  +\, T_B\bigl(x{-}{\textstyle\frac{b}{2}},y\bigr)
\biggl[1-
       \biggl(1-\frac{1-\sigma T_A\left(x{+}\frac{b}{2},y\right)}{A}\biggr)^A
       \biggr]\biggr\}.
\nonumber
\end{eqnarray}
Here $\sigma$ is the total inelastic nucleon-nucleon cross section
for which we take $\sigma\eq40$\,mb. $T_{A,B}$ is the nuclear
thickness function of the incoming nucleus $A$ or $B$, defined as
$T_A(x,y)$ $\eq\int^\infty_{-\infty} dz \rho_A(x,y,z)$;
$\rho_A(x,y,z)$ is the nuclear density given by a Woods-Saxon
profile:  $\rho_A(\bm{r})$ $\eq\frac{\rho_0}{1+\exp[(r-R_A)/\xi]}$.
We take $R_\mathrm{Cu}\eq4.2$\,fm, $\xi\eq0.596$\,fm for Cu+Cu
collisions and $R_\mathrm{Au}\eq6.37$\,fm, $\xi\eq0.56$\,fm  for
Au+Au collisions (these two parameter sets for Au and Cu nuclei
correspond to a nuclear density $\rho_0\eq0.17$\,fm$^{-3}$). The
proportionality constant $K$ does not depend on collision centrality
but on collision energy; it fixes the overall scale of the initial
energy density and, via the associated entropy, the final hadron
multiplicity to which it must be fitted as a function of collision
energy. The energy density profile for specific collision energy  is
therefore normalized by a parameter $e_0(\tau=\tau_0, r=b=0)$ giving
the peak energy density in the center of the fireball for central
collisions (impact parameter b = 0). For central Au+Au collisions at
top RHIC energies, one sets $e_0{\,\equiv\,}e(0,0;b{=}0)\eq30 \
\mathrm{GeV/fm^3}$ and $\tau_0=0.6 \ \mathrm{fm/c}$ to reproduce the
final multiplicity in ideal hydrodynamic calculations.  In the
viscous hydrodynamics comparison runs, we use the same
initialization as for ideal hydrodynamics. Viscous entropy
production then leads to slightly larger final multiplicities than
in ideal hydrodynamics. Again, we leave a corresponding retuning of
initial conditions to a careful comparison study with experimental
data. The Glauber initialization (\ref{Glauber}) does not correctly
reproduce the measured centrality dependence of
$dN_{ch}/dy$~\cite{Kolb:2001qz}. To fix this problem, more
sophisticated initialization schemes have been
developed~\cite{Kolb:2001qz}, where one uses a superposition of
wounded nucleon and binary collision densities. We also leave this
for future data comparisons.
 \\[-0.05in]

\textbf{\underline{Color glass condensate (CGC) initialization}}\\[-0.10in]

The CGC initialization based on the Kharzeev-Levin-Nardi (KLN)
approach~\cite{Kharzeev:2000ph,Kharzeev:2002ei} and its more recent
fKLN improvement~\cite{Drescher:2006pi} has been applied in earlier
ideal hydrodynamic
calculations~\cite{Hirano:2004en,Hirano:2005xf,Kuhlman:2006qp} and
recently also in viscous hydrodynamic
calculations~\cite{Luzum:2008cw,Luzum:2009sb}. Details of the CGC
initialization can be found in
Ref.~\cite{Drescher:2006pi,Hirano:2004en}.

\begin{figure}[h]
\vspace{-0.3cm} \centering
\includegraphics[width=7.5cm, height=5.8cm, angle=0]{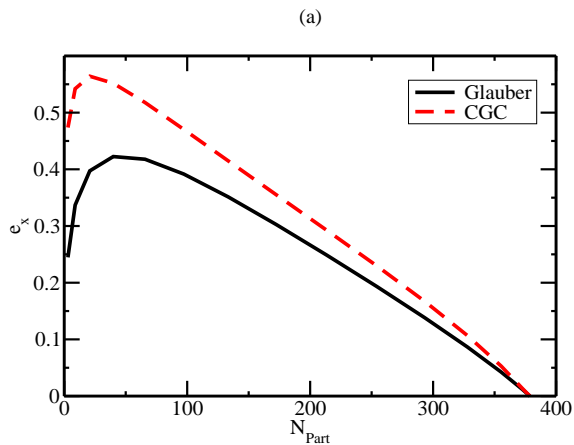}
\vspace{-0.3cm} \caption[The initial eccentricity from Glauber and
CGC initializations]{The initial eccentricity $\varepsilon_x$ from
Glauber and CGC initializations, for Au+Au collisions at different
centrality~\cite{Luzum:2008cw}. } \label{initial}
\end{figure}

Compared with the Glauber initialization, the CGC initialization
gives a more ``plateau-like" initial profile for the energy
density~\cite{Kuhlman:2006qp}, which leads to larger initial
eccentricity $\varepsilon_x=\frac{\La y^2-x^2 \Ra}{\La x^2+y^2 \Ra}$
than for the Glauber model. Fig.\ref{initial} shows the initial
eccentricity $\varepsilon_x$ from the  Glauber and fKLN
initializations~\cite{Luzum:2008cw}. One finds that the CGC
initialization predicts $20 - 50\%$ larger initial eccentricity than
the Glauber one, especially at small impact parameters, where
the eccentricity is small but the discrepancy between models is large~\cite{PRV-Scott} .\\[-0.15in]

\subsection{Initializations for $\pi^{mn}$ and $\Pi$}
Lacking a microscopic dynamical theory for the early pre-equilibrium
stage, initializing the viscous pressure tensor $\pi^{mn}$ requires
some guess-work. We here explore two options: (i) zero
initialization, which sets $\pi^{mn}_0\eq0$ at initial time
$\tau_0$~\cite{Romatschke:2007mq,Song:2007fn,Song:2007ux}.
(ii)(Navier- Stokes) N-S initialization:
$\pi^{mn}_0\eq2\eta\sigma^{mn}_0$, where the shear tensor
$\sigma^{mn}_0$ is calculated from the initial velocity profile
$u^m\eq(1,0,0,0)$
~\cite{Song:2007fn,Song:2007ux,Song:2008si,Luzum:2008cw,
Luzum:2009sb,Dusling:2007gi}. The second option is the default
choice for most of the results shown in this thesis. It gives
$\tau^2
\pi^{\eta\eta}_0\eq{-2}\pi^{xx}_0\eq{-2}\pi^{yy}_0\eq{-}\frac{4\eta}{3
\tau_0}$, i.e. a negative contribution to the longitudinal pressure
and a positive contribution to the transverse pressure.\\

\section{The equation of state (EOS)}\label{sec2d}
The equation of state (EoS), which relates the pressure $p$ to
energy density $e$ and net baryon density $n_B$: $p=p(n_B, e)$ (or
equivalently relates it to temperature $T$ and net baryon chemical
potential $\mu_B$, $p=p(T, \mu_B)$) is a necessary input for both
ideal and viscous hydrodynamic simulations. In this section, we will
describe the different EoS used in this
thesis~\cite{Song:2007ux,Song:2008si}.  Currently, all of the
existing viscous hydrodynamics calculation neglect net baryon
density and set $n_B=0$. The EoS described below thus reduces to a
simple one dimensional EOS: $p=p(e)$. \\[-0.05in]

%
\begin{figure}[t]
  \begin{center}
    \begin{minipage}[b]{0.6\linewidth}
    \includegraphics[width=0.85\linewidth,height=1.15\linewidth,clip=]{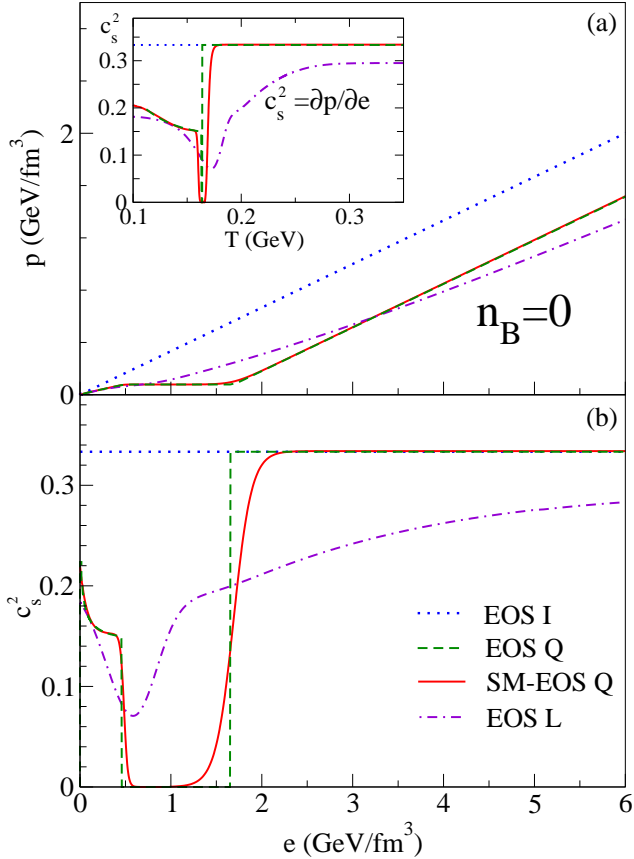}
    \end{minipage}
    \begin{minipage}[b]{0.35\linewidth}
      \caption[The equation of state]{The equation of state. Panel (a) shows the
        pressure $p$ as a function of energy density $e$ and (in the inset) the
        squared speed of sound $c_s^2=\frac{\partial p}{\partial e}$ as a function
        of temperature $T$, for EOS I, EOS Q, SM-EOS Q and EOS L.
        Panel (b) shows $c_s^2$ as a function of energy
        density $e$ for different EOS.\vspace*{3.7cm}\label{fig-EOS}}
    \end{minipage}
  \end{center}
  \vspace*{-0.8cm}
\end{figure}
%

\textbf{\underline{EOS~I}:}\\[-0.10in]

EOS~I models a non-interacting gas of massless quarks and gluons,
with $p\eq\frac{1}{3}e$. It has no phase transition. Where needed,
the temperature is extracted from the energy density via the
relation
$e\eq\left(16+\frac{21}{2}N_f\right)\frac{\pi^2}{30}\frac{T^4}{(\hbar
c)^3}$, corresponding to a chemically equilibrated QGP with
$N_f\eq2.5$ effective massless quark flavors~\cite{Kolb:1999it}. \\[-0.05in]

\textbf{\underline{EOS~Q}:} \\[-0.10in]

EOS~Q~\cite{Kolb:1999it} connects a noninteracting massless QGP gas
to a chemically equilibrated hadron resonance gas through a first
order phase transition. The EOS in the QGP phase is defined by
$p\eq\frac{1}{3}e{-}\frac{4}{3}B$ (i.e. $c_s^2\eq\frac{\partial
p}{\partial e}\eq\frac{1}{3}$). The vacuum energy (bag constant)
$B^{1/4}\eq230$\,MeV is a parameter that is adjusted to yield a
critical temperature $T_c=164$\,MeV. The hadron resonance gas below
$T_c$ can be approximately characterized by the relation
$p\eq0.15\,e$ (i.e. $c_s^2\eq0.15$)~\cite{Kolb:1999it}. The two
sides are matched through a Maxwell construction, yielding a
relatively large latent heat $\Delta
e_\mathrm{lat}\eq1.15$\,GeV/fm$^3$. For energy densities between
$e_\mathrm{H}\eq0.45$\,GeV/fm$^3$ and
$e_\mathrm{Q}\eq1.6$\,GeV/fm$^3$ one has a mixed phase with constant
pressure (i.e. $c_s^2\eq0$). \\[-0.05in]

\textbf{\underline{SM-EOS~Q}:}\\[-0.10in]

SM-EOS~Q is a smoothed version of  EOS~Q~\cite{Song:2007ux}. The
discontinuous jumps of $c_s^2$ in EOS~Q from a value of 1/3 to 0 at
$e_\mathrm{Q}$ and back from 0 to 0.15 at $e_\mathrm{H}$ generate
propagating numerical errors in VISH2+1 which grow with time and
cause problems. We avoid these by smoothing the function $c_s^2(e)$
with a Fermi distribution of width $\delta e\eq0.1$\,GeV/fm$^3$
centered at $e\eq{e_\mathrm{Q}}$ and another one of width $\delta
e\eq0.02$\,GeV/fm$^3$ centered at $e\eq{e_\mathrm{H}}$. Both the
original EOS~Q and our smoothed version SM-EOS~Q are shown in
Figure~\ref{fig-EOS}. A comparison of simulations using ideal
hydrodynamics with EOS~Q and SM-EOS~Q is given in
Appendix~\ref{appd1}. It gives an idea of the magnitude of smoothing
effects on the ideal fluid evolution of elliptic flow.\\[-0.05in]

\textbf{\underline{EOS~L}:}\\[-0.10in]

EOS~L~\cite{Song:2008si} matches the hadron resonance gas below
$T_c$ smoothly in a rapid cross-over transition to lattice QCD
data~\cite{Katz:2005br} above $T_c$. For the fit, the lattice data
were plotted in the form $p(e)$, interpolated and then smoothly
joined to the $p(e)$ curve of the HRG with a cross-over transition
near $T_c \sim 175 MeV$.  As can be seen in the upper panel of
Fig.~\ref{fig-EOS} in the inset, our procedure is not fully
thermodynamically consistent and leads to a somewhat different
temperature dependence of $c_s^2$ below $T_c$ than for EOS~Q and
SM-EOS~Q. Since this only affects the flow dynamics below our
decoupling temperature of $T_{dec}=130$\,MeV, we have not put any
effort into correcting this\footnote{\emph{The EOS L here is
different from the cross-over EOS used in
Ref~\cite{Romatschke:2007mq}, which was constructed by connecting a
hadron resonance gas EOS with the QGP EOS from a high-order
weak-coupling QCD calculation~\cite{Laine:2006cp}.}}.\\[-0.05in]
\\[-0.15in]

\underline{\textbf{PCE-EOS}} for the HRG phase:\\[-0.10in]

EOS Q assumes that chemical  and thermal freeze-out happen at the
same temperature $T_{\mathrm{th}} \sim 130 \
\mathrm{MeV}$~\cite{Kolb:2000sd,Kolb:2000fha}. However, the
experimental data indicate that chemical freeze-out happens earlier
near $T_c \sim 165 \
\mathrm{MeV}$~\cite{BraunMunzinger:2001ip,BraunMunzinger:2003zd}.
The unrealistic implementation of chemical freeze-out in EOS Q leads
to wrong predictions for the particle ratios which need to be fixed
by hand using normalization factors.

Microscopically, chemical and thermal freeze-out are related to
inelastic and total (elastic + quasi-elastic + inelastic) scattering
rates between particles, respectively. The cross sections of the
inelastic, particle number changing processes, are smaller than the
elastic and quasi-elastic ones, which leads to an earlier chemical
than thermal freeze-out~\cite{Heinz:2004qz}. A complete chemical
freeze-out corresponds to a picture where all hadron numbers are
fixed below $T_{ch}$. However, quasi-elastic resonant scatterings
constantly change resonances and their daughter particles (e.g.
$\pi\pi \rightarrow \rho \rightarrow \pi\pi$, $\pi N \rightarrow
\Delta \rightarrow \pi N$, $\pi K \rightarrow K^* \rightarrow \pi
K$, etc.), such that the resonances are in relative chemical
equilibrium with their decay products and their abundances only
freeze out at thermal decoupling. This partial chemical equilibrium
picture is naturally incorporated in any hadron cascade model, that
solves the Boltzmann equations with elastic, quasi-elastic and
inelastic cross sections for different hadron
species~\cite{Bass:1998ca,Nonaka:2006yn,Hirano:2005wx}.   In pure
hydrodynamic simulations it is realized by introducing an EOS
describing partially chemical equilibrium (PCE-EOS) with effective
chemical potentials adjusted to conserve the relative hadron
abundances after resonance decays in the HRG
phase~\cite{Kolb:2002ve,Teaney:2002aj,Huovinen:2007xh,Hirano:2002ds}.
Although the effective chemical potentials only slightly change the
equation of state $p(e)$, they affect significantly the relation
between temperature and energy density $T(e)$ in the HRG phase,
since a larger portion of the energy density is stored in particle
rest masses, reducing the thermal energy and temperature. As a
result, the typical freeze-out temperature drops from $130 \
\mathrm{MeV}$ (EOS Q) to $100 \ \mathrm{MeV}$ (PCE-EOS) for the same
freeze-out energy density $e_{dec}=0.085 \ \mathrm{GeV/fm^3}$.

Currently, the PCE-EOS has been well studied and implemented in
ideal hydrodynamic simulations, but has not been applied to any
viscous hydrodynamic simulations. However, the chemical content in
the HRG phase at freeze-out will affect the extracted value of the
QGP shear viscosity.
This will be further discussed in Chap. 8.\\

\section{Freeze-out procedure and calculation of spectra}
\label{sec2e}

The hadron spectra are computed from the hydrodynamic output via a
modified Cooper-Frye procedure \cite{Cooper:1974mv}. We here compute
spectra only for directly emitted particles and do not include
feeddown from resonance decays after freeze-out. We first determine
the freeze-out surface $\Sigma(x)$, by postulating (as common in
hydrodynamic studies) that freeze-out from a thermalized fluid to
free-streaming, non-interacting particles happens suddenly when the
temperature drops below a critical value. As in the ideal fluid case
with EOS~Q \cite{Kolb:1999it} we choose $T_\mathrm{dec}\eq130$\,MeV.
The particle spectrum is then computed as an integral over this
surface~\cite{Cooper:1974mv},
\begin{eqnarray}
\label{Cooper}
  E\frac{d^3N_i}{d^3p} &=& \frac{g_i}{(2\pi)^3}\int_\Sigma
  p\cdot d^3\sigma(x)\, f_i(x,p) \nonumber \\
  &=& \frac{g_i}{(2\pi)^3}\int_\Sigma p\cdot d^3\sigma(x)
 \left[f_{\mathrm{eq},i}(x,p)+\delta f_i(x,p) \right],
\end{eqnarray}
where $g_i$ is the spin-isospin degeneracy factor for particle
species $i$, $d^3\sigma^\mu(x)$ is the outward-pointing surface
normal vector on the decoupling surface $\Sigma(x)$ at point $x$,
which for boost-invariant freeze-out at longitudinal proper time
$\tau_f (\mathrm{r})$ reads
\begin{eqnarray}
\label{dsigma}
  p \cdot d^3\sigma(x) = \big[m_T\cosh(y{-}\eta)-\bm{p}_\perp\cdot
  \bm{\nabla}_\perp\tau_f(\bm{r})\big]
  \times
  \tau_f(\bm{r})\, r dr\, d\phi\, d\eta,
\end{eqnarray}
(with $\bm{r}\eq(x,y)\eq(r\cos\phi,r\sin\phi)$ denoting the
transverse position vector), and $f_i(x,p)$ is the local
distribution function for particle species $i$, computed from the
hydrodynamic output. Equation (\ref{Cooper}) generalizes the usual
Cooper-Frye prescription for ideal fluid dynamics
\cite{Cooper:1974mv} by accounting for the fact that in a viscous
fluid the local distribution function is never exactly in local
equilibrium, but deviates from its local equilibrium form by small
terms proportional to the non-equilibrium viscous flows
\cite{Teaney:2003kp,Song:2007ux,Baier:2006um}. Both contributions
can be extracted from hydrodynamic output along the freeze-out
surface. The equilibrium contribution is
\begin{eqnarray}
\label{f0}
  f_{\mathrm{eq},i}(p,x)
  = f_{\mathrm{eq},i}\Bigl(\frac{p{\cdot}u(x)}{T(x)}\Bigr)
  = \frac{1}{e^{p\cdot u(x)/T(x)}\pm 1},
\end{eqnarray}
where the exponent is computed from the temperature $T(x)$ and
hydrodynamic flow velocity \\  $u^\mu\eq\gamma_\perp(\cosh\eta, v_x,
v_y,\sinh\eta)$ along the surface $\Sigma(x)$:
\begin{eqnarray}
\label{pdotu}
  p\cdot u(x)\eq\gamma_\bot[m_T\cosh(y{-}\eta)
                          - p_x v_x -p_y v_y].\
\end{eqnarray}
Here $m_T\eq\sqrt{p_T^2{+}m_i^2}$ is the particle's transverse mass.

As long as only shear viscosity is implemented\footnote{\emph{For
the case with both shear and bulk viscosity, the expression for
$\delta f$ can be found in Ref.~\cite{Monnai:2009ad}.}}, the viscous
deviation from local equilibrium is given by
\cite{Teaney:2003kp,Baier:2006um}
\begin{eqnarray}
\label{deltaf}
  \delta f_i(x,p) \!\!&=&\!\!
  f_{\mathrm{eq},i}(p,x) \bigl(1{\mp}f_{\mathrm{eq},i}(p,x)\bigr)
  \frac{p^\mu p^\nu
  \pi_{\mu\nu}(x)}{2T^2(x)\left(e(x){+}p(x)\right)}.
\end{eqnarray}
The viscous correction is proportional to $\pi^{\mu\nu}(x)$ on the
freeze-out surface (normalized by the equilibrium enthalpy $e{+}p$)
and increases quadratically with the particle's momentum (normalized
by the temperature $T$). At large $p_T$, the viscous correction can
exceed the equilibrium contribution, indicating a breakdown of
viscous hydrodynamics. In that domain, particle spectra can not be
reliably computed with viscous fluid dynamics. The limit of
applicability depends on the actual value of $\pi^{\mu\nu}/(e{+}p)$
and thus on the specific dynamical conditions encountered in the
heavy-ion collision.

In Eq.~(\ref{deltaf}), the final expression for viscous correction
to the spectrum are written explicitly as:
\begin{eqnarray}
  p_\mu p_\nu \pi^{\mu\nu} &=&
  m_T^2\bigl(\cosh^2(y{-}\eta)\pi^{\tau\tau}+\sinh^2(y{-}\eta)
                 \tau^2\pi^{\eta\eta}\bigr)
\nonumber\\
  &-& 2m_T\cosh(y{-}\eta)\bigl(p_x\pi^{\tau x}+p_y\pi^{\tau y}\bigr)
\nonumber\\
  &+& \bigl(p_x^2\pi^{xx}+2p_xp_y\pi^{xy}+p_y^2\pi^{yy}\bigr).
\label{vis-correction}
\end{eqnarray}
%

Due to longitudinal boost-invariance, the integration over
space-time rapidity $\eta$ in Eq.~(\ref{Cooper}) can be done
analytically, resulting in a series of contributions involving
modified Bessel functions \cite{Heinz:2004qz,Baier:2006gy}. VISH2+1
does not exploit this possibility and instead performs this and all
other integrations for the spectra numerically~\cite{Song:2007ux}.

Once the spectrum (\ref{Cooper}) has been computed, a Fourier decomposition
with respect to the azimuthal angle $\phi_p$ yields the anisotropic flow
coefficients. For collisions between equal spherical nuclei followed by
longitudinally boost-invariant expansion of the collision fireball, only
even-numbered coefficients contribute, the ``elliptic flow'' $v_2$ being
the largest and most important one:
\begin{eqnarray}
E\frac{d^3N_i}{d^3p}(b) = \frac{dN_i}{dy\, p_T dp_T\, d\phi_p}(b) =
\frac{1}{2\pi}\frac{dN_i}{dy\, p_T dp_T}\big[1 + 2 v_2(p_T;b)
     \cos(2\phi_p) + \dots \big].
\label{eq-V2}
\end{eqnarray}
In practice it is evaluated as the $\cos(2\phi_p)$-moment of the
final particle spectrum,
\begin{eqnarray}
\label{cos2phi}
  v_2(p_T)=\langle\cos(2\phi_p)\rangle\equiv
  \frac{\int d\phi_p\,\cos(2\phi_p)\,\frac{dN}{dy\,p_T dp_T\,d\phi_p}}
                {\int d\phi_p\,\frac{dN}{dy\,p_T dp_T\,d\phi_p}}\,,\quad
\end{eqnarray}
where, according to Eq.~(\ref{Cooper}), the particle spectrum is a
sum of a local equilibrium and a non-equilibrium contribution (to be
indicated symbolically as $N\eq{N}_\mathrm{eq}+\delta N$).\\

\section[Additional viscous inputs: shear, bulk viscosity and relaxation
times]{Additional viscous inputs: shear viscosity, bulk viscosity
and relaxation times} \label{sec2c}

The shear viscosity $\eta$, bulk viscosity $\zeta$ and their
relaxation times $\tau_\pi$ and $\tau_\Pi$  are free inputs in
viscous hydrodynamic calculations.

Although the future trend for viscous hydrodynamic calculations will
be to input a temperature dependent $\eta/s$ to consider the fact
that $\eta/s$ is small in the QGP phase, reaches a minimum during
the phase transition and rises again to much larger values  in the
hadronic phase (see Chap.~1.4 for details), all of the presently
existing viscous hydrodynamic
calculations~\cite{Song:2007fn,Song:2007ux,Song:2008si,Luzum:2008cw,
Luzum:2009sb,Dusling:2007gi} (including those presented here) use a
constant $\eta/s$ as input: $\eta/s = C \times \frac 1 {4 \pi}$,
with $C=0,\ 1,\ 2,\ 3 \ ...$ ($\frac 1 {4 \pi}$ is the minimal KSS
bound from AdS/CFT ~\cite{Kovtun:2004de}).  Typically, we set
$\tau_\pi= \frac {3 \eta} {sT}$, except where mentioned otherwise.
This expression is very close to the AdS/CFT prediction
$\tau_\pi=\frac{2-\ln2}{2\pi
T}$~\cite{Bhattacharyya:2008jc,Natsuume:2007ty}, and is  half of the
kinetic theory prediction for massless Boltzmann
particles~\cite{Israel:1976tn,Baier:2006um}.

\begin{figure}[t]
\vspace{-0.0cm} \centering
\includegraphics[width=8cm,height=60mm,angle=0]{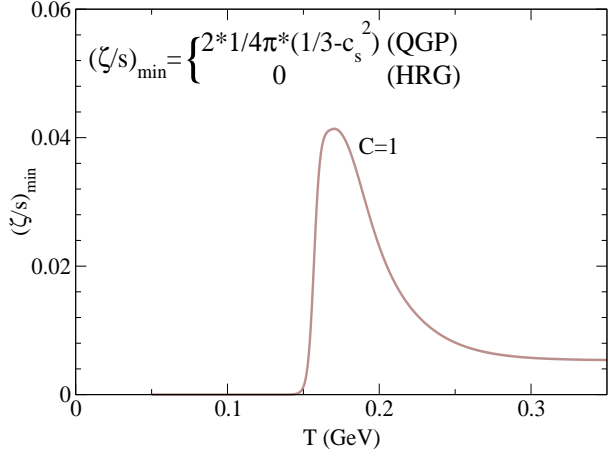}
\vspace{-3mm} \caption[Minimally constructed bulk viscosity to
entropy ratio: $\zeta/s$]{Minimally constructed bulk viscosity to
entropy ratio, $\zeta/s$, as a function of temperature.}
\vspace{0.2cm} \label{BulkVis}
\end{figure}

The bulk viscosity of the QCD matter is still under theoretical
development (see Chap.~1.4, 1.5 for details). It is generally
believed that $\zeta/s$ peaks near the phase
transition~\cite{Paech:2006st,Karsch:2007jc}. However, one finds
that the minimal peak value from AdS/CFT
predictions~\cite{Gubser:2008sz,Gubser:2008yx} is more than 10 times
smaller than the one extracted from lattice QCD data
\cite{Meyer:2007dy,Kharzeev:2007wb,Karsch:2007jc}\footnote{\emph{Refer
to Ref.~\cite{Moore:2008ws} for a critical discussion of the lattice
QCD approach.}}. Things become even more complicated for the bulk
viscosity in the hadronic phase~\cite{Kapusta:2008vb}. When
neglecting hadron masses, chiral perturbation theory predicts a
rising $(\zeta/s) (T)$ below $T_c$~\cite{Chen:2007kx}. In contrast,
a theory with non-zero hadron masses shows a decreasing $(\zeta/s)
(T)$ below $T_c$~\cite{Prakash:1993bt,Davesne:1995ms}. Considering
these theoretical uncertainties, $\zeta/s$ here is treated as a free
input. We concentrate on the bulk viscosity effects near the phase
transition, but totally neglect  bulk viscosity effects in the
hadronic matter. Fig.~\ref{BulkVis} shows the minimally constructed
$\left( \zeta/s \right)_{min}$, which is obtained by connecting the
minimal strong coupling AdS/CFT result $\zeta/s=2 (\eta/s)
(1/3-c_s^2)$~\cite{Buchel:2007mf} above $T_c$ through a Gaussian
function peaked at $T_c$ with a zero value in the hadronic phase
(the speed of sound, $c_s^2(T)$, is evaluated from the same lattice
QCD data~\cite{Katz:2005br} that are used in our EOS L). To simulate
effects from larger bulk viscosity, we multiply the entire function
$(\zeta/s)_{min}(T_c)$ by a constant $C>1$.  For the bulk relaxation
time $\tau_\Pi$, there are no standard theoretical results (see,
however, the recent work~\cite{Buchel:2009hv}). In Chap. 6.2, we set
$\tau_\Pi=\tau_\pi= \frac 3 { 4 \pi T} $ just for the purpose of a
qualitative comparison of shear and bulk viscous effects without
claim of quantitative accuracy. In Chap. 6.3 and Chap. 6.4, we will
investigate the effects of critical slowing down during the phase
transition and corresponding bulk viscosity effects. There we will
set $\tau_\Pi$ to be \textbf{a)} a temperature-independent constant
that we change it from 0.1 fm/c to 10 fm/c; \textbf{b)} a
temperature dependent function
$\tau_\Pi(T)=\max[\tilde{\tau}\cdot\frac{\zeta}{s}(T),
0.1\,\mathrm{fm}/c]$ with $\tilde{\tau} = 120$\,fm/$c$. This choice
implements phenomenologically the concept of critical slowing down;
it yields $\tau_\Pi\approx0.6$\,fm/$c$ at $T=350$\,MeV and
$\tau_\Pi\approx 5$\,fm/$c$ at $T_c$.
 \chapter[Generic Viscous Effects: Shear Viscosity]{Generic Viscous Effects\\
 \hspace{3cm} -- Shear Viscosity}
\section{Introduction}

In this chapter, we will study  generic shear viscous effects by
comparing ideal and viscous hydrodynamic runs. The results shown
here are for Au+Au collisions and complement those from our two
early viscous hydrodynamics papers~\cite{Song:2007fn,Song:2007ux}
where we presented results for the smaller Cu+Cu systems. The choice
of the Cu+Cu system was purely technical since the numerical grid in
early VISH2+1 could not accommodate the larger Au+Au collision
fireballs. This problem was solved in an up dated version of VISH2+1
with a new treatment of the boundary. However, since a similar
detailed and systematic investigation for the shear viscous effects
for Au+Au collisions have not been documented anywhere else, I
decided to redo the graphs from~\cite{Song:2007ux} for Au+Au
collisions and include them in this thesis, instead of using the old
Cu+Cu graphs.

As discussed in Ref.~\cite{Song:2008si}, the shear viscous effects
are qualitatively similar in Cu+Cu collisions and Au+Au collisions.
However, on a quantitative level, the smaller Cu+Cu system shows
larger viscous effects~\cite{Song:2008si}. This manifests itself
through a much larger viscous $v_2$ suppression for non-central
Cu+Cu collisions than for non-central Au+Au collisions.  The new
Au+Au results were calculated with the updated VISH2+1, which solves
the ``full I-S equation" rather than the simplified I-S equation
used in our early papers~\cite{Song:2007fn,Song:2007ux}, and
organized along lines similar to Ref.~\cite{Song:2008si} for
comparison and later reference. As briefly mentioned in Chap.~2.7,
the ``full" I-S equation is preferred since it preserves conformal
symmetry for a conformal fluid and reduces the dependence on the
relaxation time $\tau_\pi$. Systematic numerical comparisons between
the  ``full" and ``simplified" I-S equations were presented in
Ref.~\cite{Song:2008si}, and will be discussed later in {Chap. 7}.

Let me begin with a description of the default settings for the free
parameters (see Chap. 3) used in the calculations for this chapter.

For ideal and viscous hydrodynamic comparison runs, we use identical
initial and final conditions, with the same standard parameters
previously used in early ideal hydrodynamic simulations. Following
Refs.~\cite{Kolb:1999it,Kolb:2000fha},  we use a Glauber
initialization for the energy density and set $e_0 (\tau_0, b=0,
r=0)=30 \ \mathrm{GeV/fm^3}$ at $\tau_0=0.6 \ \mathrm{fm/c}$ for
central Au+Au collisions. The freeze-out surface is extracted using
the AZHYDRO decoupling algorithm~\cite{AZHYDRO} with a constant
decoupling temperature $T_{dec} =130 \ \mathrm{MeV}$, and the final
particle spectra are calculated with the modified Cooper-Fye
formula, given by eqs. (\ref{Cooper}-\ref{vis-correction}) in Chap
3.4. In the viscous hydrodynamic calculations, we use the KSS value
for the shear viscosity $\eta/s=1/4\pi$~\cite{Kovtun:2004de}, and
set bulk viscosity to zero: $\zeta/s=0$. For the relaxation time, we
use $\tau_\pi=3\eta/sT$, and the shear stress tensor is initialized
by the N-S initialization $\pi^{mn}(\tau_0)=2\eta\sigma^{mn}$,
unless otherwise noted (e.g. in Chap.~4.4, where we explore the
effects from varying the relaxation time and for shear stress tensor
initialization).

To isolate effects introduced by the phase transition from generic
shear viscous effects, we perform calculations with two different
EOS, using EOS I for a pure massless quark-gluon gas without phase
transition and SM-EOS~Q for a more realistic EOS that includes a
phase transition between  QGP and hadron gas. Results from the
lattice-QCD based EOS (EOS~L), can be found in Chap.~6, where we
also study the the bulk viscous effects. \\

%
\begin{figure}[t]
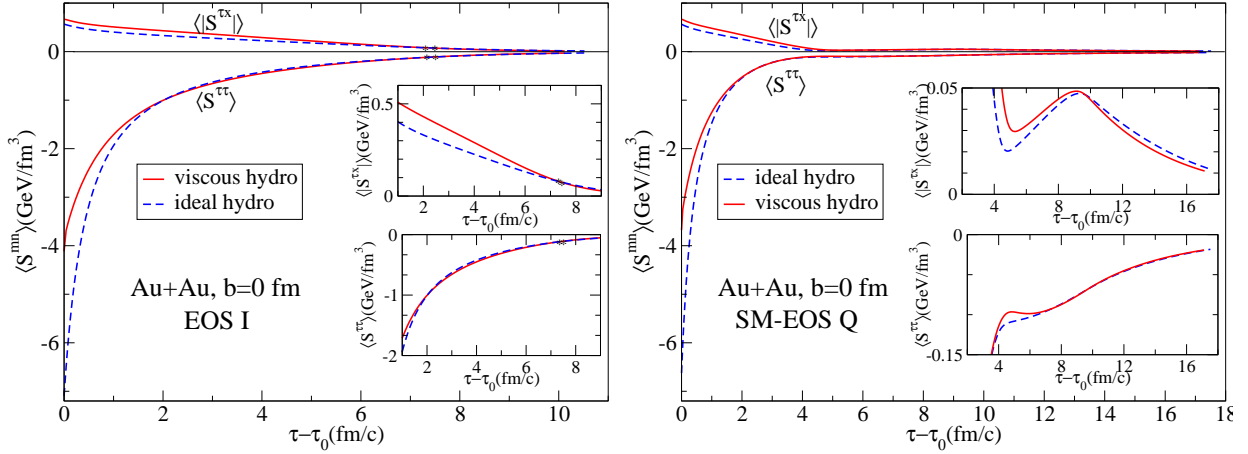

\includegraphics[bb=1 32 710 557,width=.49\linewidth,clip=]{./Figs/Chap4/EOSI/Sc30b0.eps}
\includegraphics[bb=1 32 710 557,width=.49\linewidth,clip=]{./Figs/Chap4/EOSQ/Sc30b0.eps}
\vspace{-3mm} \caption[Time evolution of the hydrodynamic source
terms]{Time evolution of the hydrodynamic source terms
(\ref{S00}-\ref{S02}), averaged over the transverse plane, for
central Au+Au collisions, calculated with EOS~I in the left panel
and with SM-EOS~Q in the right panel. The smaller insets blow up the
vertical scale to show more detail. The dashed blue lines are for
ideal hydrodynamics with $e_0\eq30$\,GeV/fm$^3$ and
$\tau_0\eq0.6$\,fm/$c$. Solid red lines show results from viscous
hydrodynamics with identical initial conditions and
$\frac{\eta}{s}\eq\frac{1}{4\pi}{\,\approx\,}0.08$,
$\tau_\pi\eq\frac{3\eta}
{sT}{\,\approx\,}0.24\left(\frac{200\,\mathrm{MeV}}{T}\right)$\,fm/$c$.
The stars in left panel indicate the freeze-out times at $T_c = 130
\ \mathrm{MeV}$. The positive source terms drive the transverse
expansion while the negative ones affect the longitudinal expansion.
} \label{source}
\end{figure}
%

\section{Hydrodynamic evolution}
\label{chap4-hydro-evo}

\textbf{\underline {Central collisions}:}\\[-0.10in]

Even without initial transverse flow, the N-S initialization
$\pi^{mn}=2\eta \sigma^{mn}$ generates three non-zero values for the
components of the shear viscous pressure tensor: $\tau^2\pi^{\eta
\eta}\eq\frac{-4\eta}{3\tau_0}$,
$\pi^{xx}\eq\pi^{yy}\eq\frac{2\eta}{3\tau_0}$, due to the
boost-invariant longitudinal expansion. Inspection of the source
terms in Eqs.~(\ref{S00}-\ref{S02}) reveals that the initially
negative $\tau^2\pi^{\eta \eta}$ effectively reduces the
longitudinal pressure, thus reducing the cooling rate, while the
initially positive values of $\pi^{xx}$ and $\pi^{yy}$ effectively
increase the transverse pressure and accelerate the development of
transverse flow in $x$ and $y$ directions. As the fireball evolves,
the shear viscous pressure tensor deviates from its N-S value
(although for small relaxation time $\tau_\pi$, the deviation
remains small~\footnote{\emph{These deviations from the N-S limit
are investigated in Sec.\ref{SimFullI-S}.}}), and the stress tensor
$\sigma^{mn}$ receives additional contributions involving the
transverse flow velocity and its derivatives (see
Eq.~(\ref{tilde-sigma})), which renders an analytic discussion of
its effects on dynamics impractical.

%
\begin{figure}[t]
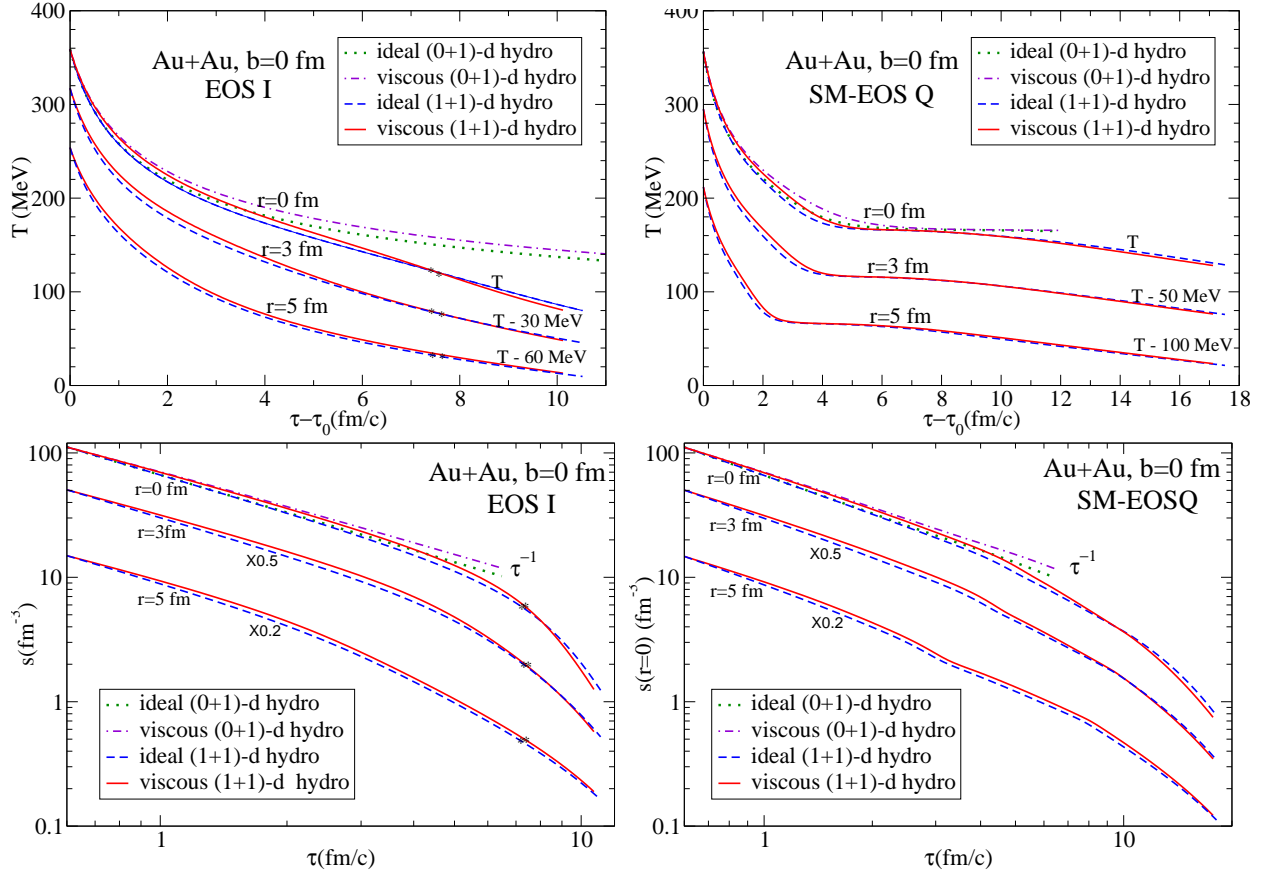

\includegraphics[bb=25 33 720 531,width=.49\linewidth,clip=]{./Figs/Chap4/EOSI/Temp.eps}
\includegraphics[bb=25 33 720 522,width=.49\linewidth,clip=]{./Figs/Chap4/EOSQ/Temp-2.eps}
\includegraphics[bb=30 42 710 524,width=.49\linewidth,clip=]{./Figs/Chap4/EOSI/entopy.eps}
\includegraphics[bb=30 42 710 522,width=.49\linewidth,clip=]{./Figs/Chap4/EOSQ/entopy.eps}
\vspace{-3mm}\caption[Time evolution of local temperature and
entropy density]{Time evolution of the local temperature  and local
entropy density (down) in central Au+Au collisions, calculated with
EOS~I  and SM-EOS~Q (right), for the center of the fireball
($r\eq0$, upper set of curves) and points at $r\eq3$\,fm and
$r\eq5$\,fm (middle and lower set of curves). For clarity, the
temperature curve for $r\eq3$\,fm and $r\eq5$\,fm have been lowered
by constant offsets, as indicated. Same parameters as in
Fig.~\ref{source}. See text for discussion. } \label{localT}
\end{figure}
%

Figure~\ref{source} shows what one gets numerically. Plotted are the
source terms (\ref{S00}) and (\ref{S01}), averaged over the
transverse plane with the energy density as weight function, as a
function of time, for evolution of central Au+Au collisions with two
different equations of state, EOS~I and SM-EOS~Q. (In central
collisions $\langle|{\cal S}^{\tau x}|\rangle\eq\langle|{\cal
S}^{\tau y}|\rangle$.) One sees that the initially strong  viscous
reduction of the (negative) source term ${\cal S}^{\tau\tau}$, which
controls the cooling by longitudinal expansion, quickly disappears.
This is due to a combination of effects: while the magnitude of
$\tau^2\pi^{\eta\eta}$ decreases with time, its negative effects are
further compensated by a growing positive contribution
$\tau\bigl(\partial_x(pv_x){+}\partial_y(pv_y)\bigl)$ arising from
the increasing transverse flow gradients. In contrast, the viscous
increase of the (positive) transverse source term ${\cal S}^{\tau
x}$ persists much longer, until about 6\,fm/$c$. After that time,
however, the viscous correction switches sign (clearly visible in
the upper inset in the right panel of Fig.~\ref{source}b) and turns
negative, thus reducing the transverse acceleration at late times
relative to the ideal fluid case. We can summarize these findings by
stating that shear viscosity reduces longitudinal cooling mostly at
early times while causing initially increased but later reduced
acceleration in the transverse direction. Due to the general
smallness of the viscous pressure tensor components at late times,
the last-mentioned effect (reduced acceleration) is not very strong.

The phase transition in SM-EOS~Q is seen to cause an interesting
non-monotonic behaviour of the time evolution of the source terms
(right panel in Fig.~\ref{source}), leading to a transient increase
of the viscous effects on the longitudinal source term while the
system passes through the mixed phase.

The viscous slowing of the cooling process at early times and the
increased rate of cooling at later times due to accelerated
transverse expansion are shown in Figure~\ref{localT} by the time
evolution of local temperatures. For comparison we also show curves
for boost-invariant longitudinal Bjorken expansion without
transverse flow, labeled ``(0+1)-d hydro''. These are obtained with
flat initial density profiles for the same value $e_0$ (no
transverse gradients). The dotted green line in the left panel shows
the well-known $T{\,\sim\,}\tau^{-1/3}$ behaviour of the Bjorken
solution of ideal fluid dynamics \cite{Bjorken:1982qr}, modified in
the right panel by the quark-hadron phase transition where the
temperature stays constant in the mixed phase.

The dash-dotted purple line shows the slower cooling in the viscous
(0+1)-dimensional case \cite{Gyulassy85}, due to reduced work done
by the longitudinal pressure. The expansion is still boost-invariant
a la Bjorken \cite{Bjorken:1982qr} (as it is for all other cases
discussed in this thesis), but viscous effects generate entropy,
thereby keeping the temperature at all times higher than for the
adiabatic case. The dashed blue (ideal) and solid red (viscous)
lines for the azimuthally symmetric (1+1)-dimensional case show the
additional cooling caused by transverse expansion. Again the cooling
is initially slower in the viscous case (solid red), but at later
times, due to faster build-up of transverse flow by the viscously
increased transverse pressure, the viscous expansion is seen to cool
the fireball center {\em faster} than ideal hydrodynamics. (Note
also the drastic reduction of the lifetime of the mixed phase by
transverse expansion; due to increased transverse flow and continued
acceleration in the mixed phase from viscous pressure gradients, it
is even more dramatic in the viscous than the ideal case.)

The lower panels of Fig.~\ref{localT} show the evolution of the
entropy density. (In the QGP phase $s{\,\sim\,}T^3$.) The
double-logarithmic presentation emphasizes the effects of viscosity
and transverse expansion on the power law
$s(\tau){\,\sim\,}\tau^{-\alpha}$: One sees that the $\tau^{-1}$
scaling of the ideal Bjorken solution is flattened by viscous
effects, but steepened by transverse expansion. As is well-known, it
takes a while (here about 6-8\,fm/$c$, depending on the EOS) until
the transverse rarefaction wave reaches the fireball center and
turns the initially 1-dimensional longitudinal expansion into a
genuinely 3-dimensional one. When this happens, the power law
$s(\tau){\,\sim\,}\tau^{-\alpha}$ changes from $\alpha\eq1$ in the
ideal fluid case to $\alpha>3$~\cite{Kolb:2003dz}. Here 3 is the
dimensionality of space, and the fact that $\alpha$ becomes larger
than 3 reflects relativistic Lorentz-contraction effects through the
transverse-flow-related $\gamma_\perp$-factor that keeps increasing
even at late times. In the viscous case, $\alpha$ changes from 1 to
3 sooner than for the ideal fluid, due to the faster growth of
transverse flow. At very late times (not shown in Fig.~\ref{localS}
) the $s(\tau)$ curves for ideal and viscous hydrodynamics are
almost perfectly parallel~\cite{Song:2007ux}, indicating that very
little entropy is produced during this late stage.

%
\begin{figure}[t]
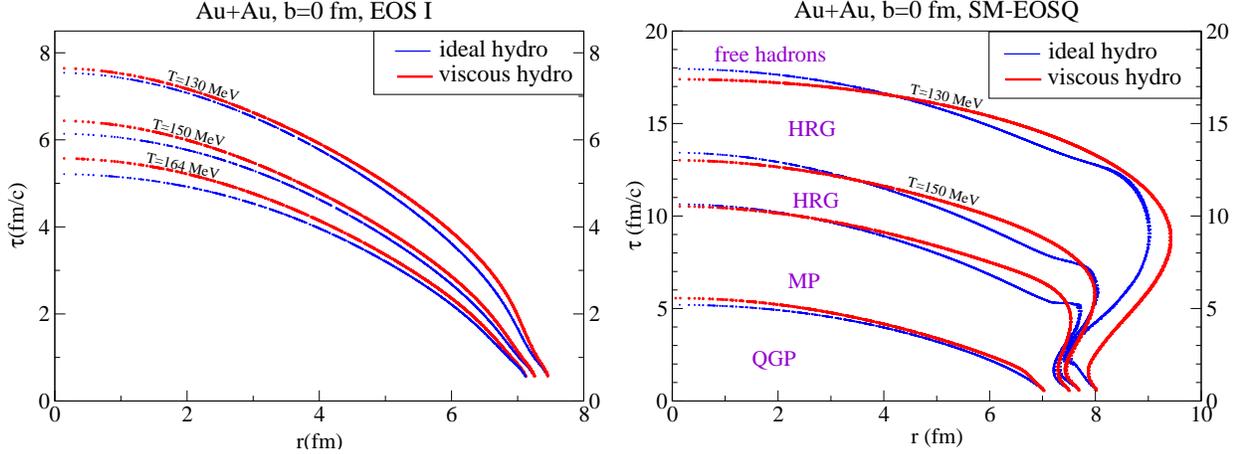

\includegraphics[bb=36 37 740 581,width=.49\linewidth,clip=]{./Figs/Chap4/Contour/EOS2_Contour-b.eps}
\includegraphics[bb=36 37 740 581,width=.49\linewidth,clip=]{./Figs/Chap4/Contour/Contour-b.eps}
\vspace*{-3mm} \caption[Constant temperature $T$ contours]{Surfaces
of constant temperature $T$  for central Au+Au collisions, evolved
with EOS~I (left panel) and SM-EOS~Q (right panel). In each panel,
results from viscous hydrodynamics (red lines) are directly compared
with the corresponding ideal fluid evolution (blue lines).  The
right panel contains two isotherms for $T_c\eq164$\,MeV, one
separating the mixed phase (MP) from the QGP at energy density
$e_\mathrm{Q}\eq1.6$\,GeV/fm$^3$, the other separating it from the
hadron resonance gas (HRG) at energy density
$e_\mathrm{H}\eq0.45$\,GeV/fm$^3$. See text for discussion. }
\label{Contourb0}
\end{figure}
%

%
\begin{figure}[t]
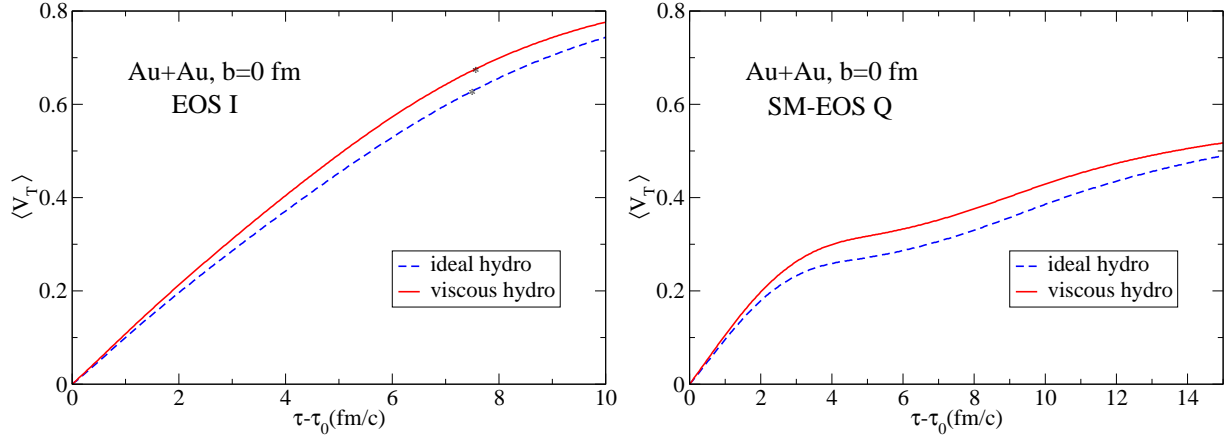

\includegraphics[bb=20 33 718 530,width=.49\linewidth,clip=]{./Figs/Chap4/EOSI/velocity.eps}
\includegraphics[bb=20 33 718 530,width=.49\linewidth,clip=]{./Figs/Chap4/EOSQ/velocity.eps}
\vspace*{-3mm}\caption[Time evolution of average radial flow
velocity]{Time evolution of the average radial flow velocity
$\langle v_T \rangle{\,\equiv\,}\langle v_\perp\rangle$ in central
Au+Au collisions, calculated with EOS~I (left panel) and SM-EOS~Q
(right panel). Solid (dashed) lines show results from ideal
(viscous) fluid dynamics. The initially faster rate of increase
reflects large positive shear viscous pressure in the transverse
direction at early times. The similar rates of increase at late
times indicate the gradual disappearance of shear viscous effects.
In the right panel, the curves exhibit a plateau from 2 to
4\,fm/$c$, reflecting the softening of the EOS in the mixed phase. }
\label{velocitySp}
\end{figure}
%

In Fig.~\ref{Contourb0} we plot the evolution of temperature in
$r{-}\tau$ space, in the form of constant-$T$ surfaces, obtained
from ideal hydrodynamics (blue lines) and viscous hydrodynamics (red
lines), respectively. Again the two panels compare the evolution
with EOS~I (left) to the one with SM-EOS~Q (right). As already
noted,  at $r\eq0$ the viscous fluid cools initially more slowly
(thereby giving somewhat longer life to the QGP phase) but later
more rapidly (thereby freezing out earlier for SM-EOS~Q). Comparing
the two sets of temperature contours shown in the right panel of
Fig.~\ref{Contourb0}, one sees that viscous effects tend to smoothen
any structures related to the (first order) phase transition in
SM-EOS~Q. The reason for this is that, with the rapid change of the
speed of sound at either end of the mixed phase, the radial flow
velocity profile develops dramatic structures at the QGP-MP and
MP-HRG interfaces~\cite{Song:2007ux,Kolb:1999it}. This leads to
large velocity gradients across these interfaces, inducing large
viscous pressures which work to reduce these gradients. In effect,
shear viscosity softens the first-order phase transition into a
smooth but rapid cross-over transition. These same viscous pressure
gradients cause the fluid to accelerate even in the mixed phase
where all thermodynamic pressure gradients vanish (and where the
ideal fluid therefore does not generate additional flow). As a
result, the lifetime of the mixed phase is shorter in viscous
hydrodynamics, as also seen in the right panel of
Figure~\ref{Contourb0}.

Fig.~\ref{velocitySp} plots the time evolution of radial velocity
$\La v_T \Ra$ (radial flow), calculated as an average over the
transverse plane with the Lorentz contracted energy density
$\gamma_\bot \ e$ as weight function, for ideal and viscous fluids.
One finds that radial flow builds up more quickly in the viscous
fluid. The shear viscosity driven acceleration is strongest in the
early and middles stages, but almost negligible in the later
hadronic stage, where the ideal and viscous fluid lines are almost
parallel with each other. Compared with the old radial flow figure
for central Cu+Cu collisions in Ref.~\cite{Song:2007ux} (see Fig. 6
there), one finds a slightly larger increase in radial flow for the
smaller Cu+Cu system, which can be interpreted as larger viscous
effects for smaller systems.  Such system size effects are more
significant for
non-central collisions, which will be discussed below.\\[-0.05in]

\textbf{\underline{Non-central collisions}:}\\[-0.10in]

\begin{figure}[t]
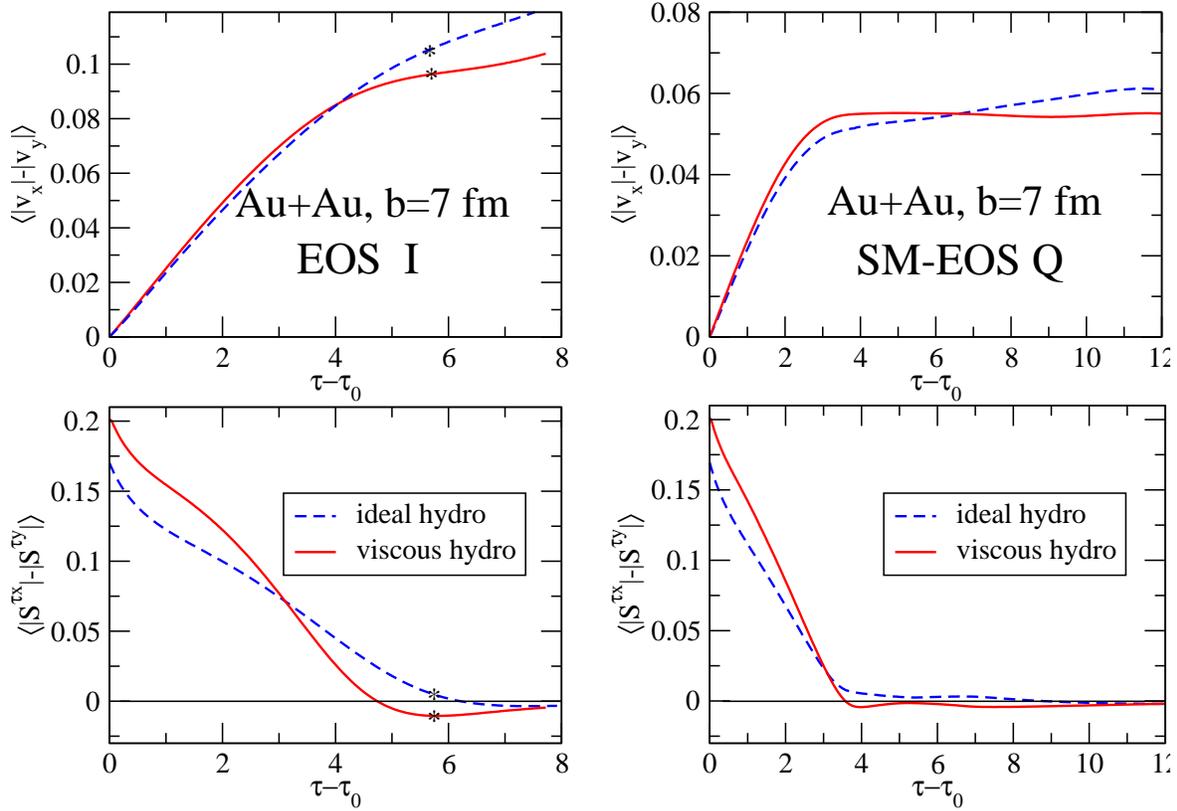

\centering
\includegraphics[bb=33 43 367 528,width=.45\linewidth,clip=]{./Figs/Chap4/EOSI/VxSc.eps}
~\hspace{2mm}
\includegraphics[bb=33 43 367 528,width=.45\linewidth,clip=]{./Figs/Chap4/EOSQ/VxSc.eps}
~\vspace{-3mm}\caption[Time evolution of transverse flow and source
term anisotropy]{Time evolution of the transverse flow anisotropy
$\La|v_x|{-}|v_y|\Ra$ (top row) and of the anisotropy in the
transverse source term $\La|{\cal S}^{\tau x}|{-}|{\cal S}^{\tau
y}|\Ra$ (bottom row). Both quantities are averaged over the
transverse plane, with the Lorentz-contracted energy density
$\gamma_\perp e$ as weight function. The left (right) column shows
results for EOS~I (SM-EOS~Q), with solid (dashed) lines representing
ideal (viscous) fluid dynamical evolution. } \label{VxSc}
\end{figure}

\begin{figure}[t]
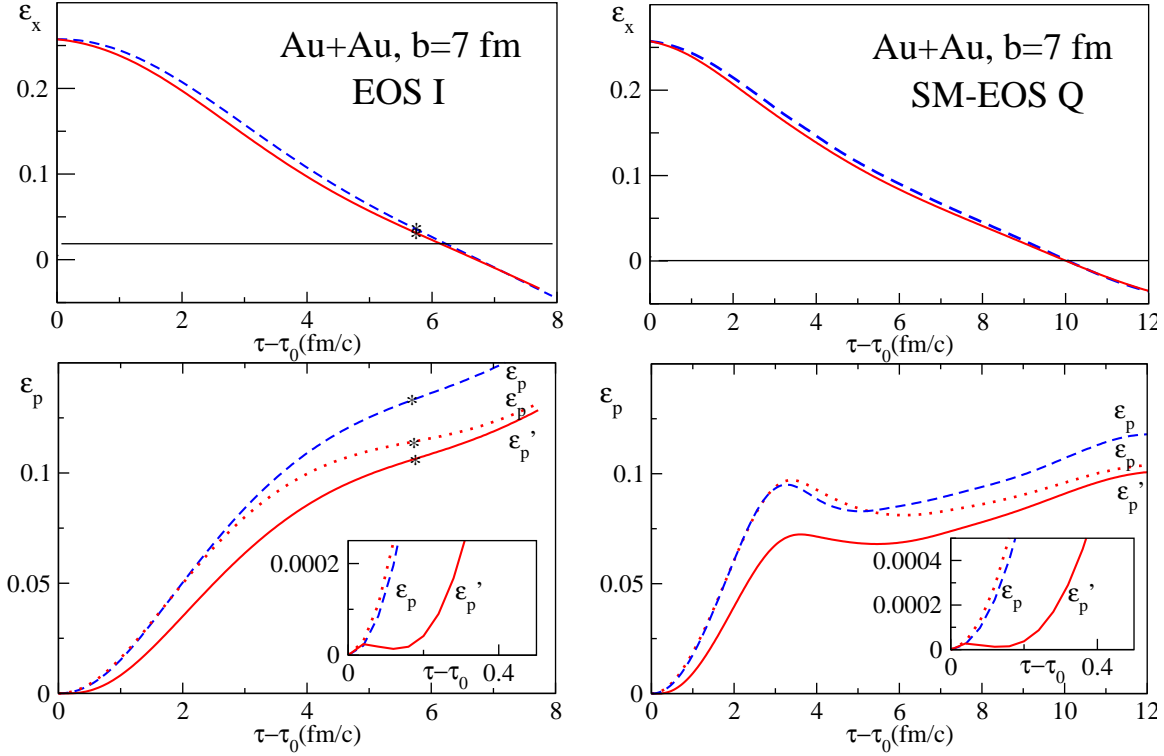

\centering
\includegraphics[bb=56 8 436 521,width=.45\linewidth,clip=]{./Figs/Chap4/EOSI/ExEp2-Eos2.eps}
\hspace{2mm}
\includegraphics[bb=55 8 436 521,width=.45\linewidth,clip=]{./Figs/Chap4/EOSQ/ExEp2.eps}
~\vspace{-3mm}\caption[Time evolution of $\epsilon_x$, $\epsilon_p$
and $\epsilon_p'$]{Time evolution for the spatial eccentricity
$\epsilon_x$, momentum anisotropy $\epsilon_p$ and total momentum
anisotropy $\epsilon_p'$ (see text for definitions), calculated for
$b\eq7$\,fm Au+Au collisions with EOS~I (left column) and SM-EOS~Q
(right column). Dashed lines are for ideal hydrodynamics while the
solid and dotted lines show results from viscous hydrodynamics. See
text for discussion. } \label{ExEp}
\end{figure}

We now turn to non-central collisions and take full advantage of the
ability of VISH2+1 to solve the transverse expansion in 2 spatial
dimensions. To evaluate the anisotropic evolution of the fireball,
we will study the time evolution of flow anisotropy
$\La|v_x|{-}|v_y|\Ra$ (Fig \ref{VxSc}) and the time evolution of
``spatial eccentricity" and ``momentum anisotropy" (Fig.\ref{ExEp}).
The spatial eccentricity~\cite{Ollitrault:1992bk, Kolb:2000sd}
characterizes the spatial deformation of the fireball in the
transverse plane; it is defined as $\epsilon_x\eq\frac{\La
x^2{-}y^2\Ra}{\La x^2{+}y^2\Ra}$ (here $\La ... \Ra$ means averaging
over the transverse plane with the energy density $e(x)$ as weight
function).  The momentum anisotropy $\epsilon_p\eq\frac{\La
T^{xx}_0{-}T^{yy}_0\Ra}{\La T^{xx}_0{+}T^{yy}_0\Ra}$ (where $\La ...
\Ra$ means average over the transverse
plane)~\cite{Ollitrault:1992bk, Kolb:2000sd} measures the anisotropy
of the transverse momentum density due to anisotropies in the
collective flow pattern, which includes only the ideal fluid part of
the energy momentum tensor. In viscous hydrodynamics, we also define
the {\em total momentum anisotropy} $\epsilon'_p\eq\frac{\La
T^{xx}{-}T^{yy}\Ra}{\La T^{xx}{+}T^{yy}\Ra}$, similarly defined in
terms of the total energy momentum tensor
$T^{\mu\nu}\eq{T}_0^{\mu\nu}{+}\pi^{\mu\nu}$, which additionally
counts anisotropic momentum contributions arising from the viscous
pressure tensor. Since the latter quantity includes effects arising
from the deviation $\delta f$ of the local distribution function
from its thermal equilibrium form which, according to
Eq.~(\ref{Cooper}), also affects the final hadron momentum spectrum
and elliptic flow, it is this {\em total momentum anisotropy} that
should be studied in viscous hydrodynamics if one wants to
understand the evolution of hadron elliptic flow. In other words, in
viscous hydrodynamics hadron elliptic flow is not simply a measure
for anisotropies in the collective flow velocity pattern, but
additionally reflects anisotropies in the local rest frame momentum
distributions, arising from deviations of the local momentum
distribution from thermal equilibrium and thus being related to the
viscous pressure.

%
%

Fig.~\ref{VxSc} shows the time evolutions of the flow anisotropy
$\La|v_x|{-}|v_y|\Ra$ and source term anisotropy $\La|{\cal S}^{\tau
x}|$  $ {-}|{\cal S}^{\tau y}|\Ra$. In central collisions these
quantities vanish. In ideal hydrodynamics, the flow anisotropy is
driven by the anisotropic gradients of the thermodynamic pressure.
In viscous fluid dynamics, the source terms (\ref{S01},\ref{S02}),
whose difference is shown in the bottom row of Fig.~\ref{VxSc},
receive additional contributions from gradients of the viscous
pressure tensor which contribute their own anisotropies.
Fig.~\ref{VxSc} demonstrates that these additional anisotropies {\em
increase} the driving force for anisotropic flow at very early times
($\tau{-}\tau_0{\,<\,}3$\,fm/$c$), but {\em reduce} this driving
force throughout the later evolution. At times
$\tau{-}\tau_0{\,>\,}5$\,fm/$c$ the anisotropy of the effective
transverse pressure even changes sign and turns negative, working to
{\em decrease} the flow anisotropy. As a consequence of this, the
buildup of the flow anisotropy stalls around
$\tau{-}\tau_0{\,\approx\,}4$\,fm/$c$ (even earlier for SM-EOS~Q
where the flow buildup stops as soon as the fireball medium enters
the mixed phase) and proceeds to slightly decrease therafter. This
happens during the crucial period where ideal fluid dynamics still
shows strong growth of the flow anisotropy. By the time the fireball
matter decouples, the average flow velocity anisotropy of viscous
hydro lags about 10-15\% behind the value reached during ideal fluid
dynamical evolution.

%
%

Figure~\ref{ExEp} correlates the decrease in time of the spatial
eccentricity $\epsilon_x$ with the buildup of the momentum
anisotropies $\epsilon_p$ and $\epsilon'_p$. In viscous dynamics the
spatial eccentricity is seen to decrease initially faster than for
ideal fluids, due to the faster radial expansion shown in
Fig.~\ref{velocitySp}. The effects of early pressure gradient
anisotropies are reflected in the initial growth rate of the flow
anisotropy (Fig.~\ref{VxSc}) and the momentum anisotropy
$\epsilon_p$ (bottom panels in Fig.~\ref{ExEp}) which are seen to
slightly exceed that observed in the ideal fluid at times up to
about 3\,fm/$c$ after the beginning of the transverse expansion.
Figure~\ref{VxSc} also shows that in the viscous fluid the flow
velocity anisotropy stalls about 2\,fm/$c$ (for SM-EOS~Q) after
start and remains about 10\% below the final value reached in ideal
fluid dynamics. This causes the spatial eccentricity of the viscous
fireball to decrease more slowly at later times than that of the
ideal fluid (top panels in Fig.~\ref{ExEp}) which, at late times,
features a significantly larger difference between the horizontal
($x$) and vertical ($y$) expansion velocities. We note that the
differences between ideal and viscous evolution were found to be
significantly larger in the smaller Cu+Cu systems studied in
Ref.~\cite{Song:2007ux}.

It is very instructive to compare the behaviour of the flow-induced
ideal-fluid contribution to the momentum anisotropy, $\epsilon_p$,
with that of the total momentum anisotropy $\epsilon'_p$. At early
times they are very different, with $\epsilon'_p$ being much smaller
than $\epsilon_p$ (see insets in the lower panels of
Fig.~\ref{ExEp}). This reflects very large {\em negative}
contributions to the anisotropy of the total energy momentum tensor
from the shear viscous pressure whose gradients along the
out-of-plane direction $y$ strongly exceed those within the reaction
plane along the $x$ direction. At early times this effect almost
compensates for the larger in-plane gradient of the thermal
pressure. The {\em negative} viscous pressure gradient anisotropy
(which is even larger in the smaller Cu+Cu systems studied
in~\cite{Song:2007ux} ) is responsible for reducing the growth of
flow anisotropies, thereby causing the flow-induced momentum
anisotropy $\epsilon_p$ to significantly lag behind its ideal fluid
value at later times. The negative viscous pressure anisotropies
responsible for the difference between $\epsilon_p$ and
$\epsilon'_p$ slowly disappear at later times, since all viscous
pressure components then become very small (see Fig.~\ref{AverPi}
below).

The net result of this interplay is a total momentum anisotropy in
Au+Au collisions (i.e. a source of elliptic flow $v_2$) that for a
``minimally'' viscous fluid with $\frac{\eta}{s}\eq\frac{1}{4\pi}$
is 20-25\% lower than for an ideal fluid, at all except the earliest
times (where it is even smaller). The origin of this reduction
changes with time: Initially it is dominated by strong momentum
anisotropies in the local rest frame, with momenta pointing
preferentially out-of-plane, induced by deviations from local
thermal equilibrium and associated with large shear viscous
pressure.
At later times, the action of these anisotropic viscous pressure
gradients integrates to an overall reduction in collective flow
anisotropy, while the viscous pressure itself becomes small; at this
stage, the reduction of the total momentum anisotropy is indeed
mostly due to a reduced anisotropy in the collective flow pattern
while momentum isotropy in the local fluid rest frame is
approximately restored.\\

\section{Final particle spectra and elliptic flow $v_2$} \label{shear-spectraV2}
\textbf{\underline{Spectra for central collisions:}}\\[-0.10in]

After obtaining the freeze-out surface, we calculate the particle
spectra from the generalized Cooper-Frye formula (\ref{Cooper}),
using the AZHYDRO algorithm \cite{AZHYDRO} for the integration over
the freeze-out surface $\Sigma$. For calculations with EOS~I which
lacks the transition from massless partons to hadrons, we cannot
compute any hadron spectra. For illustration we instead compute the
spectra of hypothetical massless bosons (``gluons''). They can be
compared with the pion spectra from SM-EOS~Q which can also, to good
approximation, be considered as massless bosons.

%
\begin{figure}[t]
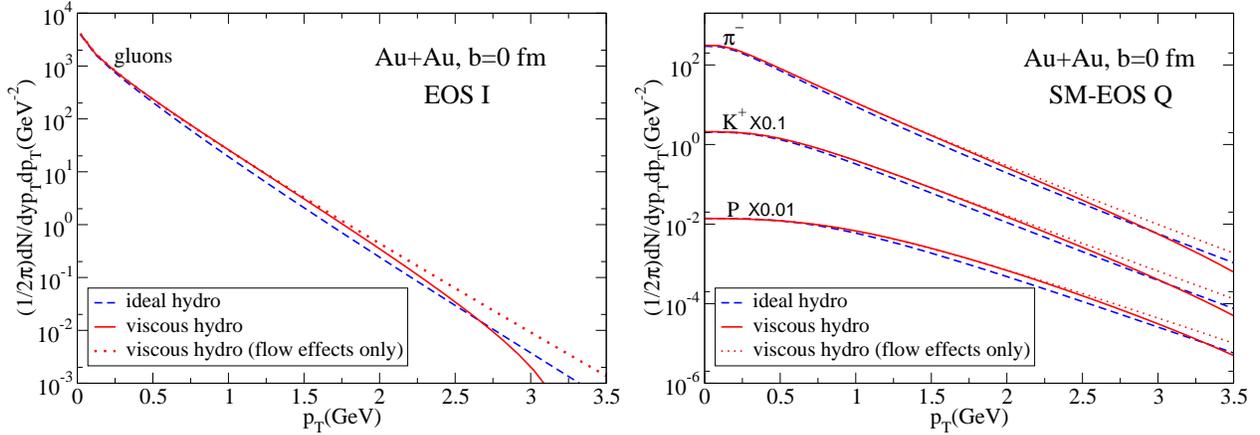

\includegraphics[width=.495\linewidth,clip=]{./Figs/Chap4/EOSI/Spectrab0.eps}
\includegraphics[width=.495\linewidth,clip=]{./Figs/Chap4/EOSQ/Spectrab0.eps}
\vspace*{-3mm} \caption[Mid-rapidity particle spectra for central
Au+Au collisions] {Mid-rapidity particle spectra for central Au+Au
collisions, calculated with EOS~I (left, gluons) and with SM-EOS~Q
(right, $\pi^-$, $K^+$ and $p$). The solid blue (red dashed) lines
are from ideal (viscous) hydrodynamics. The purple dotted lines show
viscous hydrodynamic spectra that neglect the viscous correction
$\delta f_i$ to the distribution function in Eq.~(\ref{Cooper}),
i.e. include only the effects from the larger radial flow generated
in viscous hydrodynamics. } \label{Spectra}
\end{figure}
%

The larger radial flow generated in viscous hydrodynamics, for a
fixed set of initial conditions (shown in Fig.~\ref{velocitySp}),
leads to flatter transverse momentum spectra
\cite{Chaudhuri:2005ea,Baier:2006gy,Teaney:2004qa} (at least at low
$p_T$ where the viscous correction $\delta f_i$ to the distribution
function can be neglected in (\ref{Cooper})). This is seen in
Figure~\ref{Spectra}, by comparing the dotted and dashed lines. This
comparison also shows that the viscous spectra lie systematically
above the ideal ones, indicating larger final total multiplicity.
This reflects the creation of entropy during the viscous
hydrodynamic evolution. As pointed out in
\cite{Chaudhuri:2005ea,Baier:2006gy}, this requires a retuning of
initial conditions (starting the hydrodynamic evolution later with
smaller initial energy density) if one desires to fit a given set of
experimental $p_T$-spectra. Since we here concentrate on
investigating the origins and detailed mechanics of viscous effects
in relativistic hydrodynamics, we will not explore any variations of
initial conditions. All comparisons between ideal and viscous
hydrodynamics presented here will use identical starting times
$\tau_0$ and initial peak energy densities $e_0$. The solid lines in
Figure~\ref{Spectra} show that in our calculations for
$p_T{\,\gtrsim\,}2$\,GeV/$c$ the effects from $\delta f_i$ have an
overall negative sign, leading to a reduction of the $p_T$-spectra
at large $p_T$ relative to both the viscous spectra without $\delta
f_i$ and the ideal hydrodynamic spectra. This is true for all
particle species, irrespective of the EOS used to evolve the fluid.

However, the earlier blast-wave model estimate from Teaney
\cite{Teaney:2003kp} found that the correction is positive, growing
quadratically with $p_T$. A later calculation by Dusling and Teaney,
using causal viscous hydrodynamics in the O-G formalism and  a
different (kinetic) freeze-out criterium to determine the decoupling
surface, also found a (small) positive effect from $\delta f_i$ on
the final pion spectra, at least up to $p_T\eq2$\,GeV/$c$, for
freeze-out around $T_\mathrm{dec}\sim130$\,MeV, turning weakly
negative when their effective freeze-out temperature was lowered to
below 100 MeV~\cite{Dusling:2007gi}. Inspecting
Eqs.~(\ref{Cooper},\ref{vis-correction}) reveals that the viscous
correction $\delta f_i$ in Eqs.~(\ref{Cooper},\ref{deltaf}) depends
on the signs and magnitudes of the various viscous pressure tensor
components along the freeze-out surface, weighted by the equilibrium
part $f_{\mathrm{eq},i}$ of the distribution function. Its effect on
the final $p_T$-spectra (even its sign!) is not {\em a priori}
obvious. In Appendix~\ref{hydro-blastwave}, we will explore the
origin of the discrepancy of the sign of $\delta f$ between the
different groups by investigating the effects of different
freeze-out surfaces in hydrodynamics and in the simple blast wave
model.

%
\begin{figure}[t]
\centering{\includegraphics[width=0.53\linewidth,height=62mm,clip=]{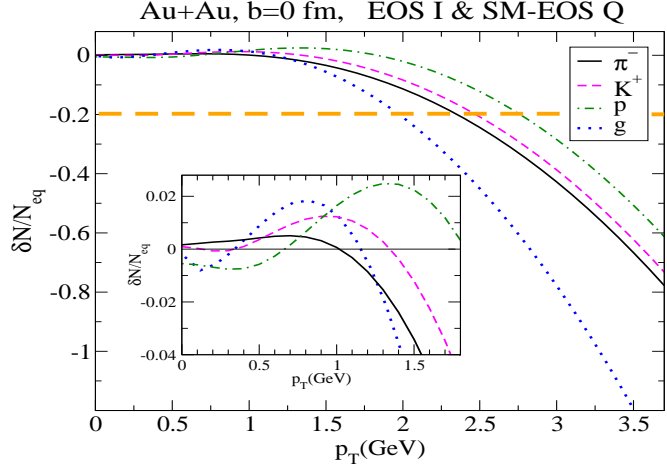}}
\vspace*{-3mm} \caption[Ratio of the viscous correction $\delta
dN/dN^0$]{Ratio of the viscous correction $\delta N$, resulting from
the non-equilibrium correction $\delta f$, Eq.~(\ref{deltaf}), to
the distribution function at freeze-out, to the equilibrium spectrum
$N_\mathrm{eq}{\,\equiv\,}dN_\mathrm{eq}/(dyd^2p_T)$ calculated from
Eq.~(\ref{Cooper}) by setting $\delta f\eq0$. The gluon curves are
for evolution with EOS~I, the curves for $\pi^-$, $K^+$ and $p$ are
from calculations with SM-EOS~Q. The horizontal dashed line at 0.2
indicates breakdown of the viscous hydrodynamic approach.}
\label{Spectra-Correction}
\end{figure}
%

In Figure~\ref{Spectra-Correction} we show the non-equilibrium
contribution to the final hadron spectra in greater detail. The
figure shows that the non-equilibrium effects from $\delta f_i$ are
largest for massless particles and, at high $p_T$, decrease in
magnitude with increasing particle mass. The assumption $|\delta
f|{\,\ll\,}f_\mathrm{eq}$, which underlies the viscous hydrodynamic
formalism, is seen to break down at high $p_T$, but to do so later
for heavier hadrons than for lighter ones. Once the correction
exceeds ${\cal O}(20\%)$ (indicated by the horizontal dashed line in
Fig.~\ref{Spectra-Correction}), the calculated spectra can no longer
be trusted.

In contrast to viscous hydrodynamics, ideal fluid dynamics has no
intrinsic characteristic that will tell us when it starts to break
down. Comparison of the calculated elliptic flow $v_2$ from ideal
fluid dynamics with the experimental data from
RHIC~\cite{Kolb:2003dz} suggests that the ideal fluid picture begins
to break down above $p_T{\,\simeq\,}2.5$\,GeV/$c$ for pions and
above $p_T{\,\simeq\,}3$\,GeV/$c$ for protons. This phenomenological
hierarchy of thresholds where viscous effects appear to become
essential is qualitatively consistent with the mass hierarchy from
viscous hydrodynamics shown in Fig.~\ref{Spectra-Correction}.

In the region $0{\,<\,}p_T{\,\lesssim\,}1.5$\,GeV/$c$, the interplay
between $m_T$- and $p_T$-dependent terms in
Eq.~(\ref{vis-correction}) is subtle, causing sign changes of the
viscous spectral correction depending on hadron mass and $p_T$ (see
inset in Fig.~\ref{Spectra-Correction}). The fragility of the sign
of the effect is also obvious from Fig.~8 in the work by Dusling and
Teaney \cite{Dusling:2007gi} where it is shown that in this $p_T$
region the viscous correction changes sign from positive to negative
when freeze-out is shifted from earlier to later times (higher to
lower freeze-out temperature). Overall, we agree with them that the
viscous correction effects on the $p_T$-spectra are weak in this
region \cite{Dusling:2007gi}. We will see below that a similar
statement does not hold for the elliptic flow. \\[-0.05in]

%
\begin{figure}[t]
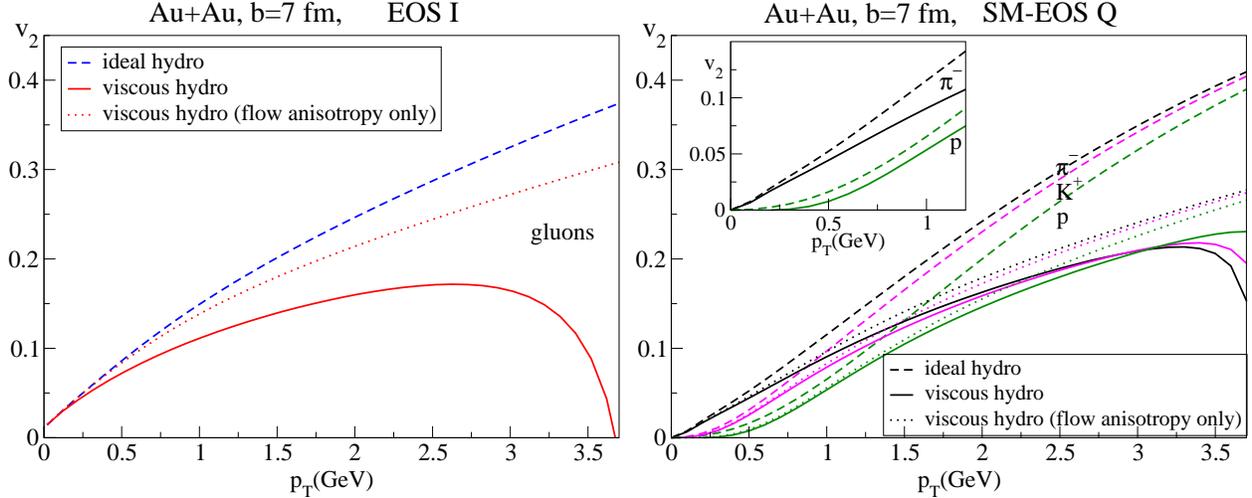
\vspace{3mm}
\includegraphics[width=.495\linewidth,clip=]{./Figs/Chap4/EOSI/V2-2.eps}
\includegraphics[width=.495\linewidth,clip=]{./Figs/Chap4/EOSQ/V2-2.eps}
\vspace*{-3mm} \caption[Differential elliptic flow $v_2(p_T)$ for
for Au+Au collisions at $b\eq7$\,fm]{Differential elliptic flow
$v_2(p_T)$ for Au+Au collisions at $b\eq7$\,fm. {\sl Left panel:}
Gluons from evolution with EOS~I. {\sl Right panel:} $\pi^-$, $K^+$,
and $p$ from evolution with SM-EOS~Q. Dashed lines: ideal
hydrodynamics. Solid lines: viscous hydrodynamics. Dotted lines:
viscous hydrodynamics without non-equilibrium distortion $\delta f$
of distribution function at freeze-out. } \label{v-2}
\end{figure}
%

\textbf{\underline{Elliptic flow for non-central collisions}}\\[-0.10in]

The shear viscous effects on the suppression of the total momentum
anisotropy $\epsilon'_p$ discussed in the last section also reflect
themselves in a suppression of the final particle elliptic flow
$v_2$, which is shown in Figure~\ref{v-2}. Even for the ``minimal''
viscosity $\frac{\eta}{s}\eq\frac{ 1}{4\pi}$ considered here one
sees a very strong suppression of the differential elliptic flow
$v_2(p_T)$ from a viscous fluid (solid lines) compared to the ideal
fluid (dashed lines). Both the viscous reduction of the collective
flow anisotropy (whose effect on $v_2$ is shown as the dotted lines)
and the viscous contributions to the anisotropy of the local
momentum distribution (embodied in the term $\delta f$ in
Eq.~(\ref{Cooper})) play big parts in this reduction. The runs with
EOS~I (which is a very hard EOS) decouple more quickly than those
with SM-EOS~Q; correspondingly, the viscous pressure components are
still large at freeze-out, and the viscous correction $\delta f$ to
the distribution function plays a bigger role. With SM-EOS~Q the
fireball doesn't freeze out until $\pi^{mn}$ has become very small
(see Fig.~\ref{AverPi} below), resulting in much smaller corrections
from $\delta f$ (difference between dashed and dotted lines in
Fig.~\ref{v-2}). On the other hand, due to the longer fireball
lifetime the negatively anisotropic viscous pressure has more time
to decelerate the buildup of anisotropic flow, so $v_2$ is strongly
reduced because of the much smaller flow-induced momentum anisotropy
$\epsilon_p$.

The net effect of all this is that, for Au+Au collisions and in the
soft momentum region $p_T{\,<\,}1.5$\,GeV/$c$, the viscous evolution
with $\frac{\eta}{s}\eq\frac{1}{4\pi}$ leads to a $\sim 30\%$
suppression of $v_2$ for pions, in both the slope of its
$p_T$-dependence and its $p_T$-integrated value. (Due to the flatter
$p_T$-spectra from the viscous dynamics, the effect in the
$p_T$-integrated $v_2$ is not quite as large as for $v_2(p_T)$ at
fixed $p_T$.)  For protons, the larger radial flow generated in
the viscous evolution also pushes the proton elliptic flow out
towards larger $p_T$ (see inserts in the right panel of Fig.\ref{v-2}),
causing an additional suppression of $v_2^{p}(p_T)$ at fixed $p_T$. This effect
is much weaker in Cu+Cu collisions in~\cite{Song:2007ux} where, due to
the small fireball size, the radial flow effects are less pronounced.  \\

\section{Sensitivity to relaxation time and $\pi^{mn}$  initialization}

\textbf{\underline{Initialization of $\pi^{mn}$}:}\\[-0.10in]

Lacking input from a microscopic model of the pre-equilibrium stage
preceding the (viscous) hydrodynamic one, one must supply initial
conditions for the energy momentum tensor, including the viscous
pressure $\pi^{mn}$. The most popular choice has been to initialize
$\pi^{mn}$ with its Navier-Stokes value, i.e. to set initially
$\pi^{mn}\eq2\eta\sigma^{mn}$. Up to this point, this has also been
our choice in Chapters 4.2 and 4.3. Ref.~\cite{Romatschke:2007mq}
advocated the choice $\pi^{mn}\eq0$ at time $\tau_0$ in order to
minimize viscous effects and thus obtain an upper limit on $\eta/s$
in a comparison with experimental data. In the present subsection we
explore the sensitivity of the final spectra and elliptic flow to
these different choices of initialization, keeping all other model
parameters unchanged.

%
\begin{figure}[t]
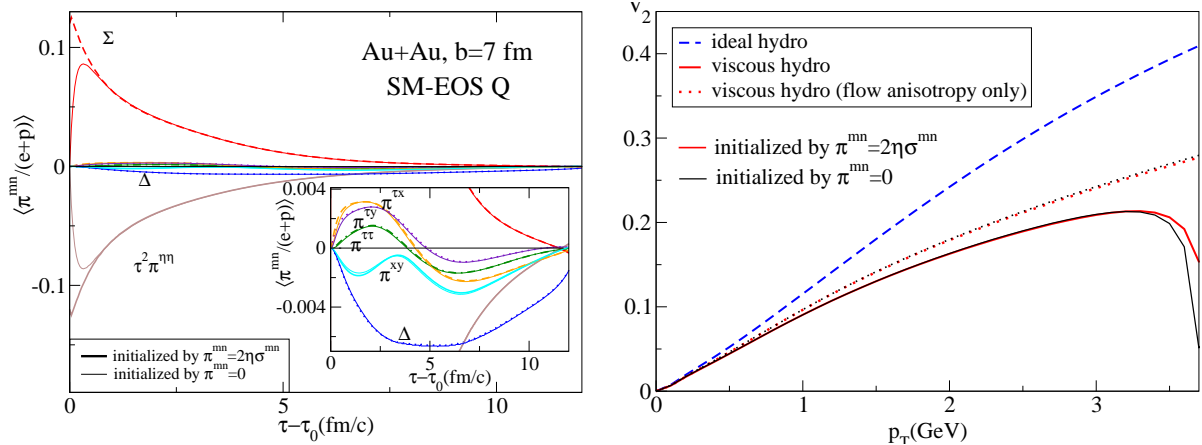

\centering \vspace{3mm}
\includegraphics[bb=19 20 692 530,width=.468\linewidth,clip=]{./Figs/Chap4/Zero-Relax/Pievo5.eps}
\includegraphics[bb=25 33 711 530,width=.49\linewidth,clip=]{./Figs/Chap4/Zero-Relax/V2-Zero.eps}
\vspace*{-3mm} \caption[Time evolution of shear viscous pressure
tensor and elliptic flow $v_2(p_T)$]{Left: Time evolution of the
various components of the shear viscous pressure tensor, normalized
by the enthalpy and averaged in the transverse plane over the
thermalized region inside the freeze-out surface (note that the
normalizing factor $e{+}p{\,\sim\,}T^4$ decreases rapidly with
time), compared with different $\pi^{mn}$ initial conditions. Right:
Differential elliptic flow $v_2(p_T)$ for pions from $b\eq7$\,fm
Cu+Cu collisions with SM-EOS~Q, compared with different $\pi^{mn}$
initial conditions. The thick lines in both figures are for N-S
initialization: $\pi^{mn}(\tau_0)\eq2\eta \sigma^{mn}$, while thin
lines are for zero initialization: $\pi^{mn}(\tau_0)\eq0$.
\label{AverPi}}
\end{figure}
%

Figure~\ref{AverPi} left  shows the time evolution of the viscous
pressure tensor, comparing the two different initializations
$\pi^{mn}(\tau_0)\eq2\eta \sigma^{mn}$ and  $\pi^{mn}(\tau_0)\eq0$.
Differences are visible only at early times
$\tau{-}\tau_0{\,\lesssim\,}5 \tau_\pi{\,\approx\,}1$\,fm/$c$. After
$\tau_\pi{\,\sim\,}0.2$\,fm/$c$, the initial differences have
decreased by roughly a factor $1/e$, and after several kinetic
scattering times $\tau_\pi$ the hydrodynamic evolution has
apparently lost all memory how the viscous terms were initialized.

Correspondingly, the final spectra and elliptic flow show very
little sensitivity to the initialization of $\pi^{mn}$, as seen in
Fig.~\ref{AverPi} right. With vanishing initial viscous pressure,
viscous effects on the final flow anisotropy are a little weaker
(dotted lines in Fig.~\ref{AverPi} right), but this difference is
overcompensated in the total elliptic flow by slightly stronger
anisotropies of the local rest frame momentum distributions at
freeze-out (solid lines in Fig.~\ref{AverPi} right). For shorter
kinetic relaxation times $\tau_\pi$, the differences resulted from
different initializations of $\pi^{mn}$ would be even smaller. \\[-0.05in]

\textbf{\underline{Relaxation time $\tau_\pi$}:}\\[-0.10in]

%
\begin{figure}[t]
  \begin{center}\vspace{7mm}
    \begin{minipage}[b]{0.56\linewidth}
   \includegraphics[height=1.3\linewidth, width=0.9\linewidth,clip=]{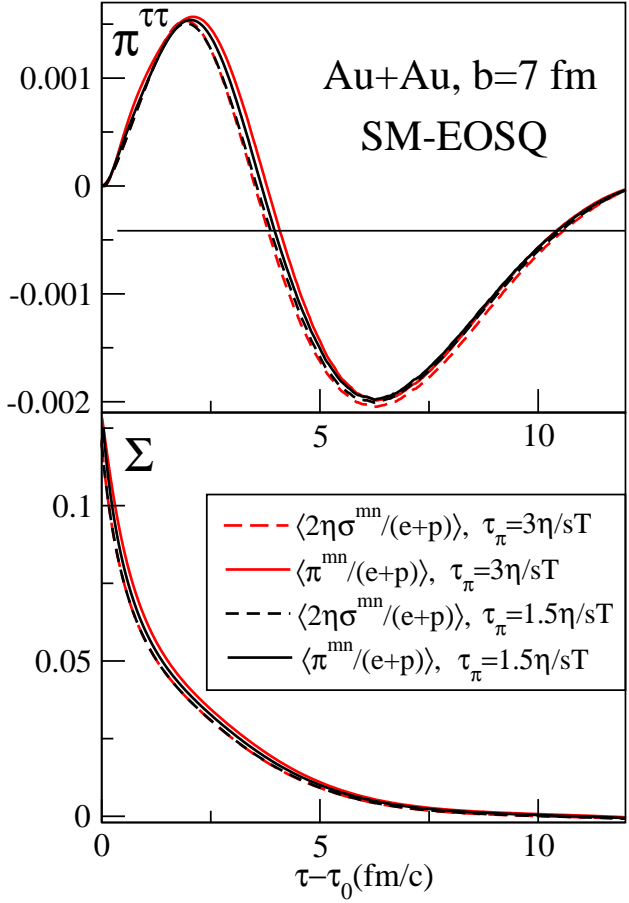}
    \end{minipage}
    \begin{minipage}[b]{0.35\linewidth}
      \caption[ Time evolution of viscous pressure tensor $\pi^{\tau\tau}$ and
         $\Sigma$ with different $\tau_\pi$]{Time evolution of the two independent
         viscous pressure tensor components $\pi^{\tau\tau}$ and
         $\Sigma\eq\pi^{xx}{+}\pi^{yy}$, for central Au+Au collisions (solid
         lines), compared with their Navier-Stokes limits
         $2\eta\sigma^{\tau\tau}$ and $2\eta(\sigma^{xx}{+}\sigma^{yy})$
         (dashed lines), for two values of the relaxation time,
         $\tau_\pi\eq3\eta/sT$ (thick red lines) and $\tau_\pi\eq1.5\eta/sT$
         (thin black lines). All quantities are scaled by the thermal equilibrium
         enthalpy $e{+}p$ and transversally averaged over the thermalized
         region inside the decoupling
         surface.\vspace*{3.2cm}\label{Relax1}}
    \end{minipage}
  \end{center}
  \vspace*{0.0cm}
\end{figure}
%

While the finite relaxation time $\tau_\pi$ for the viscous pressure
tensor in the Israel-Stewart formalism eliminates problems with
superluminal signal propagation in the relativistic Navier-Stokes
theory, it also keeps the viscous pressure from ever fully
approaching its Navier-Stokes limit $\pi^{mn}\eq2\eta\sigma^{mn}$.
In this subsection we explore how far, on average, the viscous
pressure evolved by VISH2+1 deviates from its Navier-Stokes limit,
and how this changes if we reduce the relaxation time $\tau_\pi$ by
a factor 2. (Another investigation of relaxation time effects and
the approach towards the N-S limit can be found in
Sec.~\ref{SimFullI-S}, where we compare the ``full" and
``simplified" I-S equations.)

%
\begin{figure}[t]
\centering
\includegraphics[width=0.47\linewidth,clip=]{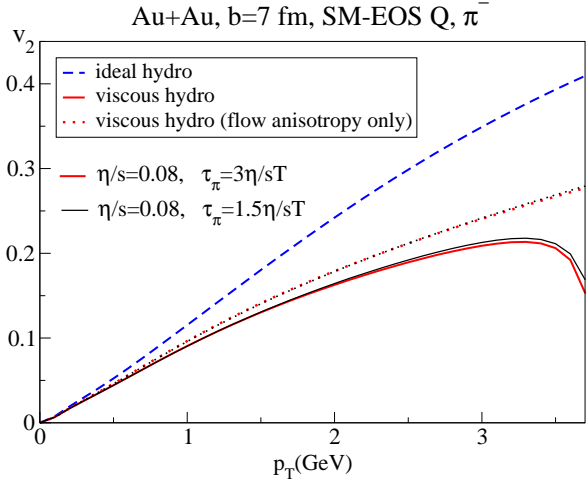}
\vspace{-3mm} \caption[$v_2(p_T)$ with different
$\tau_\pi$]{Differential elliptic flow $v_2(p_T)$ for $\pi^-$ from
$b\eq7$\,fm Au+Au collisions with SM-EOS~Q, calculated from viscous
hydrodynamics with two different values for the relaxation time
$\tau_\pi$. } \label{v-2-Beta}
\end{figure}
%

In Figure~\ref{Relax1} we compare, for non-central Au+Au collisions,
the time evolution of the scaled viscous pressure tensor, averaged
in the transverse plane over the thermalized region inside the
freeeze-out surface, with its Navier-Stokes limit, for two values of
$\tau_\pi$, $\tau_\pi\eq3\eta/sT\eq\tau_\pi^\mathrm{class}/2$ and
$\tau_\pi\eq\tau_\pi^\mathrm{class}/4$. For the larger relaxation
time, the deviation from the Navier-Stokes limit is always less than
$ 10\%$, and this fraction greatly decreases at later times. For the
twice shorter relaxation time, the fractional deviation from
Navier-Stokes decreases by somewhat more than a factor 2 and never
exceeds a value of about 5\%.

Figure~\ref{v-2-Beta} shows the corresponding elliptic flow $v_2$.
For the viscous fluid curves (both the flow anisotropy only and the
full viscous hydrodynamics curves), one finds only very slight
dependencies on relaxation times $\tau_\pi$ . This result is in
sharp contrast to what we found in~\cite{Song:2007ux} for Cu+Cu
collisions at b=7 fm, where we saw a much stronger dependence on the
relaxation time. As already pointed out in~\cite{Song:2008si}, the
main reason behind this is that the calculations presented here and
in~\cite{Song:2007ux} use different versions of the I-S equations.
The ``full" I-S equation used here greatly reduces the dependence on
$\tau_\pi$ for final observables compared to the ``simplified"  I-S
equation used in the early calculations in~\cite{Song:2007ux}, which
lead to a much stronger sensitivity to $\tau_\pi$. This will be
further
investigated in Chap.\ref{SimFullI-S}.\\

\section[Dependence on system size and collision energies]{Shear viscosity
effects: dependence on system size and collision energies}
In Ref.~\cite{Song:2008si}, we showed qualitatively that shear
viscosity effects are larger for smaller systems, by comparing the
viscous $v_2$ suppression for Au+Au collisions at b=7 fm and Cu+Cu
collisions at b=7 fm. Although the two systems have similar initial
eccentricity $\varepsilon_x$, the latter system is not a scaled
transformation of the former one due to the Woods-Saxon
parametrization in the Glauber initialization (see Chap.~3.2).
%
\begin{figure}[t]
\centering
\includegraphics[width=0.5\linewidth,clip=]{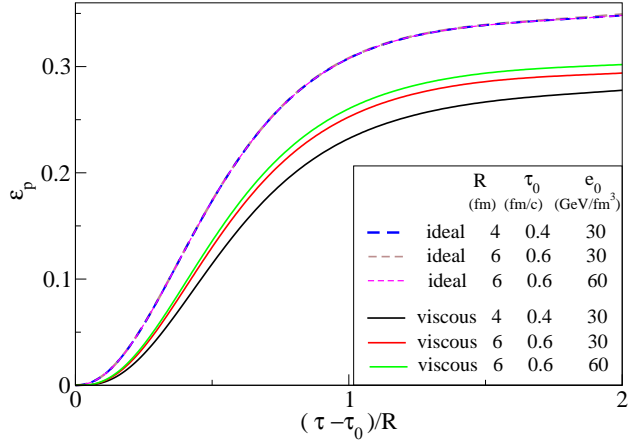}
\vspace{-3mm} \caption[$\varepsilon_p$ : dependence on system sizes
and collision energies] {Time evolution of total momentum anisotropy
$\varepsilon'_p$  from ideal and viscous hydrodynamics with
different initial conditions. For viscous hydrodynamics:
$\eta/s=0.08$ and $\tau_\pi= \frac{3\eta}{sT}$.} \label{Size-Energy}
\end{figure}
%
To quantitatively study the dependence of shear viscosity effects on
system size and collision energies, we here set the initial energy
density to a simple Gaussian function, $e(x,y, \tau_0)=e_0 \cdot
\mathrm{exp} ( - \frac{x^2}{(R/2)^2} - \frac{y^2}{R^2}) $. This
construction creates an initially elliptic fireball with
eccentricity $\varepsilon_x (\tau_0)=0.6$, independent of the
parameter $R$.  To study system size effects, we do comparison runs
with $R= 4 \ \mathrm{fm}$, $\tau_0 = 0.4 \ \mathrm{fm/c}$ and $R=6 \
\mathrm{fm}$ $\tau_0 = 0.6 \ \mathrm{fm/c}$; the second system is an
exact scale transformation of the first one in both space and time,
by a factor 1.5. To study collision energy effects, we compare
hydrodynamic evolutions starting with an initial energy density $e_0
= 30 \ \mathrm{GeV/fm^3}$ and with $e_0 = 60 \ \mathrm{GeV/fm^3}$.

Fig.~\ref{Size-Energy} shows the time evolution of total momentum
anisotropy $\varepsilon'_p$  for different initial conditions (with
ideal EOS: EOS~I). Instead of using the proper time $\tau$ for the
horizontal axis, we change it to the normalized unitless ``time"
$(\tau-\tau_0)/R$. For ideal fluid evolution, one observes exact
scale invariance, independent of system size or collision energy,
due to the scale invariance of the ideal hydrodynamic equations.
However, realistic simulations for relativistic heavy ion collisions
requires a freeze-out condition, a realistic EOS with a phase
transition, and using more complicated initial conditions (e.g. the
Glauber model). All of these introduce additional external scales
(an external time scale through the freeze-out condition, an
external energy scale through $T_c$ in the EOS, the surface
thickness in the Wood-Saxon density profile), which break the scale
invariance of ideal hydrodynamics. This will be discussed in more
detail in Chap. 5.

Microscopically, a finite shear viscosity corresponds to a finite
mean free path, which introduces an additional length scale, again
breaking the scale invariance of the hydrodynamic system. If one
examines the analytical solution (\ref{anal}) for the relativistic
Navier-Stokes equations for a (0+1)-d longitudinally boost invariant
system (see Appendix~\ref{appd2}), one finds a scale-breaking term
proportional to $\frac{\eta}{s \tau_0 T_0}$. This term shows that,
for a constant $\eta/s$, the scale breaking effect is smaller for
larger systems or higher collision energies (corresponding to larger
$\tau_0$ or $T_0$, respectively). For (2+1)-d viscous hydrodynamics,
there is no analytical solution. The numerical results in
Fig.~\ref{Size-Energy} clearly show scale breaking effects caused by
shear viscosity in the viscous $\varepsilon'_p$ suppression,
relatively to the (universal) ideal fluid lines. For identical
collision energy, one finds stronger suppression (i.e. larger scale
breaking effects) in the smaller system (comparing red and black
solid lines); for identical system size, one finds smaller viscous
$\varepsilon'_p$ suppression for higher collision energies
(comparing red and green solid lines). Both tendencies agree with
qualitative expectations based on the analytical results for the
(1+1)-d case. The system size and collision energy dependence of the
viscous $\varepsilon'_p$ suppression translates into a corresponding
dependence of the viscous $v_2$ suppression, which will be
further studied in Chap.~5.\\

\section{Breakdown of viscous hydrodynamics at high $p_T$}
\label{sec5c}

As indicated by the horizontal dashed lines in
Figs.~\ref{Spectra-Correction}, the assumption $|\delta
f|{\,\ll\,}|f_\mathrm{eq}|$ under which the viscous hydrodynamic
framework is valid breaks down at sufficiently large transverse
momenta. For a quantitative assessment we assume conservatively that
viscous hydrodynamic predictions are no longer reliable when the
viscous corrections to the particle spectra exceed 20\%.
Fig.~\ref{Spectra-Correction} shows that the characteristic
transverse momentum $p_T^*$ where this occurs depends on the
particle species and increases with particle mass. To be specific,
we here consider $p_T^*$ for pions --- the values for protons would
be about 15\% higher.

%
\begin{figure}[t]
\centering
\includegraphics[width=0.5\linewidth,clip=]{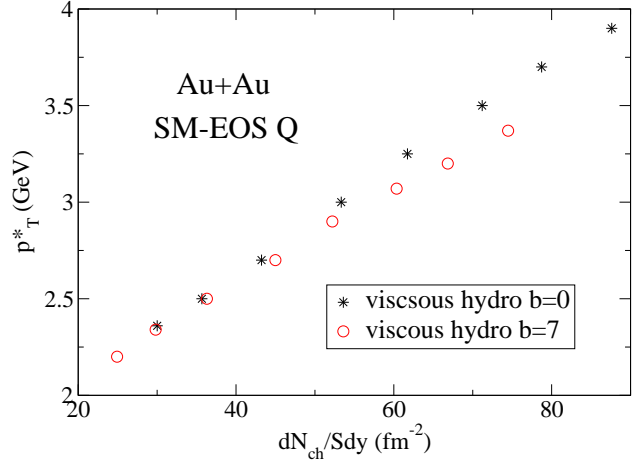}
\vspace{-3mm} \caption[Validity of viscous
hydrodynamics--characteristic transverse momentum] {Characteristic
transverse momentum $p_T^*$ where the viscous corrections to the
final pion spectrum become so large (${>\,}20\%$) that the spectrum
becomes unreliable, as a function $(1/S)dN_{ch}/dy$. Stars are for
central Au+Au collisions, open circles for semi-central Au+Au
collisions at $b\eq7$\,fm.  } \label{V2-dnN}
\end{figure}
%

In Fig.~\ref{V2-dnN} we show the breakdown momentum $p_T^*$ for
pions as a function of $(1/S)dN_{ch}/dy$ (final multiplicity
$dN_{ch}/dy$ per overlap area $S$, the details for calculating
$(1/S)dN_{ch}/dy$ can be found in Chap.~\ref{sec5}), for both
central and semi-central Au+Au collisions. (Larger $(1/S)dN_{ch}/dy$
corresponds larger collision energies. The initial time was fixed by
taking the relation $\tau_0 T_0 =\mathrm{const.}$, following
Ref.~\cite{Kestin:2008bh}.) One finds that $p_T^*$ rises with
collision energy for both central and non-central Au+Au collisions.
The rise of $p_T^*$ with increasing $e(r{=}0)$ reflects the growing
fireball lifetime which leads to smaller viscous pressure components
at freeze-out. This lifetime effect is obviously stronger for
central than for non-central collisions, leading to the faster rise
of the stars than the open circles in Fig.~\ref{V2-dnN}.  For fixed
$(1/S)dN_{ch}/dy$ (which corresponds to similar initial entropy
densities in the fireball center), we find that in central
collisions ($b=0$) the validity of viscous hydrodynamics extends to
{\em larger} values of $p_T$ than in non-central collisions ($b=7 \
\mathrm{fm}$): Viscous effects are more serious in non-central than
in central collisions.\\

\section{Conclusions}

In this chapter, we numerically studied the shear viscous effects on
the hydrodynamic evolution, final hadron spectra, and elliptic flow
$v_2$. Complementary to our early
paper~\cite{Song:2007fn,Song:2007ux}, which concentrated on Cu+Cu
collisions and used  the``simplified I-S equation", we studied Au+Au
collisions, using an updated version of VISH2+1 that solves the
``full I-S equation".

By comparing ideal and viscous hydrodynamic runs, we explored the
effects of shear viscosity for a ``minimally''~\cite{Kovtun:2004de}
viscous fluid with $\frac{\eta}{s}\eq\frac{1}{4\pi}$ in central and
non-central Au+Au collisions, comparing the evolution for two
different equations of state -- an ideal massless parton gas (EOS~I)
and an EOS with a semirealistic parametrization of the quark-hadron
phase transition (SM-EOS~Q).

We found that shear viscosity decelerates longitudinal expansion,
but accelerates the build-up of transverse flow. This slows the
cooling process initially, leading to a longer lifetime for the QGP
phase, but causes accelerated cooling at later stages by faster
transverse expansion. For SM-EOS~Q, we saw that viscous pressure
gradients during the mixed phase increase the acceleration during
this stage and slightly reduce its lifetime. They counteract large
gradients of the radial velocity profile that appear in ideal fluid
dynamics as a result of the softness of the EOS in the mixed phase,
thereby {\it de facto} smoothing the assumed first-order phase
transition of SM-EOS~Q into a rapid cross-over transition. In the
end the larger radial flow developing in viscous hydrodynamics leads
to flatter transverse momentum spectra of the finally emitted
particles, while their azimuthal anisotropy and the final elliptic
flow $v_2$ in non-central heavy-ion collisions is found to be
strongly reduced. These generic shear viscous effects are
qualitatively  similar for different systems at different collision
energies, as seen by  comparing the results for Au+Au collision
shown in this chapter with our early results for Cu+Cu collision
presented in Ref.~\cite{Song:2007ux}. On a more quantitative level,
one finds that smaller systems and lower collision energies lead to
larger viscous effects, as shown in Chap.~4.5.

Our studies show that shear viscous effects are strongest during the
early stage of the expansion phase when the longitudinal expansion
rate is largest. The discussion in Chap.~4.5 suggests that $\Gamma_s
\theta = \frac{\eta}{sT}(\partial \cdot u)$ is the parameter that
controls the strength of viscous effects.  At later times the
viscous corrections become small, although non-negligible. Small
non-zero viscous pressure components along the hadronic decoupling
surface have significant effects on the final hadron spectra that
grow quadratically with transverse momentum and thus limit the
applicability of the viscous hydrodynamic calculation to transverse
momenta below 2-3\,GeV/$c$, depending on impact parameter, collision
energy, and particle mass. Viscous effects are more important in
peripheral than in central collisions, and larger for light than for
heavy particles. Since the breakdown of viscous hydrodynamics is
signalled by the theory itself, through the relative magnitude of
the viscous pressure, the applicability of the theory can be checked
quantitatively case by case and during each stage of the expansion.

For the kinetic relaxation times $\tau_\pi$ considered in the
present work, sensitivities to the initial value of the viscous
pressure tensor were found to be small and practically negligible.
Sensitivity of the elliptic flow to the value of $\tau_\pi$ was also
found to be weak if the full I-S equation is used when solving
viscous hydrodynamics.

The most important finding here is the large $v_2$ suppression
caused by shear viscosity. While the viscous hardening of the hadron
$p_T$-spectra can be largely absorbed by retuning the initial
conditions, starting the transverse expansion later and with lower
initial entropy density \cite{Baier:2006um,Baier:2006gy}, this only
acerbates the viscous effects on the elliptic flow $v_2$ which  is
further reduced by the decreased fireball lifetime following the
retuning. The reduction of the elliptic flow $v_2$ by shear viscous
effects is therefore a sensitive and robust diagnostic tool for
shear viscosity of a fluid. This exciting finding can be used to
extract the shear viscosity from experimental data. Similar
conclusions have been obtained by several groups in the past three
years, using independently developed viscous hydrodynamic codes,
with different initial and  final conditions and different EOS input
~\cite{Song:2007fn,Song:2007ux,Romatschke:2007mq,Luzum:2008cw,Dusling:2007gi}.
The extraction of the QGP viscosity from experimental data, together
with an uncertainty analysis based on currently available  viscous
hydrodynamic studies, will be discussed in Chap. 8.
\chapter[Multiplicity Scaling of $v_2/\varepsilon$ and Shear Viscosity Effects]{Multiplicity Scaling of
$v_2/\varepsilon$\\ and Shear Viscosity Effects}
\section{Introduction}
This chapter will continue to investigate  shear viscosity effects
by studying the multiplicity  scaling of the normalized elliptic
flow $v_2/\varepsilon$ within ideal and viscous hydrodynamics. The
results compiled in this thesis are based on our
paper~\cite{Song:2008si}. The motivation for this research is
provided by the well-known systematic comparison of Voloshin {\it et
al.} \cite{Alt:2003ab,Voloshin:2007af,Voloshin:2006wi} of elliptic
flow data with ideal fluid dynamical predictions which suggests that
the elliptic flow parameter $v_2$ scaled by the initial source
eccentricity $\varepsilon$, $v_2/\ecc$, while strongly deviating
from ideal hydrodynamics at low multiplicities, still scales with
the final multiplicity $dN_{ch}/dy$ per unit overlap area $S$:
\begin{eqnarray}
\label{v2}
  \frac{v_2}{\ecc} \propto \frac{1}{S}\frac{dN_\mathrm{ch}}{dy}.
\end{eqnarray}
For ideal fluids the right hand side is a direct measure of the
initial entropy density \cite{Hwa:1985xg}. The scaling (\ref{v2})
implies that all dependence on impact parameter, collision energy
and system size can be, to good approximation, absorbed by simply
taking into account how these control parameters change the final
hadron multiplicity density. We will call this observation simply
``multiplicity scaling of the elliptic flow'', where ``elliptic
flow'' is a shorthand for the eccentricity-scaled elliptic flow
$v_2/\ecc$ and ``multiplicity'' stands for $(1/S)\dNdy$.

Such a scaling is expected for ideal fluid dynamics whose equations
of motion are scale invariant and where the eccentricity-scaled
elliptic flow is therefore predicted
\cite{Ollitrault:1992bk,Bhalerao:2005mm} to depend only on the
squared speed of sound, $c_s^2=\frac{\partial p} {\partial e}$,
which describes the stiffness of the equation of state (EOS) or
``pushing power'' of the hydrodynamically expanding matter. It has
been known, however, for many years \cite{Kolb:1999it} that this
ideal-fluid scaling is broken by the final freeze-out of the matter:
if hadron freeze-out is controlled by hadronic cross sections (mean
free paths) or simply parametrized by a critical decoupling energy
density $\edec$ or temperature $\Tdec$, this introduces an
additional scale into the problem that is independent of (or at
least not directly related to) the initial geometry of the fireball.
This breaks the above argument based on scale invariance of the
ideal fluid equations of motion. We will show here that this also
leads to a breaking of the multiplicity scaling of $v_2/\ecc$ not
only in the most peripheral or lowest energy collisions, where
freeze-out obviously cuts the hydrodynamic evolution short since the
freeze-out density is reached before the flow anisotropy can fully
build up \cite{Kolb:1999it}, but even in the most central collisions
at RHIC where freeze-out still terminates the hydrodynamic evolution
before the elliptic flow can fully saturate (see also
\cite{Heinz:2008qm,Hirano:2005xf}).

The more interesting aspect of the experimentally observed scaling
is, however, its apparent validity in regions where ideal fluid
dynamics does not work (these encompass most of the available data
\cite{Alt:2003ab}). Many years ago, simple scaling laws for the
centrality dependence of elliptic flow were derived from kinetic
theory in the dilute gas limit, where the particles in the medium
suffer at most one rescattering before decoupling
\cite{Heiselberg:1998es,Voloshin:1999gs}; these can be reinterpreted
in terms of multiplicity scaling for $v_2/\ecc$. The dilute gas
limit is expected to hold for very small collision systems, very
large impact parameters or very low collision energies. More
recently, a successful attempt was made to phenomenologically
connect the dilute gas and hydrodynamic limits with a 1-parameter
fit involving the Knudsen number \cite{Drescher:2007cd}. This fit
works very well for Au+Au and Cu+Cu data from RHIC, but predicts
that even in the most central Au+Au collisions at RHIC the ideal
fluid dynamical limit has not yet been reached and is missed by at
least 25\% \cite{Drescher:2007cd}. In this chapter, we use viscous
relativistic hydrodynamics to explore the multiplicity scaling of
$v_2/\ecc$ in the phenomenologically relevant region. We conclude
(not surprisingly since much of the available data is from regions
where the viscous hadronic phase plays a large role
\cite{Hirano:2005xf}) that the multiplicity scaling data
\cite{Alt:2003ab,Voloshin:2007af,Voloshin:2006wi} require
significant shear viscosity for the medium, especially during its
hadronic stage, but also that viscous hydrodynamics predicts subtle
scaling violations which seem to be qualitatively consistent with
trends seen in the data (even if the experimental evidence for
scaling violations is presently not statistically robust) and whose
magnitude is sensitive to the specific shear viscosity $\eta/s$.
This gives hope that future, more precise data can help constrain
the QGP shear viscosity through exactly such scaling violations.

We should caution the reader that, similar to Ref.
\cite{Drescher:2007cd} which used a constant (time-independent)
cross section for the Knudsen number, our viscous hydrodynamic
calculations are done with a constant (temperature-independent)
specific shear viscosity $\eta/s$. Neither assumption is realistic,
and we expect $\eta/s$ in particular to show strong temperature
dependence near $\Tc$ (the critical temperature for the quark-hadron
phase transition, see Chap.~1.4 for details) and emerge from the
phase transition with much larger values than in the QGP phase.
Comparisons between the results presented here and experimental data
are therefore, at best, indicative of qualitative trends, and
improved calculations, which in particular match viscous
hydrodynamics to a realistic hadron cascade below $\Tc$, are
required before quantitative extraction of $\eta/s$ from
experimental data can
be attempted.\\

\section{Multiplicity scaling of $v_2/\ecc$ in ideal and viscous
hydrodynamics} \label{sec5}

In this section we explore the multiplicity scaling (as defined in
the introduction) of the eccentricity-scaled elliptic flow
$v_2/\ecc$, comparing ideal fluid dynamics with that of
near-minimally viscous fluids with specific shear viscosity
$\frac{\eta}{s} ={\cal O}\left(\frac{1}{4\pi}\right)$ with three
different EOS: EOS I, SM-EOS Q and EOS L (see Fig.\ref{fig-EOS} and
explanation there.).  Here we consider both Au+Au collisions and
Cu+Cu collisions.  The initial conditions, final conditions and
other free inputs are described in Chap.~3. In the spectra
calculation, we neglect resonance decays and only show the elliptic
flow of directly emitted pions. To estimate the total charged hadron
multiplicity, we take the directly emitted positive pions, multiply
by 1.5 to roughly account for multiplication by resonance decays at
$\Tdec$, then multiply by another factor $2\times1.2=2.4$ to account
for the negatives and roughly 20\% of final charged hadrons that are
not pions. A proper calculation of the resonance decay chain is
computationally expensive and, for a systematic study like the one
presented here that requires hundreds of runs of  VISH2+1,
beyond our presently available resources.\\

\subsection{EOS~I: conformal fluids with $e=3p$ }
\label{sec5a}

\begin{figure}[t]
\includegraphics[width =\linewidth,height=60mm,clip=]{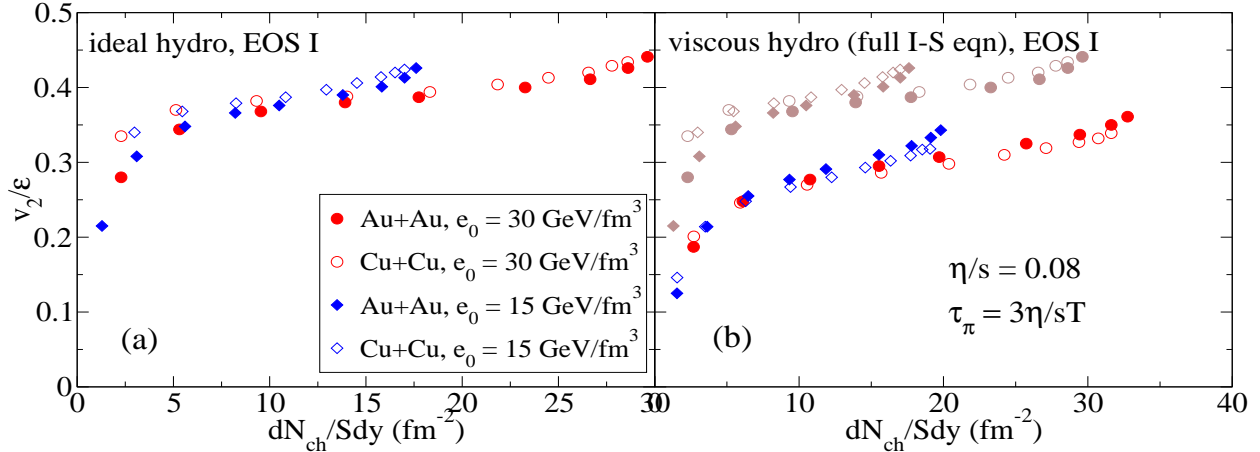}
 \caption[$v_2/\varepsilon$ as a function of
$(1/S)\dNdy$ for a conformal fluid with EOS~I]{The
eccentricity-scaled elliptic flow $v_2/\varepsilon$ as a function of
charged multiplicity density, $(1/S)\dNdy$, for a conformal fluid
with EOS~I. Results for Cu+Cu and Au+Au collisions with two
different initial energy densities at a variety of impact
parameters, as indicated in the legend, are superimposed. Results
from ideal fluid dynamics (a) are compared with those from viscous
hydrodynamics, using the full Israel-Stewart equations (b). In all
cases approximate, but not perfect multiplicity scaling is observed
(see text for discussion). In panels (b), the ideal fluid results
from the left panel are reproduced as brown symbols for comparison.
\label{F4} }
\end{figure}

We begin with the simple case of a conformal fluid with the equation
of state $e=3p$ (EOS~I), without phase transition. In this case the
speed of sound is a constant, independent of temperature $T$,
$c_s^2=\frac{1}{3}$. For the ideal fluid case, naive scaling
arguments based on the scale invariance of the ideal fluid equations
of motion would thus predict a constant $v_2/\ecc$, independent of
multiplicity density $(1/S)\dNdy$. (The nuclear overlap area $S$ is
computed as $S=\pi \sqrt{\La x^2\Ra \La y^2 \Ra}$ where
$\La\dots\Ra$ denotes the energy density weighted average over the
transverse plane, and we here use the standard the eccentricity
$\varepsilon = \frac{\La y^2- x^2 \Ra}{\La y^2+ x^2 \Ra}$, which is
slightly different from the one we used in~\cite{Song:2008si}. The
left panel in Fig.~\ref{F4} clearly contradicts this expectation.
Freeze-out at $\Tdec=130$\,MeV cuts the hydrodynamic evolution of
the momentum anisotropy $\ecc_p$ short before the elliptic flow has
fully saturated. As the left panel of the figure shows, this not
only causes a strong suppression of $v_2/\ecc$ at low multiplicity
densities, where the time between beginning of the hydrodynamic
expansion and freeze-out becomes very short, but it also breaks the
multiplicity scaling at high multiplicity density, albeit more
weakly. At a fixed value of $(1/S)\dNdy$, one sees larger $v_2/\ecc$
for more central collisions initiated at lower collision energies
(corresponding to smaller $e_0$ parameters) than for more peripheral
collisions between the same nuclei at higher beam energies, and also
for more central Cu+Cu collisions (with a rounder shape) than for
more peripheral Au+Au collisions (with a more deformed initial
shape). We find that the larger $v_2/\ecc$ values in central
compared to peripheral collisions can be traced directly to somewhat
longer lifetimes of the corresponding fireballs, i.e. to the
availability of more time to approach the saturation values of the
momentum anisotropy and elliptic flow before reaching
freeze-out~\cite{Heinz:2008qm} (see Fig.~\ref{Freeze-time} in
Appendix~\ref{Sec-cooling-rate}). These freeze-out induced scaling
violations in ideal fluid dynamics disappear at sufficiently high
collision energies (i.e. large $e_0$) where the momentum anisotropy
has time to fully saturate in {\em all} collision systems and at
{\em all} impact parameters, before freezing out (see Fig.~\ref{F5}
below).

%
%

The right panels in Fig.~\ref{F4} show the analogous results for a
minimally viscous fluid with $\frac{\eta}{s}=\frac{1}{4\pi}$ and
kinetic relaxation time $\tau_\pi=\frac{3\eta}{sT}$.  Along with the
suppression of $v_2/\ecc$ by shear viscosity, we see the appearance
of additional scale-breaking effects. Shear viscosity breaks the
multiplicity scaling of $v_2/\ecc$ because (as shown in the
preceding section) viscous effects are larger in smaller collision
fireballs. Consequently, if we compare different collision systems
that produce the same charged multiplicity density $(1/S)\dNdy$, we
find smaller $v_2/\ecc$ for Cu+Cu than for Au+Au collisions, and for
peripheral Au+Au collisions at higher collision energy than for more
central Au+Au collisions at lower collision energy. Again, the
amount of elliptic flow $v_2/\ecc$ generated for given  $(1/S)\dNdy$
correlates directly with the time  before freeze-out, as shown in
the right panel of Fig.~\ref{Freeze-time} in
Appendix~\ref{Sec-cooling-rate}.

Viscous effects also generates entropy, which increase the final
charged multiplicity $\dNdy$. Comparing in the right panels of
Fig.~\ref{F4} the brown (shaded) symbols from ideal fluid dynamics
with the colored (solid) symbols for viscous hydrodynamics, points
corresponding to the same collision system and impact parameter are
seen to be shifted to the right. This enhances the scaling
violations: for a given collision system, impact parameter and
collision energy, viscosity decreases the eccentricity scaled
elliptic flow $v_2/\ecc$, pushing the corresponding point downward
in the diagram, and simultaneously increases the entropy, pushing
the corresponding point horizontally to the right. The combination
of these two effects separates the curves for different collision
systems and energies farther in viscous hydrodynamics than in ideal
fluid dynamics.\\

\subsection{Phase transition effects: EOS~Q vs. EOS~L}
\label{sec5b}

%
\begin{figure}[t]
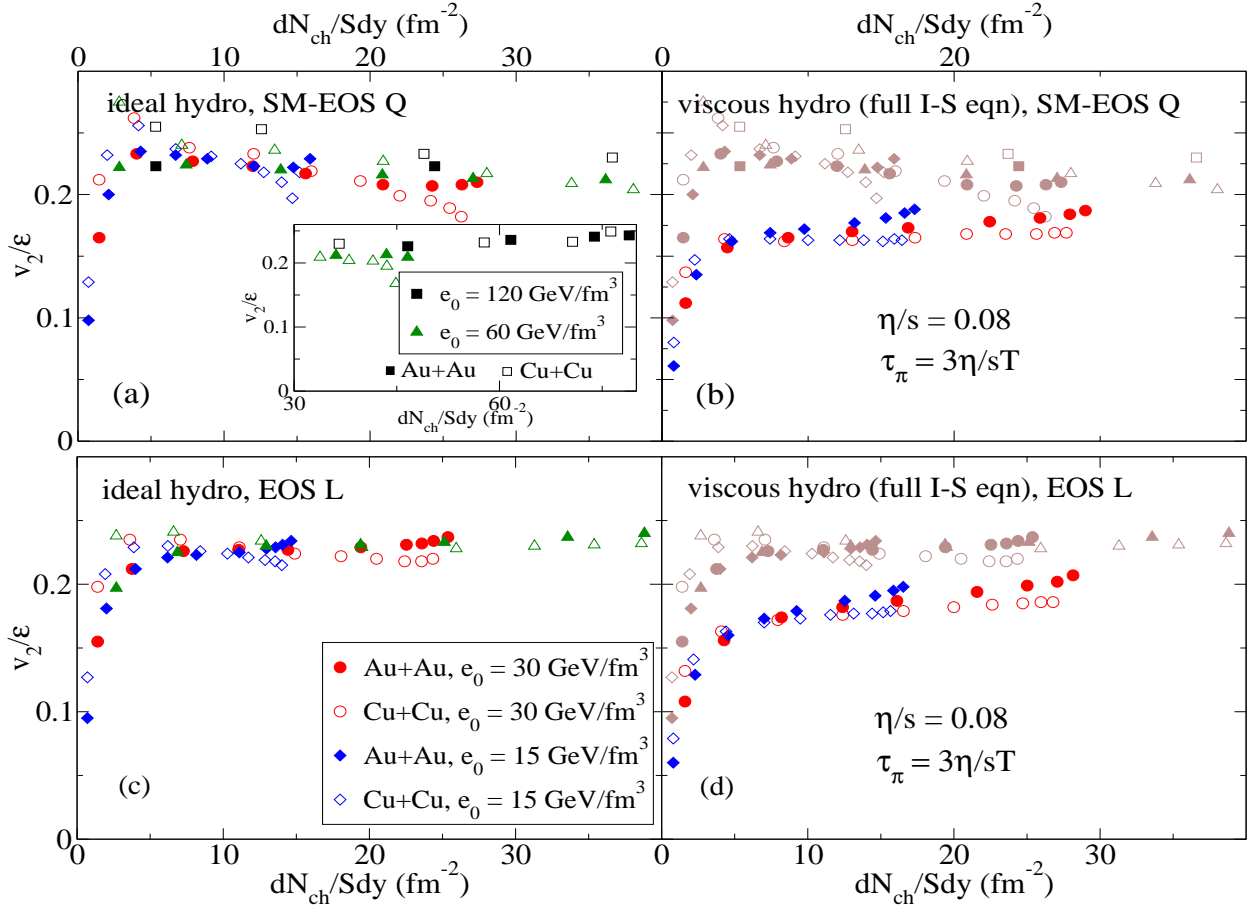

\includegraphics[width =\linewidth,height=60mm,clip=]{./Figs/Chap5/Fig5a.eps}\\
\includegraphics[width =\linewidth,height=60mm,clip=]{./Figs/Chap5/Fig5b.eps}
\caption[$v_2/\varepsilon$ as a function of $(1/S)\dNdy$ for
SM-EOS~Q and EOS~L]{Same as Fig.~\ref{F4}, but for SM-EOS~Q (top
row) and EOS~L (bottom row). For the ideal fluid case (a,c) an
extended range of $e_0$ values up to $e_0=120$\,GeV/fm$^3$ was
studied, in order to show that $v_2/\ecc$ eventually increases again
at higher collision energies \cite{Kolb:1999it}. \label{F5} }
\end{figure}
%

Figure~\ref{F5} shows the analogous results if the fluid evolves
under the influence of an equation of state with a quark-hadron
phase transition, EOS~Q (top row) or EOS~L (bottom row). Again
approximate multiplicity scaling of $v_2/\ecc$ is observed, but some
scale-breaking effects are visible in both ideal and viscous
hydrodynamics. For the equations of state with a phase transition,
the scale-breaking effects are actually larger in the ideal than in
the viscous case, i.e. {\em in viscous hydrodynamics $v_2/\ecc$
shows better multiplicity scaling than in ideal fluid dynamics!} We
interpret the large scale-breaking effects in the ideal fluid case
as a complication arising from interference between the freeze-out
process and the weak acceleration of matter in the phase transition
region. This interpretation is supported by a comparison between
SM-EOS~Q with its first-order phase transition (upper left panel in
Fig.~\ref{F5}) and the smooth crossover transition in EOS~L (lower
left panel): for ideal fluids, the scale-breaking effects are
obviously larger for SM-EOS~Q than for EOS~L. As already observed in
\cite{Song:2007fn} and in Chap.~\ref{chap4-hydro-evo}, shear
viscosity effectively smears out the phase transition and reduces
its effect on the dynamics. In Fig.~\ref{F5} this is clearly seen on
the left side of each panel (i.e. at small values of
$\frac{1}{S}\frac{dN_\mathrm{ch}}{dy}$) where for the ideal fluid
$v_2/\ecc$ shows a non-monotonic peak structure \cite{Kolb:1999it}
that is completely gone in the viscous case.

It is interesting to observe that, for ideal fluids, EOS~L leads to
about 10\% more elliptic flow under RHIC conditions than SM-EOS~Q.
The reason is that in the phase transition region EOS~L is stiffer
than SM-EOS~Q. This plays an important role at RHIC because the
softness of the EOS near $T_c$ inhibits the buildup of elliptic flow
exactly under RHIC conditions \cite{Kolb:1999it}. As a corollary we
note that, if RHIC elliptic flow data exhaust ideal fluid
predictions made with SM-EOS~Q \cite{Kolb:2003dz}, they will not
exhaust ideal fluid predictions based on EOS~L, thus leaving some
room for shear viscous effects.\\

\subsection {A look at the experimental data}
\label{sec5d}

\begin{figure}[t]
\includegraphics[bb=17 22 510 440,width=0.47\linewidth,clip=]%
                {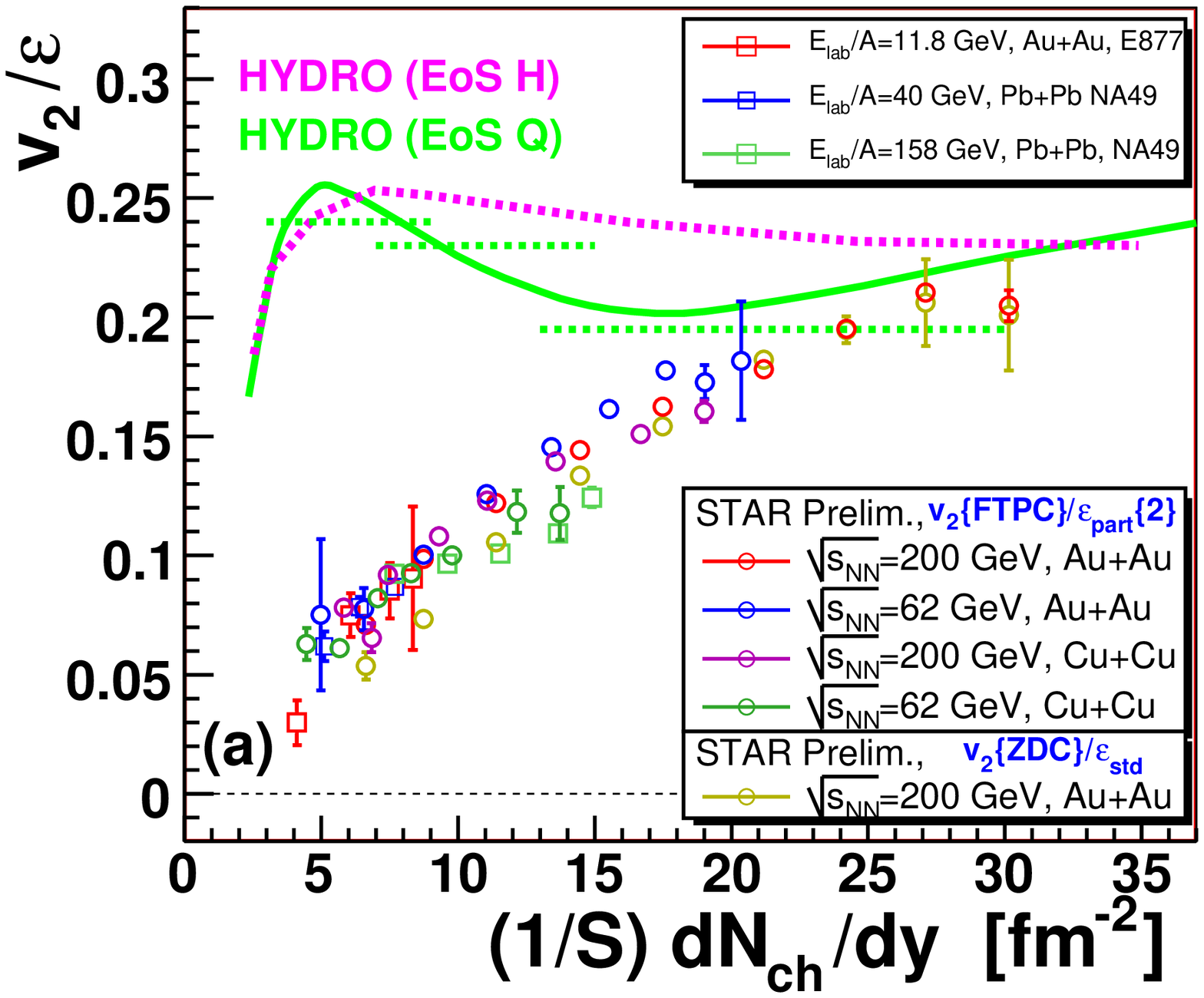}
                \hspace{2mm}
\includegraphics[width=0.49\linewidth,height=6.58cm,clip=]%
                {./Figs/Chap5/Fig7b.eps}
 \caption[$v_2/\varepsilon$ scaling: experimental data
vs. theoretical predictions]{(a) The experimental observation of
multiplicity scaling for $v_2/\ecc$, with data from Au+Au, Pb+Pb,
and Cu+Cu collisions at RHIC, SPS and AGS
\cite{Voloshin:2007af,Voloshin:2006wi}. (b) Theoretical prediction
of approximate multiplicity scaling from viscous hydrodynamics using
the full I-S equations, for three different (constant) specific
entropy values $\eta/s=0.08,\,0.16,\,0.24$. \label{F7} }
\end{figure}

Figure~\ref{F7}(a) shows the famous experimental plot by Voloshin
\cite{Voloshin:2007af,Voloshin:2006wi} which provides empirical
evidence for multiplicity scaling of $v_2/\ecc$. The lines labelled
``HYDRO'' are sketches for expectations from rough ideal fluid
dynamics, based on early calculations presented in
\cite{Kolb:1999it} for $v_2$ in Au+Au collisions at fixed impact
parameter $b=7$\,fm as a function of multiplicity (parametrized by
$e_0$). They should be replaced by the curves shown in the left
panels of Fig.~\ref{F5}.

%
%

In Fig.~\ref{F7}(b) we present multiplicity scaling curves for
$v_2/\ecc$ obtained from viscous hydrodynamics with the full I-S
equations. On a superficial level, the theoretical curves show
qualitative similarity with the experimental data, giving correct
ball-park numbers if one assumes $\eta/s \sim 0.24 \sim 3/4\pi$.
Interestingly, ignoring experimental error bars, one can see
evidence for small scaling violations in the experimental data whose
pattern agrees with the theoretical predictions from viscous
hydrodynamics (see discussion at the end of Sec.~\ref{sec5a}): the
62.5~$A$\,GeV Au+Au data lie slightly above the $200\,A$\,GeV Au+Au
points, and the $200\,A$\,GeV Cu+Cu points fall slightly below the
$62.5\,A$\,GeV Au+Au data. Of course, these fine features of the
experimental data are presently not statistically robust
(significant); much more precise data are needed to confirm or
disprove the theoretical predictions, but upcoming high-statistic
runs at RHIC should be able to deliver them.

Closer inspection of the two panels in Fig.~\ref{F7} shows, however,
that the theoretical scaling curves have the wrong slope: on the
left side of the plot, i.e. for small multiplicity densities, the
data seem to point towards larger specific shear viscosities
$\frac{\eta}{s}>3\times \frac{1}{4\pi}$ whereas on the right side of
the plot, for $\frac{1}{S}\frac{dN_\mathrm{ch}}{dy} >
20$\,fm$^{-2}$, the experimental data require smaller shear
viscosities, $\frac{\eta}{s}\lesssim(1{-}2)\times\frac{1}{4\pi}$.
But this is not at all unexpected: Collisions represented by points
in the right half of the plot correspond to high collision energies
and large initial energy densities whose expanding fireballs spend
the largest fraction of their life in the QGP phase. Fireballs
created in collisions represented by points in the left part of the
diagram have smaller initial energy densities and thus spend most of
their time in the much more viscous hadronic phase
\cite{Hirano:2005xf}. A meaningful comparison between theory and
experiment thus must necessarily account for the temperature
dependence of $\eta/s$ and its dramatic increase during the
quark-hadron phase transition \cite{Hirano:2005wx}. This would lead
to scaling curves in Fig.~\ref{F7}(b) with a larger slope that can
better reproduce the data. What one can say already now is that the
high-energy end of Fig.~\ref{F7} requires very small specific shear
viscosity $\eta/s$ for the QGP, of the same order as the minimal
value postulated in \cite{Kovtun:2004de} (unless, by using the
Glauber model eccentricity instead of the fKLN eccentricity, the
initial source eccentricity $\ecc$ was strongly underestimated in
the experimental data, see discussion in Chap.~8).\\

\section{Multiplicity scaling of entropy production in viscous hydrodynamics}
\label{sec6}

In this section, we investigate the entropy production in viscous
hydrodynamics for Au+Au and Cu+Cu systems at a variety of collision
energies and impact parameters. Similar to Ref.~\cite{Baier:2006gy},
we compute entropy production by exploiting the proportionality of
final entropy to final charged multiplicity. We compute the final
multiplicity $\dNdy$ for both ideal and viscous hydrodynamics and
then equate the fractional increase in $\dNdy$ with the fractional
increase in $d{\cal S}/dy$. This ignores a small negative correction
due to the viscous deviation of the distribution function on the
freeze-out hypersurface from local equilibrium
\cite{Israel:1976tn,Muronga:2006zw} which slightly reduces the
entropy per finally observed particle in the viscous case. The real
entropy production is thus slightly smaller than calculated with our
prescription. However, since on the freeze-out surface the viscous
pressure components are small \cite{Song:2007fn}, this correction
should be negligible.

\begin{figure}[t]
\vspace{-6mm} \centering {\includegraphics[width
=0.45\linewidth,angle=270,clip=] {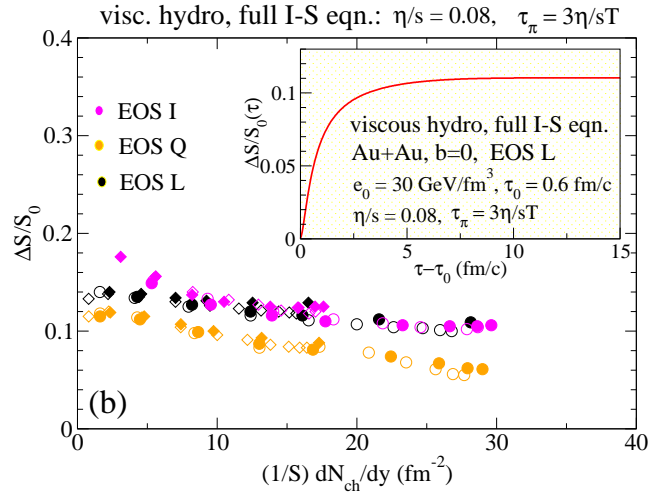} }
\caption[Entropy production as a function of charged multiplicity
density $\frac{1}{S}\frac{dN_\mathrm{ch}}{dy}$]{Entropy production
$\Delta{\cal S}$, normalized by the initial entropy ${\cal S}_0$, as
a function of charged multiplicity density
$\frac{1}{S}\frac{dN_\mathrm{ch}}{dy}$. Calculations with {\tt
VISH2+1} were performed for Au+Au and Cu+Cu collisions at various
impact parameters and collision energies, using $\eta/s=0.08$,
$\tau_\pi=3\eta/sT$, and three different equations of state (EOS~I,
SM-EOS~Q, and EOS~L). The inset in panel (b) shows the entropy
production as a function of time, for central Au+Au collisions with
parameters as indicated in the legend. \label{F8} }
\end{figure}

We checked the above procedure by also directly integrating the
viscous entropy production rate $\partial\cdot s = \pi^{\mu\nu}
\pi_{\mu\nu}/2\eta$ over the space-time volume enclosed between the
initial condition Cauchy surface and the final freeze-out surface.
This method results in slightly larger entropy production, the
relative difference amounting to about 0.7\% (or about 0.07\% in the
absolute value of $\Delta{\cal S}/{\cal S}_0$) for central Au+Au
collisions. Since the estimate from the final multiplicity gives a
lower entropy production value even without accounting for the
somewhat smaller entropy per particle in the viscous case, we
conclude that entropy production due to {\em numerical} viscosity
must be a bit smaller in the viscous fluid than in the ideal one.
This is not unreasonable, given the observation in
\cite{Song:2007fn} that, compared to the ideal fluid case, the
physical viscosity smoothens the strong velocity gradients near the
quark-hadron phase transition, thereby presumably also reducing the
effects of numerical viscosity.~\footnote{\emph{We note that our
viscous evolution starts earlier (at $\tau_0=0.6$\,fm/$c$) than that
of Ref.~\cite{Baier:2006gy} (who use $\tau_0=1$\,fm/$c$). This
earlier start results in larger entropy production fractions. As the
inset in Fig.~\ref{F8}(b) shows, most of the entropy is produced
during the early stage of the expansion. We have confirmed that the
difference between Ref.~\cite{Baier:2006gy} and the work here is
quantitatively reproduced by the entropy generated during the time
interval from 0.6 to 1.0\,fm/$c$, which can be calculated to
excellent approximation analytically \cite{Gyulassy85} (using Eq.
(D3) in Ref.~\cite{Song:2007fn}) by assuming boost-invariant
longitudinal expansion without transverse flow during this period.}}

%
%

Figure~\ref{F8} shows the viscous entropy production $\Delta{\cal
S}$, as a fraction of the initial entropy ${\cal S}_0$, for Cu+Cu
and Au+Au collisions at various impact parameters and collision
energies, as a function of multiplicity density. One observes
approximate multiplicity scaling of the fractional entropy
production, with scaling functions that depend on the  equation of
state. As for $v_2/\ecc$, we see small scale-breaking effects, but
generally the produced entropy fraction shows better multiplicity
scaling than elliptic flow. The scale breaking effects for the
viscous entropy production rate go in the same direction as with
elliptic flow insofar as, at the same value of
$\frac{1}{S}\frac{dN_\mathrm{ch}}{dy}$, larger collision systems and
more central collisions produce fractionally more entropy than
smaller or more peripheral collisions, due to their longer lifetimes
before freeze-out (see Fig.~\ref{Freeze-time}). Figure~\ref{F8} also
shows $15-25\%$ differences between the entropy production rates for
EOS~Q (first order phase transition) and EOS~L (rapid crossover
transition). The differences are largest for the most central Au+Au
and Cu+Cu collisions at top RHIC energies. The somewhat stiffer
nature of EOS~L near $T_c$ causes the fireball to expand faster and
with higher acceleration, leading to larger viscous effects than for
EOS~Q.

An important comment relates to the negative overall slope of the
scaling curves for entropy production shown in Figs.~\ref{F8}: Since
peripheral collisions produce relatively more entropy than central
collisions, and the produced entropy is reflected in the final
charged hadron multiplicity, the collision centrality dependence of
hadron multiplicities is altered by viscous effects. When viscous
effects are accounted for, the charged multiplicity $\dNdy$ will
rise more slowly as a function of the number of participant nucleons
$N_\mathrm{part}$ than for an ideal fluid with the same set of
initial conditions. In a Glauber model parametrization of the
initial conditions \cite{Kolb:2003dz} this tempering effect will
have to be compensated for by increasing the ``hard'' component in
the initial entropy production, i.e. the component that scales with
the density of binary collisions and is thus responsible for the
non-linear increase of $\dNdy$ with $N_\mathrm{part}$. In the Color
Glass Condensate approach \cite{Kharzeev:2000ph} this non-linear
rise is controlled by the centrality dependence of the saturation
momentum scale $Q_s$, with no free parameters to tune. It remains to
be seen whether the success of the CGC model in describing the
centrality dependence of $\dNdy$ \cite{Kharzeev:2001gp} survives the
inclusion of entropy (or multiplicity) producing effects resulting
from shear viscosity during the evolution from the initial CGC to
the finally observed state.\\

\section{Concluding Remarks}

The main motivation for the work presented in this chapter was
provided by the experimentally observed multiplicity scaling of the
elliptic flow, shown in Fig.~\ref{F7}a, and its deviation at low
multiplicities from ideal fluid dynamical predictions. We saw that
many of the observed features are qualitatively consistent with
viscous hydrodynamic calculations as presented here, and that the
same calculations also predict approximate multiplicity scaling for
viscous entropy production. Our studies revealed, however, that even
for ideal fluid dynamics the multiplicity scaling of the elliptic
flow is not perfect, with small scaling violations introduced by the
initial density profiles and by the freeze-out process which cuts
the evolution of elliptic flow short. Even at RHIC energies, where
the elliptic flow almost saturates before freeze-out, kinetic
decoupling truncates the momentum anisotropy at values slightly
below their asymptotic saturation value, and the deviations depend
on the size of the colliding nuclei and the deformation of the
fireball created in the collision through the time available for
building elliptic flow before freeze-out.

Shear viscosity strongly suppresses the build-up of momentum
anisotropy and elliptic flow, especially for low multiplicity
densities, i.e. at large impact parameters, low collision energies
or for small sizes of the colliding nuclei. This changes the slope
of the multiplicity scaling curve for $v_2/\ecc$ but preserves, to
good approximation, its general scaling with
$\frac{1}{S}\frac{dN_\mathrm{ch}}{dy}$. Violations of multiplicity
scaling for $v_2/\ecc$ are somewhat larger for the viscous expansion
than for the ideal fluid (especially with EOS~I), but remain small
enough to be consistent, within statistical errors, with the
experimental observation of approximate scaling. The slope of the
approximate scaling curve and the spread around this curve caused by
scaling violation increase with the value of the specific shear
viscosity $\eta/s$ and can thus be used to constrain it.

Specifically, the observed scaling violations have the following
features: At fixed multiplicity density
$\frac{1}{S}\frac{dN_\mathrm{ch}}{dy}$, viscous hydrodynamics
predicts slightly larger elliptic flow $v_2/\ecc$ for larger
collision systems or more central collisions than for smaller nuclei
colliding at similar energy or more peripheral collisions between
similar-size nuclei colliding at higher energy. Larger $v_2/\ecc$
values are associated with longer lifetimes of the corresponding
fireballs before freeze-out and thus also with larger relative
entropy production. This correlates the scaling violations for
$v_2/\ecc$ observed in Figs.~\ref{F5} and \ref{F7} with those for
the relative entropy production $\Delta{\cal S}/{\cal S}_0$ seen in
Fig.~\ref{F8}. The pattern of the predicted scaling violations shows
qualitative agreement with experiment, although higher quality data
are required to render this agreement statistically robust and
quantitative.

For a fixed (i.e. temperature independent) ratio $\eta/s$, the slope
of the multiplicity scaling curve for $v_2/\ecc$ does not agree with
experiment -- the curves predicted by viscous hydrodynamics are too
flat. The slope can be increased by allowing $\eta/s$ to increase at
lower temperatures: For small multiplicity densities (very
peripheral collisions or low collision energies), the data seem to
require $\frac{\eta}{s}>3\times \frac{1}{4\pi}$, whereas at large
multiplicity densities they appear to constrain the specific shear
viscosity to values of $\frac{\eta}{s}
\lesssim(1{-}2)\times\frac{1}{4\pi}$. While this is qualitatively
consistent with the idea that in high-multiplicity events the
dynamics is dominated by the QGP phase (whose viscosity would thus
have to be small, of order $1/4\pi$) whereas low-multiplicity events
are predominantly controlled by hadron gas dynamics (which is highly
viscous \cite{Hirano:2005xf}), much additional work is needed to
turn this observation into quantitative constraints for the function
$\frac{\eta}{s}(T)$.
\chapter[Generic Viscous Effects: Bulk Viscosity]{Generic Viscous Effects\\
\hspace{2cm} -- Bulk Viscosity}

\section{Introduction} \label{sec3}

In Chap. 4, we studied the generic \emph{shear} viscosity effects
and showed that elliptic flow $v_2$ is very sensitive to the shear
viscosity to entropy ratio $\eta/s$, and can thus be used to extract
the QGP shear viscosity from experimental
data~\cite{Song:2007fn,Song:2007ux,Song:2008si,Luzum:2008cw,
Luzum:2009sb,Dusling:2007gi}. In this chapter, we will investigate
the \emph{bulk} viscosity effects and discuss the uncertainties
introduced  by bulk viscosity when trying to constrain the QGP shear
viscosity from experimental data~\cite{Song:2008hj,Song:2009Bulk}.

We will first compare the generic shear and bulk viscosity effects
on the hydrodynamic evolution and final particle spectra and
elliptic flow, by comparison runs from \textbf{a)} ideal
hydrodynamics, \textbf{b)} viscous hydrodynamics with only shear
viscosity and \textbf{c)} viscous hydrodynamics with only bulk
viscosity. In these comparison runs, identical initial conditions
and final conditions are used, which are the same as in Chap.~4 and
explained in Chap. 3. For viscous hydrodynamics, the transport
coefficient are set as follows:  \textbf{b)}: $\eta/s= 1/4\pi$,
$\tau_\pi=3\eta/sT= 3 / 4 \pi T$ and $\zeta/s=0$ ($C=0$);
\textbf{c)}: $\eta/s= 0$, $C=1$ for $\zeta/s$ as shown in
Fig.~\ref{BulkVis}, $\tau_\Pi=3/4 \pi T$. We here use N-S
initializations for both shear stress pressure $\pi^{mn}$ and bulk
pressure $\Pi$ here. In contrast to Chap. 4, which used EOS I (ideal
EOS) and SM-EOS~Q (a quasi-first order phase transition EOS), we
employ here the lattice-based EOS L, for consistency with the bulk
viscosity, which is constructed using the squared speed of sound
$c_s^2$ from  EOS L.

Chap. 6.3 concentrates on bulk viscosity effects and studies their
sensitivity to the relaxation time and initialization for the bulk
pressure. Chap. 6.4  investigates the upper limit of $\zeta/s$,
consistent with the validity of 2nd order viscous hydrodynamics. In
Chap. 6.5 we will study how much uncertainty bulk viscous effects
might contribute to the extracted value of shear
viscosity from data. \\

\section{Shear viscosity vs. bulk viscosity effects}

\subsection{Hydrodynamic evolution}

\begin{figure}[t]
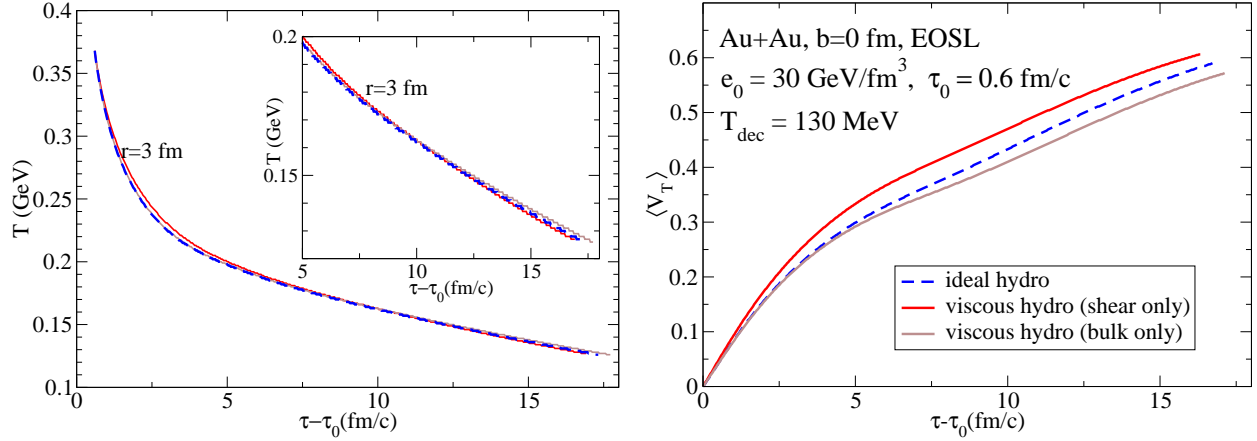

\vspace{-0.0cm}
\includegraphics[width=0.49\linewidth,clip=]{Figs/Chap6/Temp.eps}
 \hspace{0.1cm}
\includegraphics[width=0.49\linewidth,clip=]{Figs/Chap6/velocity.eps}
 \vspace{-0.3cm} \caption[Time evolution
of T and $\La v_r \Ra$: shear vs. bulk viscosity]{Time evolution of
local temperature (left) and average radial flow (right), from ideal
hydrodynamics, viscous hydrodynamics with only minimal shear or bulk
viscosity, calculated with identical initial and final conditions.}
\label{Temp-velocity-bulk}\vspace{0.2cm}
\end{figure}

The left panel of Fig.\ref{Temp-velocity-bulk} shows the time
evolution of local temperatures from ideal  and  viscous
hydrodynamics.  When compared with the ideal fluid, the shear
viscosity slows down the cooling process during the early stage, but
speeds it up during the middle and late stages. As discussed in
Ref.~\cite{Song:2007fn,Song:2007ux} and Chap.~\ref{chap4-hydro-evo},
this is caused by the competition between decelerated longitudinal
expansion and accelerated transverse expansion. During the early
stage and before the transverse flow fully builds up, the system
experiences quasi one-dimensional expansion, and the reduced work
along the longitudinal direction (caused by shear viscosity) slows
down the cooling process. In its middle and late stages, the system
is in full 3-dimensional expansion, and the faster transverse
expansion caused by the additional work done by shear viscous forces
in the transverse directions leads to faster cooling. In contrast,
bulk viscosity decelerates both the longitudinal and transverse
expansion during the entire time evolution, since the negative bulk
pressure effectively softens the EoS, especially near the phase
transition. As a result the cooling process is slowed down during
the whole evolution, with most of the effects concentrated around
the phase transition where $\zeta/s$ peaks.

The shear-viscosity-accelerated and bulk-viscosity-decelerated
transverse expansion also manifests itself  in the evolution of the
radial flow,  shown in the right panel of
Fig.\ref{Temp-velocity-bulk}. Compared with the ideal fluid, minimal
shear viscosity leads to $\sim 5\%$ more radial flow $\La v_r \Ra$
 at the end of the  evolution. The increase of $\La v_r \Ra$
mostly happens early, when the shear viscosity effects are largest.
In contrast, bulk viscosity counteracts the build-up of radial flow,
and minimal bulk viscosity reduces the final radial flow by $\sim
5\%$ relative to the ideal fluid case. Most of the bulk viscous
effects occur during the middle stage from 5-10 fm/c, which is when
a large portion of the matter passes through the  phase transition
region where $\zeta/s$ is large.

%
\begin{figure}[t]
  \begin{center}
    \begin{minipage}[b]{0.65\linewidth}
    \includegraphics[width=0.78\linewidth,height=1.12\linewidth,clip=]{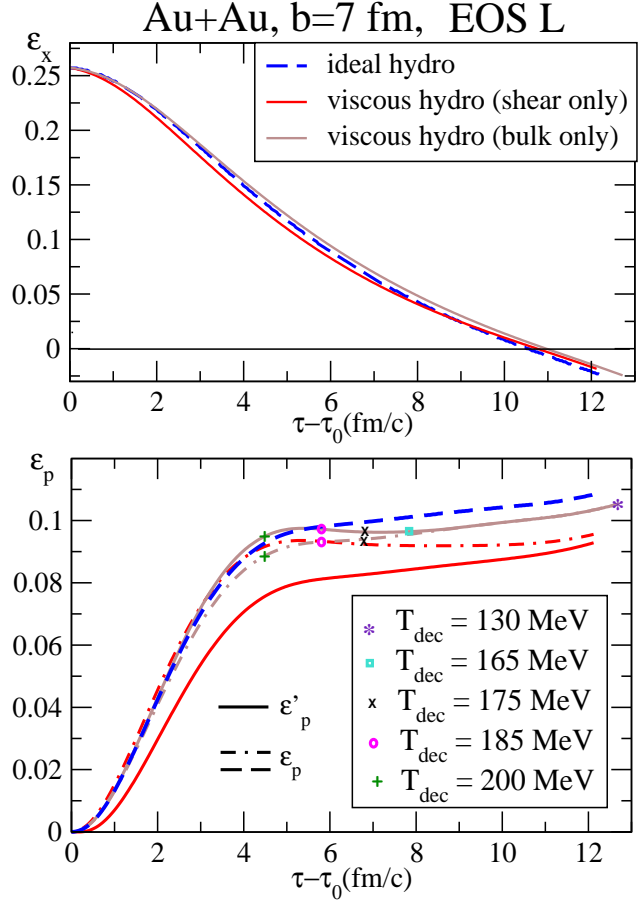}
    \end{minipage}
    \begin{minipage}[b]{0.3\linewidth}
 \caption[Time evolution for $\epsilon_x$
$\epsilon_p$ and $\epsilon_p'$: shear vs. bulk viscosity] {Time
evolution for spacial anisotropy $\epsilon_x$(upper panel) and
momentum anisotropy $\epsilon_p$, $\epsilon_p'$ (lower panel), from
ideal hydrodynamics, viscous hydrodynamics with only shear viscosity
or with only bulk viscosity The different symbols along the bulk
viscous fluid lines indicate freeze-out times for varieties of
freeze-out temperatures.} \vspace{4.3cm} \label{EXEP-bulk}
    \end{minipage}
  \end{center}
  \vspace*{-0.3cm}
\end{figure}
%

We now turn to non-central collisions.  To describe the fireball
deformation in configuration and momentum space, we use the spatial
eccentricity $\epsilon_x$ and the momentum anisotropies
$\epsilon_p$, $\epsilon_p'$, following
Ref.~\cite{Kolb:1999it,Song:2007ux} and our discussions in  Chap.~4.
Figure~\ref{EXEP-bulk}(a) shows the time evolution of the spatial
eccentricity $\epsilon_x$ for non-central Au+Au collisions at b=7
fm.  Compared with the ideal fluid, bulk viscosity slows down the
decrease of the spatial eccentricity $\varepsilon_x$ since the
negative bulk pressure decelerates both the longitudinal and
transverse expansion. In contrast, the spatial eccentricity
$\varepsilon_x$ of the shear viscous fluid without bulk viscosity
drops initially faster than that of the ideal fluid due to the
faster radial expansion shown in Fig.~\ref{Temp-velocity-bulk}
(right), but decreases more slowly than the ideal fluid at later
times, due to the smaller flow anisotropy created in the viscous
dynamics (see Fig.~\ref{VxSc} in Chap.~4). This is also reflected in
the late-time momentum anisotropy shown in Fig.~\ref{EXEP-bulk}(b),
as one sees by comparing the dashed blue ideal fluid line with the
dash-dotted red shear viscous fluid line.

It is worth discussing the time evolution of the momentum
anisotropies $\epsilon_p$ and $\epsilon_p'$ in more detail. In
viscous hydrodynamics, the difference between the total momentum
anisotropy $\epsilon_p'$ (solid lines) and flow induced momentum
anisotropy $\epsilon_p$ (dashed dotted lines) reflects the viscous
pressure anisotropy.  At freeze-out,  this difference generates
additional viscous effects on the $v_2$ suppression, arising from
the non-equilibrium part of the distribution function $\delta f$
along the freeze-out surface, which adds to or subtracts from that
caused by the viscous suppression of the flow anisotropy. Shear
viscous effects are mostly concentrated at earlier times, where they
greatly reduce $\epsilon_p'$ in comparison with $\epsilon_p$. The
negative momentum anisotropy from the shear stress pressure slows
down the growth of the flow anisotropy and causes the total momentum
anisotropy $\epsilon_p$ (solid red line) to fall behind that of the
ideal fluid (dashed blue line) at late times. Similarly  bulk
viscosity also suppresses the development flow anisotropy, since the
gradients of the effective pressure $p+\Pi$ are reduced by the
negative bulk pressure contribution. As a result, the momentum
anisotropy $\epsilon_p$ from the bulk viscous fluid (dashed dotted
brown line) stays  always below the ideal hydro one (dashed blue
line). Most of the bulk viscous effects happen during the middle
stage where $\epsilon_p'$ is seen to significantly deviate from
$\epsilon_p$ (comparing the solid and dot-dashed brown lines). In
contrast to the shear viscous fluid, for the bulk viscous fluid
$\epsilon_p'$ is larger than  $\epsilon_p$; This is mostly due to
the negative
 bulk pressure contribution in the denominator of the expression
for $\epsilon_p'$. As a result,  $\epsilon_p'$ from the bulk viscous
fluid is even slightly larger than in the ideal fluid case. But this
will not lead to an increase of hadronic $v_2$ (even if one freezes
 the system out already at the phase transition), since most of
the enhancement happens in the inner core of the QGP phase at
temperatures  $T>190 \ \mathrm{MeV}$, as the symbols along the bulk
viscous fluid lines indicate. At later times, the total
$\epsilon_p'$ from the bulk viscous fluid is always below the ideal
fluid one, indicating a bulk viscous suppression of hadronic
elliptic flow $v_2$ after freeze-out. For $T_{dec} = 130\
\mathrm{MeV}$, all of the $v_2$ suppression comes from flow effects,
since in our construction the bulk viscosity $\zeta$ is effectively
zero below $T< 150\ \mathrm{MeV}$ . If the system freezes-out near
the phase transition around $165-175 \ \mathrm{MeV}$, the difference
between $\epsilon_p'$ and $\epsilon_p$ along the bulk viscous fluid
lines indicates that the viscous corrections $\delta f$  along the
freeze-out surface will increase $v_2$; this was also pointed out by
Monnai and Hirano~\cite{Monnai:2009ad}, who calculate $\delta f$ in
terms of $\Pi$ and $\pi^{mn}$ along the freeze-out surface using
Grad's 14-moment method. However, their calculation did not include
the flow effects which (according to Fig~\ref{EXEP-bulk}) for
$T_{dec} = 165-175\ \mathrm{MeV}$ are least 2-3 times larger than
those from $\delta f$, leading to
a final overall suppression of $v_2$.\\

\subsection{Spectra and elliptic flow}

\begin{figure}[t]
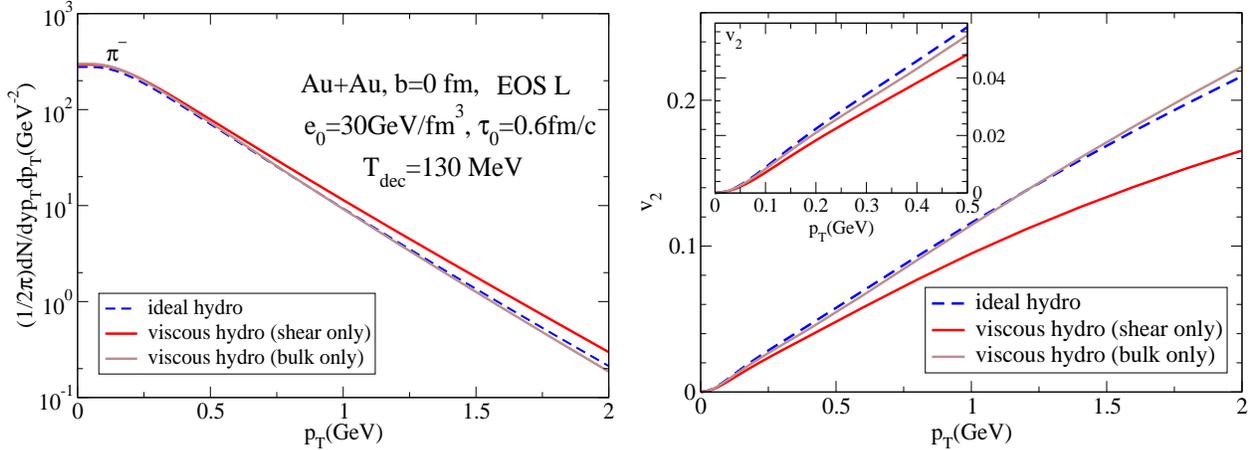

\vspace{-0.0cm}
\includegraphics[width=0.49\linewidth,height=60mm, clip=]{Figs/Chap6/Spectrab0.eps}
\hspace{0.1cm}
\includegraphics[width=0.49\linewidth,height=58.5mm, clip=]{Figs/Chap6/v2-2.eps}
\vspace{-0.3cm} \caption[ $v_2 (p_T)$: shear vs. bulk viscosity
]{$p_T$ spectra and $v_2$ for directly emitted pions (ie. without
resonance decay contribution)} \vspace{0.2cm}
\label{Spectra-v2-bulk}
\end{figure}

The left panel in Fig.~\ref{Spectra-v2-bulk} left shows the pion
$p_T$  spectra from central Au+Au collisions, calculated from ideal
hydrodynamics, viscous hydrodynamics with only minimal shear
viscosity, or viscous hydrodynamics with only minimal bulk
viscosity, using identical initial  and final conditions. We find
that shear viscosity results in flatter spectra due to larger radial
flow, as shown in Fig.~\ref{Temp-velocity-bulk} (right). In
contrast, bulk viscosity leads to steeper spectra since it decreases
radial flow build-up during the fireball expansion. At low $p_T$,
both shear and bulk viscosity lead to a larger yield, when compared
with the ones from ideal hydrodynamics. This corresponds to an
increase of the final multiplicity caused by viscous entropy
production. Even though the shear and bulk viscosity act against
each other in the slope of the final spectra, their effects add in
the viscous entropy production, and an adjustment of initial
conditions from ideal to viscous hydrodynamics can thus not be
avoided if one intents to describe a given set of experimental data.
In Ref.~\cite{Romatschke:2007jx}, it was shown that by increasing
the thermalization time $\tau_0$ and adusting  the initial energy or
entropy density, one can reduce the initial total entropy and reduce
the buildup of  the radial flow, thereby compensating for the
viscous entropy production and allowing for a successful fit of the
experimental measured final multiplicities and  spectral slopes.

In the right panel of Fig.~\ref{Spectra-v2-bulk}, we show the $p_T$
dependent $v_2$ of pions for non-central Au+Au collisions at b=7 fm.
The insert enlarges the low-$p_T$ region $p_T<0.5\ \mathrm{GeV}$. We
find that both shear and bulk viscosity suppress $v_2$ at low $p_T$.
At higher $p_T$ ($\sim 2\ \mathrm{GeV}$), shear viscosity further
suppresses $v_2$ due to a negative contribution $\delta f$ along the
freeze-out surface. In contrast, bulk viscosity increases $v_2$
above  $1\ \mathrm{GeV}$, due to its steepening effects on the $p_T$
spectra there. At $p_T=0.5 \ \mathrm{GeV}$ (approximately the mean
average transverse momentum for pions) we find that minimal bulk
viscosity suppresses $v_2$ by $\sim 5\%$ and minimal shear viscosity
suppresses $v_2$ by $\sim 20\%$. Using $v_2$ data to determine
$\eta/s$, the additional $\sim 5\%$ $v_2$ suppression from minimal
bulk viscosity can thus leads to $\sim 25\%$ uncertainties in the
extracted value of $\eta/s$. Therefore, we should not neglect bulk
viscosity effects when extracting the QGP shear viscosity from
experimental data.\\

\section[Sensitivity to relaxation time and bulk pressure initialization]
{Bulk viscosity effects: sensitivity to relaxation time and bulk
pressure initialization} \label{sec3}

We now concentrate on bulk viscosity and investigate the effects
from different relaxation times $\tau_\Pi$ and initializations of
the bulk pressure $\Pi$. For $\tau_\Pi$ we here consider three
choices: the constant values $\tau_\Pi=0.5$ and 5\,fm/$c$, and the
temperature dependent function
$\tau_\Pi(T)=\max[\tilde{\tau}\cdot\frac{\zeta}{s}(T),
0.1\,\mathrm{fm}/c]$ with $\tilde{\tau} = 120$\,fm/$c$. The last
choice implements phenomenologically the concept of critical slowing
down near a phase transition; it yields $\tau_\Pi\approx0.6$\,fm/$c$
at $T=350$\,MeV and $\tau_\Pi\approx 5$\,fm/$c$ at $T_c$. To study
memory effects, we explore two different initializations for the
bulk viscous pressure: \textbf{(a)} Navier-Stokes (N-S)
initialization, $\Pi(\tau_0)=-\zeta\partial\cdot u$, and
\textbf{(b)} zero initialization, $\Pi(\tau_0)=0$.

\begin{figure}[t]
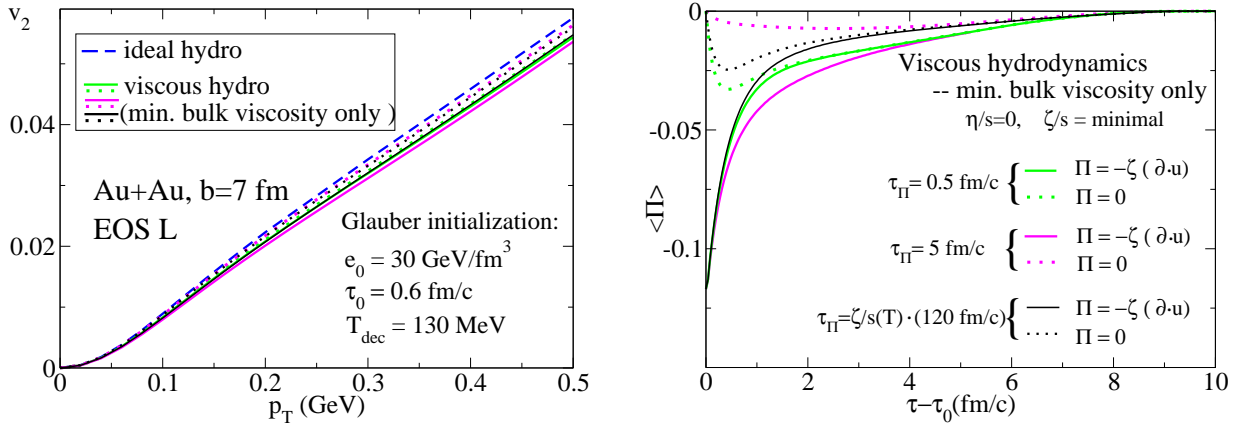

\vspace{-0.0cm}
\includegraphics[width=0.47\linewidth]{./Figs/Chap6/v2_TauPI.eps} \hspace{0.5cm}
\includegraphics[width=0.47\linewidth]{./Figs/Chap6/AvePi.eps}
\vspace{-0.3cm} \caption[$v_2(p_T)$ and $\La\Pi\Ra$: effects from
$\tau_\pi$ and initialization of $\Pi$]{{\sl Left:} Differential
pion elliptic flow $v_2(p_T)$ from ideal and viscous hydrodynamics,
including only bulk viscosity. {\sl Right:} Time evolution of the
bulk pressure $\La\Pi\Ra$ averaged over the transverse plane
(weighted by the energy density) from viscous hydrodynamics.
Different curves correspond to different initializations and
relaxation times, as indicated (see text for discussion).}
\vspace{0.2cm} \label{V2-Ave-PI}
\end{figure}

The left panel of Fig.~\ref{V2-Ave-PI} shows the differential
elliptic flow $v_2(p_T)$ of directly emitted pions (without
resonance decays) for non-central Au+Au collisions at $b\eq7$\,fm,
calculated from ideal hydrodynamics and minimally bulk viscous
hydrodynamics with identical initial and final conditions. The
different lines from viscous hydrodynamics correspond to different
relaxation times $\tau_\Pi$ and different initializations
$\Pi(\tau_0)$. One sees that these different inputs can lead to
large uncertainties for the bulk viscous $v_2$ suppression. For
minimal bulk viscosity, the $v_2$ suppression at $p_T=0.5$\,GeV
ranges from $\sim 2\%$ to $\sim 10\%$ relative  to ideal
hydrodynamics (blue dashed line in the left panel).

For the shorter relaxation time, $\tau_\Pi= 0.5 \ \mathrm{fmc/c}$,
the bulk viscous $v_2$ suppression is insensitive to the
initialization of $\Pi$, and both N-S and zero initializations show
$\sim 8\%$ $v_2$ suppression relative to ideal fluids. The reason
behind this becomes apparent in the right panel showing the time
evolution of the average bulk pressure $\La \Pi \Ra$. For short
relaxation times, $\La \Pi \Ra$ quickly loses all memory of its
initial value, relaxing in both cases to the same trajectory after
about $1-2\, \mathrm{fm/c}$ (i.e. after a few times $\tau_\Pi$).
This is similar to what we found for shear viscosity where the
microscopic relaxation times are better known and short
($\tau_\pi(T_c)\simeq 0.2-0.5\,\mathrm{fm/c}$) and where the shear
pressure tensor $\pi^{mn}$ therefore also loses memory of its
initialization after about 1\,fm/$c$ \cite{Song:2007ux}.

%

This changes if one accounts for the critical slowing down of the
evolution of $\Pi$ near $T_c$. If one simply multiplies the constant
relaxation time by a factor 10, setting $\tau_\Pi=5$\,fm/$c$, one
obtains the dotted and solid magenta lines in Fig.~\ref{V2-Ave-PI}.
Now the bulk viscous $v_2$ suppression relative to the ideal fluid
becomes very sensitive to the initialization of the bulk viscous
pressure: For zero initialization $\Pi(\tau_0)=0$, the viscous $v_2$
suppression is very small (only $\sim 2\%$ at $p_T=0.5$\,GeV/$c$).
The right panel shows that in this case the magnitude of the
(transversally averaged) bulk pressure evolves very slowly and
always stays small, leading to almost ideal fluid evolution. On the
other hand, if $\Pi$ is initialized with its Navier-Stokes value,
which initially is large due to the strong longitudinal expansion,
it decays initially more slowly than for the shorter relaxation
time. Its braking effect on the flow evolution is therefore bigger,
resulting in much stronger suppression of $v_2$ than for zero
initialization, at $\sim 10\%$ slightly exceeding even the viscous
$v_2$ suppression seen for the tenfold shorter relaxation time.

The ``critical slowing down'' scenario with temperature-dependent
$\tau_\Pi(T)$ (black lines) interpolates between the short and long
relaxation times. As for the fixed larger value
$\tau_\Pi=5$\,fm/$c$, $v_2$ depends sensitively on the
initialization of $\Pi$, but for N-S initialization the viscous
$v_2$ suppression is somewhat smaller than for both short and long
fixed relaxation times. The reasons for this are subtle since now,
at early times, the bulk viscous pressure $\Pi$ evolves on very
different time scales in the dense core and dilute edge regions of
the fireball. As a result, for N-S initialization the average value
$\La\Pi\Ra$ is {\em smaller} in magnitude than for both short and
long fixed $\tau_\Pi$, throughout the fireball evolution (right
panel, black lines).\\

\section{Validity of I-S viscous hydrodynamics and upper limits for $\zeta/s$}

Viscous hydrodynamic codes can not run with arbitrarily large bulk
viscosity. Physically, a large enough bulk viscosity can lead to a
negative value for the total effective pressure $p+\Pi$ (thermal
pressure + bulk viscous pressure) if the expansion rate $\partial
\cdot u$ is large. When this happens, the fluid becomes mechanically
unstable~\cite{Torrieri:2008ip} and will tend to break up. In
numerical simulations, this corresponds to exponential amplification
of local  numerical errors, which will eventually stop the code from
running. On the other hand,  2nd order viscous hydrodynamics applies
only for a near-equilibrium system: $\delta f \ll f_0$. This
requires that the dissipative flows are (much) smaller than the
equilibrium energy momentum tensor.

In this section, we will investigate upper limits of $\zeta/s$
allowed by  2nd order viscous hydrodynamics under dynamical
conditions encountered in heavy-ion collisions. If the input
$\zeta/s$ is small, the condition $p+\Pi (x, y, \tau)>0$ is
stratified everywhere in the transverse $(x,y)$ plane during the
entire hydrodynamic evolution. As one increases $\zeta/s$ (which
corresponds to increasing the coefficient $C$ in our parametrization
of $(\zeta/s) (T)$, see Chap.~3.5), the condition $p+\Pi>0$ will be
violated near the phase transition in certain regions of $(x,y)$ and
$\tau$. In our calculations we define the upper limit of $\zeta/s$
as the critical value of $C$ that can generate negative effective
pressure, $p+\Pi<0$,  at some position $(x,y)$ and some time $\tau$,
i.e. anywhere inside the freeze-out surface.

\begin{figure}[t]
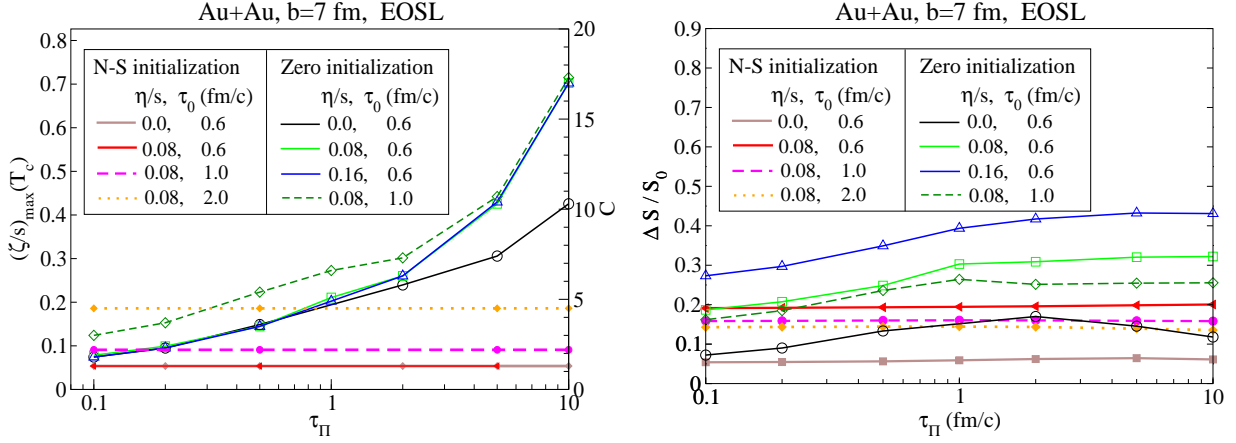

\vspace{-0.0cm}
\centering{\includegraphics[width=0.49\linewidth,height=58mm,clip=]{Figs/Chap6/Max-Zeta.eps}}
\hspace{0.1cm}
\centering{\includegraphics[width=0.47\linewidth,height=58mm,clip=]{Figs/Chap6/Max-Zeta-entropy.eps}}
\vspace{-0.3cm} \caption[upper limits of $\zeta/s$, and
corresponding entropy production]{Upper limits of $\zeta/s$, and
entropy production as a function of relaxation time, for zero
initialization $\Pi(\tau_0)=0$ and N-S initialization
$\Pi(\tau_0)=-\zeta
\partial \cdot u$, with varieties of conditions. } \ \label{Max-Zeta}
\end{figure}

Fig.~\ref{Max-Zeta} shows the upper limit  $(\zeta/s)_{max}(T_c)$ as
a function of relaxation time $\tau_\Pi$. The behavior of
$(\zeta/s)_{max}(T_c)$ as a function of $\tau_\Pi$ is seen to depend
strongly on the initialization of the bulk pressure. For N-S
initialization, $\Pi(\tau_0)=-\zeta
\partial\cdot u$,\,  $(\zeta/s)_{max}(T_c)$ is insensitive to the relaxation
time $\tau_\Pi$. In this case, the magnitude of the average bulk
pressure $\Pi$ decreases more or less monotonically  with time, as
shown in the right panel of  Fig.~\ref{V2-Ave-PI}. Violations of the
positivity condition  $p+\Pi>0$ always happen at the starting time
$\tau_0$ and at transverse positions where the matter is near the
phase transition. This leads to a $(\zeta/s)_{max}(T_c)$ that is
controlled by initial conditions and independent of the relaxation
time. If one includes shear viscosity, $(\zeta/s)_{max}(T_c)$
depends only on the initialization time $\tau_0$, but not on the
value of $\eta/s$. The dependence on $\tau_0$ arises from the strong
dependence of the initial bulk pressure $\Pi=-\zeta
\partial\cdot u$ on $\tau_0$, through the expansion rate
$\partial\cdot u=1/\tau_0$. Hence the maximal allowed
$(\zeta/s)_{max}(T_c)$ increases when the hydrodynamic evolution
starts later. This tendency is illustrated by the solid red, dashed
magenta and dotted orange lines in Fig.~\ref{Max-Zeta}: as one
increase $\tau_0$ from 0.6 fm/c to 1 fm/c and  2 fm/c, the maximal
$(\zeta/s)_{max}(T_c)$ lines increases from 0.05 to 0.09 and 0.18.

For zero initialization $\Pi(\tau_0)=0$, one finds a qualitatively
similar dependence of $(\zeta/s)_{max} (T_c)$ on $\tau_0$: The
curves $(\zeta/s)_{max} (T_c)$ move up monotonically as one
increases the starting time $\tau_0$ from 0.6 fm/c to 1.0 fm/c
(solid green and dashed dark green). In contrast to the N-S
initialization, the $(\zeta/s)_{max}(T_c)$ curves now show a strong
dependence on relaxation time  $\tau_\pi$, rising monotonically with
 $\tau_\pi$. The reason is that it takes some time for the bulk
pressure $\Pi$ to develop large enough magnitudes to violate the
positivity condition $p+\Pi>0$; again this happens typically in
regions where the matter is close to the phase transition. Smaller
relaxation times allow $\Pi$ to develop faster, resulting in a
monotonic increase of $(\zeta/s)_{max}(T_c)$ with $\tau_\Pi$. For
$\tau_\Pi< 1 \ \mathrm{fm/c}$, we find "universal"
$(\zeta/s)_{max}(T_c)-\tau_\Pi$ curves that do not depend on the
shear viscosity $\eta/s$ (solid black, green and blue curves), but
move upwards as we increase the starting time $\tau_0$. This is
because the violation of the positivity condition $p+\Pi>0$
generally happens at early times $\tau<3 \ \mathrm{fm/c}$ when the
flow profiles are not yet significantly affected by shear viscous
effects. For the two viscous fluid lines with $\eta/s=0.08$ and 0.16
(solid blue and solid green lines) one see that they continue to
overlap each other for $\tau_\pi> 1 \ \mathrm{fm/c}$, where they
break away from the $\eta/s=0$ line.  For  $\eta/s=0$, the phase
transition generates larger velocity gradients near the phase
transition, which cause instability at lower value of $\zeta/s$.
Shear viscosity smoothes out these large gradients, as discussed in
Chap.~4, allowing the fluid to evolve stably up to larger values of
$\zeta/s$. Bulk viscosity $\zeta$ alone has no smoothing influence
on  sharp structures generated by a phase transition; For zero
initialization for $\Pi$, shear viscosity is thus essential to
stabilize the evolution of the viscous fluid against mechanical
instabilities caused by strongly negative bulk viscous pressure,
especially for large relaxation time $\tau_\Pi$.\\

\section[Extracting $\eta/s$ from experimental data: uncertainties
from bulk viscosity]{Extracting $\eta/s$ from experimental data:
uncertainties introduced by bulk viscosity } \label{sec3}

\begin{figure}[t]
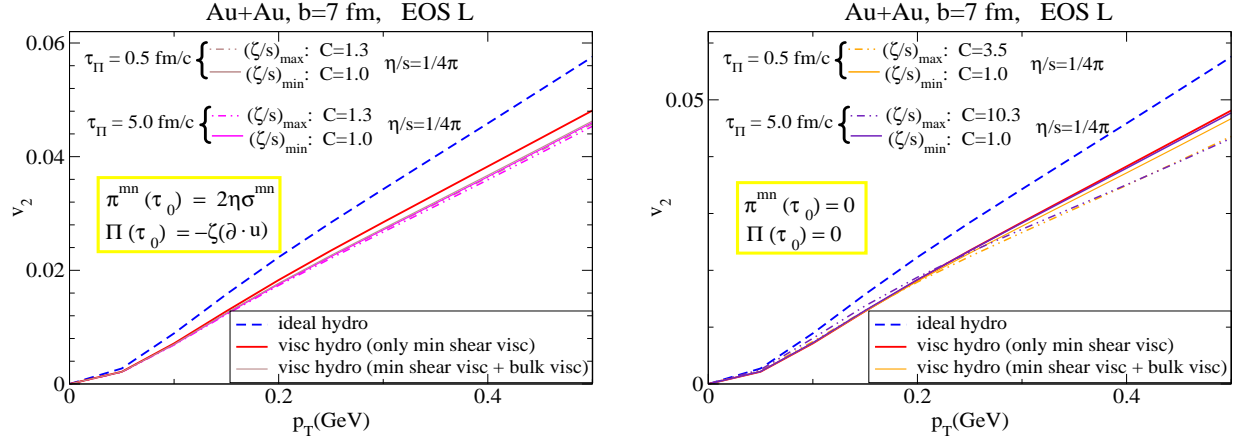

\vspace{-0.0cm}
\includegraphics[width=0.47\linewidth,height=58mm,clip=]{Figs/Chap6/v2-Supr-0.eps} \hspace{0.5cm}
\includegraphics[width=0.47\linewidth,height=58mm,clip=]{Figs/Chap6/v2-Supr-1.eps}
 \vspace{-0.3cm} \caption[$v_2$ suppression from
min shear viscosity and min/max bulk viscosity]{$v_2 (p_T)$ from
ideal hydrodynamics, viscous hydrodynamics with minimal shear
viscosity, and viscous hydrodynamics with minimal shear viscosity
and minimal or maximal bulk viscosity. Left: N-S initialization,
$\pi^{mn}(\tau_0)=2\eta \sigma^{mn}$, $\Pi(\tau_0)=-\zeta
\partial \cdot u$. Light: zero initialization,
$\pi^{mn}(\tau_0)=0$, $\Pi(\tau_0)=0$. }
\label{VisSupBulk2}\vspace{0.2cm}
\end{figure}

Similar to shear viscosity,  bulk viscosity suppresses elliptic flow
$v_2$, as shown in Fig.~\ref{Spectra-v2-bulk} (right). The bulk
viscosity effects, once included, will thus reduce the $\eta/s$
value needed to account for the suppression of a given measured
amount of $v_2/\varepsilon$ below the ideal fluid expectation.
Current theoretical uncertainties in the relaxation time $\tau_\Pi$
and the initial value of the  bulk viscous pressure $\Pi$, as well
as the unknown value of the bulk viscosity itself, thus introduce a
significant uncertainty in the extraction of $\eta/s$ from elliptic
flow data, which can not be reduced without additional theoretical
efforts. This is illustrated in Fig.~\ref{VisSupBulk2}, where we
investigate the additional viscous $v_2$ suppression from minimal
and maximal bulk viscosity. (Here, ``maximal bulk viscosity" denotes
the largest value compatible with $p+\Pi>0$ everywhere, as discussed
in Chap.~6.4). For N-S initialization (Fig.~\ref{VisSupBulk2},
left), the viscous fluid lines with both shear  and bulk viscosities
($\eta/s=1/4\pi$, C=1.0 or 1.3) lead to 20-25\% more viscous $v_2$
suppression at $p_T=0.5 \ \mathrm{GeV}$ than a fluid with shear
viscosity only. Neglecting the possible need for re-tuning the
initial conditions and other hydrodynamic parameters, this
additional $v_2$ suppression will translate into a reduction of the
extracted $\eta/s$ by \emph{O}(20\%), when fitting the same elliptic
flow data. For zero initialization (right panel), the additional
bulk viscous $v_2$ suppression ranges from 2\% to 70\%. Similarly,
this will reduce the the extracted $\eta/s$ by \emph{O}(2-70\%).
This uncertainty bound covers only the $\zeta/s$ range accessible
with our viscous hydrodynamics simulations. Larger $\zeta/s$ value
may be realized in nature but would completely destroy our ability
to extract $\eta/s$ from a comparison of $v_2$ data with viscous
hydrodynamics because the later breaks down. To remove or
significantly reduce this uncertainty, it is mandatory to place
strict theoretical limits on $(\zeta/s)(T)$ and $\tau_\Pi(T)$ as
well as on the initial value for the bulk pressure $\Pi$.\\

\section{Summary} \label{sec3}
In this chapter, we first compared shear and bulk viscous effects by
studying, for central and non-central Au+Au collisions,  simulations
based on \textbf{(a)} ideal hydrodynamics, \textbf{(b)} viscous
hydrodynamics with only minimal shear viscosity ($\eta/s=1/4\pi$)
and \textbf{(c)} viscous hydrodynamics with only minimal bulk
viscosity ($\zeta/s)_{min}$. In all comparisons, we used identical
initial and final conditions and the same EOS (EOS~L). Similarities
and differences between shear and bulk viscous effects are
summarized as follows: Shear viscosity decelerates longitudinal
expansion, but accelerates transverse expansion. When compared with
the ideal fluid case, this leads to an initial slowing down of the
cooling process, but to faster cooling during later stages, and more
radial flow generation. Bulk viscosity decelerates both longitudinal
and transverse expansion,  resulting in slower cooling throughout
and less creation of radial flow. Both shear viscosity and bulk
viscosity generate entropy during the hydrodynamic evolution, and
thus lead to larger final multiplicities. The $p_T$ spectra,
however, are flatter for a shear viscous fluid, but steeper for a
bulk viscous fluid. For non-central Au+Au collisions, we found that
both shear and bulk viscosity inhibit the development of momentum
anisotropies. Most of the shear viscous effects happen during the
early stage, while bulk viscous effects are stronger during the
middle stage when most of the matter passes through the  phase
transition. As a result, both shear and bulk viscosity suppress
elliptic flow $v_2$ at low $p_T$. At higher $p_T$, $v_2$ is further
suppressed in a shear viscous fluid due to negative contributions
from the non-equilibrium part $\delta f$ of the freeze-out
distribution function. In contrast, $v_2$ is slightly increased for
the bulk viscous fluid above $p_T>1 \ \mathrm{GeV}$, due to the
steeper $p_T$ spectra.

Effects arising from different relaxation times $\tau_\Pi$ and
different initializations of the bulk pressure $\Pi$ on the viscous
$v_2$ suppression were studied in Chap.~6.3, in order to assess
phenomenological uncertainties arising from limited theoretical
knowledge in the bulk viscous sector.  The well-known phenomenon of
critical slowing down during a phase transition suggests  that
$\tau_\Pi$ may become large near $T_c$. If one uses larger
relaxation times of the order of $\tau_\pi=5\ \mathrm{fm/c}$ (near
$T_c$ or at all temperatures), the bulk viscous $v_2$ suppression
becomes sensitive to the initialization of $\Pi$, with effects
ranging from $\sim 2\%$ (zero initialization for $\Pi$) to $\sim
10\%$ (N-S initialization for $\Pi$) for minimal bulk viscosity.
This strongly suggests a need for  more theoretical research into
the relaxation time $\tau_\Pi$ and into theories that can provide
reliable initializations for the bulk viscous pressure.

The bulk viscosity $\zeta$ itself also has theoretical
uncertainties: the peak value of $\zeta/s$ near the phase transition
from lattice QCD estimates is around 10 times larger than the
minimal AdS/CFT values, and even less is known for the bulk
viscosity of the hadronic phase. Considering these theoretical
uncertainties, we investigated the maximal peak value of $\zeta/s$
the near phase transition, $(\zeta/s)_{max} (T_c)$, that is allowed
by the the numerical applicability of 2nd order viscous
hydrodynamics with dynamical conditions encountered in heavy ion
collisions.  For N-S initialization, $(\zeta/s)_{max} (T_c)$ can be
as low as 0.05 / 0.09 / 0.18, depending on the starting time
$\tau_0$ for hydrodynamic stage, $\tau_0=0.6\ / \ 1.0 \ / \ 2.0 \
\mathrm{fm/c}$, and this limit is insensitive to the relaxation time
$\tau_\Pi$. For zero initialization, $(\zeta/s)_{max} (T_c)$
increases monotonically with the relaxation time $\tau_\Pi$, and can
reach values up to 0.7 at $\tau_\Pi= 10 \ \mathrm{fm/c}$, for
$\tau_0=0.6-1.0 \ \mathrm{fm/c}$.

The additional bulk viscous effects on the suppression of elliptic
flow $v_2$ (comparing to ideal hydrodynamics) were studied in
Chap.~6.5. We concluded that bulk viscous effects can not be ignored
when extracting the QGP shear viscosity to entropy ratio $\eta/s$
from experimental elliptic flow data. Pinning down the bulk viscous
effects is important, especially if $\eta/s$ turns out to be small
of order of the KSS bound. It urgently requires to much more tightly
constrain the allowed ranges for $\zeta/s$, $\tau_\pi$ and
$\Pi(\tau_0)$, than presently possible.
\chapter{Recent Developments in Viscous Hydrodynamics}
\section{Introduction}
During the past couple of years, several groups have independently
developed (2+1)-dimen- sional viscous hydrodynamic codes for
relativistic heavy ion collisions and published their first results
on shear viscosity
effects~\cite{Song:2007fn,Song:2007ux,Song:2008si,Romatschke:2007mq,Luzum:2008cw,
Dusling:2007gi}. In this chapter, I will put my own research results
in context with  related developments elsewhere, discuss
similarities and differences as well as the origin of the latter.
%
%

All groups that did viscous hydrodynamics simulations found that the
elliptic flow coefficient $v_2$ is very sensitive to shear
viscosity: given the large expansion rates of heavy-ion collision
fireballs, even minimal viscosity saturating the KSS bound
$\eta/s\geq1/4\pi$ \cite{Kovtun:2004de} for the viscosity to entropy
density ratio can lead to a strong (and thus easily measurable)
suppression of $v_2$. Assuming the availability of a
well-established ideal fluid dynamical baseline for $v_2$ as a
function of collision energy, centrality and system size,
measurements of this suppression could thus be used to constrain the
QGP shear viscosity from experimental data.

However, the first results for viscous $v_2$ suppression published
by the different groups seemed to show large quantitative
discrepancies, ranging from 20\% to 70\% even for 'minimal
viscosity' $\eta/s=1/4\pi$. Detailed inspection of the early papers
of each group revealed that different initial conditions, final
conditions, EOS and different versions of the 2nd order viscous
equations were used in these calculations, and that collision
systems of different size (Cu+Cu vs. Au+Au collisions) were studied.
This motivated us to systematically investigate the physical effects
of  each of these differences and to collaborate with other groups
to carefully verify our numerical codes to eliminate the possibility
of  numerical error.

In this chapter, we will summarize  recent progress in viscous
hydrodynamics.  We will show how the apparent  discrepancies on
viscous $v_2$ suppression are resolved, when one properly accounts
for physical effects arising from system size, equation of states
(EOS), and different versions of the Israel-Stewart equations (Chap.
7.2). I will also report on the  most recent code verification
results, obtained within the TECHQM collaboration~\cite{TECHQM}
(Chap. 7.4). In Chap. 7.3, we will numerically compare the different
2nd order I-S equations obtained from the different theoretical
approaches described in Chap. 2, which includes a comparison between
the ``simplified"  and ``full" I-S equations as well as some effects
from other higher order terms. In Chap. 7.4, we will
also numerically compare the I-S equation with the O-G equation. Throughout
this Chapter, we concentrated on shear viscous effects and set $\zeta=0$.\\

\section{System size, EOS and different I-S equations}
In this section, we will briefly discuss the different
manifestations of shear viscosity when one varies system size and
EOS and uses different versions of the I-S
equations~\cite{Song:2008si}. As mentioned in the Introduction, this
analysis resolves the initially puzzling differences between the
results published by different groups.

%
\begin{figure}[t]
\includegraphics[width=\linewidth,height=55mm,clip=]{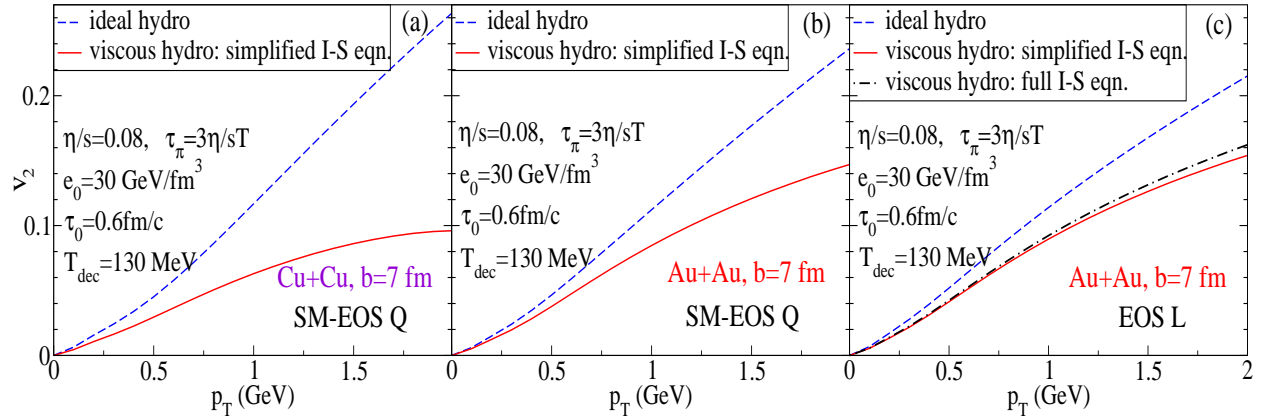}
\vspace{-3mm}\caption[$v_2(p_T)$ for different collisions systems,
EoS and I-S eqns.]{Differential elliptic flow $v_2(p_T)$ for
directly emitted pions (i.e. without resonance decay contributions),
comparing results for different collisions systems and equations of
state. (a) Cu+Cu at $b=7$\,fm with SM-EOS~Q. (b) Au+Au at $b=7$\,fm
with SM-EOS~Q. (c) Au+Au at $b=7$\,fm with EOS~L. Dashed (solid)
lines correspond to ideal (viscous) fluid dynamics, with parameters
as indicated. \label{V2-Compare} }
\end{figure}
%

Fig.~\ref{V2-Compare} shows the differential elliptic flow
$v_2(p_T)$ for directly emitted pions from ideal and viscous
hydrodynamics. Panels (a) and (b) compare two systems of different
size (Cu+Cu and Au+Au, both at $b=7$\,fm), using identical equations
of state (SM-EOS~Q), I-S equations (simplified,
eq.~(\ref{Simplifed-I-S}) in Chap.~2) and other free inputs.
Although both systems have similar initial eccentricities, the
smaller Cu+Cu system shows a much larger viscous $v_2$ suppression
(by almost 70\% below the ideal fluid value at
$p_T=2\,\mathrm{GeV}/c$ \cite{Song:2007fn,Song:2007ux}) than
observed in the larger Au+Au system where the suppression is almost
a factor two smaller. Panels (b) and (c) compare the same Au+Au
system at $b=7$\,fm for two different EOS and different I-S
equations. Changing the EOS from SM-EOS~Q to EOS~L reduces the
viscous suppression of elliptic flow by another quarter (from $\sim
40\%$ to $\sim 30\%$ below the ideal fluid value at
$p_T=2\,\mathrm{GeV}/c$). Replacing the simplified I-S equations
used in~\cite{Song:2007fn,Song:2007ux} by the full I-S equations
(eq.~(\ref{full-I-S}) in Chap.~2) used in~\cite{Romatschke:2007mq}
further reduces the $v_2$ suppression from 30\% to 25\% below the
ideal hydrodynamics value at $p_T=2\,\mathrm{GeV}/c$. This final
result is consistent with~\cite{Romatschke:2007mq}. Although for
EOS~L the additional terms in the full I-S equations only result in
a 5-10\% difference in $v_2$ suppression, its effect is much larger
for more rapidly expanding systems, such as Cu+Cu collisions driven
by a stiff conformal EOS $e=3p$ (see Chap.~7.4.1 and
Ref.~\cite{Song:2008si} for details).

We conclude that the biggest contribution to the large difference
between the results reported in Refs. \cite{Song:2007fn} and
\cite{Romatschke:2007mq} arises from the different collision systems
studied, with much larger viscous effects seen in the smaller Cu+Cu
system than in Au+Au collisions. The next most important sensitivity
is to the EOS; for the most realistic EOS studied here, EOS~L, the
differences between using the full or simplified I-S equations with
$\tau_\pi=3\eta/sT$ are only about 10\% on a relative scale, or
about 3\% on the absolute scale set by the elliptic flow from ideal
fluid dynamics. For smaller $\tau_\pi$, this last difference would
shrink even
further.\\

\section{Detailed comparison of different I-S equations}
As discussed in Chap. 2, the I-S formalism can be obtained from
different approaches: from the 2nd law of thermodynamics
(Chap.~2.3), from kinetic theory (Chap.~2.4) and from the conformal
symmetry constraint (Chap.~2.5). The common terms, obtained by all
of these approaches, form the so-called ``simplified" I-S equation
shown in eq.~(\ref{Simplifed-I-S}). The ``full" I-S equation used
here and in Ref.~\cite{Song:2008si} is defined as the simplified I-S
equation plus a term that helps to maintain the conformal symmetry
for a conformal fluid, shown in eq.~(\ref{full-I-S}). With the
identity~(\ref{full-IS-Var}), which holds in the conformal limit,
the ``full" I-S equation can be written in different forms, and
these variations are what one originally obtains from different 2nd
order theory approaches. However this ``full" I-S equation is still
not the most general 2nd order theory, and in a variety of
approaches, discussed in Chap.~2, additional 2nd order terms (a
vorticity term, a visco-elastic term, and
more~\cite{Betz:2008me,Baier:2007ix,Dusling:2009zz}), each with its
own coefficient that takes different values in different theories
(see Table~2.1), arise. To find the optimal I-S equations for
numerical implementation, one needs systematic comparisons for these
different I-S equations from different approaches. In this section,
we will give a detailed numerically comparison between  the
``simplified" and ``full" I-S equations~\cite{Song:2008si}, and then
briefly study
the effects from other higher order terms. \\

\subsection{``Simplified" and ``full" I-S equations}
\label{SimFullI-S}
%
\begin{figure}[htp]
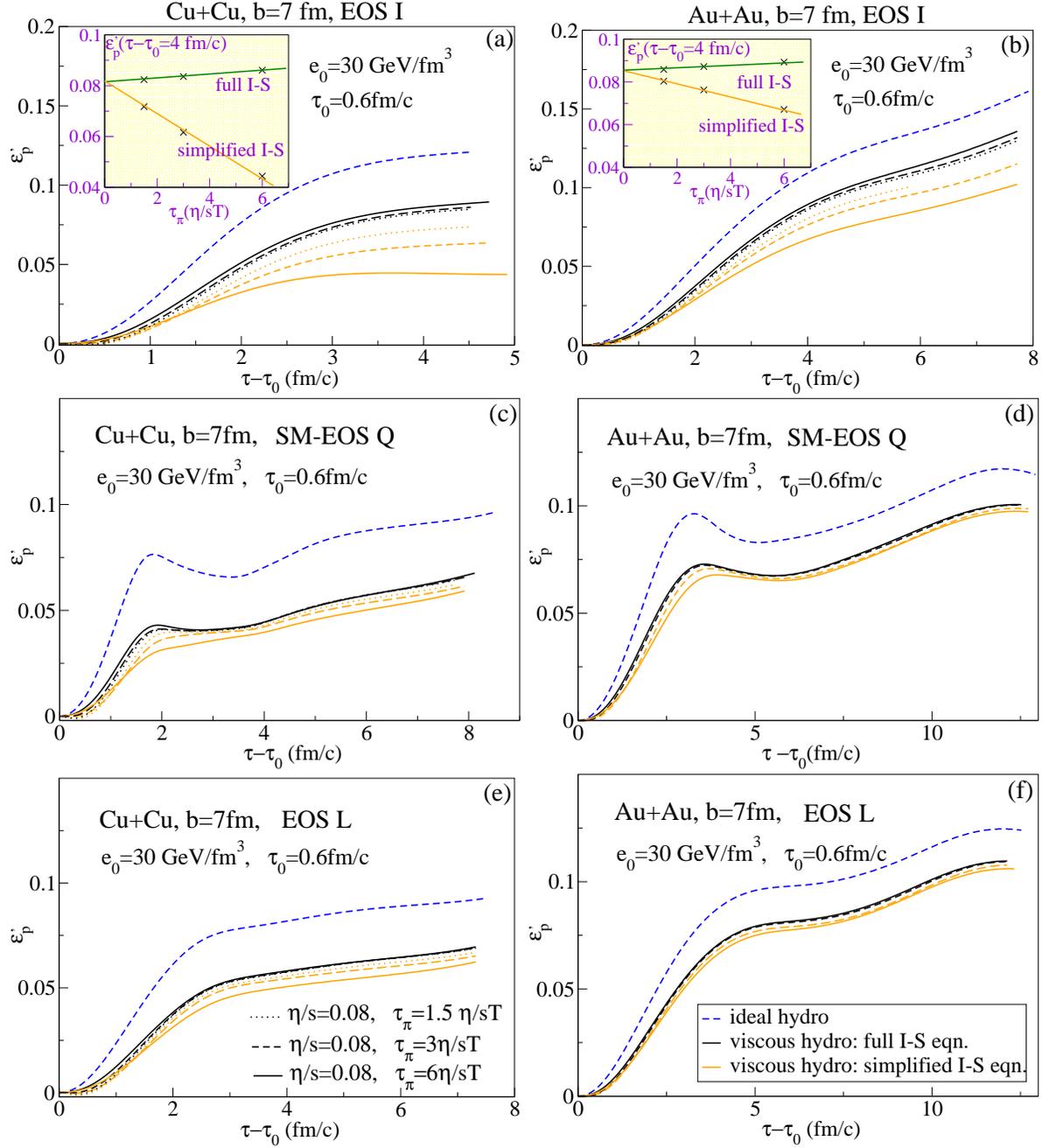

\includegraphics[width =0.47\linewidth,clip=]{./Figs/Chap7/Fig2a.eps}
\includegraphics[width =0.47\linewidth,clip=]{./Figs/Chap7/Fig2b.eps}\\
\includegraphics[width =0.47\linewidth,clip=]{./Figs/Chap7/Fig2c.eps}
\includegraphics[width =0.47\linewidth,clip=]{./Figs/Chap7/Fig2d.eps}\\
\includegraphics[width =0.47\linewidth,clip=]{./Figs/Chap7/Fig2e.eps}
\includegraphics[width =0.47\linewidth,clip=]{./Figs/Chap7/Fig2f.eps}
\caption[Time evolution of $\varepsilon'_p$: full vs. simplified I-S
eqns]{Time evolution of the total momentum anisotropy
$\varepsilon'_p$ for two collision systems (left: Cu+Cu; right:
Au+Au), three equations of state (top: EOS~I; middle: SM-EOS~Q;
bottom: EOS~L), and three values of the kinetic relaxation time
$\tau_\pi$ as indicated (dotted, dashed and solid curves,
respectively). The insets in the two top panels show the
$\tau_\pi$-dependence of the momentum anisotropy $\ecc_p$ at fixed
time $\tau-\tau_0=4$\,fm/$c$. See text for discussion. \label{F2} }
\end{figure}
%

\textbf{\underline{Evolution of momentum anisotropies}}\\[-0.15in]

Figure~\ref{F2} shows the temporal evolution of the total momentum
anisotropy $\ecc_p=\frac{\langle T^{xx}{-}T^{yy}\rangle}
{\langle T^{xx}{+}T^{yy}\rangle}$ averaged over the transverse plane%
\footnote{\emph{Note that $\ecc_p$ as defined in this section
includes the effects from both flow velocity and shear pressure
anisotropies \cite{Song:2007fn}. In Ref.~\cite{Song:2007fn} and Chap
4 we denoted it by $\ecc'_p$ in order to distinguish it from the
flow-induced momentum anisotropy $\frac{ \langle
T_0^{xx}{-}T_0^{yy}\rangle}{\langle T_0^{xx}{+}T_0^{yy}\rangle}$
which is based only on the ideal fluid part of the energy momentum
tensor and neglects anisotropies in the local fluid rest frame
caused by the shear pressure tensor $\pi^{mn}$. In the present
chapter we drop the prime for convenience.}}
for two collision systems (Cu+Cu at $b=7$\,fm on the left, Au+Au at
$b=7$\,fm on the right) and three equations of state (EOS~I (top),
SM-EOS~Q (middle), and EOS~L (bottom)). The blue dashed lines at the
top indicate the result from ideal fluid dynamics, the black and
orange lines below show viscous fluid dynamical results. The black
lines show solutions of the ``full" I-S equation, the orange ones
for the simplified I-S approach; in each case several values of the
kinetic relaxation time $\tau_\pi$ are explored. Note that our
``full" I-S equation used here is defined by eq.~(\ref{PimnPI2+1})
(already written in 2+1 dimensions), but not the full I-S equation
defined by eq.~(\ref{full-I-S}) in Chap.~3, which strictly holds
only for conformal fluids. We have, however, tested the two
expressions on the left and right side of eq.~(\ref{full-IS-Var})
against each other also for the other two equations of state
(SM-EOS~Q and EOS~L) which are not conformally invariant, and found
no discernible differences. Only for a very long relaxation time
$\tau_\pi=12\eta/sT$ (not shown in Fig.~\ref{F2}) did we see for
EOS~L a difference larger than the line width, with our result for
$\ecc_p$ lying slightly above the one obtained with the conformal
approximation eq. (\ref{full-IS-Var}).

Comparison of the black and orange lines in Fig.~\ref{F2} shows that
the sensitivity of the momentum anisotropy $\ecc_p$ to the
relaxation time $\tau_\pi$ is significantly larger for the
simplified I-S equation (orange) than for the ``full" I-S equation
(black), and that the $\tau_\pi$-dependence of $\ecc_p$ even has the
opposite sign for the two sets of equations. With the ``full" I-S
equations, $\ecc_p$ moves slowly towards the ideal fluid limit as
$\tau_\pi$ increases whereas with the simplified I-S equations
$\ecc_p$ moves away from the ideal fluid limit, at a more rapid
rate, resulting in a larger viscous suppression of the momentum
anisotropy. In the limit $\tau_\pi\to0$, both formulations approach
the same Navier-Stokes limit. The difference between ``full" and
simplified I-S equations is largest for EOS~I which is the stiffest
of the three studied equations of state, causing the most rapid
expansion of the fireball. For this EOS, the simplified I-S
equations allow for the largest excursions of $\pi^{mn}$ away from
its Navier-Stokes limit, causing a significant and strongly
$\tau_\pi$-dependent increase of all viscous effects, including the
suppression of the momentum anisotropy (Fig.~\ref{F2}) and elliptic
flow and the amount of viscous entropy production (see below).

For the other two equations of state, SM-EOS~Q and EOS~L, the
difference between ``full" and ``simplified" I-S dynamics is much
smaller, ranging from $\sim 5\%$ for Au+Au to $\sim 15\%$ for Cu+Cu
for the largest $\tau_\pi$ value of $6\eta/sT$ studied here. Note
that the viscous suppression of $\ecc_p$ is much stronger for the
smaller Cu+Cu collision system than for Au+Au. For SM-EOS~Q and
EOS~L (which yield rather similar results for $\ecc_p$, with
differences not exceeding $\sim 10\%$), the results from the ``full"
I-S equations (black lines) are almost completely independent of
$\tau_\pi$, even for the small Cu+Cu system.

The insets in the two upper panels of Fig.~\ref{F2} illustrate the
different $\tau_\pi$-dependences for $\ecc_p$ in the ``full" and
``simplified" I-S formulations, by plotting the value of $\ecc_p$
for EOS~I at a fixed time $\tau-\tau_0=4$\,fm/$c$ as a function of
$\tau_\pi$. One sees that, for the investigated range of relaxation
times, the $\tau_\pi$-dependence is linear, but that the slope has
different signs for the ``full" and ``simplified" I-S equations and
is much smaller for the ``full" I-S system. Even though VISH2+1
cannot be run for much smaller $\tau_\pi$ values, due to numerical
instabilities that develop as the Navier-Stokes limit $\tau_\pi=0$
is approached, the lines corresponding to the ``full" and the
``simplified" I-S equations are seen to nicely extrapolate to the
same Navier-Stokes point, as they should. For SM-EOS~Q and EOS~L,
the corresponding lines may no longer be linear, due to phase
transition effects, but are still characterized by opposite slopes
for the ``simplified" and ``full" I-S approaches, with almost
vanishing slope in the full I-S case. This agrees with findings
reported in
\cite{Romatschke:2007mq,Luzum:2008cw}.\\

\textbf{\underline{Elliptic flow and viscous suppression of $v_2$:
systematics}}\\[-0.15in]

%
\begin{figure}[t]
\vspace{-3mm}
\centering{\includegraphics[width=0.49\linewidth,clip=]{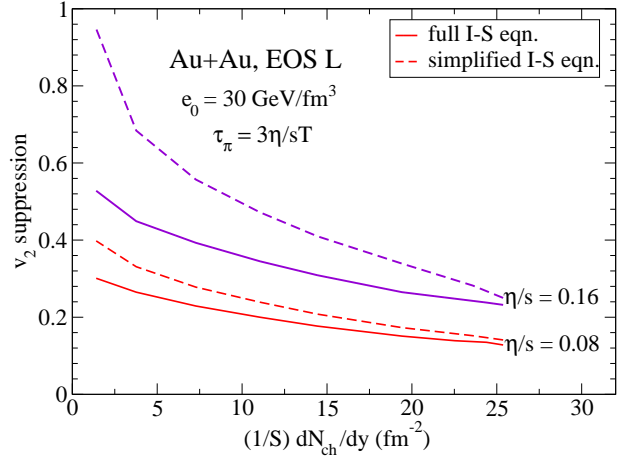}}
\vspace{-3mm} \caption[Viscous suppression of elliptic flow:  full
vs. simplified I-S eqns.]{Viscous suppression of elliptic flow,
$(v_2^\mathrm{ideal}-v_2^\mathrm{viscous})/v_2^\mathrm{ideal}$, as a
function of $(1/S)\dNdy$ for Au+Au with EOS~L, $\tau_\pi=3\eta/sT$
and two values of $\eta/s$ as indicated. Solid (dashed) lines
correspond to using the full (simplified) I-S equations,
respectively. }\label{V2suppress}
\end{figure}
%

In this subsection, we will compare the ``simplified"  and ``full"
I-S equations by studying  the viscous $v_2$ suppression with
different $\eta/s$ as inputs. As already shown in Figs.~\ref{F4} and
\ref{F5}, the slope of $v_2/\ecc$ as a function of
$\frac{1}{S}\frac{dN_\mathrm{ch}}{dy}$ remains positive for viscous
hydrodynamics curves even at the highest collision energies (or
$e_0$ values), i.e. $v_2/\ecc$ continues to increase and evolve in
direction of the asymptotic ideal fluid limit. This implies that at
higher collision energies the importance of viscous effects
decreases. This observation parallels the one made in
\cite{Song:2007fn}, namely that with increasing collision energy the
$p_T$ range increases over which viscous hydrodynamic predictions
for the single-particle momentum spectra can be trusted. The reason
is in both cases that with increasing collision energy the time
until freeze-out grows, and that (at least for constant $\eta/s$ as
assumed here and in \cite{Song:2007fn}) during the later stages of
the expansion shear viscous effects are small.

Figure~\ref{V2suppress} shows this more quantitatively. We plot the
fractional decrease of the elliptic flow relative to its ideal fluid
dynamical value,
$(v_2^\mathrm{ideal}-v_2^\mathrm{viscous})/v_2^\mathrm{ideal}$, as a
function of multiplicity density
$\frac{1}{S}\frac{dN_\mathrm{ch}}{dy}$ . Larger multiplicity
densities lead to smaller viscous suppression effects. Larger
viscosity results in stronger suppression of the elliptic flow. The
suppression effects are weaker if the ``full" I-S equations are used
than in the ``simplified" approach of Ref.~\cite{Song:2007fn} (which
also suffers from strong sensitivity to $\tau_\pi$). For minimal
viscosity, $\eta/s=1/4\pi$, the $p_T$-integrated elliptic flow $v_2$
in Au+Au collisions at RHIC is suppressed by about 20\%. The
suppression is larger at lower
energies but will be less at the LHC.\\

\textbf{\underline{Entropy production}:}\\[-0.15in]

%
\begin{figure}[t]
\centering \vspace{-8mm} {\includegraphics[height
=0.57\linewidth,angle=270,clip=]{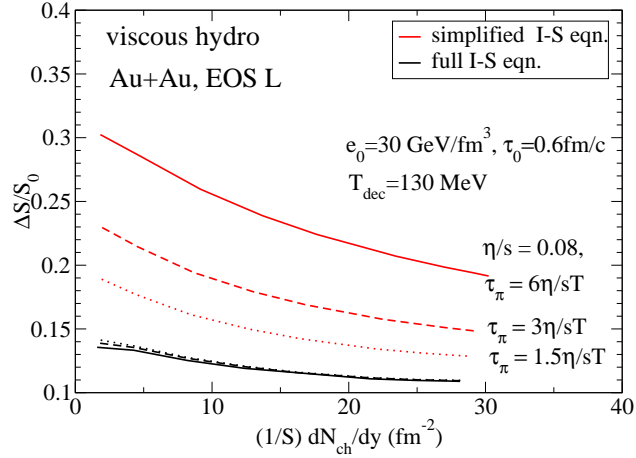}}
\vspace{-3mm} \caption[Entropy production: full vs. simplified I-S
eqns]{Sensitivity of the entropy production ratio $\Delta{\cal
S}/{\cal S}_0$ shown in Fig.~\ref{F8} to the kinetic relaxation time
$\tau_\pi$, for the Au+Au collision system with $e_0=30$\,GeV/fm$^3$
(corresponding to a collision energy of
$\sqrt{s}\approx200\,A$\,GeV). The three red curves (upper set) are
for the simplified Israel-Stewart equations, the three black curves
(lower set) for the full I-S equations. Solutions with the full I-S
equations produce less entropy and show very little sensitivity to
$\tau_\pi$. \label{entropyP} }
\end{figure}
%

In this subsection, we compare the entropy production from viscous
hydrodynamics using the  ``simplified" and ``full" I-S equations.
Figure~\ref{entropyP} shows that, when the ``simplified" I-S
equations are used, entropy production depends very sensitively on
the kinetic relaxation time $\tau_\pi$, approaching the much smaller
and almost completely $\tau_\pi$-independent entropy production
rates of the ``full" I-S framework in the limit $\tau_\pi\to0$. The
large amount of extra entropy production for non-zero $\tau_\pi$ in
the ``simplified" I-S approach must thus be considered as
unphysical. This is important because this artificial extra entropy
production (caused by unphysically large excursions of the viscous
shear pressure tensor $\pi^{mn}$ away from its Navier-Stokes value
$\pi^{mn}=2\eta\sigma^{mn}$) manifests itself as additional charged
hadron multiplicity in the observed final state. Since the final
multiplicity is used to normalize the initial energy density $e_0$,
this causes a significant distortion of the initial conditions
corresponding to a given set of experimental data, affecting their
physical interpretation.\\[-0.15in]

We conclude that using the ``full" I-S equations is mandatory if one
wants to minimize artificial effects of shear viscosity on entropy
production and elliptic flow in the realistic situation of non-zero
kinetic relaxation times. (We note that, while the value of
$\tau_\pi$ for the QGP created at RHIC is not very well known, it
can obviously not be zero).\\


\section{Code verification and comparison between I-S and O-G
equations}

Before the code verification process among different groups began,
our code VISH2+1 had already passed several tests of its numerical
accuracy: (i) in the limit of vanishing viscosity, it accurately
reproduces results obtained with the (2+1)-d ideal fluid code
AZHYDRO~\cite{AZHYDRO}; (ii) for homogeneous transverse density
distributions (\emph{i.e. } in the absence of transverse density
gradients and transverse flow) and vanishing relaxation time it
accurately reproduces the known analytic solution of the
relativistic Navier-Stokes equation for boost-invariant
1-dimensional longitudinal expansion~\cite{Gyulassy85}; (iii) for
very short kinetic relaxation times our Israel-Stewart code
accurately reproduces results from a separately coded (2+1)-d
relativistic Navier-Stokes code, under restrictive conditions where
the latter produces numerically stable solutions; and (iv) for
simple analytically parametrized anisotropic velocity profiles the
numerical code correctly computes the velocity shear tensor that
drives the viscous hydrodynamic effects. The details can be found in
Ref~\cite{Song:2007ux} and Appendix~\ref{appd-Code-Checking}.

In the rest of this section, we will  report on the recent code
verification efforts among different groups with independently
developed viscous hydrodynamic codes.

\vspace{2cm}

{\LARGE Preliminary results of the TECHQM collaboration (please
refer to https://wiki.bnl.gov/TECHQM/index.php/). I will release the
material of this part after the related paper has been made
available on-line.}

\vspace{4cm}

\section{Conclusions}
In this chapter, we first resolved the questions that arose from
several recent publications of viscous hydrodynamic calculations
(with only shear viscosity) which seemed to yield different
results~\cite{Song:2007fn,Song:2007ux,Romatschke:2007mq,Luzum:2008cw}.
After accounting for the physical effects arising from different
system sizes, different EOS and different versions of the I-S
equations, the dramatically different $v_2$ suppressions, ranging
from 70\% to 20\%,  are largely resolved~\cite{Song:2008si}. For a
realistic equation of state that implements a quark-hadron
transition (here SM-EOS~Q and EOS~L) it turns out that system size
effects play a very important role. At RHIC energies and for a
realistic EOS, the viscous suppression effects for $v_2/\ecc$ in
Cu+Cu collisions are almost twice as large as for the larger Au+Au
collision system. Non-negligible differences in the amount of
viscous $v_2$ suppression arise also from details in the EOS, with a
smooth crossover as implemented in EOS~L giving 25-30\% less
suppression than a first-order transition as in SM-EOS~Q. Compared
to system size effects and EOS uncertainties, the differences
between ``simplified'' and ``full'' I-S theory are relatively small,
affecting the viscous $v_2$ suppression at the 10\% level relative
to each other. The later code verification among different groups
 eliminated the possibility of significant numerical errors in the
independently developed codes.  Very good agreement is obtained when
testing the hydrodynamic evolution and the calculated spectra.

We also numerically studied the effects of implementing different
second order theories for causal relativistic viscous hydrodynamics.
We first compared the the ``simplified Israel-Stewart equations''
eq. (\ref{Simplifed-I-S}) used in our early
articles~\cite{Song:2007fn,Song:2007ux} and the ``full
Israel-Stewart equations'' eq. (\ref{full-I-S}) used in our later
work~\cite{Song:2008si} and in
Refs.~\cite{Romatschke:2007mq,Luzum:2008cw}. For the ``simplified''
approach,  we found a strong sensitivity of physical observables to
the presently poorly  known kinetic relaxation time $\tau_\pi$ for
the viscous shear pressure tensor $\pi^{mn}$, in contrast to a much
weaker and basically negligible $\tau_\pi$-dependence in the
``full'' approach. For non-zero $\tau_\pi$ the ``simplified I-S
equations'' allow for large excursions of $\pi^{mn}$ away from its
Navier-Stokes limit $\pi^{mn}=2\eta\sigma^{mn}$. These excursions
are artificial and disappear in the Navier-Stokes limit
$\tau_\pi\to0$ which can, however, not be stably simulated
numerically. They cause large viscous suppression effects for the
elliptic flow and large amounts of extra entropy production (i.e.
extra final hadron multiplicity). From our study we conclude that
the ``simplified I-S approach'' should be avoided, and that a
reliable extraction of $\eta/s$ from experimental data mandates the
use of the ``full Israel-Stewart
equations''~\cite{Romatschke:2007mq, Luzum:2008cw,Song:2008si}. We
also studied the effects from other higher order terms beyond the
full I-S equations,  such as the vorticity and visco-elastic terms
obtained in other approaches. The numerical evidence from our
systematic study in Chap.~7.3.2 as well as the indications from the
code verification among different groups suggest that these higher
order terms can be neglected in numerical calculations. In short, we
recommend to use the ``full" Israel-Stewart equation
(\ref{full-I-S}, \ref{full-IS-Var}) due to its relatively simple
form and its weak dependence on the microscopic relaxation time.
\chapter[Extracting $\eta/s$ from Experimental Data: Uncertainty
Analysis]{\hspace{-0cm}Extracting Shear Viscosity from Experimental
Data -- Uncertainty Analysis}

\section{Introduction}
With the efforts from different
groups~\cite{Song:2007fn,Song:2007ux,Song:2008si,Romatschke:2007mq,Luzum:2008cw,
Dusling:2007gi,Molnar:2008xj,Huovinen:2008te}, the elliptic flow has
now been widely accepted as the key observable to constrain the QGP
shear viscosity. With the availability of several independently
developed causal viscous hydrodynamic codes, we are at the threshold
for extracting the QGP shear viscosity from experimental data.
However, several issues must be clarified before we can do so. These
include: \textbf{(1)} establishing quantitative uncertainty
estimates for the hydrodynamically predicted elliptic flow related
to uncontrolled uncertainties in the  initial conditions for the
fireball eccentricity  and the remaining uncertainties in the QCD
equation of state (EOS) near and above $T_c$; {\bf (2)} inclusion of
non-equilibrium hadronic chemistry into the EOS below $T_c$ (this is
known to affect the distribution of the hydrodynamic momentum
anisotropy over the various hadron species and thus their $v_2$, but
has not yet been included in viscous hydrdynamic simulations); {\bf
(3)} exploring uncertainties related to the treatment of the final
kinetic freeze-out process; and {\bf (4)} including effects from
bulk viscosity, especially near $T_c$. In this chapter, we will try
to assess the uncertainties introduced by each of these aspects when
extracting the QGP viscosity from experimental elliptic flow
data~\cite{Song:2008hj}. Resolving these uncertainties will require
additional theoretical work that goes beyond this thesis. The
analysis laid out in this chapter can serve as a guide for such
research.\\

\section{Dynamical freeze-out and effects from late hadronic viscosity}

In ideal hydrodynamics, one usually imposes ``sudden freeze-out'',
i.e. a sudden transition from a thermalized fluid to free-streaming
particles, on a hypersurface $\Sigma(x)$ of constant temperature or
energy density~\cite{Kolb:2003dz}. The same algorithm has also been
used in most of the existing viscous hydrodynamic calculations
~\cite{Romatschke:2007mq,Song:2007fn,Song:2007ux,Luzum:2008cw,Song:2008si,
 Chaudhuri:2008sj}. Since viscous hydrodynamics is
based on an expansion in small deviations from local equilibrium,
its validity requires the microscopic relaxation time to be much
smaller than the inverse  macroscopic expansion rate,
$\tau_\mathrm{rel}
\partial{\cdot}u \ll 1$. This condition (whose long history is
discussed in Refs.~\cite{Heinz:2007in,Heinz:2006ur} where it is also
applied to ideal hydrodynamics) provides a natural criterium for a
dynamical freeze-out algorithm. Dusling and
Teaney~\cite{Dusling:2007gi} implemented it into their viscous
hydrodynamics. They find that in this case the viscous $v_2$
suppression is dominated by non-equilibrium corrections to the local
thermal distribution function along the freeze-out surface
\cite{Dusling:2007gi}. This is not a collective effect arising from
anisotropies of the flow velocity profile, but a reflection of
non-equilibrium momentum anisotropies in the local fluid rest frame.
In contrast, for isothermal freeze-out we find~\cite{Song:2007ux}
that the viscous $v_2$ suppression is dominated by the viscous
reduction of the collective flow anisotropy, while local rest frame
momentum anisotropies play a much smaller role. This comparison
shows that a careful treatment of the hadronic decoupling process
will be required for the quantitative extraction of $\eta/s$ from
elliptic flow data. Dusling also found that dynamical freeze-out can
increase the slope of the multiplicity dependence of the
eccentricity-scaled elliptic flow
$v_2/\varepsilon$~\cite{PRV-Dusling}. This is an improvement over
the scaling behavior found in~\cite{Song:2008si} for viscous
hydrodynamics with constant $\eta/s$ and isothermal freeze-out which
features a slope that is too small.

There are other reasons why a proper kinetic treatment of the late
hadronic phase is important. By matching a realistic hadron
rescattering cascade to an ideal fluid description of the QGP and
hadronization stages, Hirano {\it et al.} \cite{Hirano:2005xf}
showed that the HRG phase is highly viscous and strongly suppresses
any buildup of elliptic flow during the hadronic stage. This is
consistent with a recent analysis by Demir and Bass
\cite{Demir:2008tr} who found large shear viscosities for their
hadronic UrQMD cascade even close to $T_c$ (between 5-10 times above
the KSS bound).

To extract the QGP viscosity from elliptic flow data, one needs a
good description of the highly viscous hadronic stage in order to
properly include  the additional viscous $v_2$ suppression from the
late hadronic stage. However, viscous hydrodynamics with a
temperature dependent $\eta/s$ (using the results
from~\cite{Demir:2008tr}) can not self-consistently describe the
highly viscous hadronic stage. The HRG viscosities found
in~\cite{Demir:2008tr} are so large that the viscous hydrodynamics
description breaks down even during the early time of the fireball
evolution~\cite{PRV-Song} as one reaches the skin of the fireball
where the matter is in the HRG phase. This requires the future
development of a hybrid approach that connects viscous hydrodynamics
(for the QGP phase) with a hadron cascade. Moreover, the correct
description of the beam energy and centrality dependence of $v_2$ is
crucially affected by the changing relative weight of QGP and HRG
dynamics in building elliptic flow (in central collisions or at
higher energies the system spends more time in the QGP phase than in
peripheral collisions or at low energies). A realistic kinetic
simulation of the hadronic phase and its freeze-out thus appears to
be indispensable.\\

\section{EoS}

\textbf{\underline{Lattice EoS for the QGP phase}:}\\[-0.10in]

The EOS is a necessary input in hydrodynamic calculations. Early
ideal hydrodynamic simulations generally implemented the so-called
EOS Q, which connects the QGP phase with chemical equilibrium HRG
phase through a first order phase
transition~\cite{Kolb:2000sd,Kolb:2000fha,Huovinen:2001cy,
Teaney:2001av} (see Sec. 3.3  for details). Lattice QCD with two
light quark flavors (u and d) and one heavier quark flavor (s)
 predicts, however a smooth cross-over transition at
small net baryon chemical potential~\cite{Karsch:2003jg}. Using a
quasi-particle model,
Refs.~\cite{Huovinen:2005gy,Bluhm-Heinz:2007nu} constructed a
lattice inspired EOS (qp-EOS) and applied it to ideal hydrodynamics
simulations. From comparison runs with different EOS as inputs
(qp-EOS vs. EOS-Q etc.),  the order of magnitude of  phase
transition effects on hadron  spectra and elliptic flow were
investigated.  One found that the differential elliptic flow for
heavier particles (such as protons) at low $p_T$  is sensitive to
the order of the phase transition. However, whether experimental
data favor an EOS with  the 1st order phase transition  or a
cross-over EOS depends on the speed of sound near the phase
transition as well as the chemical compositions in the subsequent
HRG phase (as can be seen, e.g. from the different conclusions
reached in Refs.~\cite{Huovinen:2005gy}
and~\cite{Bluhm-Heinz:2007nu}, respectively).  It is thus necessary
to construct a more realistic EOS, using  the most recent lattice
QCD data for the QGP phase followed by a HRG phase in partially
chemical equilibrium, and use it as  standard input for hydrodynamic
calculations in order to eliminate uncertainties associated with
inaccurate modeling of the EOS. Recent developments in lattice QCD
simulations for the EoS in 2+1 flavor QCD with a physical strange
quark mass and almost physical light quark
masses~\cite{Cheng:2006qk,Bazavov-EOS:2009zn} make  analytical or
tabulated constructions of a Lattice QCD inspired, realistic EOS
possible, and several groups are working along this
direction~\cite{PRV-Huovinen}~\footnote{\emph{The so-called lattice
EOS (EOS~L) used in this thesis and in Ref.~\cite{Song:2008si} is
still a crude one  based on old lattice QCD data with unphysical
quark masses~\cite{Katz:2005br} and without extrapolation to the
continuum limit.}}.

The other motivation for constructing such a realistic EOS comes
from the extraction of the QGP viscosity. As pointed in
Ref.~\cite{Song:2008hj,Song:2008si} (and discussed in Chap.~7.2),
this is an  \emph{O}(25\%) effect on viscous $v_2$ suppression,
depending on whether one uses a 1st order or smooth cross-over phase
transition EOS. This translates into a \emph{O}(25\%) effect on the
extracted value of the shear viscosity to entropy ratio $\eta/s$
from experimental data. Considering that elliptic flow at present
seems to be the only variable from which $\eta/s$ can be extracted
with some accuracy, it is desirable to eliminate the uncertainties
from the EOS using our best available theoretical
knowledge based on lattice QCD.\\[-0.05in]

\textbf{\underline{Hadron resonance gas EOS with partial chemical equilibrium:}} \\[-0.10in]

Early ideal hydrodynamic
calculations~\cite{Kolb:2000sd,Kolb:2000fha,Huovinen:2001cy,
Teaney:2001av}, assuming simultaneous chemical and kinetic
freeze-out, successfully described both the shape of most
$p_T$-spectra and the integrated and differential elliptic flow data
from central to semi-central Au+Au collisions at top RHIC energies.
However, it failed to reproduce the particle yields (total final
multiplicities)  and it underpredicted the pion spectra at low
$p_T$. One example is the experimental pion/anti-proton ratio, which
could not be reproduced in these ideal hydrodynamics simulation, but
had to be fixed by hand by introducing a normalization factor.

Systematic studies of the measured particle yields for large number
of hadron species within a statistic model indicates that chemical
freeze-out happens right after the phase transition at
$T_{ch}=160-175 \
\mathrm{MeV}$~\cite{BraunMunzinger:2001ip,BraunMunzinger:2003zd}.
This   is much higher than the kinetic freeze-out temperature,
$T_{th}=90-130 \mathrm{MeV}$, which is  obtained from blast-wave
model fits to the slope of the $p_T$ spectra~\cite{Adams:2003xp}.
This separation between chemical and kinetic freeze-out motivates
the construction an EOS for the HRG phase that implements partial
chemical equilibrium among the hadron resonances as discussed in
Chap.~3.3~\cite{Kolb:2002ve,Teaney:2002aj,Huovinen:2007xh,Hirano:2002ds}.
 In this PCE-EOS, effective chemical potentials for
different hadron species (that are introduced to maintain the
measured particle ratios as the temperature decreases from $T_{ch}$
to $T_{th}$) change the relation between temperature and energy
density, reducing the kinetic freeze-out temperature for the same
decoupling energy density. Holding the energy density at the value
used in the early ideal hydrodynamic
studies~\cite{Kolb:2000sd,Kolb:2000fha,Huovinen:2001cy}, this leads
to steeper transverse momentum spectra. The experimental spectra can
be re-fitted by adding some additional flow (either from the
pre-equilibrium stage or introduced by shear viscosity), which
flattens the spectra and compensates for the lower freeze-out
temperature. However, the non-equilibrium chemical composition of
the hadronic fireball also affects the distribution of the total
momentum anisotropy among the various hadronic species at
freeze-out. Together with the additional radial flow effects, this
leads to an almost 25\% larger pion elliptic
flow~\cite{Huovinen:2007xh} compared with early ideal hydrodynamic
results and with experimental data, opening room for finite
viscosity both in the QGP and hadronic phases. It is clearly
necessary to include such a large effect also in the viscous
hydrodynamics simulations before a quantitative extraction of
$\eta/s$ from experimental  data can be attempted.

As explained in Chap.~3.3, the hadron cascade
model~\cite{Bass:1998ca}  naturally incorporates the partial
chemical equilibrium through its combination of  elastic,
quasi-elastic and inelastic cross sections among different hadron
species. Since these cross section are finite, it also accounts for
kinetic non-equilibrium (or viscous) behavior in the HRG phase, by
solving the coupled Boltzmann equations through Monte-Carlo
simulations.

Hybrid approaches, constructed by coupling (3+1)-d ideal
hydrodynamics to such a hadron cascade model just below $T_c$, show
simultaneously good descriptions of the particle yields, the
$p_T$-spectra and the differential elliptic flow $v_2(p_T)$ for
different hadron species from  Au+Au collisions at top RHIC
energies~\cite{Nonaka:2006yn,Hirano:2005wx}. Pure ideal hydrodynamic
simulations with PCE-EOS  give good descriptions of particle yields
and spectra, but over-predict the elliptic flow. Viscous suppression
of additional elliptic flow buildup in the hadronic phase is
necessary to reduce the chemical non-equilibrium enhancement of pion
elliptic flow without affecting the hadron yields. These two facts
strongly indicate that the chemical non-equilibrium $v_2$
enhancement and viscous $v_2$ suppression tend to partially balance
each other in the hadronic phase at RHIC energies. This will no
longer be true  at LHC energies, where the elliptic flow fully
saturates in the QGP phase~\cite{Hirano:2007xd} such that viscous
$v_2$ suppression in the hadronic stage is no longer significant,
while the $v_2$ enhancement caused by the non-equilibrium chemistry
in the HRG still persists since the latter reshuffles the momentum
anisotropy generated in the QGP phase among different hadron species
in the HRG phase.

Considering the above factors,  viscous hydrodynamics  requires
implementation of the PCE-EOS for the HRG phase as well as matching
to a hadron cascade. Using a partially chemically equilibrated HRG
EOS is necessary to ensure approximate insensitivity of the final
results to the exact value of the switching temperature $T_{ch}$,
where one switches from viscous hydrodynamics to the cascade
model~\cite{Teaney-Thesis}. Such a proper treatment of the HRG phase
is necessary before extracting the QGP viscosity from experimental
data. A relatively easy procedure is to directly connect viscous
hydrodynamics with a hadron cascade right after $T_c$, where the HRG
is still in chemical equilibrium. However, the PCE-EOS helps to
generate the correct hadron chemistry even at lower temperatures,
thus enlarging the temperature window where one can switch from
hydrodynamics  to the hadron cascade description, without changing
the hadron yields. This makes the implementation of a more realistic
dynamical freeze-out (see Chap. 8.2 ) possible. Such research is
important, but will have to be left
for the future.\\

\section{Bulk viscosity}
Currently, most of the existing viscous hydrodynamic calculations
concentrate on shear viscosity effects, but neglect bulk viscosity
for simplicity. Bulk viscous effects were investigated in Chap.~6
(see Ref.~\cite{Song:2009Bulk}). We found that the negative bulk
pressure ($\Pi=-\zeta \theta $ in the N-S limit) reduces the total
pressure (which effectively softens the EOS near $T_c$), decelerates
both longitudinal and transverse expansion, and suppresses the build
up of flow. As a result, the hadron spectra become steeper, and the
elliptic flow is suppressed in the low-$p_T$ region.
Fig.~\ref{Spectra-v2-bulk} in Chap.~6 showed that even the
``minimal" bulk viscosity, as constructed there, leads to $\sim
25\%$ additional viscous $v_2$ suppression if one uses the N-S
initialization. This translates into a $\sim 25\%$ uncertainty for
the extracted value of QGP shear viscosity $\eta/s$  when comparing
 experimental elliptic flow data with viscous hydrodynamics results.
 However, there are no
well-defined theoretical results for the bulk relaxation time
$\tau_\Pi$. Critical slowing down during the phase transition
suggests a larger $\tau_\Pi$ near $T_c$, but the exact analytical
form and its maximum  value are still unknown. Using a $\tau_\Pi$
with peak value of $ 5 \ \mathrm{fm/c}$ near $T_c$, we found that
bulk viscous $v_2$ suppression is very sensitive to the bulk
pressure initialization one inputs, which ranges from 2\% (or zero
initialization) to $\sim 10\%$ (for N-S initialization), if minimal
bulk viscosity as defined in Chap.~3.6 is used.

Considering the current theoretical uncertainties on bulk viscosity
itself (the peak value of  $\zeta/s$ near the phase transition and
the bulk viscosity in the hadronic phase) and our pour theoretical
control over the initialization for the bulk pressure  and the bulk
relaxation time (see Chap.~3.5 for details), it is thus unclear how
much uncertainty bulk viscosity can bring to the extraction of a
shear viscosity value of the QGP. Clearly, bulk viscous effects must
be taken into account and additional theoretical research is needed
to reduce the uncertainties of its value and relaxation time. Our
analysis showed that $\eta/s$ and $\zeta/s$ will always combine in
their effects on viscous $v_2$ suppression. At present, it is not
clear how to
sperate them, possibly by using additional observables.  \\

\section{Glauber model vs. CGC initialization}

We now come to what may turn out to be the most serious road block
for precision measurements of the QGP shear viscosity: our
insufficient knowledge of the initial source eccentricity
$\varepsilon$. It has now been known for a while that the Color
Glass Condensate (CGC) model, implemented in the initial entropy or
energy density profile via the fKLN parametrizations
(see~\cite{Hirano:2005xf,Drescher:2006pi} and Chap.~3.2 for
references), leads to $\sim 30\%$ larger initial source
eccentricities than the popular Glauber model. Ideal fluid dynamics
transforms this larger source eccentricity into $\sim30\%$ larger
elliptic flow. Since the extraction of $\eta/s$ is based on the
viscous {\em suppression} of $v_2$, obtained by comparing the
measured elliptic flow with an ideal (inviscid) fluid dynamical
benchmark calculation, a 30\% uncertainty in this benchmark can
translate into a 100\% uncertainty in the extracted value for
$\eta/s$. This was recently shown by Luzum and Romatschke (see
Fig.~8 in~\cite{Luzum:2008cw}). This uncertainty trumps most of the
other uncertainties discussed above. Worse, since the initial source
eccentricity depends on details of the shape of the fKLN profile
near the edge of the distribution where the gluon saturation
momentum scale $Q_s$ becomes small and the CGC model reaches its
limit of applicability, there is little hope that we can eliminate
this uncertainty  theoretically from first principles.

Based on their analysis of charged hadron elliptic flow data from
the STAR experiment, allowing for a 20\% systematic uncertainty of
these data, the authors of~\cite{Luzum:2008cw} found an allowed
range $0 < \eta/s < 0.1$ for Glauber and $0.08 < \eta/s < 0.2$ for
CGC initial conditions. This amount to a \emph{O}(100\%) difference
for the extracted $\eta/s$ between these two initialization models.
Since the analysis in~\cite{Luzum:2008cw} did not include a
comprehensive investigation of effects caused by permissible
variations of the EOS near $T_c$, by bulk viscosity, or by late
hadronic viscosity and non-equilibrium chemical composition at
freeze-out (see proceeding sections), one should add a significant
additional uncertainty band to these ranges. Furthermore, correcting
the experimental data for event-by-event fluctuations in the initial
source eccentricity~\cite{PRV-Poskanzer} may bring down the measured
$v_2$ values even below the range considered in~\cite{Luzum:2008cw}.
Still, we agree with Luzum and Romatschke that, even when adding all
the above effects in magnitude (ignoring the fact that several of
them clearly have opposite signs), viscous hydrodynamics with
$\eta/s > 5 \times (1/4\pi)$ would suppress the
elliptic flow too much to be incompatible with experiment.\\

\section{Numerical viscosity}

To study the effects from shear and bulk viscosity one must ensure
that numerical viscosity is under control and sufficiently small.
Simply speaking, numerical viscosity comes from the discretization
of the hydrodynamic equations for numerical calculation. It causes
entropy production even in ideal hydrodynamics without shocks and
can never be fully avoided. To minimize numerical viscosity, the
flux-corrected transport algorithm  SHASTA~\cite{SHASTA} employed by
 VISH2+1 (and by its ideal fluid ancestor AZHYDRO~\cite{AZHYDRO})
 implements an ``antidiffusion step'' involving a
parameter $\Lambda$ called ``antidiffusion constant''~\cite{SHASTA}.
For a given grid spacing, numerical viscosity is maximized by
setting $\Lambda=0$. In standard situations, the default value
$\Lambda=\frac{1}{8}$ minimizes numerical viscosity
effects~\cite{SHASTA}. With $\Lambda=\frac{1}{8}$ and typical grid
spacing $\Delta x=\Delta y=0.1$\,fm, $\Delta \tau=0.04$\,fm/$c$,
AZHYDRO generates only 0.3\% additional entropy in central Au+Au
collisions. This is negligible when compared with the ${\cal
O}(10\%)$ entropy production by  VISH2+1 for a fluid with real shear
viscosity $\eta/s=1/4\pi$.

By increasing the grid spacing in  AZHYDRO  and/or changing
$\Lambda$, we can explore the effects of numerical viscosity on
radial and elliptic flow. We find that numerical viscosity has
little effect on the development of radial flow but reduces $v_2$ in
very much the same way as does real shear viscosity. Since we gauge
the effects of $\eta/s$ on $v_2$ by comparing results from VISH2+1
for $\eta/s\ne 0$ to those for $\eta/s=0$, we should explore how
much in the latter case $v_2$ is already suppressed by numerical
viscosity. We can do this by setting $\eta/s=0$ and reducing the
grid spacing until $v_2$ stops changing (i.e. until we have
completely removed all numerical viscosity effects on $v_2$). In
this way we have ascertained that for our standard grid spacing
numerical viscosity suppresses the differential elliptic flow
$v_2(p_T)$ by less than 2\%.\\

\section{Conclusions}
While the elliptic flow $v_2$ generated in non-central heavy-ion
collisions is very sensitive to the shear viscosity to entropy ratio
$\eta/s$ of the QGP, it is also significantly affected by (i)
details of the initialization of the hydrodynamic evolution, (ii)
bulk viscosity and sound speed near the quark-hadron phase
transition, and (iii) the chemical composition and non-equilibrium
kinetics during the late hadronic stage. Not all of these effects
are presently fully under control. Recent attempts to extract the
specific shear viscosity $\eta/s$ phenomenologically, by comparing
experimental elliptic flow data with viscous hydrodynamics, have
established a robust upper limit~\cite{Luzum:2008cw,Song:2008si}
\begin{eqnarray}
\label{etas2}
  \left.\frac{\eta}{s}\right|_\mathrm{QGP} < 5\times \frac{1}{4\pi},
\end{eqnarray}
tantalizingly close to the conjectured KSS
bound~\cite{Kovtun:2004de}, but further progress requires
elimination of the above systematic uncertainties. Since some of
these influence the build-up of elliptic flow in opposite
directions, it is quite conceivable that the QGP specific viscosity
is in fact much closer to the KSS bound
$\left.(\eta/s)\right|_\mathrm{KSS} =1/4\pi$ than suggested by the
upper limit (\ref{etas2}). Ongoing improvements on the theory side
should help to reduce or eliminate most of the above uncertainties,
bringing us closer to a quantitative extraction of $\eta/s$ for the
quark-gluon plasma. The single largest uncertainty, however, is
caused by our poor knowledge of the initial source eccentricity
which varies by about 30\% between models. As shown in
\cite{Luzum:2008cw}, this translates into an ${\cal O}(100\%)$
uncertainty for $\eta/s$. It seems unlikely that theory can help to
eliminate this uncertainty from first principles. It thus appears
crucial to develop experimental techniques that may help us to pin
down the initial source eccentricity phenomenologically, with
quantitative precision at the percent level.
\chapter[$\eta/s$ from Other Considerations and Extraction Methods]
{$\eta/s$ from Other Considerations\\ \hspace{3cm} and Extraction
Methods}

\section{Shear viscosity and Knudsen number}

The Knudsen number approach starts with fitting the experimental
multiplicity scaling curve for  $v_2/\varepsilon$  as a function of
$\frac{1}{S}\frac{dN}{dy}$ by a simple formula:
\begin{eqnarray}
\frac{v_2}{\varepsilon}=\frac{v_2^{hydro}}{\varepsilon}\frac{1}{1+K/K_0}\,
. \label{Ollitrault-eqn}
\end{eqnarray}
The above empirical formula are phenomenologically obtained from a
two-dimensional parton cascade model
simulations~\cite{Ollitrault:PLB2005,Ollitrault:PRC08-hydro-transport,
Ollitrault:PRC07-v2scaling} from which one can directly measure the
elliptic flow $v_2$ by sampling the particle distribution, and also
calculate the Knudsen number $K$\footnote{\emph{Knudsen number $K$
is defined as the ratio of the mean free path and the system size,
$K\equiv\frac{\lambda}{R}$. For a parton cascade model simulated
within a two-dimensional $R_x \times R_y$ box, the system size is
estimated by $R = (\frac{1}{R_x^2}+\frac{1}{R_y^2})^{-1/2}$. The
mean free path $\lambda$ is inversely proportional to transport
cross section $\sigma$ and particle density $n$,
$\lambda=\frac{1}{\sigma n}$, with $n=\frac{N}{4\pi R_x R_y}$ (N is
the number of test particles in a Monte Carlo simulation).}}. $K_0$
is the free parameter in this formula, and from two-dimensional
parton cascade model simulation runs one finds $K_0=0.7$ for a
dilute system.

Extracting the QGP viscosity requires an estimate of the Knudsen
number from experimental data first. It is known that elliptic flow
$v_2$ is developed at an early stage, roughly at a time scale
$\bar{R}/c_s$ for an EoS with constant speed of sound
$c_s$~\cite{Ollitrault:PRC07-v2scaling}. At the time $\tau \sim
\bar{R}/c_s$, the particle density can be estimated by $c \tau n
\sim \frac{1}{S} \frac{dN}{dy}$, assuming that total particle number
is conserved and the transverse size of the system does not vary
significantly. Then one can connect the Knudsen number with the
final multiplicity per unit area $(1/S) dN/dy$ as
follows~\cite{Ollitrault:PLB2005}:
\begin{eqnarray}
\frac{1}{K} \equiv \frac{R}{\lambda} = \frac{\sigma n R}{1}=
\frac{\sigma}{S}\frac{dN}{dy} c_s \, .\label{Kndsen}
\end{eqnarray}

After fitting the experimental multiplicity scaling curve for
$v_2/\varepsilon$ (for Glauber initial eccentricity and CGC initial
eccentricity, respectively) with eqs.(~\ref{Ollitrault-eqn}) and
(~\ref{Kndsen}), one finds $v_2^{hydro}/\varepsilon =0.30 \pm 0.02$
and $\sigma = 4.3 \ \mathrm{mb}$ for Glauber initialization, and
$v_2^{hydro}/\varepsilon =0.22 \pm 0.01$ and $\sigma = 7.6\
\mathrm{mb}$ for CGC initialization (using
$c_s^2=1/3$)~\cite{Ollitrault:PRC07-v2scaling}.

The shear viscosity to entropy ratio is estimated using the formula
for a classical gas of massless particles with isotropic
differential cross section~\cite{Kox:1976}:
\begin{eqnarray}
\frac{\eta}{s}=0.316\frac{T}{c\sigma n} \, . \label{eta-cross}
\end{eqnarray}
One finds $\eta/s=0.19$ for Glauber initialization and $\eta/s=0.11$
for CGC initialization, with the inputs $T= 200\ \mathrm{MeV}$ and
$n= 3.9 \
\mathrm{fm^{-3}}$~\cite{Ollitrault:PRC07-v2scaling}\footnote{\emph{$n=
3.9\ \mathrm{fm^{-3}}$ is estimated at the time when $v_2$ is
developed~\cite{Ollitrault:PLB2005}.}}.  Since the particle density
evolves with time, eq.~(\ref{eta-cross}) shows that a constant cross
section and constant $\eta/s$ throughout the fireball evolution are
mutully exclusive. The $\eta/s$ value extracted from this method
thus represents at best an average value, taken over the history of
the medium. Detailed hydrodynamic simulations in our
group~\cite{PRV-Scott} show that the evolution history (and thus
this average) depends on the initialization (Glauber vs. CGC) and
the EOS. The estimated $\eta/s$ values from
Ref.~\cite{Ollitrault:PRC07-v2scaling} thus have to be considered
with caution.
\\

\begin{figure}[t]
\label{Xu-Fig}
\begin{center}
\includegraphics[width=.495\linewidth,height=60mm,clip=]{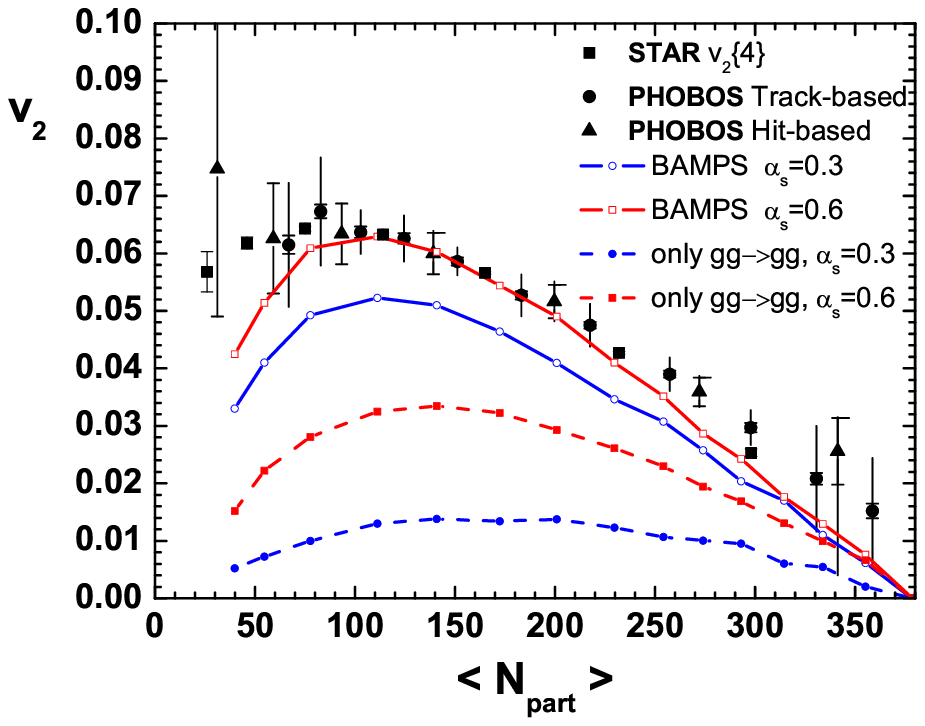}
\includegraphics[width=.495\linewidth,height=60mm,clip=]{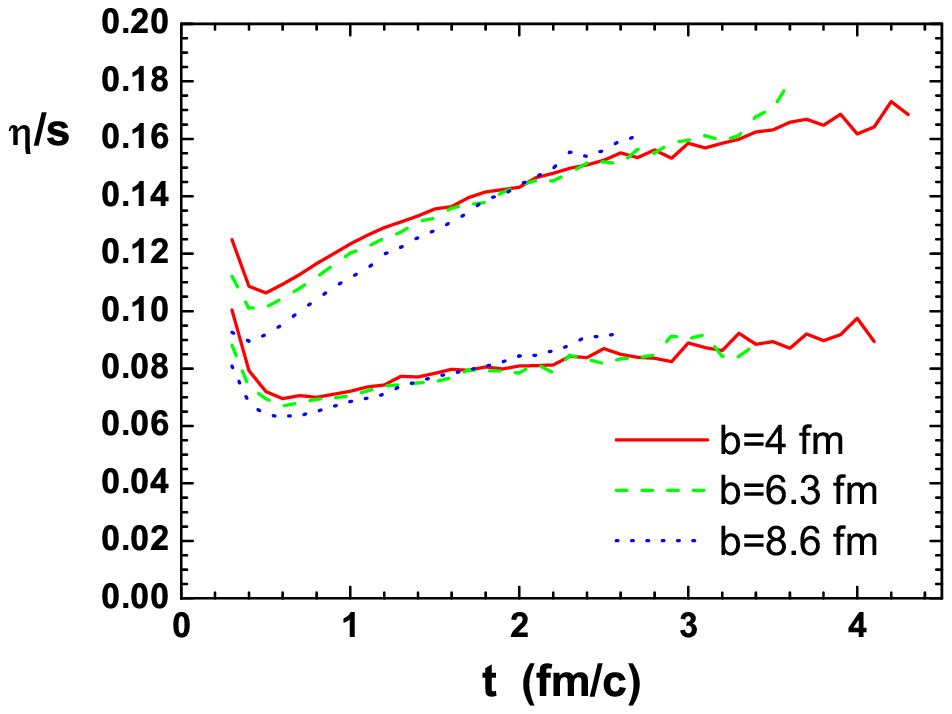}
\vspace{-3mm}
 \caption[elliptic flow and shear viscosity from BAMPS]{Left panel:
 elliptic flow $v_2(|y|<1)$ from BAMPS~\cite{Xu:PRL08-V2shear}(Boltzmann Approach of Multi-Parton Scattering) using
$\alpha_s=0.3$ and $0.6$, compared with the
PHOBOS~\cite{Back:2004mh-PHOBOS} and STAR~\cite{Adams:2004bi-STAR}
data. Right panel: The shear viscosity to entropy density ratio
$\eta/s$ at the central region during the entire
expansion~\cite{Xu:PRL08-V2shear}. $\eta/s$ values are extracted
from the simulations at impact parameter $b=4$, $6.3$, and $8.6$
fm.The upper band shows the results with $\alpha_s=0.3$ and the
lower band the results with $\alpha_s=0.6$}
\end{center}
\end{figure}

\section{Shear viscosity from the parton cascade model}
In contrast to the macroscopic hydrodynamic approach, the QGP
fireball evolution can also be microscopically described by a Parton
Cascade Model (PCM), solving the Boltzmann equation for colliding
gluons. The earlier parton cascade
model~\cite{Molnar:PRC00-PCM,Zhang:1997ej} that only included
$2\rightarrow 2$ elastic gluon collisions was not be able to achieve
thermal equilibrium and to reproduce the large experimentally
measured elliptic flow, unless unrealistically large scattering
cross sections, far beyond the expectations of PQCD on which the
parton cascade is based, were assumed~\cite{Molnar:2001ux}.
Recently, a radiative parton cascade model including radiative $2
\leftrightarrow 3$ processes and their inverse has been developed to
study, on a semiclassical level, the dynamics of gluon matter
produced at RHIC
energies~\cite{Xu:PRC05-Thermal,Xu:2007aa,Xu:2008av,Zhang:2008zzu,Zhang:2009rk}.
This approach has shown that some well-known strongly coupled QGP
phenomena (such as  fast thermalization of the initial
non-equilibrium system, large elliptic flow and a small shear
viscosity to entropy ratio) can be achieved within a framework that
is based on weakly coupled QCD~\cite{Xu:2008dv}.

The left panel of Fig.~\ref{Xu-Fig} shows the elliptic flow $v_2$,
calculated from the parton cascade model with only two body
interactions ($gg\rightarrow gg$) and with multi-parton interactions
($ gg\rightarrow gg$, $gg\rightarrow ggg$ and $ggg\rightarrow gg$,
denoted as BAMPS\footnote{\emph{Boltzmann Approach of Multi-Parton
Scattering}} in the figure), using different values of the coupling
constant $\alpha_s$ as input~\cite{Xu:PRL08-V2shear}. One finds that
large elliptic flow $v_2$,  comparable to the experimental data, can
be achieved within the perturbative QCD description if the radiative
collision processes $gg \leftrightarrow ggg$ are included.

The shear viscosity $\eta$ can be extracted from the Parton Cascade
Simulation through the following formula (using the N-S
approximation)~\cite{Xu:PRL08-shear}:
\begin{equation}
\label{shv} \eta \cong \frac{1}{5} n \frac{\langle
E(\frac{1}{3}-v_z^2) \rangle} {\frac{1}{3}-\langle v_z^2 \rangle}
\frac{1}{\sum R^{\rm tr}+ \frac{3}{4} n
\partial_t (\ln \lambda)}\,.
\end{equation}
Here $n$ is gluon density, $E$ is gluon energy, $v_z=\frac{p_z}{E}$
is the gluon velocity along the beam direction and $\lambda$ is the
gluon fugacity, $\lambda=n/n_{eq}$. $\sum R^{\rm tr}$ denotes the
total transport collision rate, the expression of which can be found
in Ref~\cite{Xu:PRL08-shear}.

The right panel of Fig.~\ref{Xu-Fig} shows the shear viscosity to
entropy ratio $\eta/s$ extracted from BAMPS, for different impact
parameters and $\alpha_s$ values. One finds that $\eta/s$ depends
weakly on the gluon density or temperature (indicated by the time
evolution on the horizontal axis), but depends strongly on
$\alpha_s$. For $\alpha_s=0.6$, which best fits the experimental
elliptic flow data, $\eta/s \simeq 0.08$, a value that is very close
to the minimal KSS bound $1/4\pi$, is reached. The authors stated,
however, that $\eta/s$ may increase once  hadronization and a
subsequent hadron cascade are included~\cite{Xu:2008dv}.

The authors of Ref.~\cite{Xu-El:2008yy} derived an improved
expression for shear viscosity within the second order
Israel-Stewart formalism, for a one-dimensionally expanding particle
system. One finds that although the improved I-S expression
increases the extracted $\eta/s$, when compared to eq.~(\ref{shv}),
the difference is within 20\% for $\alpha_s =0.3-0.6$~\cite{Xu:PRL08-shear}.\\

\section{Shear viscosity and radiative energy loss}

The estimation of the shear viscosity from radiative energy loss is
inspired by the kinetic theory expression of shear viscosity
$\eta$~\cite{Kox:1976},
\begin{eqnarray}
\eta = C \rho \La p \Ra  \lambda_{mfp}=C \La p \Ra
\frac{1}{\sigma_{tr}} \, ,\label{shear-kinetic}
\end{eqnarray}
where $\rho$ is the particle density, $\La p \Ra$ is the average
particle momentum and $\lambda_{mfp}$ is the mean free path. Again,
the last expression replaces the mean free path by
$\lambda_{mfp}=(\rho \sigma_{tr})^{-1}$ with particle density $\rho$
and transport cross section $\sigma_{tr}$. In perturbative QCD and
QED, the transport cross section $\sigma_{tr}$ is dominated by small
angle scattering and can be expressed by
\begin{eqnarray}
\sigma_{tr}=\frac{4}{\La E^2_{cm} \Ra}\int d k^2_\bot k^2_\bot
\frac{d \sigma}{dk^2_\bot} \, , \label{cross-Sec}
\end{eqnarray}
where $\La E^2_{cm} \Ra$ is the center of mass energy, with $\La
E^2_{cm} \Ra \approx 18 T^2 $ for a thermal medium with temperature
$T$.

On the other hand, the transport parameter $\hat{q}$ (which
describes the per unit length energy loss of a fast parton traveling
through the QGP by gluon bremsstrahlung), has a theoretical
definition associated with the transverse momentum broadening of a
fast parton through interaction with the medium:
\begin{eqnarray}
\hat{q}=\rho \int d k^2_\bot k^2_\bot \frac{d \sigma}{d k^2_\bot} \,
. \label{q-hat}
\end{eqnarray}
Here $\frac{d \sigma}{dk^2_\bot}$ is the differential cross section
for elastic scattering in the medium, and $\rho$ is the medium
particle density.

Comparing eqs. (\ref{shear-kinetic}) and (\ref{q-hat}), one finds
that the shear viscosity to the entropy density ratio $\eta/s$ can
be expressed as~\cite{Majumder:PRL07-shear}:
\begin{eqnarray}
\frac{\eta}{s} \approx 1.25 \frac{T^3}{\hat{q}}\, , \qquad \qquad
\mathrm{(for \ weak \ coupling)}.\label{shear-qhat}
\end{eqnarray}
Here one set $C=1/3$~\cite{Kinetic-Theory80,Gyulassy85} and used
$\La p \Ra \approx 3T$, and $s \approx 3.6\rho$ for a thermal
ensemble consisting of massless bosons.

In Ref.~\cite{Majumder:PRL07-shear} the relationship between shear
viscosity $\eta$ and jet quenching parameter $\hat{q}$ was formally
derived within the weakly coupled QGP picture, for two different
circumstances: \textbf{1)} the distribution of quasi-particles
satisfies a linearized Boltzmann equation with a soft scattering
collision term in leading-log approximation for a pure gluon
gas~\cite{Arnold:2000dr}; \textbf{2)} the evolution of the
quasi-particle distribution function satisfies a Fokker-Planck
equation, which does not contain a collision term but includes the
effects of a random color field. The latter contributes a small
anomalous shear viscosity~\cite{Asakawa:2006tc,Asakawa:2006jn}. One
finds that the coefficient 1.25 found in eq. (\ref{shear-qhat})
holds in both of these cases.

In the above weak coupling scenario (which describes the quark gluon
plasma by nearly massless quasi-particles), one estimates that
$\eta/s=0.12-0.24$ for a lower value of $\hat{q}^{HT}=1-2 \
\mathrm{GeV^2/fm}$, and $\eta/s=0.008-0.024$ for a higher value of
$\hat{q}^{ASW}=10-30\ \mathrm{GeV^2/fm}$, with $T\approx 340 MeV$.
The uncertainty of the estimated $\eta/s$ mainly comes from the
different values of $\hat{q}$ obtained from different schemes, and
the error bands in each estimate comes from whether one takes
$\hat{q}$ to be proportional to $\varepsilon^{4/3}$ or to $T^3$. The
authors of~\cite{Majumder:PRL07-shear} pointed out that for the
larger $\hat{q}$ value, the $\eta/s$ ratio derived
from(~\ref{shear-qhat}) strongly violates the KSS Bound, which they
interpreted as a sign of break-down of the weakly-interacting
quasi-particle picture underlying the relation(~\ref{shear-qhat}).

However, eq. (\ref{shear-qhat}) is no longer true in the case of
strong coupling. After checking the shear viscosity and $\hat{q}$
from strongly coupled N=4 superymmetric Yang-Mills (SYM) theory and
for pion gases (where the strong coupling limit of QCD is exhibited
by confinement), Majumder, Muller and
Wang~\cite{Majumder:PRL07-shear} also found that
\begin{eqnarray}
\frac{\eta}{s} \gg 1.25   \frac{T^3}{\hat{q}} \, , \qquad \qquad
\mathrm{(for \ strong \ coupling}). \label{shear-qhat2}
\end{eqnarray}

So in the strong coupling limit, $\eta/s$ can not be usefully
derived from a measurement of $\hat{q}$. Considering
eqs.(\ref{shear-qhat}) and (\ref{shear-qhat2}) together, they
conclude instead that \emph{``The determination of both shear
viscosity $\eta$ and jet quenching parameter $\hat{q}$ from
experimental data would give a quantitative assessment of the
strongly coupled nature of the quark gluon plasma produced in heavy
ion collisions"}~\cite{Majumder:PRL07-shear}.\\

\section{Shear viscosity and heavy quark diffusion}

Heavy quarks are  unique probes to study the transport properties of
the QGP. Based on the property that the masses ($M_c = 1.3 \
\mathrm{GeV}$ and $M_b= 4.2\ \mathrm{GeV}$) of heavy quarks are much
larger than the medium temperature ($ T\sim \ 300 \mathrm{MeV}$),
Moore and Teaney proposed a simple model to study the motion of
heavy quarks by the following stochastic
equations~\cite{Moore:2004tg}:
\begin{eqnarray}
\frac{d p_i}{dt}=\xi_i(t) - \eta_D p_i,
 \qquad \qquad \La \xi_i(t) \xi_j(t') \Ra = k \delta_{ij} \delta
 (t-t').
\end{eqnarray}
Here $\eta_D$ is the drag coefficient, which measures the momentum
loss per unit time, and $\xi_i$ characterizes the uncorrelated
random momentum kicks from the medium.

The diffusion coefficient $D$  is defined as the probability of
``starting a heavy quark at $x=0$ at $t=0$ and finding it at the
mean squared position $Dt$ at a later time":
\begin{eqnarray}
\La x_i(t) x_j(t)\Ra =2D t\delta_{ij}\  \longrightarrow  \ 6 Dt= \La
x^2(t)\Ra.
\end{eqnarray}
One finds that $D$ is related to the drag coefficient $\eta_D$ at
zero momentum~\cite{Moore:2004tg}:
\begin{eqnarray}
D=\frac{T}{M \eta_D (0)}\,.
\end{eqnarray}
After perturbatively calculating the heavy quark diffusion
coefficient $D$ and comparing it with the weakly-coupled QGP shear
viscosity $\eta$ at leading order, Moore and Teaney found that the
heavy quark diffusion coefficient is related to the hydrodynamic
diffusion coefficient $\frac{\eta}{e+p}$ by a factor of $\sim
6$~\cite{Moore:2004tg},  For later convenience, we rewrite this as
follows (using $e+p=Ts$):
\begin{eqnarray}
\frac{\eta}{s} \approx \frac{1}{6} TD \qquad \qquad \qquad
\mathrm{(wQGP)}\,. \label{diffu-weak}
\end{eqnarray}

%
\begin{figure}[t]
  \begin{center}
    \begin{minipage}[h]{0.6\linewidth}
    \includegraphics[width=0.9\linewidth,height=0.9\linewidth]{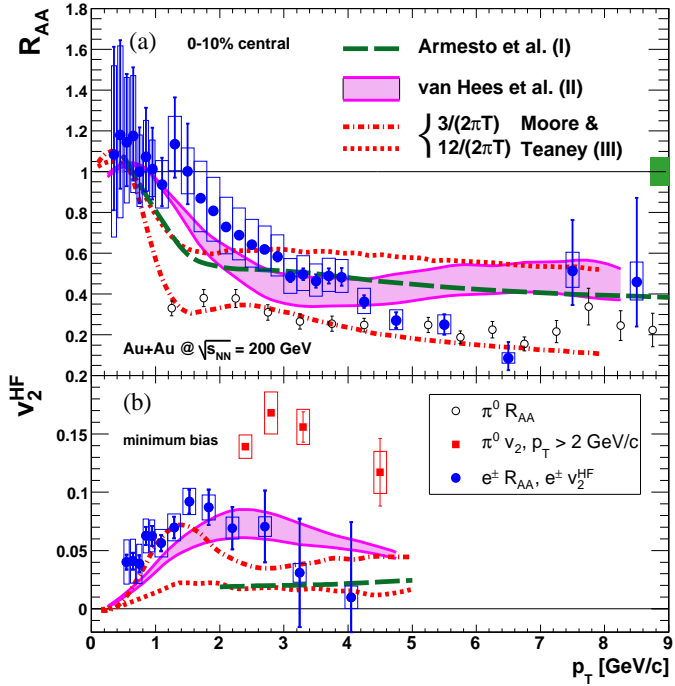}
    \end{minipage}
    \begin{minipage}[h]{0.35\linewidth}
      \caption[$R_{\rm AA}$ and $v_2$of heavy-flavor electrons]
      {\label{Heavy-quark}(a) $R_{\rm AA}$ of heavy-flavor electrons in
0-10\% central collisions compared with $\pi^0$
data~\cite{Adler:2005ab} and model calculations (curves
I~\cite{Armesto:2005mz}, II~\cite{vanHees:2005wb}, and
III~\cite{Moore:2004tg}). (b) $v_2^{\rm HF}$ of heavy-flavor
electrons in minimum bias collisions compared with $\pi^0$
data~\cite{Adler:2003cb} and the same models. (Taken from the PHENIX
data~\cite{Adare:2006nq})}
     \vspace{0.3cm}
    \end{minipage}
  \end{center}
\end{figure}
%

The heavy quark diffusion coefficient can also be calculated in
strongly coupled $N=4$  Super-Yang-Mills theory, using the AdS/CFT
correspondence, which gives $D \simeq 1/(2\pi
T)$~\cite{CasalderreySolana:2006rq}. Recalling the remarkable result
$\eta/s = 1/4\pi$ for a large class of strongly coupled theories
that have a gravity dual ($N=4$ SYM theory is one example among
them), one finds the following simple relation between the heavy
quark diffusion constant and the shear viscosity for a strongly
coupled system:
\begin{eqnarray}
\frac{\eta}{s} \approx \frac{1}{4\pi} D(2\pi T) = \frac{1}{2} TD
\qquad  \mathrm{(N=4 \ SYM)}\,. \label{diffu-strong}
\end{eqnarray}

Fig.~\ref{Heavy-quark} shows the experimental quenching ratio
$R_{AA}$ (see Chap.~1.13) and $v_2$ for ``non-photonic electrons"
from semi-leptonic decays of heavy flavors (i.e.  of hadrons
carrying charm and botton quarks) for Au+Au collisions at top RHIC
energies, together with the theoretical predictions from different
groups. The model that describes both $R_{AA}$ and $v_2$ well is
from van Hees et al.~\cite{vanHees:2005wb}, leading to an estimate
of the heavy quark diffusion constant $D \simeq (4-6) (2 \pi T)$.
Using the relations (\ref{diffu-weak}) and (\ref{diffu-strong}) for
weakly and strongly coupled systems, respectively, one finds:
\begin{eqnarray}
\frac{\eta}{s}= \left\{ \begin{array}{l@{~~~}l}
  (1.33-2)/4\pi & \mathrm{for \  weak \ coupling, }\\
 (4-6)/4\pi & \mathrm{for \ strong \ coupling}.
          \end{array} \right.
\end{eqnarray}

A more detailed assessment of estimating $\eta/s$ from the heavy
quark diffusion constant $D$ with different models can be found in
the recent review article~\cite{Rapp:2008qc}, to which we refer the interested reader.\\

\section{Summary and comments}

\begin{figure}[t]
\begin{center}
\includegraphics[width=0.9\linewidth,height=70mm]{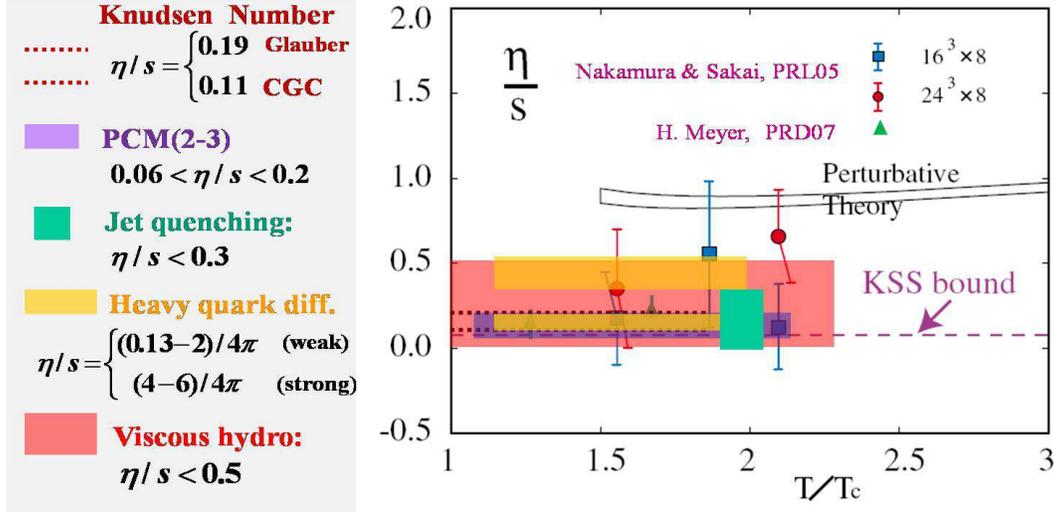}
\vspace{-3mm}
 \caption[$\eta/s$: estimated from other methods]{A summary for the estimated
 $\eta/s$ from different method: \textbf{a)} from Knudsen number and multiplicity $v_2$ scaling
 (Ref.~\cite{Ollitrault:PRC07-v2scaling} and Chap.~9.1);
 \textbf{b)} from parton cascade model with $2 \rightarrow 2$ and $2 \rightarrow 3$ process
 (Ref.~\cite{Xu:PRL08-V2shear} and Chap.~9.2); \textbf{c)} from  transport parameter $\hat{q}$ of radiative energy
 loss (Ref.~\cite{Majumder:PRL07-shear} and Chap.~9.3); \textbf{d)} from heavy quark diffusion constant $D$
 (Ref.~\cite{Rapp:2008qc} and Chap.~9.4); \textbf{e)} from viscous hydrodynamics
 (Ref.~\cite{Luzum:2008cw,Song:2008si} and Chap.~8);
 \textbf{f)} SU(3) gluconic lattice QCD results from
 Nakamura and Sakai~\cite{Nakamura:2004sy} and Meyer~\cite{Meyer:2007ic},
 respectively.
} \label{estimate-etaS}
\end{center}
\end{figure}

In Fig.~\ref{estimate-etaS}, we summarize the estimated $\eta/s$
from different methods, together with the lattice QCD estimates for
a pure gluon plasma. One finds that all of these results fall in a
relatively narrow band near the KSS bound at $1/4\pi$. This strongly
indicates that the QGP created at RHIC energies is an almost perfect
liquid.

When interpreting the error bands in the figure, one must however
pay attention to the assumptions and limitations of each method. For
example, the narrow purple band is from the estimate based on parton
cascade model, which simulates the dynamics of the pure SU(3)
gluons~\cite{Xu:PRL08-V2shear}. To compare with the experimental
data, it directly converts partons to hadrons, but totally neglects
the effects from the phase transition and the succeeding hadronic
expansion. Including these, the actual error band from the parton
cascade model based estimate is very likely much larger than the
current one. Although in the future the parton cascade model can be
connected with the hadron cascade model to account for the highly
viscous hadronic stage, this will not resolve generic difficulties
of cascade models in treating the change of vacuum structure that
occurs during the phase transition and reproducing a realistic EOS
(see Chap.~1.12 for further discussion). As a result, the parton
cascade model based approach also has difficulties to properly
account for bulk viscous effects, which, as we showed in Chap.~6,
should not be neglected when extracting the QGP shear viscosity.

The Knudsen number approach provides a simple way to estimate the
shear viscosity to entropy ratio $\eta/s$ from the experimental
multiplicity  scaling data for $v_2/\varepsilon$. However,  it only
gives an effective viscosity averaged over the whole evolution of
the fireball, rather than the QGP viscosity,  since the Knudsen
number formula derived in Ref.~\cite{Ollitrault:PRC07-v2scaling}
only deals with time averaged variables. On the other hand, the
framework of this approach has also been challenged. For example,
questions were raised about the basic assumption of this approach,
which assumes a constant transport cross section $\sigma$ during the
whole fireball evolution and then extracts a constant $\eta/s$ for a
particular collision system. However, the Parton Cascade Model shows
that a constant cross section $\sigma$ corresponds an $\eta/s$ that
increases monotonically with time, while a constant $\eta/s$
corresponds to a monotonically increasing
$\sigma$~\cite{Molnar:2008jw}. This raises questions about the
self-consistency of the Knudsen number approach.

The radiative energy loss and heavy quark diffusion approaches
investigated both weakly and strongly coupled media and showed that
the corresponding formulas for $\eta/s$ are dramatically different.
As emphasized in Ref.~\cite{Majumder:PRL07-shear}, instead of using
the different formulas to extract $\eta/s$, it is more preferable to
use them to distinguish the strongly or weakly coupled nature of the
QGP medium after both $\eta/s$ and $\hat{q}$ have been extracted
from experimental data.

Simulations from current viscous hydrodynamic models show $\eta/s<
5/4\pi$. This error band is almost the largest, when compared with
other estimates. However it is a pretty safe results, since it takes
into account all kinds of uncertainties that currently persist (i.e.
initial conditions, EOS, bulk viscosity, viscosity of the hadronic
stage -- see Chap. 8 for details). Compared with other methods,
viscous hydrodynamics is a tool that directly attacks the problem
for extracting the QGP viscosity. In the future it can be
systematically improved through a controlled analysis of all
presently persisting uncertainties. This is expected to further
narrow down the uncertainty in $\eta/s$ by a significant amount.
\chapter{Summary and Concluding Remarks}
Ideal hydrodynamics has been a great success in describing and (in
the case of elliptic flow) even predicting relativistic heavy ion
collision data at RHIC energies. This has led to the well-known
announcement that ``RHIC scientists serve up the perfect liquid" --
the quark-gluon plasma (QGP). To answer the question \emph{"How
perfect is the QGP fluid?"}quantitatively, one needs a dynamical
framework which allows to include dissipative effects accurately and
consistently. Causal viscous hydrodynamics is such a tool that not
only improves the description of heavy-ion fireball dynamics in
regions of parameter space where ideal hydrodynamics begins to fail,
but also allows to extract even small values of $\eta/s$ through
detailed comparison with precise experimental data.

This thesis focused on causal viscous hydrodynamics for relativistic
heavy ion collisions. The work included developing a numerical code
to solve the causal viscous hydrodynamic equations and searching for
possible signals for extracting the QGP viscosity. Based on the
explicit form of the general 2nd-order Israel-Stewart (I-S)
equations in 2+1-dimensions (derived by my advisor U. Heinz and
myself~\cite{Heinz:2005bw}), assuming longitudinal boost-invariance
but arbitrary dynamics in the 2-dimensional transverse plane, I
subsequently developed a (2+1)-d viscous hydrodynamic code, called
VISH2+1 (for ``Viscous Israel-Stewart Hydrodynamics in 2+1
dimensions'')~\cite{Song:2007fn,Song:2007ux}. This code has not only
passed extensive tests against the well-established (2+1)-d ideal
fluid code AZHYDRO and analytically known solutions for problems
with reduced dimensionality during the early stage of my
work~\cite{Song:2007ux} (see Appendix A.~2), but recently also
passed detailed code verification tests (within the TECHQM
Collaboration) against two other viscous hydrodynamics codes that
were developed independently at the University of Washington and at
Stony Brook later on (see Chap.~7.4).

Using VISH2+1, we numerically studied the effects from shear
viscosity~\cite{Song:2007fn,Song:2007ux} (Chap.~4) and bulk
viscosity~\cite{Song:2009Bulk} (Chap.~6) on the hydrodynamic
evolution of the QGP fireball, the final hadron spectra, and their
elliptic flow coefficient $v_2$. We found that shear viscosity
reduces the longitudinal expansion, but accelerates transverse
expansion, which leads to larger radial flow and flatter spectra
when compared with an ideal fluid with identical initial and final
conditions. Bulk viscosity reduces both longitudinal and transverse
expansion, resulting in smaller radial flow and steeper spectra.
Both shear and bulk viscosity suppress elliptic flow $v_2$ at low
$p_T$. It turns out that $v_2$ is very sensitive to the QGP shear
viscosity, and that  in heavy ion collisions at RHIC energies even
the conjectured lower bound, $\eta/s=1/4\pi$, leads to a large
suppression of $v_2$. We explored the scaling behavior of $v_2$ with
the initial source eccentricity $\varepsilon_x$, by computing
$v_2/\varepsilon_x$ as a function of charged hadron multiplicity in
both ideal and viscous hydrodynamics, for Cu+Cu and Au+Au collisions
at a variety of impact parameters and collision energies. We found
that smaller systems and collisions at lower energies feature
stronger shear viscosity effects, and that non-zero shear viscosity
breaks the scaling of $v_2/\varepsilon_x$ with charged multiplicity
by small amounts that can perhaps be used to further constrain the
QGP shear viscosity experimentally~\cite{Song:2008si} (Chap.~5).

The first attempt to extract the QGP viscosity from RHIC elliptic
flow data using viscous hydrodynamics comes from the work of Luzum
and Romatschke. Their results
indicate~\cite{Luzum:2008cw,Song:2008hj}
\begin{eqnarray}
\label{etas}
  \left.\frac{\eta}{s}\right|_\mathrm{QGP} < 5\times
  \frac{1}{4\pi}\,.
\end{eqnarray}
A more precise extraction of the QGP viscosity, however, requires to
at least consider the following aspects in the
future~\cite{Song:2008hj} (Chap.~8):

\begin{itemize}
\item \underline{Better understanding of the initialization}:
the largest contribution to the present uncertainty range for
$\eta/s$ comes from uncontrolled uncertainties in the initial
conditions for the fireball eccentricity, more specifically, the
Glauber initialization vs. the Color Glass Condensate
initialization. This issue can not be solved by hydrodynamics
itself, but requires new experimental techniques to help us pin down
the initial source eccentricity phenomenologically. In addition, it
is important to implement  event-by-event fluctuations in the
initial source eccentricity in viscous hydrodynamic simulations to
account for additional contributions to the elliptic flow $v_2$.

\item \underline{Constructing a better EOS}: The non-equilibrium
chemistry in the hadronic phase is known to affect the distribution
of the hydrodynamic momentum anisotropy over the various hadron
species and thus their elliptic flow, but has not yet been included
in viscous hydrodynamic simulations. With the availability of new,
more accurate lattice QCD data on the EOS, using  almost physical
quark masses, it is now possible and necessary to construct a new
EOS that connects this latest lattice EOS above $T_c$ with the
non-equilibrium hadronic EOS below $T_c$.

\item \underline{Viscous hydrodynamics + hadron cascade}: The
hadronic stage is highly viscous, to the extent that the framework
of viscous fluid dynamics may break down. It is therefore necessary
to match viscous hydrodynamics to a microscopic hadronic cascade
below $T_c$. It is also important to investigate how well the late
non-equilibrium hadronic stage can be reasonably simulated by
viscous hydrodynamics with a temperature-dependent viscosity/entropy
ratio that increases rapidly as the matter approaches freeze-out,
and explore possibilities of extracting the viscosity of hadronic
matter.

\item \underline{Including effects from bulk
viscosity}: My recent research has shown that bulk viscosity also
suppresses $v_2$, thus adding to the effects from shear viscosity.
It is therefore important to include  bulk viscosity effects when
extracting the QGP shear viscosity. However, relaxation times and
initial values for the bulk pressure are required inputs in viscous
hydrodynamic calculations, in addition to the transport coefficients
and the EOS. Near $T_c$, the bulk viscosity $\zeta$ can exceed the
shear viscosity $\eta$ of the strongly interacting matter. If the
relaxation time $\tau_\Pi$ for the bulk viscous pressure $\Pi$ is
short, it quickly loses memory of its initial valaue, but the
relatively large peak value of $\zeta/s$ near $T_c$ can lead to a
significant viscous suppression of the elliptic flow $v_2$,
competing with shear viscous effects. If $\tau_\Pi$ grows rapidly
near $T_c$, due to critical slowing down, the bulk viscous
suppression effects on $v_2$ depend crucially on the initial value
of $\Pi$: If $\Pi$ is zero initially, bulk viscous effects on $v_2$
are almost negligible; if $\Pi$ is initially large, however, as for
the case of the N-S initialization, it remains relatively large
throughout the evolution, suppressing the buildup of elliptic flow
at a level that again competes with shear viscous effects.
Additional research on initial conditions and relaxation times for
the bulk viscous pressure is therefore necessary for a quantitative
extraction of $\eta/s$ from measured data.
\end{itemize}

In short, the field of viscous relativistic fluid dynamics is
experiencing a fast and healthy development, with  contributions
from and collaborations among several different groups. As a result,
we have reached a new threshhold: we are in the process of
quantitatively extracting the transport properties of the QGP, a new
type of  matter that we knew almost nothing about even a decade ago.
What has been covered in this thesis is only the beginning of the
story.


%
%

\appendix
\chapter{}
\vspace{0cm}

\section{Coordinates and Transformations}
The three spatial coordinates and  time  form a four dimensional
coordinate system $x^\mu=(t,x,y,z)$, called \emph{Cartesian
coordinates}. Throughout this thesis, we use the metric tensor
$g_{\mu\nu}=g^{\mu\nu}= \mathrm{diag} (1, -1, -1, -1)$, such that
four vectors ($x^\mu$, for example)  transform as follows:
\begin{eqnarray}
x^\mu=(t,x,y,z), \qquad  x_\mu=g_{\mu\nu}x^\nu =(t,-x,-y,-z).
\end{eqnarray}

One generally sets the $z$-axis parallel to the beam direction, and
correspondingly calls the $(x,y)$ plane  the transverse plan (with x
pointing in the direction of the impact parameter). Within the
forward light-cone $|z|<t$, $\eta-\tau$ coordinates $x^m=(\tau, x,
y, \eta)$ (with $\tau = \sqrt{ t^2-z^2}$ and $\eta=\frac{1}{2} \ln
\frac{t+z}{t-z}$) prove more useful in high energy particle and
nuclear physics. The metric in this coordinate system reads $g^{mn}=
\mathrm{diag} (1, -1, -1, -1/\tau^2)$, $g_{mn}= \mathrm{diag} (1,
-1, -1, -\tau^2)$.

Here we list the transformation between Cartesian and $\eta-\tau$
coordinates:

%
\begin{figure}[h]
  \begin{center}
    \begin{minipage}[b]{0.55\linewidth}
    \includegraphics[width=0.8\linewidth,height=0.5\linewidth,clip=]{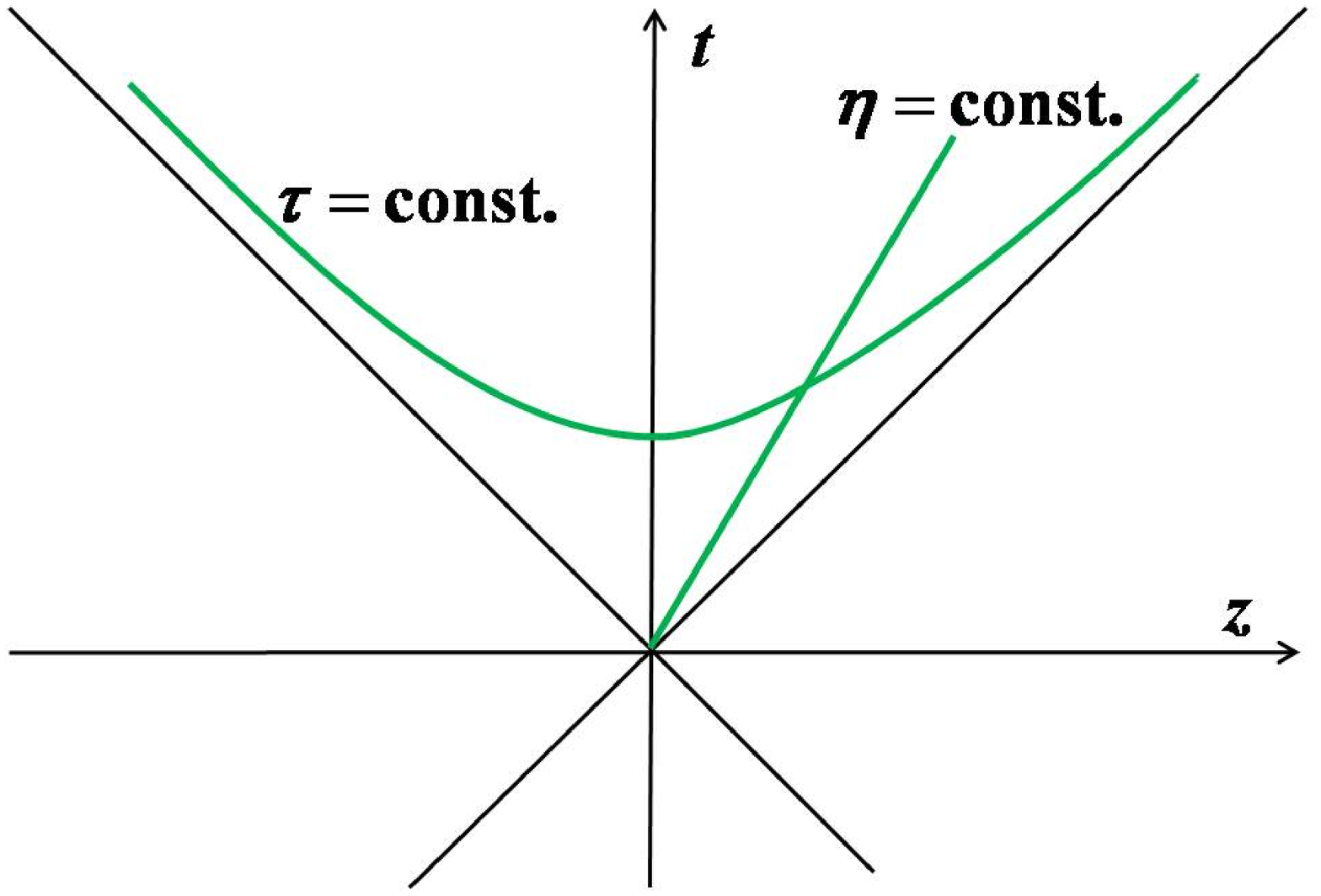}
    \end{minipage}
    \begin{minipage}[b]{0.4\linewidth}
\begin{eqnarray*}
x^{\mu}=(t,x,y,z) &&  x^m=(\tau, x, y, \eta)         \nonumber \\
t=\tau \cosh \ \eta   \ \ \ &\qquad&   \tau=\sqrt{t^2-z^2} \nonumber \\
z=\tau \sinh \ \eta    \ \ \ &\qquad&  \eta=\arctan (z/t)
\end{eqnarray*}
 \vspace*{0.7cm}
    \end{minipage}
  \end{center}
  \vspace*{+0.3cm}
\end{figure}
%

\section{Details of the viscous hydro code VISH2+1}

\subsection{Expressions for $\tilde{\pi}^{mn}$ and
$\tilde{\sigma}^{mn}$} \label{appa1}

The expressions for $\tilde{\pi}^{mn}$ and $\tilde{\sigma}^{mn}$ in
eq.~(\ref{PimnPI2+1}) are
\begin{eqnarray}
\label{tilde-pi}
  \tilde{\pi}^{mn} =
  \begin{pmatrix} \pi^{\tau\tau}&\pi^{\tau x}&\pi^{\tau y}&0\\
                  \pi^{\tau x}&\pi^{xx}&\pi^{x y}&0\\
                  \pi^{\tau y}&\pi^{xy}&\pi^{y y}&0\\
                  0 & 0& 0& \tau^2\pi^{\eta\eta}
  \end{pmatrix},
\end{eqnarray}
\begin{eqnarray}
\label{tilde-sigma}
  \tilde{\sigma}^{mn} &=&
  \begin{pmatrix} \partial_\tau u^\tau &
                  \frac{\partial_\tau u^x{-}\partial_x u^\tau}{2} &
                  \frac{\partial_\tau u^y{-}\partial_y u^\tau}{2} & 0\\[0.5ex]
                  \frac{\partial_\tau u^x{-}\partial_x u^\tau}{2} &
                  -\partial_x u^x &
                  -\frac{\partial_x u^y{+}\partial_y u^x}{2} & 0\\[0.5ex]
                  \frac{\partial_\tau u^y{-}\partial_y u^\tau}{2} &
                  -\frac{\partial_x u^y{+}\partial_y u^x}{2} &
                  -\partial_y u^y & 0\\[0.5ex]
                  0 & 0 & 0 & -\frac{u^\tau}{\tau}
  \end{pmatrix}
\nonumber\\
  && -\frac{1}{2}
  \begin{pmatrix}
        D\left((u^\tau)^2\right) & D(u^\tau u^x) & D(u^\tau u^y) & 0\\[0.5ex]
        D(u^\tau u^x)    & D\left((u^x)^2\right) & D(u^xu^y)     & 0\\[0.5ex]
        D(u^\tau u^y)    & D(u^x u^y)    & D\left((u^y)^2\right) & 0\\[0.5ex]
        0                & 0             & 0                     & 0
  \end{pmatrix}
\\
  && +\frac{1}{3}(\partial\cdot u)
  \begin{pmatrix} (u^\tau)^2{-}1 & u^\tau u^x  & u^\tau u^y  & 0\\[0.5ex]
                  u^\tau u^x     & (u^x)^2{+}1 & u^x u^y     & 0\\[0.5ex]
                  u^\tau u^y     & u^x u^y     & (u^y)^2{+}1 & 0\\[0.5ex]
                  0              & 0           & 0           & 1
  \end{pmatrix}.
\nonumber
\end{eqnarray}
Here
$D\eq{u^\tau}\partial_\tau{\,+\,}u^x\partial_x{\,+\,}u^y\partial_y$
and $\partial\cdot u\eq\partial_\tau u^\tau + \partial_x u^x +
\partial_y u^y +\frac{u^\tau}{\tau}$.\\

\subsection{Velocity finding} \label{appb}

As shown in~\cite{Heinz:2005bw}, since we evolve all three
components $\pi^{\tau\tau}$, $\pi^{\tau x}$, and $\pi^{\tau y}$ (one
of which is redundant due to the constraint $\pi^{\tau m}u_m\eq0$),
the flow velocity and energy density can be found from the
energy-momentum tensor components with the same efficient
one-dimensional zero-search algorithm employed in ideal
hydrodynamics~\cite{Rischke:1995ir,Rischke:1995mt}. This is
important since this step has to be performed after each time step
at all spatial grid points in order to evaluate the EOS $p(e)$.

Using the output from the numerical transport algorithm, one defines
the two-dimensional vector $\bm{M}=(M_x, M_y){\,\equiv\,}(T^{\tau
x}{-}\pi^{\tau
 x}, T^{\tau y}{-}\pi^{\tau y})$. This is (up to the
substitution $p+\Pi \rightarrow p$ ) the ideal fluid part of the
transverse momentum density vector; as such it is parallel to the
tranverse flow velocity $\bm{v}_\bot\eq(v_x,v_y)$. Introducing
further $M_0{\,\equiv\,}T^{\tau\tau}{-}\pi^{\tau\tau}$, one can
write the energy density as
\begin{eqnarray}
\label{edensity}
  e=M_0-\bm{v}_\bot \cdot \bm{M} = M_0 - v_\bot M,
\end{eqnarray}
where $v_\perp\eq\sqrt{v_x^2{+}v_y^2}$ is the transverse flow speed
and $M\equiv\sqrt{M_x^2{+}M_y^2}$. One sees that solving for $e$
requires only the magnitude of $\bm{v}_\perp$ which is obtained by
solving the implicit
relation~\cite{Rischke:1995ir,Rischke:1995mt,Heinz:2005bw}
\begin{eqnarray}
\label{vperp}
  v_\perp = \frac{M}{M_0+p(e{=}M_0{-}v_\perp M) + \Pi}.
\end{eqnarray}
by a one-dimensional zero-search. The flow velocity components are
then reconstructed using
\begin{eqnarray}
\label{vi}
  v_x=v_\perp\frac{M_x}{M}, \qquad  v_y=v_\bot\frac{M_y}{M}.
\end{eqnarray}

Note that this requires direct numerical propagation of all three
components ($\pi^{\tau \tau}$, $\pi^{\tau x}$ and $\pi^{\tau y}$)
since the flow velocity is not known until after the velocity
finding step has been completed. Hence the transversality constraint
$\pi^{\tau m}u_m\eq0$ cannot be used to determine, say,
$\pi^{\tau\tau}$ from $\pi^{\tau x}$ and $\pi^{\tau y}$. However, it
can be used after the fact to test the numerical accuracy of the
transport code.\\

\section {Tests of the viscous hydro code VISH2+1} \label{appd-Code-Checking}

%
\begin{figure}[t]
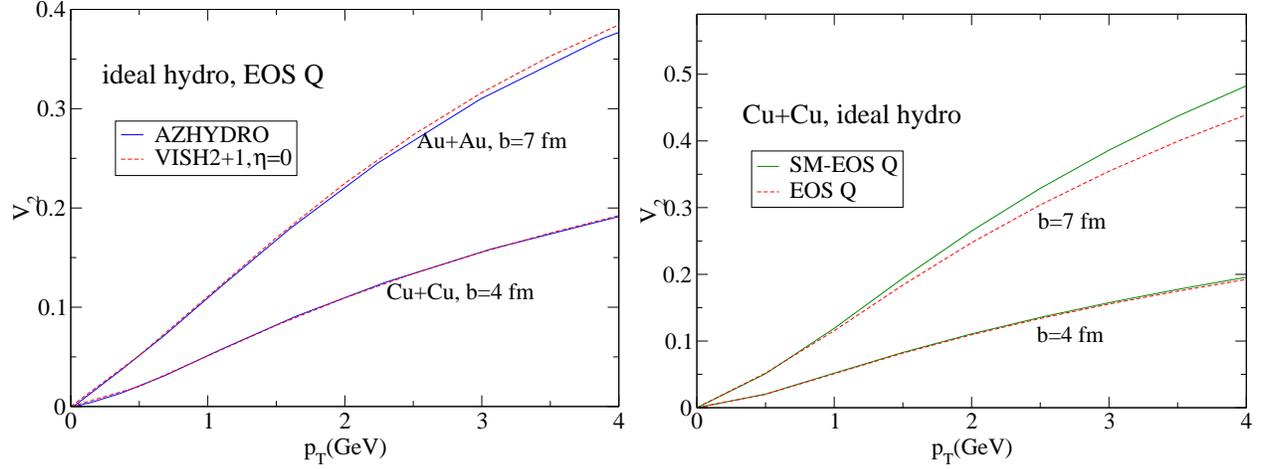

\includegraphics[width=.495\linewidth,clip=]{./Figs/Chap10/Fig22a.eps}
\includegraphics[width=.495\linewidth,clip=]{./Figs/Chap10/Fig22b.eps}
\vspace*{-3mm} \caption[$v_2(p_T)$ for ideal fluid: comparison
between VISH2+1 and AZHYDRO]{{\sl Left:} Differential elliptic flow
$v_2(p_T)$ for $\pi^-$ from $b\eq4$\,fm Cu+Cu collisions and
$b\eq7$\,fm Au+Au collisions, using EOS~Q. Results from VISH2+1 for
$\eta\eq0$ and $\pi^{mn}\eq0$ (dashed lines) are compared with the
ideal fluid code AZHYDRO (solid lines). {\sl Right:} $v_2(p_T)$ for
$\pi^-$ from Cu+Cu collisions at impact parameters $b\eq4$ and
7\,fm, comparing VISH2+1 evolution with EOS~Q (dashed) and SM-EOS~Q
(solid) in the ideal fluid limit $\eta\eq0$, $\pi^{mn}\eq0$. }
\label{v2-CompEOS4}
\end{figure}
%

\subsection{Testing the ideal hydro part of VISH2+1}\label{appd1}

When one sets $\pi^{mn}\eq0$ initially and takes the limit
$\eta\eq0$, VISH2+1 simulates the evolution of an ideal fluid, and
its results should agree with those of the well-tested and publicly
available (2+1)-dimensional ideal fluid code AZHYDRO \cite{AZHYDRO}.
Since VISH2+1 was written independently, using only the
flux-corrected SHASTA transport algorithm from the AZHYDRO package
\cite{AZHYDRO,SHASTA} in its evolution part, this is a useful test
of the code. The left panel in Fig.~\ref{v2-CompEOS4} shows that,
for identical initial and final conditions as described in Chap. 3,
the two codes indeed produce almost identical results. The small
difference in the Au+Au system at $b\eq7$\,fm is likely due to the
slightly better accuracy of AZHYDRO which, in contrast to VISH2+1,
invokes an additional timesplitting step in its evolution algorithm.

When comparing our VISH2+1 results with AZHYDRO we initially found
somewhat larger discrepancies which, however, could be traced back
to different versions of the EOS used in the codes (EOS~Q in
AZHYDRO, the smoothed version SM-EOS~Q in VISH2+1). In the left
panel of Fig.~\ref{v2-CompEOS4} this difference has been removed, by
running also VISH2+1 with EOS~Q. In the right panel we compare
VISH2+1 results for EOS~Q and for SM-EOS~Q, showing that even the
tiny rounding effects resulting from the smoothing procedure used in
SM-EOS~Q (which renders the EOS slightly stiffer in the mixed phase)
lead to differences in the elliptic flow for peripheral collisions
of small nuclei which exceed the numerical error of the code.\\

\subsection[Comparison with analytical results]{Comparison with analytical results for (0+1)-d boost-invariant
viscous hydrodynamics}\label{appd2}

For boost-invariant longitudinal expansion without transverse flow,
the relativistic Navier-Stokes equations read \cite{Gyulassy85} (for
zero bulk viscosity)
\begin{eqnarray}
 &&\frac{\partial e}{\partial \tau} +\frac{e+p+\tau^2 \pi^{\eta\eta}}{\tau}=0,
\\
 &&\tau^2 \pi^{\eta\eta}=-\frac{4}{3} \frac{\eta}{\tau}.
\end{eqnarray}

\begin{figure}[t]
\centering{\includegraphics[width=0.50\linewidth,clip=]{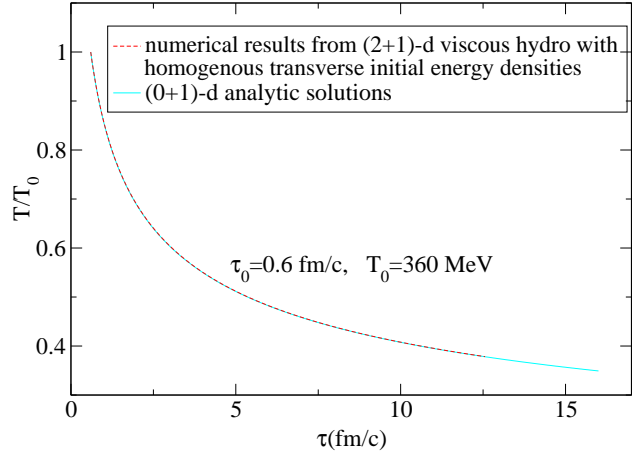}}
\vspace*{-3mm} \caption[(0+1)-d N-S viscous hydrodynamics:
analytical vs. numerical results]{Comparison between the analytical
temperature evolution for (0+1)-d boost-invariant Navier-Stokes
viscous hydrodynamics (solid line) and numerical results from
VISH2+1 with homogeneous transverse initial energy density profiles
(dashed line). } \label{CompAnaly}
\end{figure}

For an ideal gas EOS $p\eq\frac{1}{3}e{\,\sim\,}T^4$ this leads to
the following analytic solution for the evolution of the temperature
\cite{Gyulassy85}:
\begin{eqnarray}
\label{anal}
  \frac{T(\tau)}{T_0}=\Bigl(\frac{\tau_0}{\tau}\Bigr)^{1/3}
  \Bigl[1+\frac{2\eta}{3 s \tau_0 T_0}
        \Bigl(1-\Bigl(\frac{\tau_0}{\tau}\Bigr)^{2/3}\Bigr)\Bigr].
\end{eqnarray}
To test our code against this analytical result we initialize
VISH2+1 with homogeneous tranverse density distributions (no
transverse pressure gradients and flow) and use the Navier-Stokes
identification $\pi^{mn}\eq2\eta\sigma^{mn}$ in the hydrodynamic
part of the evolution algorithm, sidestepping the part of the code
that evolves $\pi^{mn}$ kinetically. It turns out that in this case
the relativistic Navier-Stokes evolution is numerically stable.
Fig.~\ref{CompAnaly} compares the numerically computed temperature
evolution from VISH2+1 with the analytic formula (\ref{anal}), for
$\eta/s\eq0.08$ and $T_0\eq360$\,MeV at $\tau_0\eq0.6$\,fm/$c$. They
agree perfectly.\\

%
%

\subsection[Reduction of VISH2+1 to relativistic Navier-Stokes theory]
{Reduction of VISH2+1 to relativistic Navier-Stokes theory for small
$\eta$ and $\tau_\pi$} \label{appd3}

Having tested the hydrodynamic part of the evolution algorithm in
Appendix~\ref{appd1}, we would like to demonstrate also the accuracy
of the kinetic evolution algorithm that evolves the viscous pressure
tensor components. A straightforward approach would be to take
VISH2+1, set the relaxation time $\tau_\pi$ as close to zero as
possible, and compare the result with a similar calculation as in
Appendix~\ref{appd1} where we sidestep the kinetic evolution
algorithm and instead insert into the hydrodynamic evolution code
directly the Navier-Stokes identity $\pi^{mn}\eq2\eta\sigma^{mn}$.
Unfortunately, this naive procedure exposes us to the well-known
instability and acausality problems of the relativistic
Navier-Stokes equations. The suggested procedure only works if a set
of initial conditions and transport coefficients can be found where
these instabilities don't kick in before the freeze-out surface has
been reached.
%
\begin{figure}[t]
\centering{\includegraphics[width=0.50\linewidth,clip=]{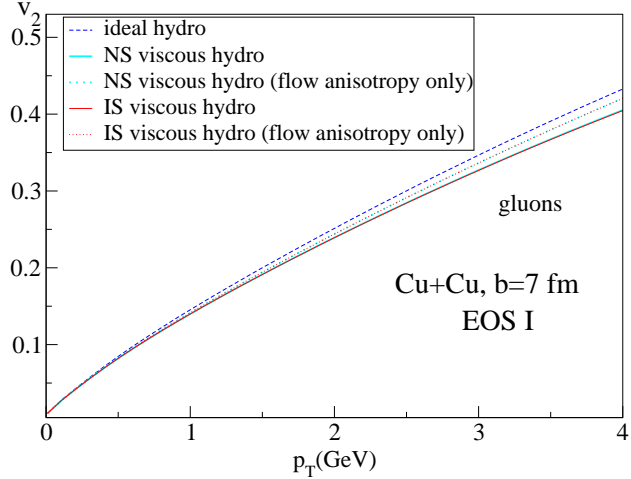}}
\vspace*{-3mm} \caption[Reduction of VISH2+1 to N-S limits with
small $\tau_\pi$]{Differential elliptic flow $v_2(p_T)$ for gluons
from $b\eq7$\,fm Cu+Cu collisions, calculated with ideal
hydrodynamics (blue dashed line), relativistic Navier-Stokes (NS)
hydrodynamics (light blue lines), and Israel-Stewart (IS) viscous
hydrodynamics with $\frac{\eta}{s}\eq\frac{T}{2 \,\mathrm{GeV}}$ and
$\tau_\pi\eq0.03$\,fm/$c$ (red lines), using EOS~I. The lines for NS
and IS viscous hydrodynamics are almost indistinguishable. Solid
lines show the full results from viscous hydrodynamics, dotted lines
neglect viscous corrections to the spectra and take only the flow
anisotropy effect into account. }
\label{v2-2}
\end{figure}
%
%

We found that sufficiently stable evolution of the relativistic
Navier-Stokes algorithm (i.e. of VISH2+1 with the identification
$\pi^{mn}\eq2\eta\sigma^{mn}$) can be achieved for standard initial
density profiles in Cu+Cu collisions and the simple ideal gas
equation of state EOS~I by choosing a very small and temperature
dependent specific shear viscosity
$\frac{\eta}{s}\eq0.01\,\frac{T}{200\,\mathrm{MeV}}\eq\frac{T}
{2\,\mathrm{GeV}}$. For the Israel-Stewart evolution we use a
relaxation time which is correspondingly short:
$\tau_\pi\eq\frac{3\eta}{sT}\eq0.03$\,fm/$c$.

Figure~\ref{v2-2} shows the differential elliptic flow $v_2(p_T)$
for gluons in $b\eq7$\,fm Cu+Cu collisions evolved with these
parameters. The dashed line gives the ideal fluid result. The solid
and dotted lines show the total elliptic flow and the anisotropic
flow contribution to $v_2(p_T)$, respectively, similar to the left
panel Fig.~\ref{v-2}. There are two solid and dotted lines with
different colors, corresponding to Israel-Stewart and Navier-Stokes
evolution; they are indistinguishable, but clearly different from
the ideal fluid result. We conclude that, for small shear viscosity
$\eta/s$ and in the limit $\tau_\pi{\,\to\,0}$, the second-order
Israel-Stewart algorithm reproduces the Navier-Stokes limit and
that, therefore, VISH2+1 evolves the kinetic equations for
$\pi^{mn}$ accurately.\\

\section {Hydrodynamics vs. blast wave model} \label{hydro-blastwave}

As discussed in Chap.~\ref{shear-spectraV2}, the viscous corrections
to the final pion spectra from the hydrodynamic model have a
different sign (at least in the region $p_T{\,>\,}1$\,GeV) than
those originally obtained by Teaney \cite{Teaney:2003kp}. In this
Appendix we try to explore the origins of this discrepancy. We will
see that the sign and magnitude of viscous corrections to the
(azimuthally averaged) particle spectra are fragile and depend on
details of the dynamical evolution and hydrodynamic properties on
the freeze-out surface. Fortunately, they same caveat does not seem
to apply to the viscous corrections to elliptic flow where
hydrodynamic and blast wave model calculations give qualitatively
similar answers.

%
\begin{figure}[t]
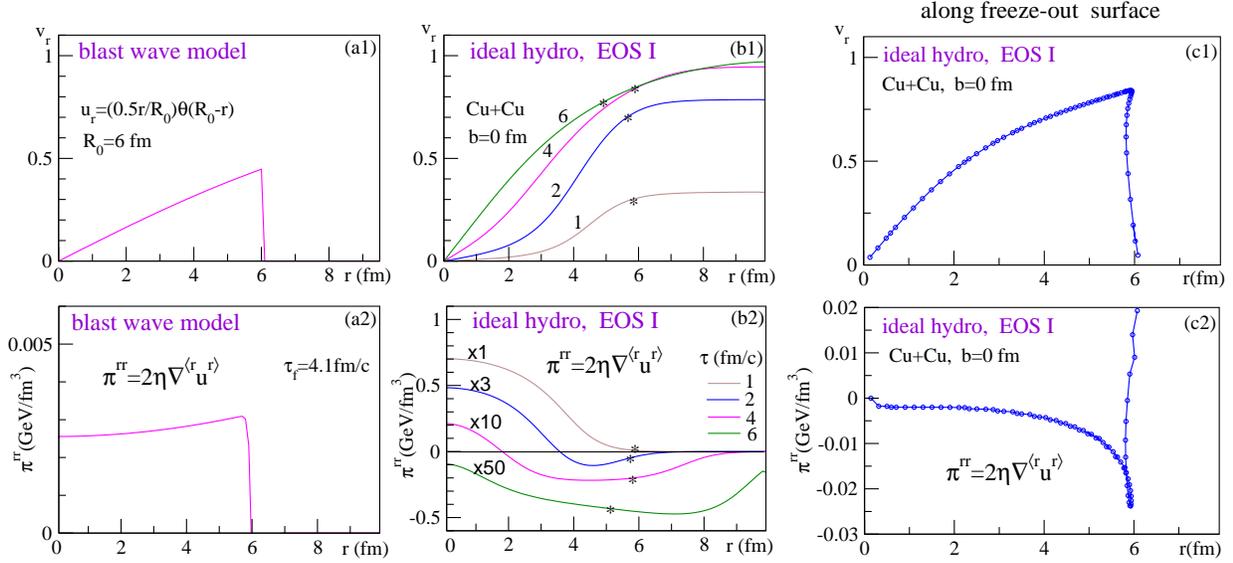

\includegraphics[width=.62\linewidth,clip=]{./Figs/Chap10/Fig25a.eps}
\includegraphics[width=.355\linewidth,clip=]{./Figs/Chap10/Fig25b.eps}
\caption[Velocity and decoupling surface: hydrodynamics vs.
blastwave model]{{\sl Top row:} Velocity profiles from the blast
wave model (left) and from the hydrodynamic model with EOS~I at
fixed times (middle) and along the decoupling surface (right). {\sl
Bottom row:} The corresponding profiles for the transverse shear
viscous pressure $\pi^{rr}$ in the Navier-Stokes limit,
$\pi^{rr}\eq2\eta\nabla^{\left\langle\mu\right.}u^{\left.\nu\right\rangle}$.
Calculations are for central Cu+Cu collisions, and the curves in the
middle panels correspond to the times $\tau\eq1$, 2, 4, and
6\,fm/$c$. See text for discussion. } \label{BW-dyn}
\end{figure}
%

Following Teaney's procedure, we calculate $\pi^{mn}$ in the
Navier-Stokes limit $\pi^{mn}$ $\eq2\eta\sigma^{mn}$. We do this
both in the blast wave model and using the results for $\sigma^{mn}$
from VISH2+1. For the blast wave model we assume like Teaney
freeze-out at constant $\tau$ with a box-like density profile
$e(r)\eq{e}_\mathrm{dec} \theta(R_0{-}r)$, where
$e_\mathrm{dec}\eq0.085$\,GeV/fm$^3$ is the same freeze-out energy
density as in the hydrodynamic model for EOS~I, and $R_0\eq6$\,fm.
The velocity profile in the blast wave model is taken to be linear,
$u_r(r)\eq{a}_0\frac{r}{R_0}\theta(R_0{-}r)$, with $a_0=0.5$;
freeze-out is assumed to occur at $\tau_\mathrm{dec}\eq4.1
$\,fm/$c$. $R_0$, $a_0$ and $\tau_\mathrm{dec}$ are somewhat smaller
than in Ref.~\cite{Teaney:2003kp} since we study Cu+Cu instead of
Au+Au collisions. We concentrate here on a discussion of $\pi^{rr}$
for illustration; the expression for $\sigma^{rr}$ is found in
Ref.~\cite{Heinz:2005bw}, Eq.~(A11c). While $\pi^{rr}$ from VISH2+1
differs from $2\eta\sigma^{rr}$ due to the finite relaxation time
$\tau_\pi$ (see Sec.~\ref{sec5c}), we have checked that the signs of
these two quantities are the same on the freeze-out surface so that
our discussion provides at least a qualitatively correct analysis of
the viscous spectra corrections in the two models.

In Fig.~\ref{BW-dyn} we compare the freeze-out profiles for the
radial flow velocity and $2\eta\sigma^{rr}$ from the blast wave
model. In spite of qualitative similarity of the velocity profiles,
the freeze-out profiles of $2\eta\sigma^{rr}$ are entirely different
and even have the opposite sign in the region where most of the
hydrodynamic particle production occurs (left and right columns in
Fig.~\ref{BW-dyn}). The middle column shows that at fixed times
$\tau$, the hydrodynamic profile for $2\eta\sigma^{rr}$ shows some
similarity with the blast wave model in that $2\eta\sigma^{rr}$ is
positive throughout most of the interior of the fireball. What
matters for the calculation of the spectra via Eq.~(\ref{Cooper}),
however, are the values of $2\eta\sigma^{rr}$ on the freeze-out
surface $\Sigma$ where they are negative, mostly due to radial
velocity derivatives. This explains the opposite sign of the viscous
correction to the spectra in the hydrodynamic model and shows that,
as far as an estimate of these viscous corrections goes, the blast
wave model has serious limitations.\\

\section{Cooling rates for ideal and viscous heavy-ion fireballs}
\label{Sec-cooling-rate}
\begin{figure}[t]
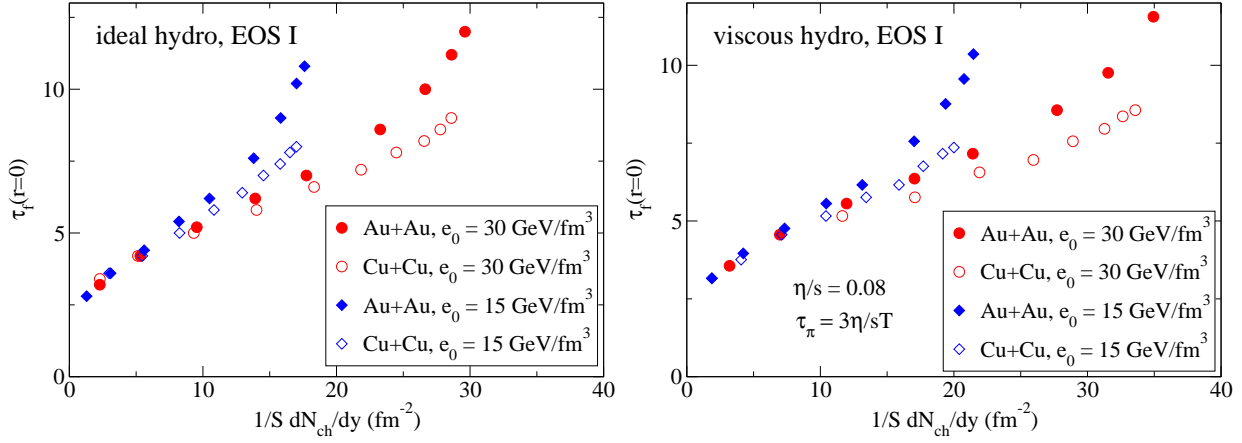

  \begin{center}
 \includegraphics[bb=21 22 718 522,width=0.49\linewidth,clip=]{Figs/Chap10/3tf_Id_EOSI.eps}
 \includegraphics[bb=21 22 718 522,width=0.49\linewidth,clip=]{Figs/Chap10/3tf_Vis_EOSI.eps}
  \end{center}
  \hspace{-3mm}
  \caption[freeze-out times for Au+Au and Cu+Cu collisions]{\label{F1}
  Central freeze-out times for Au+Au and Cu+Cu collisions with
  differential initial peak energy densities $e_0\equiv e(0,0;b{=}0)$
  as a function of final charged hadron multiplicity density.
  Points belonging to one series correspond to different impact
  parameters, with central (peripheral) collisions corresponding
  to large (small) multiplicity densities. Left: ideal fluid dynamics.
  Right: viscous fluid dynamics with $\eta/s=0.08$ and $\tau_\pi=
  3\eta/sT$.
  }
  \label{Freeze-time}
 \end{figure}

In the early expansion stage of a heavy-ion collision, shear
viscosity leads to a reduction of the longitudinal and an increase
of the transverse pressure. The reduced longitudinal pressure
decreases the work done by longitudinal expansion, thereby reducing
the initial cooling rate of the fireball. This leads to a somewhat
increased lifetime of the quark-gluon plasma phase. The increased
transverse pressure, on the other hand, causes larger transverse
acceleration and stronger radial flow of the matter than in ideal
fluid dynamics. Due to the larger transverse flow, the center of the
viscous fireball cools more quickly during the late stages than an
ideal fluid, thereby slightly reducing the total fireball lifetime
until freeze-out for central and near-central collisions (see
Fig.~\ref{F1}). The viscous fireballs created in peripheral
collisions don't live long enough for this mechanism to manifest
itself; they live longer than their ideal counterparts, due to the
decreased initial cooling rate arising from the smaller longitudinal
pressure (left points in Fig.~\ref{F1}).

%
%

For an equation of state (EOS) with a first-order quark-hadron phase
transition, the mixed phase matter is free of pressure gradients and
hence not accelerated. Velocity gradients in the mixed phase
generate, however, viscous pressure components whose gradients
continue to accelerate the viscous fluid even while it passes
through the mixed phase where the thermal pressure gradients vanish.
As a result, the viscous fluid spends less time in the mixed phase
than the ideal one.

In ideal hydrodynamics, a first-order phase transition with a mixed
phase (MP) generates large velocity gradients near the QGP-MP and
MP-HG (HG = hadron gas) interfaces. In viscous fluids such gradients
generate viscous pressures that act against building up large
velocity gradients. In consequence, all prominent structures in
ideal hydrodynamics that arise from discontinuities in the speed of
sound in ideal hydrodynamics are washed out by viscous effects.
Shear viscosity thus effectively turns a first-order phase
transition into a smooth cross-over transition.

\chapter{Glossary}

RHIC: Relativistic Heavy Ion Collider

LHC: Large Hadron Collider

CGC : Color Glass Condensate

QGP: the quark gluon plasma

HRG:  hadron resonance gas

MP: Mixed phase

EOS : equation of state

PCE-EOS: partially chemical equilibrium EOS (for the HRG phase)
\\[-0.05in]

N-S equation: Navier-Stokes equation

I-S equation: Israel-Stewart equation

O-G equation: \"Ottinger-Gremla equation

VISH2+1: Viscous Israel-Stewart Hydrodynamics in 2+1 dimensions

AZHYDRO:  AZimuthally  asymmetric (ideal) Hydrodynamics in 2+1
dimensions
\\[-0.05in]

$g^{\mu\nu}$: metric tensor

$\tau$: longitudinal proper time

$\eta$: space time rapidity

$\gamma$: Lorentz contraction factor

$u^\mu$: four velocity
\\[-0.05in]

$A$: atomic number

$b$: impact parameter

$R$: nuclear radius

$e_0(r=0, b=0)$: initial peak energy density value at $r=0$

$e_{dec}$: decoupling energy density

$\tau_0$: initial time

$\tau_f$: freeze-out time

$c_s$: speed of sound

$T_{cr}$: critical temperature

$T_{dec}$: decoupling temperature
\\[-0.05in]

$\varepsilon_x$: spatial eccentricity

$\varepsilon_p$: momentum anisotropy

$v_T$: transverse flow velocity

$N_{ch}$: produced charged hadrons

$y$: momentum rapidity

$p_T$: particle transverse momentum

$v_2$: elliptic flow coefficient

$S$: overlap area
\\[-0.05in]

$T^{\mu\nu}$: energy momentum tensor

$N^{\mu}$: conserved charge flow

$f(x,p)$: distribution function

$f_0(x,p)$: equilibrium distribution function

$\delta f(x,p)$: non-equilibrium part in the distribution function
$f=f_0+ \delta f$
\\[-0.05in]

$e$: energy density

$p$: pressure

$T$: temperature

$s, \mathcal{S}$: entropy density, total entropy

$\Delta \mathcal{S}$ entropy production
\\[-0.05in]

$\pi^{\mu\nu}$: shear pressure tensor

$\Pi$: bulk pressure

$q^\mu$: heat flow

$\zeta$: bulk viscosity

$\eta$: shear viscosity

$\lambda$: heat conductivity

$\tau_\pi$: relaxation time for shear pressure tensor

$\tau_\Pi$: relaxation time for bulk pressure

$\lambda_1, \lambda_2, \lambda_3$: other 2nd order transport
coefficients in the viscous equation for $\pi^{\mu\nu}$

$\sigma^{\mu\nu}$, $\nabla^{\La \mu} u^{\nu \Ra}$ : velocity stress
tensor

$\Omega^{\mu\nu}$: vorticity tensor



%
%

\singlespacing
\footnotesize
\bibliographystyle{utphysnew}
\bibliography{2-songbib}
\end{document}